\documentclass[preprint]{elsarticle}

\usepackage{amsmath}
\usepackage{amsfonts}
\usepackage{amssymb}
\usepackage{amsthm}
\usepackage{graphicx}
\usepackage{hyperref}
\usepackage{color}
\usepackage{flafter}
\usepackage[normalem]{ulem}

\usepackage[toc,page]{appendix}
\usepackage{makecell}
\usepackage{slashed}
\usepackage{tikz-cd}
\usepackage{makecell}
\usepackage{subcaption}
\usepackage{graphicx}

\allowdisplaybreaks

\definecolor{OliveGreen}{rgb}{0,0.6,0}
\definecolor{Orange}{rgb}{1.00, 0.65, 0}
\definecolor{Grey}{rgb}{0.43, 0.5, 0.5}

\newcommand{\B}[1]{\textcolor{blue}{#1}}
\newcommand{\R}[1]{\textcolor{red}{#1}}

\newcommand{\Fig}[1]{Fig.~\ref{#1}}
\newcommand{\Eq}[1]{Eq.(\ref{#1})}
\newcommand{\nn}{\nonumber\\}

\newcommand{\be}{\begin{eqnarray}}
\newcommand{\ee}{\end{eqnarray}}
\newcommand{\bpm}{\begin{pmatrix}}
\newcommand{\epm}{\end{pmatrix}}

\newcommand{\p}{\partial}

\newcommand{\Tr}{{\rm Tr}}
\newcommand{\tr}{{\rm tr}}

\renewcommand{\Im}{{\rm Im}}
\renewcommand{\Re}{{\rm Re}}
\newcommand{\ua}{\uparrow}
\newcommand{\cG}{\mathcal{G}}
\newcommand{\da}{\downarrow}
\newcommand{\ra}{\rightarrow}

\renewcommand{\v}[1]{{\boldsymbol{#1}}}
\renewcommand{\a}{\alpha}
\renewcommand{\b}{\beta}

\newcommand{\e}{\epsilon}
\newcommand{\s}{\sigma}
\renewcommand{\t}{\tau}
\newcommand{\g}{\gamma}
\newcommand{\G}{\Gamma}

\newcommand{\Avg}[1]{\langle #1 \rangle}

\begin{document}

\title{Non-abelian bosonization in two and three spatial dimensions and applications}

\author[ucb,lbl]{Yen-Ta Huang}
\ead{yenta.huang@berkeley.edu}
\author[ucb,lbl]{Dung-Hai Lee}%\corref{cor1}
\ead{dunghai@berkeley.edu}
\address[ucb]{Department of Physics, University of California, Berkeley, California 94720, USA}
\address[lbl]{Materials Sciences Division, Lawrence Berkeley National Laboratories, Berkeley, California 94720, USA}

\setlength{\parindent}{0pt}

\setlength{\parindent}{0pt}

\date{\today}

\begin{abstract}
	In this paper, we generalize Witten's non-abelian bosonization in $(1+1)$-D to two and three spatial dimensions. Our theory applies to fermions with relativistic dispersion. The bosonized theories are non-linear sigma models with level-1 Wess-Zumino-Witten terms. We apply the bosonization results to the $SU(2)$ gauge theory of the $\pi$-flux phase, critical spin liquids in 1,2,3 spatial dimensions, and twisted bilayer graphene.
	
\end{abstract}

\maketitle

\tableofcontents

\newpage
\part*{Introduction} \label{sec:intro}
\addcontentsline{toc}{part}{Introduction}

Bosonization in $(1+1)$-D has been a very useful theoretical tool.  It allows one to map a theory, where the fundamental degrees of freedom are fermionic, to a theory with bosonic degrees of freedom.  Often, things that can be seen easily in one picture are difficult to see in the other. The best-known bosonization is the abelian bosonization \cite{Mattis1965,Coleman1975,Mandelstam1975,Luther1975}, where fermions are  solitons in the Bose field. A shortcoming of the abelian bosonization, when fermions have flavor (e.g., spin) degrees of freedom, is that the flavor symmetries are hidden. This problem was solved by Witten's non-abelian bosonization \cite{Witten1984}. In this paper we generalize Witten's non-abelian  bosonization to $(2+1)$ and $(3+1)$ space-time dimensions.\\

The limitation of our theory is that it only applies to fermions with relativistic dispersion. (However, we do not restrict the Fermi velocity to be  the speed of light.) In the absence of a mass gap, such theories have Dirac-like dispersion relation. In one space dimension, massless fermions are generically relativistic at low energies. In two and three space dimensions,  relativistic massless fermions have  been discovered in many experimental condensed matter systems. Examples include graphene and twisted bilayer graphene, Dirac and Weyl semi-metal,...etc. Moreover,  relativistic massless fermions can appear in the mean-field theory of strongly correlated systems. Such theory serves as the starting point of a more rigorous treatment. For example,  the ``spinon $\pi$-flux phase'' mean-field theory sets the stage for a gauge theory description of the Mott insulating state of the half-filled Hubbard model.\\

Another important area where relativistic massless fermions appear is at the boundary of topological insulators or superconductors, which are simple examples of symmetry-protected topological (SPT) phases. The classification of topological insulator/superconductor\cite{Kitaev2009,Ryu2010} can be viewed as asking how many copies of the massless fermion theories on the boundary are required to couple together before a symmetry-allowed mass term emerges. \\ %simplest representatives of such phases are massive relativistic fermion theories. Kitaev's classification  is based on such theories. When a relativistic fermion topological state is cut open, the boundary is a massless fermion theory. \\

The paper contains two major parts: I. bosonization and II. applications. Each of them contains several sections, namely, 14 sections in part I and 3 sections in part II.  In each section of part I, we discuss an important step or input of the bosonization. We shall illustrate the relevant concepts with examples in the lowest spatial dimension where it first appears. For higher spatial dimensions, we simply present the result while leaving the details to the appendices. Together, the 14 sections in part I provide the readers with the idea and technical details of the bosonization. In part II there are 3 sections, each gives an example of how this bosonization can be applied.  The topics include the $SU(2)$ gauge theory of the $\pi$-flux phase of half-filled Hubbard model, the critical spin liquid of ``bipartite-Mott insulators'' in spatial dimensions 1, 2, 3, and the twisted bilayer graphene. Finally,  the 11 appendices provide the details omitted in the main text. \\

\part{Bosonization}

\section{The idea}\label{theidea}
\hfill

In this paper, by ``Bosonization'', we mean to construct bosonic theories that are equivalent to theories of massless relativistic fermions. In the rest of the paper, unless otherwise stated, ``massless fermion'' always refers to massless relativistic fermion. Here we stress again  that ``relativistic massless fermion'' does not imply the Fermi velocity is the speed of light. As mentioned in the introduction, in several $(2+1)$ and $(3+1)$ dimensional condensed matter systems, relativistic massless fermions have been encountered. \\ 

We look at the massless fermion theories from two points of view. On one hand, as $d$ (spatial) dimensional theories, the massless fermion theories have  emergent symmetries and symmetry anomalies. On the other hand, the massless fermions can be realized on the boundary of $d+1$ topological insulators/superconductors where the emergent symmetries are the protection symmetries.\\

If the emergent symmetries are to be respected, the (massless) fermions can not develop an energy gap. But what if we introduce mass terms (or the bosonic order parameters), at the expense of breaking the emergent symmetries, then fluctuate the order parameters {\it smoothly} (in both space and time) until the symmetries are restored ? Since the order parameter fluctuations are smooth, we expect the fermion gap to remain intact. Under such conditions, we can integrate out the fermions to yield  bosonic non-linear sigma models governing the dynamics of the order parameters. From the perspective of the boundary of topological insulators/superconductors, after integrating out the fermions, what's left are fluctuating order parameters and the non-linear sigma models. Because the protection symmetries are restored by the order parameter fluctuations, the non-linear sigma models are either gapless or possess topological order.  It turns out that the non-linear sigma models have a special type of topological term: the level-one Wess-Zumino-Witten (WZW) term. Such term encodes the 't Hooft anomaly\footnote{The 't Hooft anomaly refers to the obstruction in gauging the continuous part of the emergent global symmetries. Under such conditions, once gauge field is introduced, the partition function fails to be gauge invariant.} associated with the boundary of topological insulators/superconductors. %In the anomaly inflow picture, this is due to the  Noether current flowing from the higher dimensional topological insulators/superconductors to the boundary. Because the bosonic non-linear sigma model is attached to the same bulk SPT, it must manifest the same anomaly. The WZW term is the manifestation of such an anomaly. 
Due to the WZW term, the non-linear sigma models are gapless, hence potentially can be equivalent to the massless fermion theories. This equivalence is supported by the fact that the fermion and boson theories have (1) the same symmetries, (2) the same anomalies, and (3) the boson theories have fermionic solitons.% Using the above ideas we have successfully reproduced Witten's bosonization in $(1+1)$-D.
\\

%Another important area where relativistic massless fermions naturally appear is at the boundary of topological insulators or superconductors, which are simple examples of symmetry-protected topological (SPT) phases. The simplest representatives of such phases are massive relativistic fermion theories. Kitaev's classification \cite{Kitaev2009} is based on such theories. When a relativistic fermion topological state is cut open, the boundary is a massless fermion theory. \\

%There is a folklore that so long as the protection symmetry of a fermionic SPT is not broken, the boundary should be either gapless or exhibit topological order for $d\ge 2$.  Our bosonization idea is based on this folklore. In particular, we shall {\it assume} that any relativistic gapless bosonic theory that can be attached to the boundary of a free fermion SPT while respecting all symmetries of the bulk fermions is equivalent to the massless fermion theory. Using this idea we successfully reproduced Witten's non-abelian bosonization in 1D. In subsequent discussions, we shall explain when a boson theory is attachable to the boundary of a free fermion SPT.  \\   

As to the question of why do we bother to bosonize? One reason is it allows us to determine the low energy physics of a non-trivial boson theory by solving the theory of free massless fermions, and often what is subtle in one picture can become clearer in the other. Of course, we will not stop at the massless free fermion theories, the goal of bosonization is to enable one to go further. This will become clear in the applications.\\

\section{Emergent symmetries of the massless fermion theory}\label{ES}
\hfill

A necessary condition for two theories to be equivalent is that they have the same symmetry.  Thus it is important to determine the symmetry of massless fermion theories.    It turns out the symmetries of such theories are rather rich. Because the massless fermion theories are low energy {\it effective} theories,  we shall refer to their symmetries as the {\it emergent symmetries}.  \\

In the following, we shall consider massless $n$-flavor Dirac (or Majorana) fermion theories in spatial dimensions 1, 2, and 3. Such theories can be split into two main categories, namely, complex class and real class.  A theory in complex classes can be solely written in terms of {\it complex} Dirac fermion fields. Moreover, in the presence of a cutoff, its Hilbert space is the {\it eigenspace} of certain ``charge'' operator $Q$. In the following we shall focus on the $Q=0$ eigenspace, i.e.,  the ``charge neutral point'' in condensed matter physics. The charge operator is the generator of a (continuous) global U(1) symmetry. In contrast, a theory in the real class is expressed in terms of Majorana fermion fields. In this class, there is no requirement for a conserved charge operator.\\

%We should mention that the boundary theories of free fermion SPTs, constructed by the Lorentz invariant representatives discussed in Ref. \cite{Kitaev2009}, precisely fall into the two categories described above, and are $n$-flavor Dirac (or Majorana) fermion theories. They have the same emergent symmetry as the massless fermion theories we are about to describe. We stress that these emergent symmetries contain the so-called ``protection symmetries'' generated by, e.g., time reversal, charge, and charge conjugation operators. However, the emergent symmetry group is usually larger than the protection symmetry group. \\ 
\subsection{Complex class}\hfill

Now, as an example, let's determine the emergent symmetry group of a one dimensional massless fermion theory. To this end let's first consider a complex class, $n$-flavor, massless Dirac fermion theory described by the following action

\begin{align}
\label{S01C}
&S_0 = \int dx^0 dx^1 \psi^\dagger (\partial_0  - i \Gamma_1 \partial_1 )\psi  ~~{\rm where}\\
& \Gamma_1 = Z I_n \notag
\end{align}

Here $I_n$ denotes $n\times n$ identity matrix. In the following we shall use the shorthand $I,X,Y,Z,E$ to denote the Pauli matrix $\s_{0,x,y,z},i\s_y$, and when two matrix symbols stand next to each other, e.g., $Z I_n$, it means tensor product $Z \otimes I_n$.  
For complex fermion field $\psi$, the possible unitary transformations include
\be
&&\psi \ra U \cdot \psi\nn
&&\psi \ra C \cdot (\psi^\dagger)^T
\nonumber
\ee
where $U$ and $C$ are unitary matrices. Note that as a discrete transformation (the second line of the above equations), the charge conjugation transformation  does leave the $Q=0$ eigenspace invariant \footnote{However, we do not allow the charge conjugation operator to generate {\it continuous} transformations, since under such transformations $\psi$ will go into the superposition of $\psi$ and $\psi^\dagger$. This violates the requirement that the Hilbert space is the eigenspace of the charge operator.}.  \\

One can easily show that the full emergent symmetries of the action in \Eq{S01C} are 
\be
&&\text{Chiral $U(n)$ symmetry:}\nn
&&~~U(n)_+ \times U(n)_-:\psi\ra \Big(P_+ \otimes g_+ + P_- \otimes g_-\Big)\psi
\text{~where } g_\pm \in U(n) \nn
&&\text{Charge conjugation  symmetry:}\nn
&&~~ C:\psi \rightarrow \left( Z \otimes I_n \right) (\psi^\dagger)^T \nn
&&\text{Time reversal  symmetry (anti-unitary):}\nn
&&~~ T:\psi \rightarrow \left( X \otimes I_n \right) \psi 
\label{symm1C}\ee
\noindent Here 
\begin{align}
\label{Ppm}
P_\pm := \frac{I \pm Z}{2}
\end{align}
are the projection operators with the subscript $\pm$ denoting the ``right/left'' moving fermions, respectively. Note that any other anti-unitary symmetry can be written in terms of the composition of a unitary symmetry and the time reversal transformation above.  \\

\subsection{Real class}
\hfill

Next, we consider the one-dimensional massless theory in the real class. In this case, we write the action in terms of the n-component Majorana fermion field
\begin{align}
\label{S01R}
& S_0 = \int dx^0 dx^1 \, \chi^T \left[\partial_0 - i \Gamma_1 \partial_1 \right]\chi~~{\rm where} \\
&\Gamma_1 := Z I_n \notag
\end{align}

\noindent For Majorana fermion field, the possible unitary transformations are of the form 
$$\chi\ra O \cdot\chi $$
\noindent where $O$ is an orthogonal matrix. The full emergent symmetries of the action in \Eq{S01R} are
\be
&&\text{Chiral $O(n)$ symmetry:}\nn
&&O(n)_+ \times O(n)_-:\chi\ra\Big(P_+ \otimes g_+ + P_- \otimes g_-\Big)\chi~
\text{where } g_\pm \in O(n) \nn
&&\text{Time reversal symmetry (anti-unitary):}\nn
&&T: \chi \rightarrow (X\otimes I_n) \chi. 
\label{symm1R}
\ee
\\

In $D=d+1$ space-time dimension, the massless fermion actions are 

\begin{align}
\text{Complex class:} ~~&S_0 = \int d^D x \, \psi^\dagger \left[\partial_0 - i \sum\limits_{i=1}^d \Gamma_i \partial_i \right]\psi \notag\\
\text{Real class:} ~~&S_0 = \int d^D x\, \chi^T\left[\partial_0 - i \sum\limits_{i=1}^d \Gamma_i \partial_i \right]\chi
\label{s0}
\end{align}
where $\psi$ and $\chi$ are complex and Majorana fermion fields, respectively. In table \ref{tab:emergentSymm} we summarize the emergent symmetries of massless fermion theories in 1,2 and 3 dimensions. See the detailed derivation in appendix \ref{appendix:emergentSymm}. Here the discrete symmetries, such as charge conjugation or time-reversal,  should be viewed as the generators of more general charge conjugation and time-reversal transformations. For example, compounding the time-reversal transformation with an arbitrary unitary symmetry yields another anti-unitary symmetry. The reason for the particular choice of the discrete symmetry generators in table \ref{tab:emergentSymm} will be discussed in subsection \ref{choice} of appendix\ref{appendix:anomaliesf}.

\begin{table}
	\small
\begin{tabular}{ |c|c|c| }
\hline
$(1+1)$-D 		&	Real class	& Complex class	\\
\hline
$\Gamma_i$		&  	$Z\otimes I_n $ 	&	$Z \otimes I_n$		\\
\hline
	Emergent symmetries		
	&	\thead{$T = X \otimes I_n$ \\ 		
		$O_+(n) \times O_-(n): P_+ \otimes g_+ + P_- \otimes g_-$	\\ 
		where $g_+ \in O_+(n)$ and $g_- \in O_-(n)$}		
	&	\thead{ $T = X \otimes I_n$ \\ $C=Z \otimes I_n$ \\		
		$U_+(n) \times U_-(n): P_+ \otimes g_+ + P_- \otimes g_-$	\\ 
		where $g_+ \in U_+(n)$ and $g_- \in U_-(n)$}		\\
	\Xhline{3\arrayrulewidth}
		$(2+1)$-D 		&	Real class	& Complex class	\\
	\hline
	$\Gamma_i$		&  	$Z \otimes I_n$	, $X \otimes I_n$ 	&	$Z \otimes I_n$	, $X \otimes I_n$	\\
	\hline
	Emergent symmetries			
	&	\thead{$T = E \otimes I_n$ \\ 		
		$O(n) : I \otimes g$	\\ 
		where $g \in O(n)$}		
	&	\thead{$T = Y \otimes I_n$ \\ 	
		$C=I \otimes I_n$ \\		
		$U(n) : I \otimes g$	\\ 
		where $g \in U(n)$}		\\
	\Xhline{3\arrayrulewidth}
		$(3+1)$-D 		&	Real class	& Complex class	\\
	\hline
	$\Gamma_i$		&  	$ZI \otimes I_n$	, $XI \otimes I_n$, $YY \otimes I_n$ 	&	$ZI \otimes I_n$	, $XI \otimes I_n$, $YZ \otimes I_n$	\\
	\hline
	Emergent symmetries			
	&	\thead{$T = EZ \otimes I_n$ \\ 		
		$U(n) : II \otimes g_1 - IE \otimes g_2$	\\ 
		where $u=g_1 + i g_2 \in U(n)$}		
	&	\thead{$T = YZ \otimes I_n$ \\ 	
		$C=IX \otimes I_n$ \\		
		$U_+(n)\times U_-(n) : IP_+ \otimes g_+ + IP_- \otimes g_-$	\\ 
		where $g_+ \in U_+(n)$ and $g_- \in U_-(n)$}		\\
	\hline
	\end{tabular}
\caption{A summary of the emergent symmetries of massless fermions in  $(1+1)$-D, $(2+1)$-D, and $(3+1)$-D. Here $P_\pm:=(I\pm Z)/2$ as in \Eq{Ppm}.} 
\label{tab:emergentSymm}
\label{emsymm}\end{table}

\section{Mass terms and mass manifolds}
\label{massManifold1d}
\hfill

Mass terms, or order parameters, are fermion bilinears, namely,  \be &&\psi^\dagger M\psi,~~{\rm  or}\nn&&\chi^T M\chi,\label{mtm}\ee which opens an energy gap when added to \Eq{s0}. To achieve that, the {\it hermitian} mass matrix $M$ must anti-commute with all the gamma matrices, i.e., 
\begin{align}
\label{mass}
&\{M,\G_i\}=0~{\rm for}~i=1,...,d 
\end{align}  
\noindent  We will further require that the gap is flavor independent by imposing
\begin{align}
\label{constantGap}
M^2 = m^2 \cdot 1
\end{align}
\noindent Here $1$ means the identity matrix of appropriate size. 
%(which is denoted by $m$) 
%(denoted by $m$ here)
The mass matrices satisfying \Eq{mass} and \Eq{constantGap} form a topological space -- the %when the associated fermion bilinear is added to \Eq{s0}, form the 
{\it mass manifold}.  In the simplest case, it can be a k-dimensional sphere. In general, it is a closed $k$-dimensional manifold. If, in addition to \Eq{mass}, the mass terms are required to be invariant under certain unitary or anti-unitary transformations, the mass manifold will be affected. In the classification of the free fermion SPTs, it is important to know what is the homotopy group of the mass manifold \cite{Kitaev2009}.  
\\

In the following we give two examples in one spatial dimension, to let the readers get a feeling of what's involved in figuring out the mass manifold.\\

\subsection{Complex class}
\hfill

Let the U(1) symmetry transforms the field field according to
$$\psi\ra e^{i\theta}\psi.$$ Then all mass terms in the form 
\begin{align*}
	\psi^\dagger M^{\mathbb{C}} \psi,
\end{align*}
are invariant under U(1).  Here the superscript $\mathbb{C}$ is to remind us that this is  a mass matrix in the complex fermion class. 
$M^{\mathbb{C}}$ is a $2n\times 2n$ ($2n$ is the number of component of $\psi$) satisfying

\begin{align*}
&M^{\mathbb{C}} = \left( M^{\mathbb{C}}\right)^\dagger\\
	&\{ M^{\mathbb{C}}, \Gamma_i \} = 0 \\
		&\left( M^{\mathbb{C}} \right)^2 = m^2  I_{2n}
\end{align*}

Here $I_{2n}$ is the $2n\times 2n$ identity matrix. Associated with the massless fermion action given in \Eq{S01C}, the first two conditions require $M^{\mathbb{C}}$ to be of the form

\be
	M^{\mathbb{C}} = m \left(X \otimes H_1 + Y \otimes H_2\right)
	\label{mass1C}
\ee

\noindent where $H_1$ and $H_2$ are $n\times n$ hermitian matrices. If we define 
\be Q^{\mathbb{C}}:= H_1+ i H_2, \label{qcdef}\ee 
\noindent it can be easily shown that the third condition requires $$Q^{\mathbb{C}} \cdot \left( Q^{\mathbb{C}}\right)^\dagger = I_n.$$  Therefore the mass manifold for one dimension, in complex class, is the topological space formed by $n\times n$ unitary matrices.\\

\subsection{Real class}
\hfill

In this case, the mass term is the Majorana fermion bilinear

\begin{align*}
	\chi^T M^{\mathbb{R}} \chi
\end{align*}

\noindent where the matrix $M^{\mathbb{R}}$ is an anti-symmetric matrix satisfying

\begin{align*}
&M^{\mathbb{R}} = \left( M^{\mathbb{R}}\right)^\dagger\\
	&\{ M^{\mathbb{R}}, \Gamma_i \} = 0 \\
		&\left( M^{\mathbb{R}} \right)^2 = m^2  I_{2n}
\end{align*}
The first two conditions require

\begin{align*}
	M^{\mathbb{R}} = m \left(Y\otimes S + X\otimes (i A) \right)
\end{align*}

\noindent where $S$ and $A$  are real symmetric and anti-symmetric matrix, respectively. If we define $$Q^{\mathbb{R}}:= S+A$$ 
the last condition requires $$Q^{\mathbb{R}} \cdot \left( Q^{\mathbb{R}}\right)^T = I_n.$$ Thus, the mass manifold is the space of $n\times n$ orthogonal matrices.\\

In table \ref{tab:massManifold} we summarize the mass manifolds for 1,2 and 3 dimensions. The detailed derivations are left in appendix \ref{appendix:massManifold}.

\begin{table}
	\small
	\begin{tabular}{ |c|c|c| }
		\hline
		$(1+1)$-D 		&	Real class & Complex class	\\
		\hline
		$\Gamma_i$		&  	$Z\otimes I_n $ 	&	$Z \otimes I_n$		\\
		\hline
		Mass manifold
		&	\thead{$M=Y \otimes S + X \otimes (iA)$ \\ 
			where $Q^{\mathbb{R}} = S+A \in O(n)$}
		&	\thead{$M=X \otimes H_1 + Y \otimes H_2$ \\ 
			where $Q^{\mathbb{C}} = H_1+i H_2 \in U(n)$}	\\
			\Xhline{3\arrayrulewidth}
		$(2+1)$-D 		&	Real class & Complex class	\\
		\hline
		$\Gamma_i$		&  	$Z \otimes I_n$	, $X \otimes I_n$ 	&	$Z \otimes I_n$	, $X \otimes I_n$	\\
		\hline
		Mass manifold
		&	\thead{$M= Y \otimes S $ \\ 
			where $Q^{\mathbb{R}} = S \in \bigcup_{l=0}^n\frac{O(n)}{O(l) \times O(n-l)}$}	
		&	\thead{$M= Y \otimes H $ \\ 
			where $Q^{\mathbb{C}} = H \in \bigcup_{l=0}^n\frac{U(n)}{U(l) \times U(n-l)}$}	\\
			\Xhline{3\arrayrulewidth}
		$(3+1)$-D 		&	Real class & Complex class	\\
		\hline
		$\Gamma_i$		&  	$ZI \otimes I_n$	, $XI \otimes I_n$, $YY \otimes I_n$ 	&	$ZI \otimes I_n$	, $XI \otimes I_n$, $YZ \otimes I_n$	\\
		\hline
		Mass manifold
		&	\thead{$M= YX \otimes S_1 + YZ \otimes S_2 $ \\ 
			where $Q^{\mathbb{R}} = S_1 + i S_2 \in \frac{U(n)}{O(n)}$}	
		&	\thead{$M= YX \otimes H_1 + YY \otimes H_2 $ \\ 
			where $Q^{\mathbb{C}} = H_1 + i H_2 \in U(n)$}	\\
		\hline
	\end{tabular}
	\caption{A summary of the mass manifolds for the real and complex class fermions in $(1+1)$-D, $(2+1)$-D, and $(3+1)$-D.} 
	\label{tab:massManifold}
\end{table}

\section{The symmetry anomalies of the fermionic theories}
\label{fermionAnomalies}
\hfill

Emergent symmetries of a low-energy effective theory can be broken when a cutoff is imposed. 
In this section, we review the symmetry anomalies of the massless fermion theories.\\ %In section \ref{tHooftWZW}, we will see that the same anomalies can be reproduced in the bosonized non-linear sigma models.

\subsection{The 't Hooft anomaly of continuous symmetry}\label{fthooft}
\hfill

The emergent symmetries discussed in the section \ref{ES} can suffer the ``'t Hooft anomaly''. A theory is said to have the 't Hooft anomaly with respect to global  symmetry group $G$ if there are obstructions against gauging $G$ \cite{Hooft1980}. In the following we shall use the $(1+1)$-D complex class to illustrate the ideas.\\

The simplest example is the chiral anomaly associated with the $(1+1)$-D complex class theory defined in \Eq{s0}. This theory has emergent global $U_+(n) \times U_-(n)$ symmetry. However, when one tries to gauge this symmetry, an anomaly is encountered. Namely, in the presence of gauge field with non-zero curvature, the theory can not be made to conserve the  Noether's current associated with the {\it full}~  $U_+(n) \times U_-(n)$ symmetry.\\

Starting from the massless fermion theory, we can introduce the $U_+(n) \times U_-(n)$ gauge field (i.e., ``gauging'' $U_+(n) \times U_-(n)$) via minimal coupling. Moreover, we can define the effective gauge action after integrating out fermions,
\begin{align}
	&W[A_+, A_-] = -\ln \left[ \int D\psi \, D\bar{\psi} e^{-S[\psi,\bar{\psi},A_+,A_-]} \right], ~~	\text{where }\notag \\
	&S[\psi,\bar{\psi},A_+,A_-] = \int d^2 x~ \, \bar{\psi} \, \left[ i \gamma^\mu \left( \partial_\mu + i  P_+\otimes  A_{+,\mu}  + i P_-  \otimes A_{-,\mu}   \right) \right] \psi. 
\label{weffA}\end{align}

Here $A_\pm$ are the $n\times n$ matrix value gauge fields associated with $U_\pm(n)$, and $P_\pm$ are the projection operators selecting the chiral fermion modes defined in \Eq{Ppm}. Adler\cite{Adler1969}, Bell, and Jackiw\cite{Bell1969} first showed that in the presence of a diagonal (i.e., $A_+=A_-$) $U(1)$ gauge field, the axial current is not conserved. Shortly after, this was generalized by Bardeen \cite{Bardeen1969} who showed that under infinitesimal gauge transformation, $W$ in \Eq{weffA} is not gauge invariant, namely,
\be
&&\delta W := W[A_+ + d\epsilon_+ , A_- + d\epsilon_-] - W[A_+, A_-] \notag\nn
&&=  -\frac{i}{4\pi} \int\limits_{\mathcal{M}}~ {\rm tr} \left[ A_+  d\epsilon_+ - A_- d\epsilon_-   \right].
\label{Bardeen1c}
\ee
\noindent This is the 't Hooft anomaly. \\

This phenomenon is also connected to the physics of SPT. In odd space dimension, this connection constitutes the so-called ``anomaly inflow picture'' \cite{Callan1985}. In fact, each of the emergent symmetry groups in table \ref{tab:emergentSymm} protects a $D+1$ dimensional $\mathbb{Z}$-classified free fermion SPT. The $D$-dimensional massless free fermion theories in \Eq{s0} describe the boundary of these SPTs.  We shall discuss this point further in appendix \ref{appendix:SPT}.\\

\begin{figure}[h]
	\begin{center}
		\includegraphics[scale=0.3]{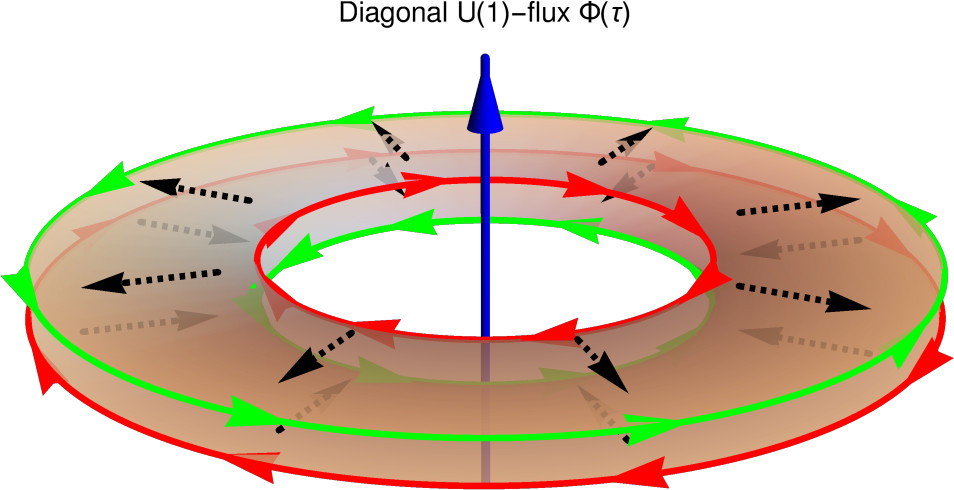}
		\caption{Two layers of annulus shape Chern insulators with $\sigma_{xy}=\pm 1$ stacked together. The outer edge harbors the 1D $n=1$ non-chiral massless fermion modes. The green and red arrows represent the opposite chiralities. When a time-dependent diagonal $U(1)$ flux pierces the inner hole, the induced electric field in the azimuthal direction causes a Hall current (dashed arrows) flowing from inner to outer boundary in the top layer and from outer to inner boundary in the bottom layer. As the result,  the chiral current $J_+-J_-$ is not conserved viewed from the outer edge alone. This system is realized as the ``spin Hall insulator'' experimentally.}
		\label{DCI}
	\end{center}
\end{figure}

The most familiar anomaly inflow example is for $n=1$ in 1D. In this case, we can view the 1D (non-chiral) massless fermions  as the edge modes of two Chern insulators stacked together, with each Chern insulator having Hall conductivity $\sigma_{xy}=\pm 1$ (see \Fig{DCI}). In the presence of a time-dependent flux associated with the diagonal gauge field, there will be  the electric fields in the azimuthal direction. This induces a Hall current causing the charge to flow from the outer to the inner edge on one layer, and from the inner to the outer edge on the other layer.  Viewing from the edge (one-dimensional world), the chiral current $J_+-J_-$ is not conserved. This manifests the chiral anomaly, namely gauging the diagonal U(1) symmetry breaks axial U(1) symmetry - an example of the 't Hooft anomaly.
\\

Although the $U_+(n) \times U_-(n)$ anomaly makes it impossible to gauge the whole group consistently, it's possible to gauge a subgroup of it. For example, if we only gauge the diagonal subgroup $U(n) $ within $U_+(n) \times U_-(n)$, i.e., if 

\begin{align*}
	&A:=A_+ = A_-\\
	&\epsilon := \epsilon_+ = \epsilon_-
\end{align*}
then the two terms in \Eq{Bardeen1c} cancel out, hence the theory is anomaly free with respect to  diagonal $U(n)$ subgroup.\\

\subsection{A heuristic way to determine the 't Hooft anomaly }\label{Anomalyheuristic}
\hfill

The discussions presented above require rather involved field theory calculations. However, there is a heuristic way to get the correct answer. The basis of this heuristic argument is the fact that {\it if a theory can be defined on a lattice with all its (continuous) symmetry, then these symmetries can be gauged without anomaly.}  Again, the above statement is suggested by the SPT physics, namely, the boundary modes (which has 't Hooft anomaly) of an SPT can not be defined on a lattice in the dimension of the boundary. In the following we shall again use the $(1+1)$-D complex class to illustrate the ideas.
\\

Under Wilson's regularization\cite{Wilson1977}(see later), whether a theory with global symmetry group $G$ can be defined on a lattice, is determined by whether there is a mass term that respects $G$
\footnote{Using Wilson's regularization method \cite{Wilson1977}, the existence of such a mass term is a sufficient condition for the theory to be regularizable on a lattice. However, it is more involved to show that it is the necessary condition \cite{Huang2020}.
}. 
Thus, a theory with the $U_+(n) \times U_-(n)$ anomaly, means no mass term is $U_+(n) \times U_-(n)$ symmetric. Again, this is the condition that the gaplessness of the boundary modes %of a  $U_+(n) \times U_-(n)$ -protected SPT 
is symmetry protected.  
\\

First, we show that no mass term is allowed if  $U_+(n) \times U_-(n)$ symmetry is to be respected. Under $U_+(n) \times U_-(n)$ the fermion field transform as
\begin{align*}
	\psi \rightarrow  \left( P_+ \otimes g_+ + P_- \otimes g_- \right) \psi~~{\rm where}~~P_\pm={I\pm Z\over 2}.
\end{align*}
Under such transformation, there is, e.g., no mass term preserving the axial $U_A(1)$ generated by $ZI_n$. This is because according to table \ref{tab:massManifold} the mass terms have the form $$\psi^\dagger \left( X\otimes H_1 + Y \otimes H_2\right) \psi.$$ In fact, the anomaly is not only in the axial $U_A(1)$ part. To see that, let's consider $n>1$. The diagonal $U(n)$ symmetry requires that both $H_1$ and $H_2$ be proportional to the identity matrix. However, such mass term would break $U_+(n)$.\\

Now we show that if we relax the condition to only demanding the diagonal  $U(n)$ symmetry, there is a mass term. For example, 
\begin{align*}
	M_{\rm reg} = X \otimes I_n. 
\end{align*}
This means that we can then write down a lattice model in momentum space using Wilson's regularization\cite{Wilson1977}
\begin{align*}
	{\rm \hat{H}} = \sum_{k\in{\rm  BZ}} \psi_k^\dagger \left[ \sin k \, \Gamma_1 + (1- \cos k ) M_{\rm reg}\right] \psi_k
\end{align*}
where ``BZ'' stands for the Brillouin zone. We can Fourier transform the above hamiltonian back to the real space which gives us a lattice tight-binding model.  The  diagonal $U(n)$ gauge field can then be introduced via Peierls' substitution 
\begin{align*}
	\psi_j^\dagger \psi_i \rightarrow \psi_j^\dagger e^{i A_{i,j}} \psi_i
\end{align*}
\noindent for two adjacent sites $i,j$. Here $A_{i,j}$ is the gauge connection from site $i$ to $j$.\\

\subsection{Discrete global symmetry anomaly }
\label{TRAnoamly2Cfree}
\hfill

A (global) discrete symmetry in a fermion theory can also be broken by regularization. In this subsection, we shall review the simplest example -- the ``parity anomaly"\cite{Redlich1984,Redlich1984b} of the $(2+1)$-D Dirac fermions in the complex class.\\

When the anomaly-free $U(n)$  symmetry is gauged, 
the low energy fermion action  is given by
\begin{align}
\label{actGauged2C}
S =& \int d\tau \, d^2{\v x} \, \psi^\dagger \left[ \left(\partial_0 + i\, I \otimes A_0 \right) - i \Gamma_i \left(\partial_i + i \, I\otimes A_i\right)\right] \psi \\
&\text{ where } \Gamma_1 = ZI_n ,~~~ \Gamma_2=XI_n \notag
\end{align}
\noindent Here $A_\mu$ is the $n\times n$ matrix-valued  $U(n)$ gauge field. Under the global emergent symmetries listed in table \ref{tab:emergentSymm}, the gauged field transforms as
\begin{align}
\label{gaugeTrans2C}
&\text{$U(n)$:~} 	A_\mu \rightarrow g \cdot A_\mu \cdot g^\dagger\notag\\
&\text{Time reversal:~} A_\mu \rightarrow  -\left(A_\mu\right)^* \notag\\
&\text{Charge conjugation:~} 	A_\mu \rightarrow  -\left(A_\mu\right)^T 
\end{align}
\noindent It's easy to check that the low energy action \Eq{actGauged2C} is invariant under the combined transformation of the fermion and the gauge field. \\

As we saw in the preceding subsection, the condition for a symmetry to be anomaly-free is the theory can be regularized while preserving the symmetry. In the present case, to preserve  $U(n)$  we need to choose a regularization that is $U(n)$ invariant. In Wilson's regularization\cite{Wilson1977} this amounts to choose a $U(n)$ invariant regularization mass. The most general mass term is given by $$M=m \, Y\otimes H,$$ where $H$ is an $n\times n$ hermitian matrix with $H^2=I_n$. When acted upon by the global $U(n)$, $$M\ra (I\times g)^\dagger \cdot M \cdot (I\times g)$$ (see table \ref{tab:emergentSymm}). Requiring it to be invariant forces us to choose \be M_{\rm reg}= m \,Y \otimes I_n.\label{mregg}\ee Under Wilson's regularization the momentum space Hamiltonian of the massless Dirac fermion (without gauge field) read, 
\begin{align}
{\rm \hat{H}} = \sum_{{\v k} \in BZ} \psi^\dagger_{\v k} \left[ \sin k_1 \Gamma_1 + \sin k_2 \Gamma_2 + \left(2 - \cos k_1 - \cos k_2 \right)M_{\rm reg} \right]\psi_{\v k}
\end{align}
To incorporate the gauge field, we Fourier transform the above equation back to real space and introduce the gauge field by Peierls' substitution. This is all good as far as regularizing \Eq{actGauged2C} is concerned.\\

 Under the action of the discrete symmetries, however
\begin{align*}
\text{Charge conjugation: }& M_{\rm reg} \rightarrow -\left( I\otimes I_n \right) \cdot M_{\rm reg}^T \cdot \left( I\otimes I_n \right) = M_{\rm reg} \\
\text{Time reversal: }& M_{\rm reg} \rightarrow \left( Y\otimes I_n \right) \cdot M_{\rm reg}^* \cdot \left( Y\otimes I_n \right) = -M_{\rm reg} 
\end{align*}
Therefore charge conjugation is respected by the regularization, however, time-reversal symmetry is not.\\

It was first shown by Redlich \cite{Redlich1984,Redlich1984b} that one can detect the time-reversal anomaly through the effective $U(n)$ gauge action after integrating out the fermions. We reproduce his argument in the following. In momentum space (the Brillouin zone) we have four low energy Dirac fermions, each around a time-reversal invariant ${\v k}$ points:
\begin{align}
\label{DiracNodes2C}
{\v k} = (0,0) + {\v q}:  ~~~ & {\rm \hat{H}}_{(0,0)} \approx \sum_{\text{small }{\v q}} \psi^\dagger_{(0,0)+ {\v q}} \left[ q_1 \, \Gamma_1 + q_2 \, \Gamma_2  \right]\psi_{(0,0)+ {\v q}}\\
{\v k} = (\pi,0) + {\v q}:  ~~~ & {\rm \hat{H}}_{(\pi,0)} \approx \sum_{\text{small }{\v q}} \psi^\dagger_{(\pi,0)+ {\v q}} \left[ -q_1 \, \Gamma_1 + q_2 \, \Gamma_2  + 2m \, M_{\rm reg} \right]\psi_{(\pi,0)+ {\v q}}\notag\\
{\v k} = (0,\pi) + {\v q}:  ~~~ & {\rm \hat{H}}_{(0,\pi)} \approx \sum_{\text{small }{\v q}} \psi^\dagger_{(0,\pi)+ {\v q}} \left[ q_1 \, \Gamma_1 - q_2 \, \Gamma_2  + 2m \, M_{\rm reg} \right]\psi_{(0,\pi)+ {\v q}}\notag\\
{\v k} = (\pi,\pi) + {\v q}:  ~~~ & {\rm \hat{H}}_{(\pi,\pi)} \approx \sum_{\text{small }{\v q}} \psi^\dagger_{(\pi,\pi)+ {\v q}} \left[ -q_1 \, \Gamma_1 - q_2 \, \Gamma_2  + 4m \, M_{\rm reg} \right]\psi_{(\pi,\pi)+ {\v q}}\notag
\end{align}
\noindent Among the four, the first is massless and preserves the time-reversal symmetry. The remaining three, however, acquire a large regularization mass, which is time-reversal breaking. In the presence of the $U(n)$ gauge field, these massive Dirac fermions would each contribute a Chern-Simons effective gauge action after the fermions are integrated out\cite{Redlich1984b}. In particular, for each massive fermion the effective gauge action is $$\frac{1}{2}\times (\pm 1)\times {i\over 4\pi}\int  AdA,$$ where the sign depends on the product of the signs in front of $q_1 \Gamma_1$, $q_2 \Gamma_2$, and $M_{\rm reg}$. Combing them, the massive fermions contribute the following breaking effective action 
\be
\left( -\frac{1}{2}-\frac{1}{2}+\frac{1}{2} \right) \frac{m}{|m|} \frac{i}{4\pi} \int  \tr\left[A \, dA + \frac{2i}{3}A^3 \right] = -\frac{i}{8\pi}\int \tr\left[A \, dA + \frac{2i}{3}A^3 \right].
\label{panom}\ee
This is time-reversal odd, as can be explicitly shown by replacing $A_\mu \rightarrow  -\left(A_\mu\right)^*$ and complex conjugating the action. As to the massless fermions near $\v k=(0,0)$, based on the fact that the first line of \Eq{DiracNodes2C} is time reversal invariant so should their effective gauge action. Thus after regularization, the time-reversal symmetry of \Eq{actGauged2C} is broken!
As expected, charge conjugation is not broken by the regularization. Since $T$ is broken while $C$ is not, based on the $CPT$ invariance, the parity should also be broken 
\footnote{
In two space dimension, the "parity" transformation $P$ is realized by spatial reflection. Take the reflection in $x$-direction as an example, the fermion field transforms according to $$\psi(\tau,x,y) \xrightarrow{P} XI_n \cdot \psi(\tau,-x,y).$$ It is easy to see the that the regularization mass defined in \Eq{mregg} changes sign under $P$. However, the combined $CPT$ transformation leaves it invariant. % transformation alone, but respects the combined transformation $CPT$. 
Thus, there is no $CPT$ anomaly. The same conclusion can be drawn by looking at the parity transformation of the effective gauge action. Under $P$ the gauge field transforms as
\begin{align*}
	A_\tau(\tau,x,y) \xrightarrow{P}& A_\tau(\tau,-x,y)  \\
	A_x(\tau,x,y) \xrightarrow{P}& -A_x(\tau,-x,y)  \\
	A_y(\tau,x,y) \xrightarrow{P}& A_y(\tau,-x,y)  
\end{align*}
\noindent Again, \Eq{panom} changes sign under $P$, but is invariant under $CPT$.
}.
\\

In table \ref{tab:anomalies}, we summarize the maximal anomaly-free continuous symmetry and the discrete symmetry  that is broken after regularization. The only discrete symmetry which possesses anomaly occurs in $(2+1)$-D for the time-reversal symmetry. More detailed discussions are left to appendix \ref{appendix:anomaliesf}.

\begin{table}
	\centering
	\begin{tabular}{ |c|c|c| }
		\hline
		$(1+1)$-D 		&	Real class & Complex class	\\
		\hline
		Global Symmetry	
		&\thead{{\bf Discrete}\\ Anti-unitary: $T^2=+1$\\
			{\bf Continuous unitary}\\Chiral $O(n) \times O(n)$}	
		&	\thead{{\bf Discrete}\\ Anti-unitary: $T^2=+1$\\
			Unitary: $C^2=+1$\\ {\bf Continuous unitary}\\Chiral $U(n) \times U(n)$}		\\
		\hline
		Aanomaly free part
		&Diagonal $O(n)$, $T$ & Diagonal $U(n)$, $T$, $C$	\\
		%\hline
		%Symmetry broken by gauging
		%&${O(n)\times O(n)\over O(n)}$ & ${U(n)\times U(n)\over U(n)}$\\
		\Xhline{3\arrayrulewidth}
		$(2+1)$-D 		&	Real class & Complex class	\\
		\hline
		Global Symmetry	
		&\thead{{\bf Discrete}\\ Anti-unitary: $T^2=-1$\\{\bf Continuous unitary}\\$O(n)$}	
		&	\thead{{\bf Discrete}\\ Anti-unitary: $T^2=-1$\\
			Unitary: $C^2=+1$\\ {\bf Continuous unitary}\\$U(n)$}		\\
		\hline
		Anomaly free part
		&$O(n)$ & $U(n)$, $C$	\\
		%\hline
		%Symmetry broken by gauging
		%&$T$ & $T$	\\
		\Xhline{3\arrayrulewidth}
		$(3+1)$-D 		&	Real class & Complex class	\\
		\hline
		Global Symmetry	
		&\thead{{\bf Discrete}\\ Anti-unitary: $T^2=-1$\\{\bf Continuous unitary}\\$U(n)$}	
		&	\thead{{\bf Discrete}\\ Anti-unitary: $T^2=-1$\\
			Unitary: $C^2=+1$\\ {\bf Continuous unitary}\\ Chiral $U(n)\times U(n)$}		\\
		\hline
		Anomaly free part
		&$O(n)$, $T$ & Diagonal $U(n)$, $T$, $C$	\\
		%\hline
		%Symmetry broken by gauging
		%&${U(n)\over O(n)}$ & ${U(n)\times U(n)\over U(n)}$	\\
		\hline
	\end{tabular}
	\caption{The summary of the global symmetry groups and the anomaly-free parts of the symmetry groups of the massless fermions (and the bosonized non-linear sigma models) in $(1+1)$-D,$(2+1)$-D, and $(3+1)$-D. } 
	\label{tab:anomalies}
\end{table}

\section{Breaking the emergent symmetry by the mass terms}
\label{symmBreaking}
\hfill

The mass terms discussed in the last section {\it necessarily} break some of the emergent symmetries in table \ref{tab:emergentSymm}.  This is because so long as the full emergent symmetries remain unbroken, the fermions will remain massless. In the rest of this section, we use one-dimensional examples to illustrate this.

\subsection{Complex class}
\hfill

The mass terms for the complex class in $(1+1)$-D can be written as

\begin{align*}
\psi^\dagger \left( X\otimes H_1 + Y \otimes H_2\right) \psi 
= \psi^\dagger \begin{bmatrix} 0 & \left( Q^{\mathbb{C}} \right)^\dagger \\ Q^{\mathbb{C}} & 0 \end{bmatrix} \psi.
\end{align*}
When acted upon by the emergent symmetries in \Eq{symm1C},  $Q^{\mathbb{C}}$ transforms as

\begin{align}
	\label{symmTransform1c}
	&U_+(n) \times U_-(n): Q^{\mathbb{C}}\ra 	g_-^\dagger \cdot Q^{\mathbb{C}} \cdot g_+ \\
	&\text{Charge conjugation}: Q^{\mathbb{C}}\ra							 \left(Q^{\mathbb{C}}\right)^* 	\notag\\
	&\text{Time reversal}: Q^{\mathbb{C}}\ra						 \left(Q^{\mathbb{C}}\right)^T. \notag
\end{align}

Thus a space-time constant $Q^{\mathbb{C}}$ breaks the emergent symmetry because both $g_+$ and $g_-$ can be arbitrary unitary matrices.

\subsection{Real class}
\hfill

For the real class in $(1+1)$-D, the mass term  can be written as

\begin{align}
	\chi^\dagger \left[ Y\otimes S + X \otimes (iA)\right] \chi
	= \chi^T \begin{bmatrix} 0 & -i\left( Q^{\mathbb{R}} \right)^T \\ i Q^{\mathbb{R}} & 0 \end{bmatrix} \chi.
	\label{mass1R}
\end{align}

When the emergent symmetries in \Eq{symm1R} acts on it $Q^{\mathbb{R}}$ transforms as

\begin{align*}
	&O_+(n) \times O_-(n): Q^{\mathbb{R}}\ra 	g_-^T \cdot Q^{\mathbb{R}} \cdot g_+ \\
	&\text{Time reversal}: Q^{\mathbb{R}}\ra						 \left(Q^{\mathbb{R}}\right)^T.\\
\end{align*}

Therefore a space-time non-zero $Q^{\mathbb{R}}$  breaks the emergent symmetry because both $g_+$ and $g_-$ can be arbitrary orthogonal matrices.
\\

\section{Restoring the emergent symmetries}
\label{restoreEmergent}
\hfill

So far we have seen that space-time constant  $Q^{\mathbb{C}}$ or $Q^{\mathbb{R}}$ breaks the emergent symmetry. But what if $Q^{\mathbb{C}}$ and $Q^{\mathbb{R}}$ fluctuates in space-time? As in statistical mechanics, when the order parameters fluctuate, the broken  symmetry can be restored. Likewise, if we fluctuate $Q^{\mathbb{C}}$ and $Q^{\mathbb{R}}$ over the appropriate mass manifold we expect the emergent symmetry to be restored. \\

Our approach is conceptually similar to that in Ref.\cite{Wang2015,Metlitski2016} where, on the surface of the topological insulator, the fluctuating superconducting order parameters restore the symmetries of the massless fermions. The important difference is that the required order parameter fluctuation in Ref.\cite{Wang2015,Metlitski2016} is not smooth, because it involves the proliferation of superconducting vortices. Since the structure of vortex cores is important in that approach, and such structure depends on the short-distance physics, this approach is constrained to the surface of SPTs where regularization is not an issue. In contrast, our goal is to bosonize the low energy effective theory, where the emergent symmetry is necessarily broken at short distances (due to anomaly).  As the result, we restrict our order parameter to be smooth in space and time, so that they act on the low energy theory only. \\ %We shall further discuss this in section \ref{condbos}.}\\

But what does ``appropriate mass manifold'' mean? For complex class in $(1+1)$-D,  $Q^{\mathbb{C}}$  needs to fluctuate over the space formed by  $n\times n$ unitary matrices, or $U(n)$. Such a space is connected and has a single component. On the other hand for the real class in 1D, $Q^{\mathbb{R}}$ needs to fluctuate in the space formed by $n\times n$ orthogonal matrices, or $O(n)$.  This space has two disconnected  components, corresponding to $\det [ Q^{\mathbb{R}}] = \pm 1$. It's only when $Q^{\mathbb{R}}$ fluctuates in both components with the equal statistical weight we can restore the emergent symmetry.\\

In $(3+1)$-D the mass manifold consists of a single component, in which $Q^{\mathbb{C,R}}$ fluctuate.  However, in $(2+1)$-D the mass manifold in complex class is $\cup_{l=0}^n\frac{U(n)}{U(l) \times U(n-l)}$ which contains $n+1$ disconnected components. Here $Q^{\mathbb{C}}$ needs to fluctuate in the component $l=n/2$ in order to restore the time reversal symmetry\footnote{Of course this requires $n$ to be even.}. In real class, the mass manifold in two space dimension is $\cup_{l=0}^n\frac{O(n)}{O(l) \times O(n-l)}$, and  $Q^{\mathbb{R}}$ needs to fluctuate in the $l=n/2$ component in order to restore the time reversal symmetry. We summarize the results for higher dimensions in table \ref{tab:symmRestore} and leave the detail in appendix \ref{appendix:massManifold}.\\

\begin{table}
	\centering
	\begin{tabular}{ |c|c|c| }
		\hline
		$(1+1)$-D 		&	Real class & Complex class	\\
		\hline
		\thead{Symmetry transformations \\ of $Q^{\mathbb{C},\mathbb{R}}$}
		&	\thead{$T: Q^{\mathbb{R}} \rightarrow \left( Q^{\mathbb{R}} \right)^T $ \\ 
			$O_+(n) \times O_-(n): Q^{\mathbb{R}} \rightarrow g_-^T \cdot Q^{\mathbb{R}} \cdot g_+$}
		&	\thead{$ T: Q^{\mathbb{C}} \rightarrow \left( Q^{\mathbb{C}} \right)^T $ \\ 
			$C: Q^{\mathbb{C}} \rightarrow \left( Q^{\mathbb{C}} \right)^* $ \\ 
			$U_+(n) \times U_-(n): Q^{\mathbb{C}} \rightarrow g_-^\dagger \cdot Q^{\mathbb{C}} \cdot g_+$}\\
		\hline
		\thead{The mass manifold required to\\restore the full emergent symmetries}
		&	$O(n)$
		&	$U(n)$\\
		\Xhline{3\arrayrulewidth}
		$(2+1)$-D 		&	Real class & Complex class	\\
		\hline
		\thead{Symmetry transformations \\ of $Q^{\mathbb{C},\mathbb{R}}$}
		&	\thead{$T: Q^{\mathbb{R}} \rightarrow - Q^{\mathbb{R}}  $ \\ 
			$O(n) : Q^{\mathbb{R}} \rightarrow g^T \cdot Q^{\mathbb{R}} \cdot g$}
		&	\thead{$T: Q^{\mathbb{C}} \rightarrow -\left( Q^{\mathbb{C}} \right)^* $ \\ 
			$C: Q^{\mathbb{C}} \rightarrow \left( Q^{\mathbb{C}} \right)^T $ \\ 
			$U(n) : Q^{\mathbb{C}} \rightarrow g^\dagger \cdot Q^{\mathbb{C}} \cdot g$}\\
		\hline
		\thead{The mass manifold required to\\restore the full emergent symmetries}
		&	\thead{$\frac{O(n)}{O(n/2) \times O(n/2)}$ \\ for $n \in $ even}
		&	\thead{$\frac{U(n)}{U(n/2) \times U(n/2)}$ \\ for $n \in $ even}\\
		\Xhline{3\arrayrulewidth}
		$(3+1)$-D 		&	Real class & Complex class	\\
		\hline
		\thead{Symmetry transformations \\ of $Q^{\mathbb{C},\mathbb{R}}$}
		&	\thead{$T: Q^{\mathbb{R}} \rightarrow \left( Q^{\mathbb{R}} \right)^* $ \\ 
			$U(n): Q^{\mathbb{R}} \rightarrow u^T \cdot Q^{\mathbb{R}} \cdot u$}
		&	\thead{$T: Q^{\mathbb{C}} \rightarrow \left( Q^{\mathbb{C}} \right)^* $ \\ 
			$C: Q^{\mathbb{C}} \rightarrow \left( Q^{\mathbb{C}} \right)^T $ \\ 
			$U_+(n) \times U_-(n) : Q^{\mathbb{C}} \rightarrow g_-^\dagger \cdot Q^{\mathbb{C}} \cdot g_+$}\\
		\hline
		\thead{The mass manifold required to\\ restore the full emergent symmetries}
		&	$\frac{U(n)}{O(n)}$
		&	$U(n)$\\
		\hline
	\end{tabular}
	\caption{The summary of the symmetry transformations of $Q^{\mathbb{R},\mathbb{C}}$, and the mass manifolds in which the $Q^{\mathbb{R},\mathbb{C}}$ fluctuations can restore the full emergent symmetries.} 
	\label{tab:symmRestore}
\end{table}

\section{The conditions for the effective theory being bosonic}\label{condbos}
\hfill

In order to achieve bosonization, the fermions in \Eq{S01C} and \Eq{S01R} must not appear in the low energy theory. %the ``elementary'' particles are fermions, analogous to quarks. On the other hand, the matrix fields $Q^{\mathbb{C}}$ and $Q^{\mathbb{R}}$ are bosonic. They describe ``mesons''  made of a ''quark'' (particle) and an ''anti-quark'' (hole). 
To ensure that, we need to impose some conditions on the space-time dependence of $Q^{\mathbb{C}}$ and $Q^{\mathbb{R}}$. Namely, as functions of $\v x$ and $\t$,  $Q^{\mathbb{C}}(\t,\v x)$ and $Q^{\mathbb{R}}(\t,\v x)$ needs to fluctuate smoothly (comparing with the length and time scale set by $m$).  Under such conditions, the original fermions can be integrated out, yielding a non-linear sigma model for the order parameters. The idea is similar to that encountered in magnetism, where electrons  form local moments. After integrating out the electrons we arrive at an effective theory -- a non-linear sigma model describing the fluctuations of the local moments in space and time.\\

\section{Fermion integration}\label{NLSM}
\hfill

In this section, using $(1+1)$-D as an example, we shall describe how to integrate out the fermions. In higher spatial dimensions we shall present the results while leaving the details in appendix \ref{appendix:fermionInt}.\\

\subsection{Complex class}\label{NLSM1}
\hfill

The fermion action with a space-time dependent mass term reads
\be
	S =& \int d\tau \, d{\v x} \, \psi^\dagger \left[ \partial_0 - i\Gamma_1 \partial_1 + m \hat{M}(\tau, \v x) \right] \psi 
	\label{S1C}
\ee
where and $\G_1=ZI_n$, and  \be\{\G_1,\hat{M}(t,\v x)\}=0,~~{\rm and}~~\hat{M}(\t,\v x)^2=I_{2n}.\label{mcondit}\ee
The $\hat{M}(\t,\v x)$ that satisfies \Eq{mcondit} is given by
$$
	\hat{M}(\t,\v x) = m \left[X \otimes H_1 (\t, \v x) + Y \otimes H_2 (\t,\v x)\right],
$$
\\

For smooth order parameter configurations $\hat{M}(\t,\v x)$, the fermion integration can be done via gradient expansion (See \cite{Abanov2000} for example. We shall convert the action to a Lorentz invariant form and present the general formalism applicable for all spatial dimensions in appendix \ref{appendix:fermionInt}). The resulting effective action consists of two types of terms: the non-topological and topological terms. For the non-topological term (the stiffness term) we shall keep the one with the smallest number of space-time derivatives  (they are the most relevant in the renormalization group sense). The topological term is dimensionless. In $(1+1)$-D, explicit fermion integration yields (see appendix \ref{appendix:fermionInt} for details)
\begin{align}
W[Q^{\mathbb{C}}]  = \frac{1}{8\pi}   \int\limits_{\mathcal{M}} d^2x \, \tr\left[\p_\mu Q^{\mathbb{C}\dagger}  \partial^\mu Q^{\mathbb{C}}\right] -  \frac{2\pi i}{24\pi^2}    \int\limits_{\mathcal{B}}   \, {\rm tr} \Big[   \left( \tilde{Q}^{\mathbb{C}\dagger}   d \tilde{Q}^{\mathbb{C}}  \right)^3    \Big],  \label{wzw1C}
\end{align}
where $Q^{\mathbb{C}} $ is given in \Eq{mass1C} and \Eq{qcdef}. The first term in \Eq{wzw1C} is the stiffness term and the second is the Wess-Zumino-Witten (WZW) topological term. \Eq{wzw1C} reproduces the  level-1 $U(n)$ (abbreviated as $U(n)_1$) WZW model in Witten's non-abelian bosonization \cite{Witten1984}. Note that the symbol ``$\tr$'' means tracing over the $n\times n$ portion of the matrix. (In doing fermion integration, we have already traced out the matrix part involving $\g^\mu$'s). In \Eq{wzw1C} $\mathcal{M}$ is the space-time manifold, and $\mathcal{B}$ is the extension of the space-time manifold $\mathcal{M}$ so that $$\partial \mathcal{B} = \mathcal{M}.$$ In addition, $\tilde{Q}^{\mathbb{C}}(u,x)$ is an extension field of $Q^{\mathbb{C}}(x)$ so that \be &&\tilde{Q}^{\mathbb{C}}(u=1,x)= Q^{\mathbb{C}}(x)~~{\rm and}\nn &&\tilde{Q}^{\mathbb{C}}(u=0,x)= {\rm constant}\nonumber\ee In the equation above,  ``constant'' means a space-time independent matrix.\\

For simplicity we shall focus on the space-time manifold $\mathcal{M}=S^D$ so that $\mathcal{B}$ is a
$D+1$-dimensional disk. The reason for this choice is to ensure the extension $\tilde{Q}^{\mathbb{C}}(u,x)$ exists. Because we require a smooth evolution from $Q^{\mathbb{C}}(u=0,x)$ to $\tilde{Q}^{\mathbb{C}}(u=1,x) $ ($x$ denotes $(\t,\v x)$), it means the mapping  $$Q^{\mathbb{C}}: (u=1,x)\ra \text{mass manifold}$$ is homotopically equivalent to the mapping $$Q^{\mathbb{C}}: (u=0,x)\ra \text{mass manifold}.$$ Since $\tilde{Q}^{\mathbb{C}}(u=0,x)= {\rm constant}$ is homotopically trivial, a necessary condition for the smooth extension to exist is 
$$ \pi_D(\text{mass manifold})=0,$$ i.e., all smooth mappings from the space-time manifold to the mass manifold are homotopically trivial. It turns out this condition is met for sufficiently large $n$ in all spatial dimensions. We shall return to this point in appendix \ref{appendix:massManifold}, \ref{appendix:fermionInt}, and \ref{appendix:enlarge}. For $(1+1)$-D, $ \pi_2(U(n))=0$ for any $n$.
\\

For the WZW term to be well defined, it had better not depend on the extension. When there are two different extensions on the $D+1$ dimensional disk, say one defined by $\tilde{Q}_1^{\mathbb{C}}$ on $\mathcal{B}_1$ and the other by $\tilde{Q}_2^{\mathbb{C}}$ on $\mathcal{B}_2$, the difference in the WZW term associated with these two extensions is given by

\be
	\Delta W_{WZW}[\tilde{Q}^{\mathbb{C}}] = -\frac{2\pi i}{24\pi^2}    \int\limits_{\mathcal{B}_1\cup (-\mathcal{B}_2)}    \, {\rm tr} \Big[   \left( \tilde{Q}^{\mathbb{C}\dagger}   d \tilde{Q}^{\mathbb{C}}  \right)^3    \Big] 
	\label{wzwdiff}
\ee

\noindent where $-\mathcal{B}_2$ is the mirror reflection of $\mathcal{B}_2$. Since $\mathcal{B}_1 \cup (-\mathcal{B}_2)=S^{D+1}$, removing the factor $2\pi i$, \Eq{wzwdiff} is the topological invariant associated with $\pi_{2+1}(\text{mass manifold})$. It turns out that for all relevant cases, $\pi_{D+1}(\text{mass manifold})$ $=$$\mathbb{Z}$ (see appendix \ref{appendix:massManifold}). In $(1+1)$-D, $\pi_3(U(n))=\mathbb{Z}$ for $n\ge 2$ ($n=1$ corresponds to flavorless or spinless fermion where the bosonization is abelian.). The coefficient of the  WZW term renders $\Delta W_{WZW}=2\pi i \times \text{integer}$. The fact that the WZW term is $2\pi i$ times the topological invariant implies the level ($k$) is 1.  After the exponentiation, the phase factor associated with the WZW term is well-defined.\\

\subsection{Real class}\label{NLSM2}
\hfill

The 1+1-D Majorana fermion action with a space-time dependent mass read

\begin{align}
	S =& \int d\tau \, d{\v x} \, \chi^T \left[ \partial_0 - i\Gamma_1 \partial_1 +m \hat{M}(\tau, \v x) \right] \chi 
	\label{S1R}
\end{align}
\noindent where 
\begin{align*}
\Gamma_1 = ZI_n \text{ and } \hat{M}(\t,\v x)=	\left[Y\otimes S + X\otimes (i A) \right].
\end{align*}

Following the same steps discussed in the last subsection, fermion integration yields the following effective action (see appendix \ref{appendix:fermionInt})
\begin{align}
	W[Q^{\mathbb{R}}] &=  \frac{1}{16\pi}   \int\limits_{\mathcal{M}} d^2x \, {\rm tr}\left[ \p_\mu Q^{\mathbb{R}T} \partial^\mu Q^{\mathbb{R}} \right] -\frac{2\pi i}{48\pi^2}    \int\limits_{\mathcal{B}}   {\rm tr} \Big[   \left(\tilde{Q}^{\mathbb{R}T}   d \tilde{Q}^{\mathbb{R}}\right)^3\Big].
	\label{wzw1R} 
\end{align}
\Eq{wzw1R} is the $O(n)_{k=1}$ WZW model. Again, $\tilde{Q}^{\mathbb{R}}(u,x)$ is extension field of $Q^{\mathbb{R}}(x)$, which exists if $\pi_D(\text{mass manifold})=0$. In $(1+1)$-D, $\pi_2(O(n))=0$ for $n\ge 3$.  Here the difference in the WZW term associated with two different extension is the topological invariant associated with $\pi_3(O(n))=\mathbb{Z}$ for relevant $n$ (see appendix \ref{appendix:massManifold}). The coefficient of the WZW term renders $\Delta W_{WZW}=2\pi i\times\text{integer}$ hence yields the same phase factor upon exponentiation. Again,  the fact that the WZW term is $2\pi i$ times the topological invariant implies the level ($k$) is 1.\\

Thus, for both complex and real classes, the bosonization of massless fermion is the non-linear sigma model with WZW term. This reproduces  Witten's non-abelian bosonization results, which was obtained using a totally different method (the current algebra).\\

The above bosonization scheme can be straightforwardly   generalized to higher dimensions. One thing that needs some care is the fact that the homotopy group of the mass manifold depends on $n$. For $n$ exceeds certain value $\pi_{D+1}(\text{mass manifold}) $ $=$ $\mathbb{Z}$. In that case fermion integration does lead to a nonlinear sigma model with $k=1$ WZW term. However, for small $n$ (before the ``homotopy stabilization'') sometimes, e.g., $\pi_{D+1}(\text{mass manifold})=0$. We shall discuss one such instance in appendix \ref{appendix:enlarge}. Fortunately, for the vast majority of applications $n$ is sufficiently big so that $\pi_{D+1}(\text{mass manifold})=\mathbb{Z}$.

\section{Non-linear sigma models in $(2+1)$-D and $(3+1)$-D}
\label{NLSigma2D3D}
\hfill

As mentioned, the bosonization strategy described in the preceding section can be applied to two and three spatial dimensions. To facilitate later discussions, including the applications in $(2+1)$-D and $(3+1)$-D, the explicit form of the nonlinear sigma models in table \ref{tab:WZWTable} are given here. For briefness, we shall only include the results for sufficiently large $n$ so that $\pi_{D+1}({\rm mass manifold})=\mathbb Z$. As discussed earlier, under such conditions the non-linear sigma model possesses a WZW term.\\ 

\subsection{Complex class in $(2+1)$-D}
\label{bosonization2C}
\hfill

For Dirac fermions with $n$ flavors in the complex class, after bosonization the sigma model matrix field (or the order parameter)  lives in the space of complex Grassmannian, namely, $$Q^{\mathbb{C}}(x)\in\frac{U(n)}{U(n/2) \times U(n/2)}.$$ This means that at any space-time point $x$, $Q^{\mathbb{C}}(x)$ is an $n \times n$ hermitian matrix with half of the eigenvalues  $+1$, and the other half $-1$. One can specify $Q^{\mathbb{C}}(x)$ by the unitary matrix, $C(x)$, which renders $Q^{\mathbb{C}}(x)$ diagonalized upon similarity transformation, i.e., 
$$Q^{\mathbb{C}}(x)=C(x) \cdot \begin{pmatrix}I_{n/2}&0\\0 &-I_{n/2}\end{pmatrix} \cdot C^\dagger(x).$$
Obviously two different $C(x)$s related by

\begin{align*}
	C^\prime(x) =
	C(x) \cdot \begin{pmatrix} g_1(x) & 0 \\ 0 & g_2(x) \end{pmatrix},
\end{align*}
\noindent where $g_1(x),g_2(x)\in U(n/2)$, will lead to identical $Q^{\mathbb{C}}(x)$. Due to this redundancy, the order parameter lives in the quotient space $\frac{U(n)}{U(n/2) \times U(n/2)}$.\\

Explicit fermion integration yields the following non-linear sigma model
\begin{align}
	W[Q^{\mathbb{C}}] ={1\over 2\lambda_3} \int\limits_{\mathcal{M}} d^3 x \, \tr\Big[\Big(\p_\mu Q^{\mathbb{C}}\Big)^2\Big] -  \frac{ 2 \pi i }{256 \pi^2}      \int\limits_{\mathcal{B}} \, \tr \Big[\tilde{Q}^{\mathbb{C}}    \,\left( d \tilde{Q}^{\mathbb{C}} \right)^4 \Big], 
	\label{wzw2C}
\end{align}
\noindent where $\lambda_3$ is a parameter having the dimension of length. In the limit where the short distance cutoff is zero, \be\lambda_3={8\pi\over m}\label{lambda2D}\ee where $m$ is the fermion energy gap.\\

 The first term in \Eq{wzw2C} is the stiffness term and the second is the level-1 ($k=1$) Wess-Zumino-Witten term. $\tilde{Q}^{\mathbb{C}}(x,u)$ is the extended field of $Q^{\mathbb{C}}(x)$, which exist because $\pi_3(\frac{U(n)}{U(n/2) \times U(n/2)})=0$ for $n\geq 4$. The difference in the WZW term associated with two different extensions is $2\pi i$ times the topological invariant associated with $\pi_4(\frac{U(n)}{U(n/2) \times U(n/2)})=\mathbb{Z}$. Consequently upon exponentiation, different extensions yield the same phase factor. (To recapitulate the explanation, the readers are referred to subsection \ref{NLSM1}.)

\subsection{Real class in $(2+1)$-D}
\label{bosonization2R}
\hfill

For massless $n$-flavor Majorana fermions in the real class,  the fluctuating order parameters $Q^{\mathbb{R}}(x)$  lives in the space of real Grassmannian, namely,  $$Q^{\mathbb{R}}(x)\in \frac{O(n)}{O(n/2) \times O(n/2)}.$$ This means that at any space-time point $x$, $Q^{\mathbb{R}}(x)$ is an $n \times n$ real symmetric matrix, with half of the eigenvalues  $+1$, and the other half $-1$. One can specify $Q^{\mathbb{R}}(x)$ by  the orthogonal matrix, $R(x)$, required to render $Q^{\mathbb{R}}(x)$ diagonalized, namely,  $$Q^{\mathbb{R}}(x)=R(x) \cdot  \begin{pmatrix}I_{n/2}&0\\0 &-I_{n/2}\end{pmatrix} \cdot R^T(x).$$ Two different $R(x)$s related by 
\begin{align*}
R^\prime(x)=
	R(x) \cdot \begin{pmatrix} g_1(x) & 0 \\ 0 & g_2(x) \end{pmatrix},
\end{align*}
\noindent where $g_1(x),g_2(x)\in O(n/2)$, will lead to identical $Q^{\mathbb{R}}(x)$. Due to this redundancy, the order parameter lives in the quotient space $\frac{O(n)}{O(n/2) \times O(n/2)}$.\\

Explicit fermion integration leads to the following non-linear sigma model
\begin{align}
	W[Q^{\mathbb{R}}] ={1\over 4\lambda_3} \int\limits_{\mathcal{M}} d^3 x \, \tr\Big[\Big(\p_\mu Q^{\mathbb{R}}\Big)^2\Big] -  \frac{ 2 \pi i }{512 \pi^2}      \int\limits_{\mathcal{B}} \tr \Big[\tilde{Q}^{\mathbb{R}}    \,\left( d \tilde{Q}^{\mathbb{R}} \right)^4 \Big] .
	\label{wzw2R}
\end{align}
Again, $\lambda_3$ has the dimension of length, and in the limit where the short-distance cutoff is zero $\lambda_3$ is given by \Eq{lambda2D}. \\

The first term in \Eq{wzw2R} is the stiffness term and the second is the Wess-Zumino-Witten topological term of level $k=1$. $\tilde{Q}^{\mathbb{R}}(x,u)$ is the extended field of $Q^{\mathbb{R}}(x)$, which exist because $\pi_3(\frac{O(n)}{O(n/2) \times O(n/2)})=0$ for $n\geq 6$.  The difference in the WZW term associated with two different extensions is $2\pi i$ times the topological invariant associated with $\pi_4(\frac{O(n)}{O(n/2) \times O(n/2)})=\mathbb{Z}$. Consequently upon exponentiation 
different extensions yield the same phase factor. (Again, to recapitulate the explanation, the readers are referred to subsection \ref{NLSM2}.) \\

\subsection{Complex class in $(3+1)$-D}
\label{bosonization3C}
\hfill

For the $n$-flavor massless Dirac fermions in the complex class,
the fluctuating order parameters $Q^{\mathbb{C}}(x)$ lives in the space of $n\times n$ unitary matrices, namely, 
$$Q^{\mathbb{C}}(x)\in U(n).$$
Explicit fermion integration leads to the following non-linear sigma model
\begin{align}
	W[Q^{\mathbb{C}}] = &  \frac{1}{2\lambda_4^2}   \int_\mathcal{M} d^4 x \, \tr\left[  \partial_{\mu} Q^{\mathbb{C}} \partial^{\mu}Q^{\mathbb{C}\dagger}\right] 
	-\frac{2\pi}{480\pi^3}    \int\limits_{\mathcal{B}}    \, \tr \Big[\left(\tilde{Q}^{\mathbb{C}\dagger} d\tilde{Q}^{\mathbb{C}} \right)^5 \Big],
	\label{wzw3C}
\end{align}
where $\lambda_4$ has the dimension of length. Using dimensional regularization  $\lambda_4$ is given by
\be
{1\over \lambda_4}= \left[ \frac{\Gamma(0^+)m^{2}}{8 \pi^{2}} \right]^{1/2},\label{lambda3D}\ee  
signifying that $\lambda_4$ is cutoff-dependent.  Here $\Gamma(0^+)$ is the gamma function evaluated at $0^+$ from dimensional regularization (see appendix \ref{appendix:fermionInt} for the details). \\

The first term in \Eq{wzw2C} is the stiffness term and the second is the level $k=1$ Wess-Zumino-Witten term. $\tilde{Q}^{\mathbb{C}}(x,u)$ is the extended field of $Q^{\mathbb{C}}(x)$, which exist because $\pi_4(U(n))=0$ for $n\geq 3$. The difference in the WZW term associated with two different extensions is $2\pi i$ times the topological invariant associated with $\pi_5(U(n))=\mathbb{Z}$. Consequently upon exponentiation different extensions yield the same phase factor. (Again, to recapitulate the explanation, the readers are referred to subsection \ref{NLSM1}.) \\

\subsection{Real class in $(3+1)$-D}
\label{bosonization3R}
\hfill

For the $n$-flavor massless Majorana fermions in the complex class, the fluctuating order parameters  $Q^{\mathbb{R}}(x)$ lives in the space of ``real Lagrangian Grassmannian'', namely, $$Q^{\mathbb{R}}(x)\in \frac{U(n)}{O(n)}.$$ This means that at any space-time point $x$, $Q^{\mathbb{R}}(x)$ is an $n \times n$ symmetric unitary matrix. According to the Autonne decomposition (e.g., corollary 2.6.6 of \cite{Horn2012}), any symmetric unitary matrix can be decompose into $$Q^{\mathbb{R}}(x)= W(x) \cdot W^T(x),$$ where $W(x)$ is unitary. Hence, two different $W(x)$s related by
\begin{align*}
	W^\prime(x)= W(x) \cdot g(x),
\end{align*}
where $g(x)\in O(n)$, will lead to identical $Q^{\mathbb{R}}(x)$. Due to this redundancy, the order parameter lives in the quotient space $\frac{U(n)}{O(n)}$.\\

Explicit fermion integration yields the following non-linear sigma model 
\begin{align}
	W[Q^{\mathbb{R}}] = &  \frac{1}{4\lambda_4^2}  \int_\mathcal{M} d^4 x \, \tr\left[  \partial_{\mu} Q^{\mathbb{R}} \partial^{\mu}Q^{\mathbb{R}\dagger}\right] 
	-\frac{2\pi}{960\pi^3}    \int\limits_{\mathcal{B}}   \, \tr \Big[\left(\tilde{Q}^{\mathbb{R}\dagger} d\tilde{Q}^{\mathbb{R}} \right)^5 \Big].
	\label{wzw3R}
\end{align}

The first term in \Eq{wzw2R} is the stiffness term and the second is the  level $k=1$ Wess-Zumino-Witten topological term. $\tilde{Q}^{\mathbb{R}}(x,u)$ is the extended field of $Q^{\mathbb{R}}(x)$, which exist because $\pi_4(U(n)/O(n))=0$ for $n\geq 5$. The difference in the WZW term associated with two different extensions is $2\pi i$ times the topological invariant associated with $\pi_5(U(n)/O(n))=\mathbb{Z}$.  Consequently upon exponentiation different extensions yield the same phase factor. (Again, to recapitulate the explanation, the readers are referred to subsection \ref{NLSM2}.) \\

In table \ref{tab:WZWTable} we summarize the $n$ values above which $\pi_{D+1}(\text{mass manifold})$ is stabilized. We shall discuss some of the small $n$ cases which are relevant to our applications in appendix \ref{appendix:enlarge}.

\begin{table}
	\centering
	\begin{tabular}{ |c|c|c| }
		\hline
		 		&	Real class & complex class 	\\
		\hline
		$(1+1)$-D
		&	\thead{$O(n)_1$ WZW term \\ stabilized for $n\geq 3$}
		&	\thead{$U(n)_1$ WZW term\\ stabilized for $n\geq 2$}\\
		\hline
		$(2+1)$-D 	
		&	\thead{$\left[\frac{O(n)}{O(n/2)\times O(n/2)}\right]_1$ WZW term\\stabilized for $n\geq 6$}
		&	\thead{$\left[\frac{U(n)}{U(n/2)\times U(n/2)}\right]_1$ WZW term\\stabilized for $n\geq 4$}\\
		\hline
		$(3+1)$-D 	
		&	\thead{ $\left[U(n)/O(n)\right]_1$ WZW term \\stabilized for $n\geq 5$}
		&	\thead{$U(n)_1$ WZW term\\stabilized for $n\geq 3$}\\
		\hline
	\end{tabular}
	\caption{The $n$ values above which the $\pi_{D+1}(\text{mass manifold})$ is stabilized.} \label{tab:WZWTable}
\end{table}

\subsection{The value of the stiffness constant and the phases of non-linear sigma models}
\hfill

Unlike in $(1+1)$-D, the stiffness constants of the non-linear sigma models in $(2+1)$-D and $(3+1)$-D are dimensionful parameters. A natural question then arises, how does the values of these parameters determine the phase of the non-linear sigma models ?  For small $\lambda_3$ and $\lambda_4^2$ the action costs of space-time varying $Q^{\mathbb{R}, \mathbb{C}}$ is large, hence we expect spontaneous symmetry breaking to occur.  Quantum disorder sets in for large $\lambda_3$ and $\lambda_4^2$. In the presence of the WZW term, the quantum disordered phase is gapless. It is in the latter phase do the non-linear sigma models represent the massless free fermions.

\section{Non-linear sigma models as the effective theories of interacting fermion models}\label{IntFermion}
\hfill

As we have seen in section \ref{NLSigma2D3D}, while the coefficient in front of the stiffness term in the non-linear sigma model is dimensionless in $(1+1)$-D, those in $(2+1)$-D and $(3+1)$-D are dimensionful parameters. This begs the question of what are these parameters? and for what values of these parameters are the non-linear sigma models equivalent to the massless fermion theories? In addition, for $D=2+1$  the mass manifold consists of more than one connected components. What kind of model can realize phases correspond to different components of the mass manifold? In the following, we answer these questions by focusing on the complex class. It is straightforward to generalize the result to the real class. \\

%So far we have argued that when the order parameter, $Q^{\mathbb{C},\mathbb{R}}$, of the non-linear sigma models fluctuate to restore the emergent symmetry, the results are massless fermion phases. On the other hand,  $Q^{\mathbb{C},\mathbb{R}}$ could spontaneously break the emergent symmetries and lead to massive phases. 
%More interestingly, in $(2+1)$-D where the mass manifold of $Q^{\mathbb{C},\mathbb{R}}$ consists of more than one connected components. What kind of interacting fermion model can have the phases correspond to different components of the mass manifold?}\\% 

As listed in table \ref{tab:massManifold}, the mass terms correspond to  $Q^{\mathbb{C}}$ are given by
\be
&& (1+1)\text{-D}:~~M[Q^\mathbb{C}]= X\otimes {1\over 2}\left[Q^\mathbb{C}+\left(Q^\mathbb{C}\right)^\dagger\right]+ Y\otimes {1\over 2i}\left[Q^\mathbb{C}-\left(Q^\mathbb{C}\right)^\dagger\right]\nn
&& (2+1)\text{-D}:~~M[Q^{\mathbb{C}}]=Y\otimes Q^{\mathbb{C}}\nn
&& (3+1)\text{-D}:~~M[Q^{\mathbb{C}}]=YX\otimes {1\over 2}\left[Q^\mathbb{C}+\left(Q^\mathbb{C}\right)^\dagger\right]+YY\otimes {1\over 2i}\left[Q^\mathbb{C}-\left(Q^\mathbb{C}\right)^\dagger\right]\nn
\label{mass1to3}\ee

Let's consider the four-fermion interacting %respecting the full emergent symmetry we  consider 
generated by the  following inverse Hubbard-Stratonovich transformation,
\begin{align}
\label{invhs}
&\exp\left\{-S_{\rm I}\left[\psi^\dagger,\psi\right]\right\}:=\notag\\&\int D\left[\mathcal{Q}(x)\right] \, \exp\left\{-\int d^Dx \left[ \psi^\dagger~M[\mathcal{Q}(x)] ~\psi + \frac{1}{2\lambda_I} \tr\left[\mathcal{Q}(x)^\dagger \mathcal{Q}(x) \right] \right]\right\}
\end{align}
\noindent where $\mathcal{Q}(x)$ is an $n \times n$  matrix-valued function of space-time. We note that the strength of the four fermion interaction in \Eq{invhs} is proportional to $\lambda_I$.\\

The emergent global symmetries transform $\mathcal{Q}(x)$ in exactly the same way as $Q^{\mathbb{C}}$ (see table \ref{tab:emergentSymm}). This is because $\mathcal{Q}(x)$ and $Q^{\mathbb{C}}$ couple to the same fermion bi-linears. Such transformation can be absorbed by the redefinition of the integration variable $\mathcal{Q}(x)$. Therefore as long as the integration measure in \Eq{invhs} is symmetric under the symmetry transformations, $S_I$ is invariant under the action of emergent symmetries. \\

When $\lambda_I$ is sufficiently large, it is energetically favorable for $$\tr \left[\Avg{\mathcal{Q}^\dagger (x) \mathcal{Q}(x)}\right]$$ to acquire a non-zero expectation value. Assuming such expectation value doesn't spontaneously break the continuous symmetry\footnote{The possible symmetry breaking phases are captured by the non-zero expectation value $\Avg{\mathcal{Q}(x)}$.} it must satisfy
\begin{align*}
\Avg{\mathcal{Q}^\dagger (x) \mathcal{Q}(x)} \rightarrow g^\dagger \cdot \Avg{\mathcal{Q}^\dagger (x) \mathcal{Q}(x)} \cdot g = \Avg{\mathcal{Q}^\dagger (x) \mathcal{Q}(x)}
\end{align*}
\noindent for all $g \in U(n)$ (for $(1+1)$-D and $(3+1)$-D $g\in U_+(n)$). This requires the expectation value of $\mathcal{Q}^\dagger (x) \mathcal{Q}(x)$ to be proportional identity matrix,
\begin{align*}
\Avg{\mathcal{Q}^\dagger (x) \mathcal{Q}(x)} = \kappa^2 I_n
\end{align*}
where  $\kappa^2$ should grow monotonically with $\lambda_I$. At low energy and long wavelength, the dynamics of $\mathcal{Q}$ is governed by the Goldstone modes $Q^\mathbb{C}(x)$, where
%In the above equation we have assumed $\kappa^2$ to be space-time independent, because we expect the space-time fluctuation of $m^2$ (the Higgs mode) to be massive. Under such condition the low energy and long wavelength physics is captured by the Goldstone modes 
$$\mathcal{Q}(x) \rightarrow \kappa Q^\mathbb{C}(x),~~{\rm and}~~\left(Q^\mathbb{C}\right)^\dagger Q^\mathbb{C} = I_n.$$ The manifold in which $Q^\mathbb{C}(x)$ fluctuates is exactly  the mass manifold given in table \ref{tab:massManifold}.\\

The effective action governing the fluctuations of $Q^\mathbb{C}(x)$ is given by the results of section \ref{NLSigma2D3D}, where the stiffness term coefficients ${1\over 2\lambda_3}$ and ${1\over 2\lambda_4^2}$ should grow with $\kappa^2$ which, in turn, monotonically increases with $\lambda_I$. As the result, strong four-fermion interaction implies small $\lambda_3$ and $\lambda_4^2$, while weak four fermion-interaction implies large   $\lambda_3$ and $\lambda_4^2$. Thus, we obtain a duality-like relation, namely, strong coupling fermion theory corresponds to weak coupling non-linear sigma model, and weak coupling fermion theory corresponds to strong coupling non-linear sigma model. Since, by dimension counting, local four-fermion interaction is an irrelevant perturbation to the massless theory in $(2+1)$- and $(3+1)$-D, we expect there is a range of large $\lambda_3$ and $\lambda_4^2$ where the non-linear sigma model is massless.\\

Now we come to $(2+1)$-D, where according to table \ref{tab:massManifold} , the mass manifold has $n+1$ components, namely, $$Q^\mathbb{C} \in \bigcup_{l=0}^n \frac{U(n)}{U(l) \times U(n-l) }.$$ (Here $l$ corresponds to the number positive eigenvalues of $Q^\mathbb{C}(x)$, the readers are referred to appendix \ref{appendix:massManifold} for details.) The condition that the order parameter is a smooth function of space-time confines $Q^\mathbb{C}(x)$ to fluctuate in one of the mass manifold components. If such fluctuation is to restore the time-reversal symmetry, it further restricts $l=n/2$ (we focus on $n=$ even). However, if we allow the possibility of spontaneous time-reversal symmetry breaking, then  $Q^\mathbb{C}(x)$ can fluctuate in the $l\ne n/2$ mass manifold. It is  interesting whether the order parameter fluctuation in the $l\ne n/2$ mass manifolds can restore the unitary part of the emergent symmetry, and if it does can the resulting phase be gapless.\\
 
%Regardless of which component of the mass manifold $Q^\mathbb{C}(x)$ fluctuates in, the effective theory is governed by the following non-linear sigma model in appendix \ref{appendix:fermionInt}.\\

\section{Global symmetries of the non-linear sigma models}
\label{symmNLSigma}
\hfill

Up to this point, we have derived the non-linear sigma model. The bosonic partition function is given by %By fluctuating the $Q^{\mathbb{C,R}}$ field over the space-time, we hope to restore the emergent symmetry. To be explicit the partition function we consider is given by
$$
\mathcal{Z}=\int D[Q^{\mathbb{C,R}}] ~e^{-S_{\rm{NL}\s}\left[Q^{\mathbb{C,R}}\right]}.$$
Here $Q^{\mathbb{C,R}}\in \text{mass manifolds}$, and the integration measure is defined so that at every space-time point 
$Q^{\mathbb{C,R}}$ and the symmetry transformed $Q^{\mathbb{C,R}}$ (see table \ref{tab:WZWTable}) have the same weight.\\

Now, using the  complex class in $(1+1)$-D as an example, we demonstrate that the non-linear sigma model in \Eq{wzw1C} respects the emergent symmetries of the massless free fermion theory. 
 Under the  action of the global emergent symmetries, a configuration $Q^{\mathbb{C}}(\tau, \bf x)$  transforms by  \Eq{symmTransform1c}, namely,

\begin{align*}
	&U_+(n) \times U_-(n): Q^{\mathbb{C}}(\tau, {\bf x})\ra  g_-^\dagger \cdot Q^{\mathbb{C}}(\tau, {\bf x}) \cdot g_+ \\
	&\text{Charge conjugation}: Q^{\mathbb{C}}(\tau, {\bf x})\ra  	\left(Q^{\mathbb{C}}(\tau, {\bf x})\right)^* 	\\
	&\text{Time reversal}: Q^{\mathbb{C}}(\tau, {\bf x})\ra		 	 \left(Q^{\mathbb{C}}(\tau, {\bf x})\right)^T.
\end{align*}
Under the action of $U_+(n) \times U_-(n)$ 
\be &&Q^{\mathbb{C}\dagger}\partial_{\mu} Q^{\mathbb{C}}\ra g_+^\dagger\cdot\left( Q^{\mathbb{C}\dagger}\partial_{\mu} Q^{\mathbb{C}}\right) \cdot g_+
\nonumber\ee
Due to the cyclic invariance  of trace, the similarity transformations cancel out and the action \Eq{wzw1C} is invariant.\\

Under  charge conjugation, the stiffness term transforms as

	\begin{align*}
	& -\frac{1}{8\pi}   \int\limits_{\mathcal{M}} d^2x \, \tr\left[  \left(Q^{\mathbb{C}\dagger}  \partial^\mu Q^{\mathbb{C}}\right)^2 \right]
	\rightarrow  -\frac{1}{8\pi}   \int\limits_{\mathcal{M}} d^2x \, \tr\left[  \left(Q^{\mathbb{C}T}  \partial^\mu Q^{\mathbb{C}*}\right)^2 \right] \\
	=& -\frac{1}{8\pi}   \int\limits_{\mathcal{M}} d^2x \, \tr\left[  \left(  \partial^\mu Q^{\mathbb{C}\dagger}Q^{\mathbb{C}}\right)   \left(  \partial^\mu Q^{\mathbb{C}\dagger}Q^{\mathbb{C}}\right)\right]=-\frac{1}{8\pi}   \int\limits_{\mathcal{M}} d^2x \, \tr\left[  \left(  Q^{\mathbb{C}\dagger} \partial^\mu Q^{\mathbb{C}}\right)   \left(   Q^{\mathbb{C}\dagger}\partial^\mu Q^{\mathbb{C}}\right)\right]
	\end{align*}

hence is invariant. Here the first equality in the second line is due to the invariance of trace under transposing, and the second equality is due to $ \partial^\mu Q^{\mathbb{C}\dagger}Q^{\mathbb{C}}=-Q^{\mathbb{C}\dagger}\partial^\mu Q^{\mathbb{C}}.$ A similar argument applies to the WZW term,
	\begin{align*}
	&\frac{2\pi i}{24\pi^2}    \int\limits_{\mathcal{B}}   \tr \Big[   \left( \tilde{Q}^{\mathbb{C}\dagger}   d \tilde{Q}^{\mathbb{C}}  \right)^3    \Big]
	\rightarrow  \frac{2\pi i}{24\pi^2}    \int\limits_{\mathcal{B}}   \tr \Big[   \left( \tilde{Q}^{\mathbb{C}T}   d \tilde{Q}^{\mathbb{C}*}  \right)^3    \Big]\\
	= & -\frac{2\pi i}{24\pi^2}    \int\limits_{\mathcal{B}}   \, {\rm tr} \Big[   \left(d \tilde{Q}^{\mathbb{C}\dagger}  \tilde{Q}^{\mathbb{C}}     \right)^3    \Big]
	= \frac{2\pi i}{24\pi^2}    \int\limits_{\mathcal{B}}   \,{\rm tr} \Big[   \left( \tilde{Q}^{\mathbb{C}\dagger}   d \tilde{Q}^{\mathbb{C}}  \right)^3    \Big].
	\end{align*}
\noindent The extra minus sign in the second line is because transposing causes an odd number of crossing of the  differential 1-forms. This negative sign is canceled out in the last term due to the odd number of negative signs arising from $d Q^{\mathbb{C}\dagger}Q^{\mathbb{C}}=-Q^{\mathbb{C}\dagger}d Q^{\mathbb{C}}.$  Therefore \Eq{wzw1C} is invariant under charge conjugation.\\

Under the action of time-reversal transformation, the stiffness term transforms as

	\begin{align*}
		& \frac{1}{8\pi}   \int\limits_{\mathcal{M}} d^2x \,{\rm  tr}\left[\p_\mu Q^{\mathbb{C}\dagger}  \partial^\mu Q^{\mathbb{C}}\right]
		\rightarrow  \frac{1}{8\pi}  \int\limits_{\mathcal{M}} d^2x \,{\rm  tr}\left[\p_\mu Q^{\mathbb{C}*}  \partial^\mu Q^{\mathbb{C}T}\right] \\
		=&\frac{1}{8\pi}  \int\limits_{\mathcal{M}} d^2x \,{\rm  tr}\left[ \partial^\mu Q^{\mathbb{C}}\p_\mu Q^{\mathbb{C}\dagger} \right]
		=  \frac{1}{8\pi}   \int\limits_{\mathcal{M}} d^2x \,{\rm  tr}\left[\p_\mu Q^{\mathbb{C}\dagger}  \partial^\mu Q^{\mathbb{C}}\right]
		\end{align*}

\noindent As for the WZW term, note that the $i$ in front becomes $-i$ due to the complex conjugation involved in the time-reversal transformation \footnote{The time-reversal symmetry in Euclidean space-time requires a complex conjugation on the Boltzmann weight. It is important to check whether a term is real or complex before deciding how time-reversal transformation acts.}. Thus the WZW term transforms as

	\be
	&&\frac{2\pi i}{24\pi^2}    \int\limits_{\mathcal{B}}   \,{\rm tr} \Big[   \left( \tilde{Q}^{\mathbb{C}\dagger}   d \tilde{Q}^{\mathbb{C}}  \right)^3    \Big]
	\rightarrow  -\frac{2\pi i}{24\pi^2}    \int\limits_{\mathcal{B}}   \, {\rm tr} \Big[   \left( \tilde{Q}^{\mathbb{C}*}   d \tilde{Q}^{\mathbb{C}T}  \right)^3    \Big]\nn
	&&=  \frac{2\pi i}{24\pi^2}    \int\limits_{\mathcal{B}}   \, {\rm tr} \Big[   \left(d \tilde{Q}^{\mathbb{C}}  \tilde{Q}^{\mathbb{C}\dagger}     \right) \left(d \tilde{Q}^{\mathbb{C}}  \tilde{Q}^{\mathbb{C}\dagger}     \right) \left(d \tilde{Q}^{\mathbb{C}}  \tilde{Q}^{\mathbb{C}\dagger}     \right)    \Big]\nn
	&&=  \frac{2\pi i}{24\pi^2}    \int\limits_{\mathcal{B}}   \,{\rm tr} \Big[   \left( \tilde{Q}^{\mathbb{C}\dagger}   d \tilde{Q}^{\mathbb{C}}  \right)^3    \Big]
	\label{wzwtr}\ee

The disappearance of the minus sign in the second line is because transposing causes an odd number of crossings of  differential 1-forms. The passing to the third line follows from the cyclic invariance of trace.\\

In summary, the non-linear sigma model is invariant under the action of the global emergent symmetries. The same conclusion applies to the real and complex classes nonlinear sigma models in other space-time dimensions. %  in table \ref{tab:WZWTable}, namely, they are invariant under the global transformations listed in table \ref{tab:symmRestore}. 
The details is left in appendix \ref{appendix:fermionInt}.\\

\section{The symmetry anomalies of the nonlinear sigma models}
\label{bosonAnomalies}
\hfill

A necessary condition for the bosonized non-linear sigma model to be equivalent to the massless fermion theory is that the former  has the same symmetry anomalies as the original massless fermion theories. In this section, we will show this is indeed the case.\\

\subsection{Gauging the non-linear sigma models and the 't Hooft anomalies}\label{tHooftWZW}
\hfill

In table \ref{tab:anomalies}  we see that in $(1+1)$-D and $(3+1)$-D, the massless free fermion theories have the 't Hooft anomalies (with respect to the continuous symmetries). In this subsection, we first gauge the non-linear sigma models and then determine their 't Hooft anomalies. 
\\

Again, taking the complex class $(1+1)$-D example, under an infinitesimal $U_+(n) \times U_-(n)$ transformation, $Q^{\mathbb{C}}$ and gauge fields transformed as 
\be
&&Q^{\mathbb{C}} \rightarrow  e^{-i \epsilon_-} Q^{\mathbb{C}} e^{i \epsilon_+}\nn
&&A_\pm \rightarrow  A_\pm + d\epsilon_\pm + i [A_\pm, \epsilon_\pm]
\label{gtr}\ee
\noindent where we let $g_\pm = e^{i\epsilon_\pm}$ in the symmetry transformation. For the stiffness term, the usual minimal coupling guarantees the gauge invariance
\begin{align*}
W_{\rm stiff}[Q^{\mathbb{C}},A_{+},A_{-}] = -\frac{1}{8\pi} \int\limits_{\mathcal{M}} d^2 x \, \tr\left[\left(Q^{\mathbb{C} \dagger} \left( \partial_\mu Q^{\mathbb{C}} - i Q^{\mathbb{C}} A_{+,\mu} + i A_{-,\mu} Q^{\mathbb{C}} \right) \right)^2 \right].
\end{align*}
However, it is less clear how to gauge the WZW term. %, due to its extension nature. 
Here we follow Witten's ``trial-and-error'' method \cite{Witten1983b}, which we shall explain in the following. \\

First, we determine the variation of the WZW term when $Q^{\mathbb{C}}$ undergoes  space-time dependent transformation given by the first line of \Eq{gtr}
\begin{align*}
&\delta \Big[ - \frac{i}{12\pi} \int\limits_{\mathcal{B}} {\rm tr} \left[\left(Q^{\mathbb{C} \dagger} d Q^{\mathbb{C}} \right)^3 \right]\Big] \\
=& \frac{1}{4\pi} \int\limits_{\mathcal{M}}  {\rm tr} \left[d\epsilon_+ \left( Q^{\mathbb{C} \dagger} d Q^{\mathbb{C} } \right) + d\epsilon_- \left(d Q^{\mathbb{C} } Q^{\mathbb{C} \dagger}  \right) \right]
\end{align*}
Here we remark that although writing down the action requires the extended space-time manifold $\mathcal{B}$, the variation of the action can be expressed solely in the space-time manifold $\mathcal{M}$, which is $(1+1)$-D in the example.\\

In an attempt to make the theory gauge invariant, we subtract a term with $d\epsilon_\pm $ replaced by $A_\pm$. Together, the gauge variant part becomes
\begin{align*}
&\delta \Big[ - \frac{i}{12\pi} \int\limits_{\mathcal{B}} {\rm tr} \left[\left(Q^{\mathbb{C} \dagger} d Q^{\mathbb{C}} \right)^3 \right] - \frac{1}{4\pi} \int\limits_{\mathcal{M}} {\rm tr} \left[ A_+ \left( Q^{\mathbb{C} \dagger} d Q^{\mathbb{C} } \right) + A_- \left(d Q^{\mathbb{C} } Q^{\mathbb{C} \dagger}  \right)\right]\Big] \\
=&-\frac{i}{4\pi} \int\limits_{\mathcal{M}} {\rm tr} \left[ A_+ \left( d\epsilon_+ - Q^{\mathbb{C} \dagger} d\epsilon_- Q^{\mathbb{C} } \right) + A_- \left( -d\epsilon_- + Q^{\mathbb{C} } d\epsilon_+ Q^{\mathbb{C} \dagger}   \right)   \right]
\end{align*}
\\

Last, we repeat the previous step by adding another term with $d\epsilon_\pm $ in the above equation replaced by $A_\pm$. After some work we obtain

\begin{align*}
&\delta \Big[  -\frac{i}{12\pi} \int\limits_{\mathcal{B}} {\rm tr} \left[\left(Q^{\mathbb{C} \dagger} d Q^{\mathbb{C}} \right)^3 \right] - \frac{1}{4\pi} \int\limits_{\mathcal{M}} {\rm tr} \left[ A_+ \left( Q^{\mathbb{C} \dagger} d Q^{\mathbb{C} } \right) + A_- \left(d Q^{\mathbb{C} } Q^{\mathbb{C} \dagger}  \right)+ i A_+ Q^{\mathbb{C} \dagger}  A_- Q^{\mathbb{C} } \right]\Big] \\
=&-\frac{i}{4\pi} \int\limits_{\mathcal{M}} {\rm tr} \left[ A_+  d\epsilon_+ - A_- d\epsilon_-   \right]
\end{align*}

Now the gauge variant part contains no $Q^{\mathbb{C}}$ anymore. Hence we cannot find any term to cancel the remaining non-gauge-invariance. This result reproduces Bardeen's result in \Eq{Bardeen1c}. \\

In summary, the gauged WZW model is given by
\begin{align*}
W[Q^{\mathbb{C}},A_+,A_-]=&-\frac{1}{8\pi} \int\limits_{\mathcal{M}} d^2 x \, \tr\left[\left(Q^{\mathbb{C} \dagger} \left( \partial_\mu Q^{\mathbb{C}} - i Q^{\mathbb{C}} A_{+,\mu} + i A_{-,\mu} Q^{\mathbb{C}} \right) \right)^2 \right]\\
-&\frac{i}{12\pi} \int\limits_{\mathcal{B}} {\rm tr} \left[\left(Q^{\mathbb{C} \dagger} d Q^{\mathbb{C}} \right)^3 \right] \\
-& \frac{1}{4\pi} \int\limits_{\mathcal{M}} {\rm tr} \left[ A_+ \left( Q^{\mathbb{C} \dagger} d Q^{\mathbb{C} } \right) + A_- \left(d Q^{\mathbb{C} } Q^{\mathbb{C} \dagger}  \right)+ i A_+ Q^{\mathbb{C} \dagger}  A_- Q^{\mathbb{C} } \right]\Big] .
\end{align*}
Moreover, we have shown that it has the same 't Hooft anomaly for the continuous symmetry as the original massless fermion. In appendix \ref{appendix:anomaliesb}  we summarize the gauged non-linear sigma model in $d=1,2,3$.
\\

\subsection{Discrete symmetry anomalies}\label{dsan}
\hfill

%The principle we use to determine whether there is an anomaly for discrete symmetries is analogous. For example, after determining the anomaly-free part of $U_+(n)\times U_-(n)$ (the diagonal $U(n)$), we can ask whether such gauging will break discrete symmetries. This is achieved by asking whether the diagonal $U(n)$ -invariant mass term breaks the discrete symmetries. \\

In section \ref{TRAnoamly2Cfree}, we saw that massless fermion theory has a time-reversal anomaly for the complex class in $(2+1)$-D.  This anomaly originates from the massive Dirac fermion at time reversal invariant $\v k$ points other than $\v k=(0,0)$ where the mass breaks time-reversal.  We would like to see the same phenomenon in the nonlinear sigma model.\\

In the following, we focus on the complex class in $(2+1)$-D. First, let's focus on the vicinity of $\v k = 0$ (under Wilson's regularization). The bosonized model is given by \Eq{wzw2C}. Following Witten's trial-and-error method discussed in the preceding subsection (see appendix \ref{appendix:anomaliesb} for the detail), we obtain  the following gauged nonlinear sigma model,
\begin{align}
\label{MaingaugedWZW2C}
W[Q^{\mathbb{C}}, A] =&{1\over 2\lambda_3} \int\limits_{\mathcal{M}} d^3 x \, \tr\Big[\Big(\p_\mu Q^{\mathbb{C}}+i [A_\mu,Q^{\mathbb{C}}]\Big)^2\Big]  \\
-&  \frac{ 2 \pi i }{256 \pi^2}   \Big\{   \int\limits_{\mathcal{B}} \tr \Big[\tilde{Q}^{\mathbb{C}}    \,\left( d \tilde{Q}^{\mathbb{C}} \right)^4 \Big] \notag\\
+& 8 \int\limits_{\mathcal{M}} \tr \Big[ i A Q^{\mathbb{C}} (dQ^{\mathbb{C}})^2 - (A Q^{\mathbb{C}})^2 dQ^{\mathbb{C}} \notag\\
-& \frac{i}{3} (A Q^{\mathbb{C}})^3 + i A^3 Q^{\mathbb{C}} -AQ^{\mathbb{C}}F - AFQ^{\mathbb{C}} \Big]
\Big\} \notag
\end{align}
\noindent This action  is invariant under global symmetry transformations where the gauge field and $Q^{\mathbb{C}}$ are transformed according to \Eq{gaugeTrans2C} and table \ref{tab:symmRestore}. This is expected, given the low energy fermion theory near $\v k=0$ respects these symmetries.\\

For the Dirac fermions near $\v k = (\pi,0), (0, \pi)$, and $(\pi,\pi)$, there are time reversal breaking masses, namely,  $M = 2m \, Y\otimes I_n$ for $\v k=(\pi,0), (0,\pi)$ and $M=4m \, Y\otimes I_n$ for $(\pi,\pi)$. The non-linear sigma model describes these massive fermions is again given by
\Eq{MaingaugedWZW2C} except that now $l=n$ or $0$. Due to the signs in front of $q_1\G_1$ and $q_2\G_2$ at $\v k = (\pi,0), (0, \pi)$, and $(\pi,\pi)$ the effective  mass sign for these massive fermions are given by $$\eta_{\v k}:=\text {sign of~} (q_1\G_1)\times\text {sign of~} (q_2\G_2)\times\text {sign of~}(m).$$ Consequently the $Q^{\mathbb{C}}$ associated with the massive fermions obeys  
\be Q^{\mathbb{C}}=\eta_{\v k}I_n.\label{regf}\ee

We can thus use the gauged nonlinear sigma model in appendix \ref{appendix:anomaliesb} to predict the Chern-Simons term due to the massive fermions at $\v k = (\pi,0), (0, \pi)$, and $(\pi,\pi)$ by plug in \Eq{regf}.
For these space-time constant $Q^{\mathbb{C}}$ we can drop all the terms with derivatives on $Q^{\mathbb{C}}$. The remaining can be combined into the Chern-Simons term. Summing the contribution from ${\v k}$ around $(\pi,0)$, $(0,\pi)$, and $(\pi,\pi)$, we get
\begin{align*}
W_{(\pi,0)}+W_{(0,\pi)}+W_{(\pi,\pi)} =&\left( -\frac{1}{2}-\frac{1}{2}+\frac{1}{2} \right) \frac{m}{|m|} \frac{i}{4\pi} \int \tr\left[A \, dA + \frac{2i}{3}A^3 \right] \\
=& -\frac{i}{8\pi}\int \tr\left[A \, dA + \frac{2i}{3}A^3 \right]
\end{align*}
\noindent which agrees with \Eq{panom}.\\

As for other discrete symmetry anomalies, with the input of how $Q^\mathbb{C,R}$ and the gauge field transform under discrete symmetries, it's simple to show that in $(1+1)$-D and $(3+1)$-D, there is no discrete-symmetry-anomaly after gauging the anomaly-free part of the continuous symmetries. In $(2+1)$-D, gauging the continuous symmetry breaks the time-reversal symmetry as discussed in subsection \ref{dsan}.\\

In appendix \ref{appendix:anomaliesb}, we show that all the symmetry anomalies of massless fermions in table \ref{tab:anomalies} are reproduced by the corresponding gauged nonlinear sigma models. This lends strong support to the idea that the nonlinear sigma models are equivalent to the original massless fermion theories.\\

\section{Soliton of the non-linear sigma models and the Wess-Zumino-Witten terms}
\label{solitonAndWZW}
\hfill

In order for the bosonization to hold, somehow the bosonic non-linear sigma model must possess fermion degrees of freedom.  In this section, we show that due to the WZW term, the solitons of the non-linear sigma model are fermions.\\

%The Wess-Zumino-Witten term fundamentally impacts the properties of the non-linear sigma models. For example, in $(1+1)$-D, Witten \cite{Witten1984} showed that, due to the WZW term, the current algebra of the fermion theory can be reproduced by the bosonic model. Also, as we saw in section \ref{tHooftWZW}, upon gauging the WZW term records the information of the `t Hooft anomaly. In this section we show another important effect of the WZW term, namely, it alters the exchange statistics of the soliton in the non-linear sigma model. 
%\\

\subsection{Soliton classification}\label{solitonClass}
\hfill

Soliton is a spatial texture of the ``order parameter'' ($Q^{\mathbb{C,R}}$). Such texture represents a non-trivial mapping from the spatial space to the mass manifold, i.e., the space where the order parameter lives.  In $d$ spatial dimension,  solitons are classified by the $d$-th homotopy group of the mass manifold, namely,
\begin{align*}
	\pi_d \left( \text{mass manifold} \right).
\end{align*}
\noindent In appendix \ref{appendix:massManifold}, we list the relevant homotopy groups. Since exchange  statistics only make sense for spatial dimension greater than one,  in the following we shall focus on $d\ge 2$. For the nonlinear sigma models considered in section \ref{NLSigma2D3D}, when $n$ is sufficiently large so that there is a WZW term, the soliton classifications are $\mathbb{Z}$ for the complex classes, and are $\mathbb{Z}_2$ for the real classes, namely, 

\begin{align*}
	&\pi_2 \Big( \frac{U(n)}{U(n/2) \times U(n/2)} \Big) = \mathbb{Z} &\text{  for } n\geq 4 \\
	&\pi_2 \Big( \frac{O(n)}{O(n/2) \times O(n/2)} \Big) = \mathbb{Z}_2 &\text{  for } n\geq 6 \\
	&\pi_3 \Big( U(n) \Big) = \mathbb{Z} &\text{  for } n\geq 3 \\
	&\pi_3 \Big( \frac{U(n)}{O(n)} \Big) = \mathbb{Z}_2 &\text{  for } n\geq 5 \\
\end{align*} 
\noindent This means that for the complex classes, we can define a topological quantum number, namely, the ``soliton charge'' $Q_{\rm sol}$. When we fuse two solitons of different charges, $Q_{\rm sol}$ adds; for the real classes, on the other hand, this soliton charge is defined $mod~ 2$  so that two solitons with unit soliton charges can fuse into zero soliton charge. 
\\

\subsection{Soliton charge and the conserved $U(1)$ charge $Q$}
\label{solitonCharge}
\hfill

For the complex classes, it is natural to ask what is the relation between the soliton charge $Q_{\rm sol}$ and the conserved charge $Q$. The conserved charge $Q$ is associated with a global U(1) symmetry. In $(3+1)$-D such U(1) symmetry belongs to a diagonal subgroup of  $U_+(n) \times U_-(n)$. As shown in table \ref{tab:anomalies}, it is anomaly-free. For $(2+1)$-D the $U(1)$ symmetry is a subgroup of the global symmetry group $U(n)$, which is also anomaly-free according to table \ref{tab:anomalies}. 
\\

In appendix \ref{appendix:anomaliesb} we present the gauged non-linear sigma model. In particular, by focusing on the term linear in the gauge field (associated with the anomaly-free $U(1)$ subgroup) derived from the WZW term, we can extract the $U(1)$ current. The answer is \footnote{The same result can be derived by fermion integration.}

\begin{align}
\label{solitonCurrent}
	&\text{$(2+1)$-D}:~ J^\mu = -\frac{i}{16 \pi} \epsilon^{\mu\nu\rho}\tr\left[Q^\mathbb{C} \partial_\nu Q^\mathbb{C} \partial_\rho Q^\mathbb{C}  \right] \\
	&\text{$(3+1)$-D}:~J^\mu = -\frac{1}{24 \pi^2} \epsilon^{\mu\nu\rho\sigma}\tr\left[\left( Q^{\mathbb{C}\dagger} \partial_\nu Q^\mathbb{C}\right) \left( Q^{\mathbb{C}\dagger} \partial_\rho Q^\mathbb{C}\right)\left( Q^{\mathbb{C}\dagger} \partial_\sigma Q^\mathbb{C}\right)  \right]. \notag
\end{align}

Thus the $U(1)$ charge given by

\begin{align*}
	&\text{$(2+1)$-D}:~Q =  -\frac{i}{16 \pi} \int d^2 {\v x} \, \epsilon^{ij}\tr\left[Q^\mathbb{C} \partial_i Q^\mathbb{C} \partial_j Q^\mathbb{C}  \right] \\
	&\text{$(3+1)$-D}:~Q = -\frac{1}{24 \pi^2} \int d^3 {\v x} \, \epsilon^{ijk}\tr\left[\left( Q^{\mathbb{C}\dagger} \partial_i Q^\mathbb{C}\right) \left( Q^{\mathbb{C}\dagger} \partial_j Q^\mathbb{C}\right)\left( Q^{\mathbb{C}\dagger} \partial_k Q^\mathbb{C}\right)  \right].\\
\end{align*}

\noindent These are, in fact, exactly the same expression as the topological invariant corresponding to $\pi_2(\frac{U(n)}{U(n/2) \times U(n/2)})=\mathbb{Z}$ in $(2+1)$-D and $\pi_3(U(n))=\mathbb{Z}$ in $(3+1)$-D (see appendix \ref{appendix:massManifold} for the details). Thus, for both cases
\begin{align}
	\label{solitonU(1)charge}
Q=Q_{\rm sol}.
\end{align}

\subsection{Statistics of soliton}
\hfill

One way to derive the statistics of soliton is to calculate the topological spin by comparing Berry's phase difference between the following two processes. In the first process, we have a static soliton. In the second process, the spatial soliton configuration is adiabatically rotated by $2\pi$ in time. Following Witten \cite{Witten1983a}, we show in appendix \ref{appendix:solitonStat} that such Berry's phase difference is $e^{-i k \pi}$, where $k$ is the level of the WZW term (see appendix \ref{appendix:solitonStat} for the details).  Since all nonlinear sigma models in section \ref{NLSigma2D3D} have $k=1$ WZW term, their solitons are fermion.\\

%Combining \Eq{solitonU(1)charge} and Fermi statistics of solitons motivates us to conjecture that the soliton in the complex class is the original fermion. Under this conjecture, the quantum disordered phases of the nonlinear sigma models are due to the proliferating the fermionic soliton.

\section{A summary of part I}
\hfill

So far, we have established the fact that the fermion and boson theories have the same global symmetries and anomalies. In addition, we have shown that the solitons of the bosonic theories are fermions.  All these support the equivalence between the fermion and boson theories. Now we present a brief summary of part I.\\

We begin in section \ref{theidea} by presenting the essential idea underlying the present work. Prior to performing the fermion integration, we first identify the emergent symmetries in section \ref{ES}, and the mass manifolds in section \ref{massManifold1d}. For a given massless fermion theory, the mass manifold is the topological space formed by all mass terms that can fully gap out the fermions. We then work out the anomalies with respect to the emergent symmetries in section \ref{fermionAnomalies}. Afterward, we introduce mass terms at the expense of breaking the emergent symmetries in section \ref{symmBreaking} and fluctuate the mass terms {\it smoothly} to regain the emergent symmetries in section \ref{restoreEmergent}. As discussed in section \ref{condbos}, the smoothness of the mass fluctuations is to ensure that the original fermions remain gapped, hence can be integrated out to yield non-linear sigma models in section \ref{NLSM} and section \ref{NLSigma2D3D}.\footnote{The procedure can be easily applied to higher dimensions, though we shall not pursue it in the present paper.}. The level-1 WZW term  resulting from the fermion integration is checked against the prediction of homotopy groups in the appendix, which is referred to in sections \ref{NLSM} and \ref{NLSigma2D3D}. In section \ref{IntFermion}, we present local interacting fermion theories that have duality-like relationships with the bosonized non-linear sigma models. In section  \ref{symmNLSigma}, we analyze the symmetries of the non-linear sigma models. A comparison with the results obtained in section \ref{ES} leads to the conclusion that the fermion and boson theories have the same symmetry. Using the method of reference \cite{Witten1983b} we determine the anomalies of the non-linear sigma models in section \ref{bosonAnomalies}. A comparison with the results obtained in section \ref{fermionAnomalies} leads to the conclusion that the fermion and boson theories have the same anomalies. Finally, in section \ref{solitonAndWZW}, we show the bosonized theories have fermionic degrees of freedom, namely the solitons of the non-linear sigma models.\\

\part{Applications}\label{appi}

\section{The SU(2) gauge theory of the $\pi$-flux phase of the half-filled Hubbard model}
\label{Mott}

\subsection{The ``spinon'' representation of the spin operator}
\hfill

The paradigmatic model describing a Mott insulator is the Hubbard model in the large $U$ limit. At half-filling, every site is occupied by one electron. Below the Mott-Hubbard gap, the active degrees of freedom are those of spins. Through Anderson's super-exchange \cite{Anderson1950}, the dynamics of the spins is governed by the anti-ferromagnetic Heisenberg interaction 
\begin{align*}
\hat{H} =& \sum_{\Avg{ij }} J_{ij} \vec{S}_i \cdot \vec{S}_j. 
\end{align*}
In the ``spinon'' treatment \cite{Coleman1984,Lee2006} one decomposes a spin-$1/2$ operator into auxiliary fermion (spinon) operators
\be
S_i^a =\frac{1}{2} f_{i\a}^\dagger \sigma^a_{\a\b} f_{i\b}, 
\label{stspn}
\ee
\noindent and supplement it with the single occupation constraints \be &&f_{i\ua}^\dagger f_{i\ua}+f_{i\da}^\dagger f_{i\da}=1\nn
&&f^\dagger_{i\ua}f^\dagger_{i\da}=0\nn&&f_{i\da}f_{i\ua}=0.
\label{constraint}\ee
In the following we shall refer to the above constraints as the ``Mott constraint''.
The decomposition in \Eq{stspn}, where one separates the physical spin degrees of freedom into the auxiliary ``spinon'' degrees of freedom, is an example of the so-called ``slave particle'' approach. %To restore the original spin Hilbert space, it is important to respect the constraints \Eq{constraint}. 
\\

In terms of the spinon operators the Heisenberg Hamiltonian read
\begin{align*}
\hat{H} =&\frac{1}{4} \sum_{\Avg{ij }} J_{ij}\left( f_{i\alpha}^\dagger \sigma^a_{\alpha \beta} f_{i\beta} \right) \left(  f_{j\gamma}^\dagger \sigma^a_{\gamma \delta} f_{j\delta} \right) \\
=&\frac{1}{4} \sum_{\Avg{ij }} J_{ij}\left( -f_{i\alpha}^\dagger f_{i\alpha} f_{j\beta}^\dagger f_{j\beta} - 2 f_{i \alpha}^\dagger f_{j \alpha} f_{j \beta}^\dagger f_{i \beta}  \right)  \\
= &-\frac{1}{2} \sum_{\Avg{ij }} J_{ij}\left(\frac{1}{2}f_{i\alpha}^\dagger f_{i\alpha} f_{j\beta}^\dagger f_{j\beta} +f_{i \alpha}^\dagger f_{j \alpha} f_{j \beta}^\dagger f_{i \beta}  \right)  
\end{align*}
Upon Hubbard-Stratonovich transformation, we express
\begin{align}
&\exp\Big\{-\int_0^\b d\t \Big[\sum_i f^\dagger_{i\a}\p_0f_{i\a}+H\Big]\Big\}=\nn
&\int D[U]\exp\Big\{-\int_0^\b d\t \Big[\sum_i\ \psi^\dagger_{i}\p_0 \psi_{i}+\sum_{\Avg{ij}} \frac{3}{8} J_{ij}\Big( - \left(\psi^\dagger_{i}U_{ij} \psi_{j} + h.c.\right) + \frac{1}{2}\Tr \left[U_{ij}^\dagger U_{ij} \right] \Big)\Big]\Big\}.\nn
\end{align}
\noindent where 
\begin{align}
\psi_i=\begin{pmatrix}f_{i\ua}\\f^\dagger_{i\da}\end{pmatrix}, U_{ij}=  
\begin{bmatrix}
\chi_{ij}^*		&	\Delta_{ij}	\\
\Delta_{ij}^*	&	-\chi_{ij}
\end{bmatrix}.\label{mfan}
\end{align}\\

For later convenience, we  rewrite the spinon operator in terms of Majorana fermions 
\begin{align*}
f_{i \alpha} := F_{i ,1 \alpha} + i F_{i, 2 \alpha},
\end{align*}
in terms of which, the  spin operators are represented as
\be
&&S_i^a =\frac{1}{2} F_i^\dagger \Sigma^a F_i,~~{\rm where}\nn
&&\Sigma^a = \left(YX, IY, YZ \right).
\label{sop}
\ee
\noindent In the last line, the first and second Pauli matrices carry  the Majorana and spin indices, respectively. \\

The spin operators in \Eq{sop} are invariant under the following local ``charge-SU(2)''
transformation
$$F_i\ra W_i F_i$$
\noindent where $W_i$ is generated by 
$$T^b =( XY, YI, ZY).$$
In terms of The Majorana fermion operators, the Mott constraint in \Eq{constraint} becomes 
\be
&&f^\dagger_{i \alpha} f_{i \alpha} -1=  F^T_i \left( YI \right) F_i =F^T_i T^2 F_i=0 \nn
&&\epsilon^{\alpha \beta} \left( f_{i \alpha} f_{i \beta} +f^\dagger_{i\beta} f^\dagger_{i \alpha}  \right)=  F^T_i \left( XY \right) F_i =F^T_i T^1 F_i=0 \nn
&&i\epsilon^{\alpha \beta} \left(  f_{i \alpha} f_{i \beta} - f^\dagger_{i\beta} f^\dagger_{i \alpha}  \right)=  F^T_i \left( ZY \right) F_i =F^T_i T^3 F_i=0
\label{constr2}\ee
These constraints are implemented via the Lagrange multipliers in the  path integral
$$
Z=\int D[F] D[U]D[a_0] \exp{\left(-S\right)} 
$$
with
\begin{align}
S=&\int_0^\b d\t \Big\{\sum_i F^T_{i}\p_0 F_{i}+ \sum_{\Avg{ij}} \frac{3}{8} J_{ij} \Big[F_i^T\Big(  Re[\chi_{ij} ] YI + i \, Im[\chi_{ij}] II  + Re[\Delta_{ij}] XY\notag\\&- \, Im[\Delta_{ij}] ZY \Big) F_j 
+|\chi_{ij}|^2 + |\eta_{ij}|^2 \Big] + i\sum_i   a^b_{i0}  \left( F^T_i T^b F_i\right)\Big\}.
\label{spap}
\end{align}
\\

\subsection{The $\pi$-flux phase mean-field theory and the $SU(2)$ gauge fluctuations}\label{su2chargeconf}
\hfill

In treating the path integral, \Eq{spap}, one often starts from a mean-field theory where $U_{ij}$ and $a^b_{i0}$ are assumed to be space-time independent. To see the many possible mean-field ansatzes we refer the readers to, e.g., Ref.\cite{Lee2006}. In the following, we shall focus on the so-called ``$\pi$-flux phase mean-field theory''\cite{Affleck1988} for the nearest neighbor Heisenberg model.\\

\begin{figure}[h]
	\begin{center}
		\includegraphics[scale=0.3]{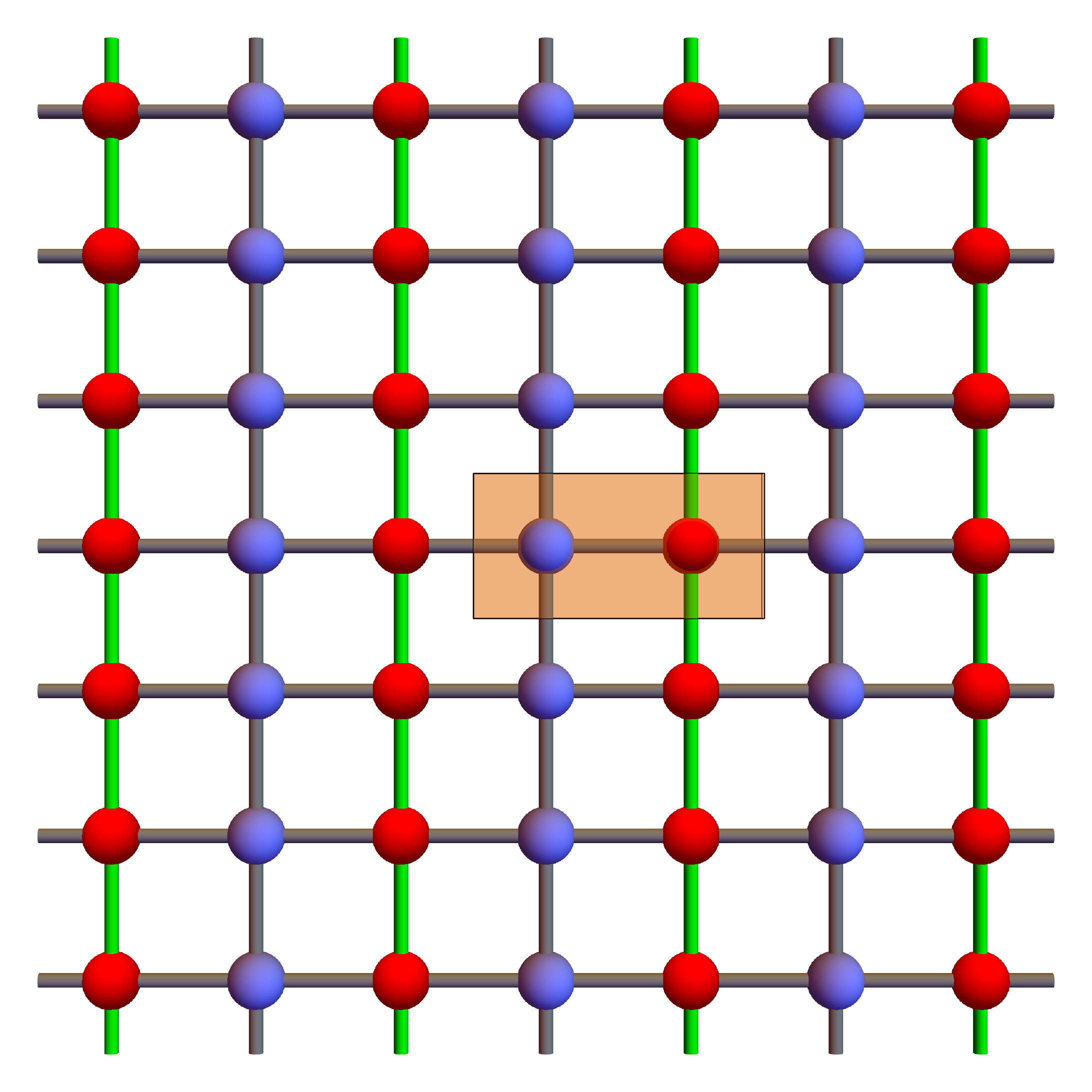}
		\caption{The $\pi$-flux mean-field theory. Here the black bonds represent hopping amplitude $i\chi$ in the positive x- or y-direction and the green bonds represent $-i\chi$. The unit cell is enclosed by the orange rectangle. }
		\label{piflux}
	\end{center}
\end{figure}

The $\pi$-flux mean field theory assumes the following mean-field $\bar{U}_{ij}$ and $\bar{a}^b_{i0}$
\be
&&\bar{\Delta}_{ij}=0,~~\bar{a}^b_{i0}=0,\nn
&&\bar{\chi}_{i,i+\hat{x}} = i \chi, \nn
&&\bar{\chi}_{i,j+\hat{y}} = i (-1)^{i_x} \chi
\label{anchi}
\ee
\noindent where $\chi$ is a real parameter (see \Fig{piflux}).
\noindent This leads to the 
following fermion mean-field Hamiltonian,
\begin{align*}
\hat{H}_{\rm MF}= -\frac{3}{4} J \sum_{i} \Big\{ i \, \chi \left[F_{i+\hat{x}}^T \left(II\right) F_{i }\right]+ i \, (-1)^{i_x} \chi \left[F_{i+\hat{y}}^T \left(II\right) F_{i }\right]+ h.c.  \Big\}
\end{align*}

\noindent Because the Pauli matrices are identity in both the Majorana and spin spaces, this mean-field Hamiltonian enjoys both global spin- and charge-SU(2) symmetries generated by 
\be
&&\text{Spin-SU(2) generators:}~~\Sigma^a = \left(YX, IY, YZ \right)\nn
&&\text{Charge-SU(2) generators:}~~T^a = \left(XY, YI, ZY \right).
\label{cssu2}\ee
\\

Using the eigenvalues $\pm 1$ of the ``sub-lattice Pauli matrix'' $Z$ to label the blue and red sub-lattices in \Fig{piflux},  and performing Fourier transform we obtain the following momentum-space mean-field Hamiltonian

\be
&&\hat{H}_{\rm MF} = - \frac{3}{4}J \chi \, \sum_{\bf k} F_{\bf-k}^T \, \left[ II \otimes 
\begin{bmatrix}
i \left( e^{i k_2} - e^{-i k_2} \right) 	&	-i + i e^{2i k_1} 	\\
i - i e^{-2ik_1}								&	-i\left( e^{ik_2} - e^{-ik_2} \right)
\end{bmatrix}\right] F_{\bf k}\nn
&&=- \frac{3}{4}J \chi \, \sum_{\bf k} F_{\bf-k}^T \, \left[ II \otimes \left( - \sin 2k_1 \, X + \left( 1-\cos 2k_1 \right)Y -2 \sin k_2 \, Z \right)\right] F_{\bf k}.\nn
\label{mjmf}
\ee
In the above equation the tensor product of Pauli matrices are ordered according to
$$\text{Majorana}\otimes \text{spin}\otimes \text{sub-lattice}.$$ In \Eq{mjmf} the (halfed) Brillouin zone is $$-\pi/2\le k_1< \pi/2, ~~-\pi\le k_2<\pi$$ and the Dirac nodes are situated at ${\bf k}_0= (0, 0)$ and $(0, \pi)$, which are referred to as two ``valleys'' in the following.\\

Expand $\bf k = \v k_0 + \v q$ around these two Dirac nodes, and Fourier transform (w.r.t. $\v q$) back to the real space, we obtain the following low energy mean-field Hamiltonian

\begin{align*}
\hat{H}_{\rm MF} = \int d^2 x  \, \tilde{F}^T \left( -i \Gamma_i \partial_i  \right) \tilde{F},
\end{align*}

where

\be
&&\Gamma_1 = IIXI \nn
&&\Gamma_2 = IIZZ.
\label{gmmt}
\ee

The tensor product of four Pauli matrices in \Eq{gmmt} are arranged according to  $$\text{Majorana}\otimes \text{spin}\otimes \text{sub-lattice}\otimes \text{valley}.$$ Including the sub-lattice and valley Pauli matrices the generators of the charge and spin SU(2) transformations are given by 
\be
&&\text{Spin-SU(2) generators:}~~\Sigma^a = \left(YXII, IYII, YZII \right)\nn
&&\text{Charge-SU(2) generators:}~~T^a = \left(XYII, YIII, ZYII \right).
\label{cssu3}\ee
\\

Because the local charge-SU(2) gauge degrees of freedom is a redundancy in the original half-filled Mott insulator, we expect the field theory in \Eq{spap} to have the local charge-SU(2) symmetry. This motivates one to think the low energy theory, including fluctuations in $U_{ij}$ and $a_{i0}^b$, is a charge-SU(2) gauge theory with 
$$U_{ij}=\bar{U}_{ij}e^{i a_{ij}}$$
where $a_{ij}=a_{ij}^b T^b$ is the spatial component of the charge-SU(2) gauge field. According to Ref.\cite{Lee2006,Dagotto1988}, because the mean-field $\bar{U}_{ij}$ commutes with the {\it global} charge-SU(2) transformations, the low theory is a charge-SU(2) gauge theory, with  $a_0$ and $a_{ij}$ playing the roles of the time and spatial components of the gauge field, respectively.  \\

The partition function of the charge-SU(2) gauge theory reads
\be
&&\mathcal{Z}=\int D[\tilde{F}] D[a_\mu] e^{-S[\tilde{F},a_\mu]}\nn
&&S=\int d^3 x \Big\{\tilde{F}^T\left[\left( \partial_0 + i a_0^a T^a \right)-i \sum_{i=1}^2 \Gamma^i(\partial_i + i a_i^a T^a)\right]\tilde{F}+ {1\over 2g} f_{\mu\nu}^2\Big\}.
\label{SU2Higgs}
\ee
In \Eq{SU2Higgs} the ${1\over 2g} f_{\mu\nu}^2$ is generated by integrating out the higher energy fermion degrees of freedom.
\noindent The theory in \Eq{SU2Higgs} describes the $n=8$ real class fermion theory coupled to a dynamic charge-SU(2) gauge field. \\

According to the bosonization in section \ref{bosonization2R}, the bosonized theory is a gauged $\frac{O(8)}{O(4) \times O(4)}$ nonlinear sigma model with the $k=1$ WZW term \footnote{Note that although for $n=8$, the homotopy group are not yet stabilized, fermion integration still gives a WZW term. When $\mathcal{B}$ is a closed manifold, and after division by $2\pi i$, the WZW term is the topological invariant of one of the $\mathbb{Z}$ factor of the $\pi_4$.}. Here the charge-SU(2) subgroup  of the fermion (emergent) global symmetry group $O(8)$ is gauged. \\

In the following let's {\it assume} that the effect of the SU(2) gauge field is to cause confinement (note, however, we are not implying the deconfined phase does not exist)
\footnote{
	Note that unlike the compact U(1) gauge field, here the confinement can be not caused by the proliferation of monopoles. This is based on the following homotopy argument. The $SU(2)$ gauge configurations on the space-time surface $S^2$ surrounding the location of the monopole are classified by the mapping  classes of $S^2\ra BSU(2)$, where $BSU(2)$ is the classifying space of $SU(2)$. Using the following identity in algebraic topology,
	\begin{align*}
	[S^2, BSU(2) ]_* = [\Sigma S^1, BSU(2)]_* = [S^1,SU(2)]_* = \pi_1(SU(2))=0,
	\end{align*}
	it follows that there is no topologically non-trivial gauge field configuration on $S^2$, hence there is no monopole. Here $\Sigma$ denotes ``reduced suspension'', and $[X_1,X_2]_*$ is the homotopy class of base-point-preserving maps $X_1\rightarrow X_2$. Physically speaking, assuming the  $SU(2)$ monopole exists, we can take the northern and southern hemispheres as the patches to define the gauge connection so that on each patch, the gauge field configuration is non-singular. On the equator, $S^1$, where the two patches overlap, a gauge transformation must relate the gauge fields originated from the two patches. At each point of $S^1$ the gauge transformation is an element in $SU(2)$. 
Therefore the monopole classification is given by the homotopy class of gauge transformation on the $S^1$, i.e., $\pi_1(SU(2))$.
}. 
Under such condition, the fermion-antifermion pair oder parameter (analogous to mesons in QCD) must be a charge-SU(2) singlet. 
Since $Q^\mathbb{R}$ is precisely the ``meson'' field, it follows that in the charge-SU(2) confined phase the finite energy $Q^\mathbb{R}$ are restricted to a sub-manifold of $\frac{O(8)}{O(4)\times O(4)}$ which are invariant under the charge-SU(2) transformation \footnote{In addition to restricting $Q^\mathbb{R}$ to be invariant under charge-SU(2) transformations, the charge-SU(2) gauge fluctuations can also generate four-fermion interactions, the effects of which are not studied in the current work.}\footnote{Our result is analogous to Witten's non-linear sigma model description of QCD in the color SU(3) confined phase\cite{Witten1983a}.}. This sub-manifold is the $S^4$ spanned by the following $5$ mutually anti-commuting masses,
\be
&&S^4=\Big\{\sum_{i=1}^5 n_i M_i; ~~\sum_{i=1}^5 n_i^2=1\Big\}, ~~{\rm where}\nn
&&M_i = YXZX, IYZX, YZZX, IIYI, IIZY.
\label{NV}
\ee
\\

In order to match the gamma matrices and mass matrices convention in table \ref{tab:emergentSymm} and \ref{tab:massManifold} (based on which the non-linear sigma models in subsection \ref{NLSigma2D3D} and appendix \ref{appendix:anomaliesf} and \ref{appendix:anomaliesb} are derived), we will make the following change the basis. We first exchange the order of the third and the fourth (i.e., sub-lattice and valley) Pauli matrices, followed by the orthogonal transformation,
\begin{align*}
II\otimes \begin{bmatrix} I & 0 \\ 0 & X \end{bmatrix}
\end{align*}
In the new basis the gamma matrices and the mass terms become
\be
&&\tilde{\Gamma}_1 	=IIIX \nn
&&\tilde{\Gamma}_2 	=IIIZ \nn
&&\tilde{M_i}			=YXYY, IYYY, YZYY, IIXY, IIZY
\label{fmss}\ee
These are consistent with the matrices shown in table table \ref{tab:emergentSymm} and table \ref{tab:massManifold}, except a trivial exchange of the first and the last Pauli matrices. In this basis, the order parameter $Q^\mathbb{R}$ is defined by $\tilde{M} =m\, Q^\mathbb{R} \otimes Y$.

\subsection{Antiferromagnet, Valence bond solid, and the ``deconfined'' quantum crtical point}
\hfill

For the mass manifold in \Eq{NV}, we expect the non-linear sigma model to have a WZW term because $\pi_4(S^4)=\mathbb{Z}$.
 Substituting 
  \be
&&Q^\mathbb{R}=n_i N_i~\text{where}\nn
&&N_i =\left(YXY, IYY, YZY, IIX, IIZ \right) 
\label{fmsg}\ee
\noindent into the non-linear sigma model given by \Eq{wzw2R} in subsection \ref{bosonization2R} 
we obtain 
\be
W[\hat{n}] ={2\over\lambda_3}\int\limits_{\mathcal{M}} d^3 x \, \left( \partial_\mu n_i \right)^2 - \frac{2\pi i }{64 \pi^2} \int\limits_{\mathcal{B}} \epsilon^{ijklm} \tilde{n}_i \, d\tilde{n}_j  \, d\tilde{n}_k \, d\tilde{n}_l \, d\tilde{n}_m.
\label{o5}\ee
This model has O(5) global symmetry generated by the pair-wise product of the matrices in $\tilde{M_i}$, which are also the generators of $O(8)$ that commutes with the charge-SU(2) generators. Hence \Eq{o5} is often referred to as the ``O(5)'' non-linear sigma model in the literature \cite{Senthil2003,Tanaka2005,Senthil2006,Wang2017} (a recent related work can be found in \cite{Zou2021}).
\\

Now we address the physical meaning of the five masses given in \Eq{NV} (or equivalently the physical meaning of  $\tilde{M}_i$ in \Eq{fmss}). 
The first three of the masses in \Eq{NV} correspond to the N\'eel order parameters, while the last two to the valence bond solid (VBS) orders. To see this, we first note that the first three masses rotate into each other under spin-SU(2),

\begin{align*}
\Sigma^a = \left( YXII, IYII, YZII \right)
\end{align*}

\noindent while the last two are invariant.\\

\begin{figure}[h]
	\begin{center}
		\includegraphics[scale=0.5]{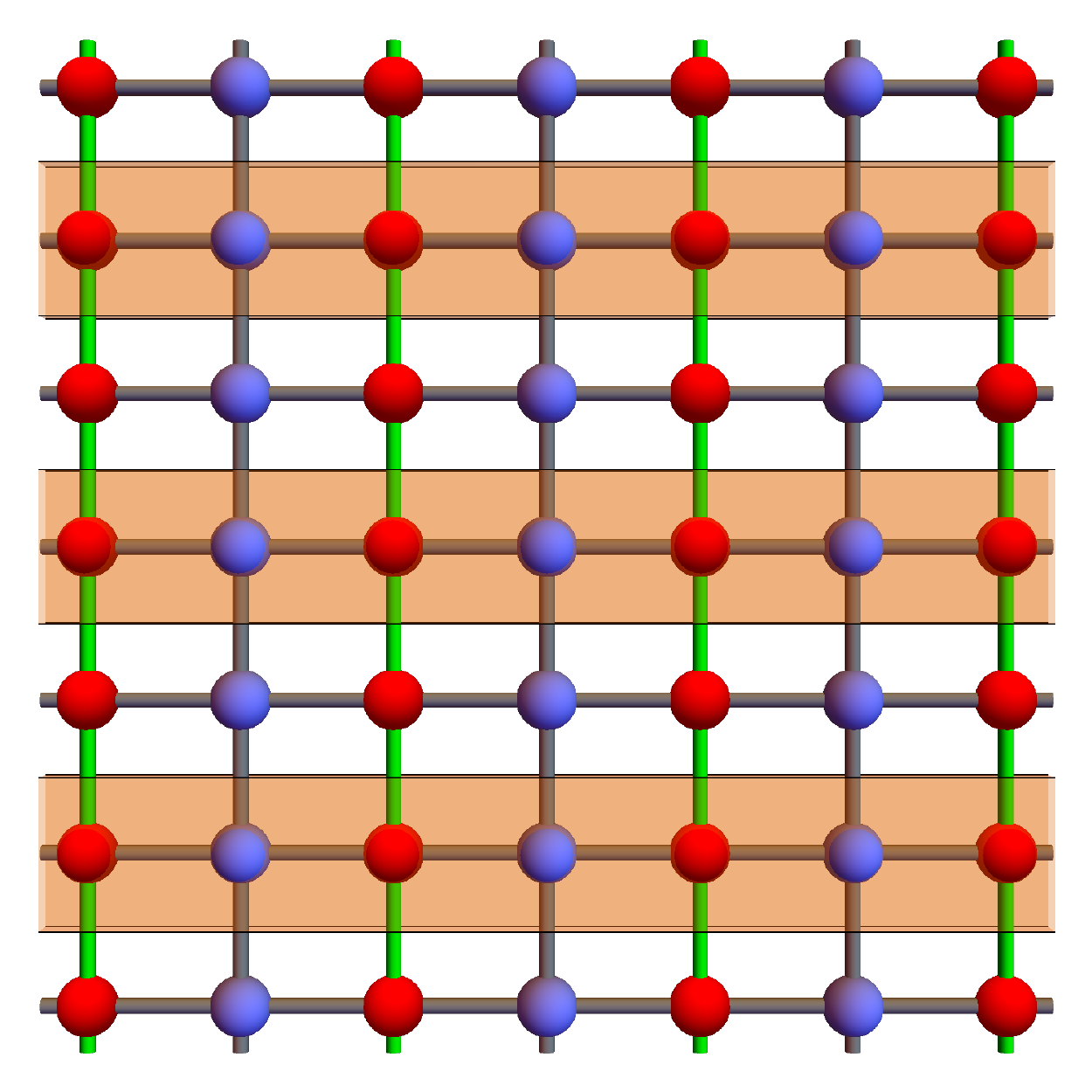}
		\caption{Translation by one lattice constant in the x-direction compounded with the gauge transformation which multiplies the fermion operators on sites in the orange rows by -1 leaves the mean-field Hamiltonian invariant.  }
		\label{PSG}
	\end{center}
\end{figure}

We can also deduce the effect of translation by one lattice constant on these mass terms. In writing down the mean-field Hamiltonian we have chosen a particular charge-SU(2) gauge that explicitly breaks the symmetry associated with x-translation by one-lattice spacing. However, this is an artifact of gauge choice. The compounded transformation where the x-translation is followed by the  gauge transformation  which multiplies the fermion operator located on the orange rows in \Fig{PSG} by $-1$
\be
&&\left( F_{\mathcal{I},1}, F_{\mathcal{I},2} \right) \xrightarrow{\hat{T}_{\hat{x}}} (-1)^{\mathcal{I}_y} \times
\left( F_{\mathcal{I},2} , F_{\mathcal{I}+\hat{x},1} \right) \nn
&&\left( F_{\mathcal{I},1}, F_{\mathcal{I},2} \right) \xrightarrow{\hat{T}_{\hat{y}}}  \left( F_{\mathcal{I}+\hat{y},1} , F_{\mathcal{I}+\hat{y},2} \right), 
\label{tsg1}
\ee
leaves the mean-field ansatz invariant. This is an example of ``projective transformation''. In \Eq{tsg1} $\mathcal{I}$ label the unit cell in \Fig{piflux} 
, and we have omitted the Majorana and spin indices because they are unaffected by the translation.\\

In the following, we derive the effects of the ``projective translation'' on the fermion operator $\tilde{F}$ which is related to $F$ via

\begin{align*}
	F_{\mathcal{I},l}=\sum\limits_{\text{small } \v q} \left( \tilde{F}_{q,(l,1)} e^{i {\v q} \cdot \mathcal{I}} +\tilde{F}_{q,(l,2)} e^{i ((0,\pi)+{\v q}) \cdot \mathcal{I}}\right)
\end{align*}
\noindent where the $(l,v)$  are the indices for sub-lattice and valleys respectively. Doing the inverse Fourier transform, the above projective translation transforms $\tilde{F}$ according to

$$\tilde{F}\ra T_{\hat{x},\hat{y}}\tilde{F}$$ 
where
\be
&&T_{\hat{x}} = IIXX\nn
&&T_{\hat{y}} = IIIZ.
\label{tro}\ee
Here we have put back the Majorana and spin  (i.e., the first two) Pauli matrices.\\

Under $T_{\hat{x},\hat{y}}$ the mean-field Hamiltonian is invariant, but the first three mass terms change sign under $\hat{T}_{\hat{x}}$ and  $\hat{T}_{\hat{y}}$ (as should the N\'eel order parameter) while the remaining two masses each breaks $\hat{T}_{\hat{x}}$ or $\hat{T}_{\hat{y}}$. These are the expected transformation properties of the VBS order parameters. \\

In appendix \ref{appendix:chargeSU(2)Decoupling} we show that the order parameters in \Eq{fmsg} completely decouple from the charge-SU(2) gauge field.  Thus even in the presence of such gauge field the non-linear sigma model preserves the form in \Eq{o5}. 
Before moving on, there is one additional thing worth mentioning, namely,  $$\pi_2(S^4)=0.$$ Hence there is no soliton in the order parameter associated with \Eq{o5}. 
\\

In summary, we have found that after the charge-SU(2) confinement  \Eq{o5} describes the critical point between the AFM and VBS phases the so-called ``deconfined quantum critical point''\cite{Senthil2003,Tanaka2005,Sandvik2007,Xu2008}. It is important to note that in the treatment so far, we have assumed that the sole effect of the charge-SU(2) confinement is to restrict $Q^\mathbb{R}$ to the appropriate submanifold of $\frac{O(8)}{O(4)\times O(4)}$. However, the SU(2) gauge field fluctuations can also induce four-fermion interactions, the effects of which are not studied in the current work. %  These interactions can favor the N\'eel order once spontaneous symmetry breaking sets in. Such order is well-known to exist in the half-filled Hubbard model. 
It is very satisfying that aside from the N\'eel order parameter, \Eq{fmsg} captures the best-known VBS order parameter once the N\'eel is destabilized by quantum fluctuation\cite{Haldane1988,Read1989,Figueirido1990,Jiang2012,Gong2014}.  \\

\section{The critical spin liquid of ``bipartite Mott insulators'' in  $D=1+1,2+1$ and $3+1$.}
\label{Mott_1D_2D}
\hfill

The idea explained in the preceding section can be generalized to the insulating phase of ``bipartite Mott insulators''. \\

A bipartite Mott insulator is a Mott insulator whose lattice consists of two sub-lattices, and hoppings only occur between different sub-lattices. The nearest-neighbor spin-$1/2$ antiferromagnetic Heisenberg model in one spatial dimension  describes the dynamics of spin degrees of freedom in a one-dimensional bipartite Mott insulator. It realizes the $SU(2)_1$ WZW non-linear sigma model, where the emergent symmetries are realized in a non-onsite (e.g., translation) fashion. This model serves as a paradigm of, e.g., quantum number fractionalization, and has profoundly influenced theoretical physics. It is natural to ask what is the generalization of this non-linear sigma model in the Mott insulating phase of higher dimensions. In the present section, we answer this question.\\

In a Mott insulating phase, the low energy degrees of freedom are the spins.  Since the spin operators are invariant under the charge-SU(2) transformation, there are lots of choices in fractionalizing the spin into spinons. Different choices are related by the spinon charge-SU(2) gauge transformation. The spin-spin interaction is generated by Anderson's super-exchange,  the spinon mean-field theory amounts to choosing a spinon tight-binding model which reproduces the spin-spin interaction after super-exchange. Since the spin-spin interaction is independent of which charge-SU(2) gauge we choose, we shall choose the gauge so that the hoppings are purely imaginary in the following. The reason for doing so is because in such a gauge, the mean-field spinon Hamiltonian is charge-SU(2) invariant. This gauge choice exists when the Mott insulator is bipartite.\\

In section \ref{su2chargeconf}, we saw that the Mott insulating condition is imposed by the constraint that the order parameter  $Q^{\mathbb{R}}$ is a charge-SU(2) singlet.  %This constraint originates from the fact that in a Mott insulator, the low energy states are annihilated by $\sum_\s c^\dagger_{i\s}c_{i\s}-1$, $c^\dagger_{i\ua}c^\dagger_{i\da}$ and $c_{i\da}c_{i\ua}$. 
In this section we show that imposing such constraints allows us to derive the spin effective theory in  bipartite Mott insulators in spatial dimensions 1,2 and 3 \footnote{Although we will not pursue it in the present paper, the discussion in the following can be generalized to the cases with larger flavor number or in higher dimensions.}.\\
%We first show that by implementing the Mott constraint this way, we obtain the correct effective theory for spin degrees of freedom in the case of $n=2$ in 1+1 D. Afterward we apply this procedure to the strongly correlated single-layer graphene, which potentially can be regarded as the $\nu=2$ filling for the twisted bilayer graphene \cite{Po2019}.

\subsection{(1+1)-D}
\hfill

\subsubsection{The analog of the $\pi$-flux phase}
\hfill

\begin{figure}
	\begin{subfigure}{0.5\textwidth}
		\includegraphics[width=\linewidth]{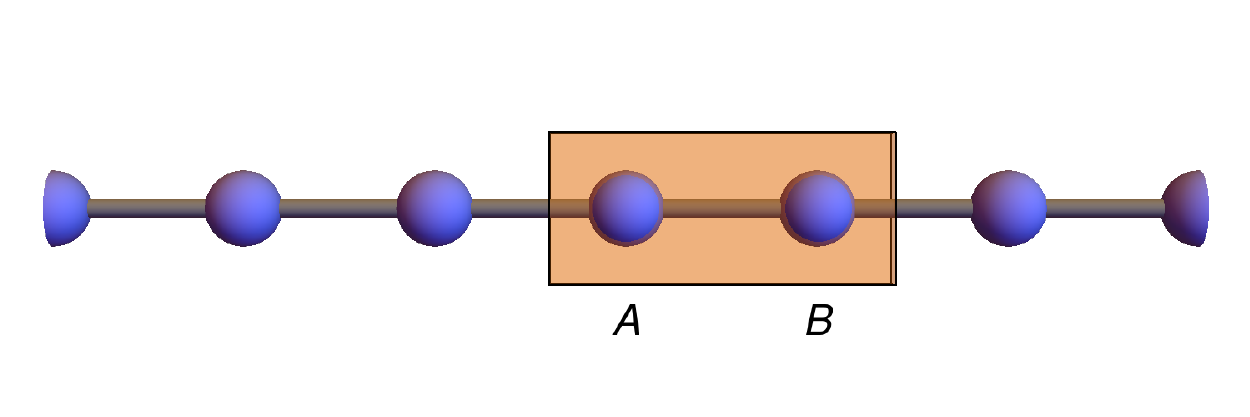}
		\caption{Real hopping} \label{1DTB_fig:1a}
	\end{subfigure}%
	\hspace*{0.3in}    % maximize separation between the subfigures
	\begin{subfigure}{0.5\textwidth}
		\includegraphics[width=\linewidth]{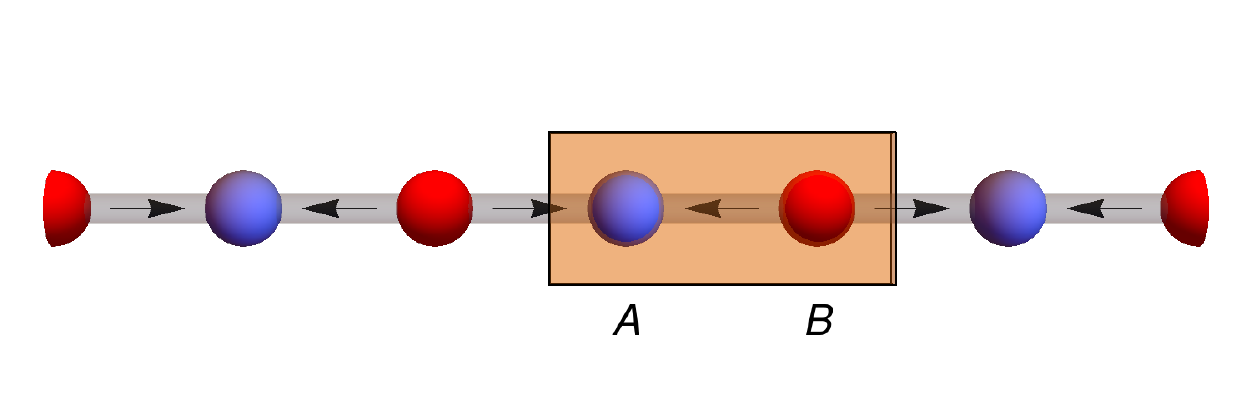}
		\caption{Imaginary hopping} \label{1DTB_fig:1b}
	\end{subfigure}%
	\caption{
		(a) The usual 1D nearest neighbor tight-binding with real hopping. (b) Upon the gauge transformation $(c^A_j ,c^B_{ j }) \rightarrow ( c^A_j, i \,c^B_{ j })$, hoppings become purely imaginary with alternating sign. The arrows point in the direction where the hopping is $+it$. The hopping Hamiltonian in panel (b) is charge-SU(2) invariant.
	}
	\label{1DTB}
\end{figure}

For the nearest neighbor tight-binding model with real hopping in 1D, one can break the lattice into $A,B$ sub-lattice and do the transformation $(c^A_j ,c^B_{ j }) \rightarrow ( c^A_j, i \,c^B_{ j })$ to make the hopping purely imaginary (see figure \ref{1DTB}). This leads to the lattice model
\begin{align*}
	\hat{H} = &
		t \sum_k c_k^\dagger\left[  I \otimes \begin{pmatrix} 
			0	&	i + i e^{-ik} 	\\
			-i-ie^{+ik}	&	0		\\
		\end{pmatrix} \right] c_k \\
	=& -t  \sum_k c_k^\dagger \left[ -(\sin k) IX +(1+\cos k) IY  \right] c_k
\end{align*}
\noindent Here identity matrix $I$ part acts on the spin. After linearizing around $k_F=\pi$,  the low energy effective theory in Majorana fermion basis reads
$$
H=\int dx ~\chi^T(x)~ \left[ -i \G_1 \partial_1 \right]~\chi(x)$$
where  
\begin{align}
	\Gamma_1 = I I X. 
\end{align}
To comply with the gamma matrix notation in table \ref{tab:emergentSymm}, we perform a basis transformation $\chi \rightarrow e^{-i\frac{\pi}{4}IIY} \chi$, so that the gamma matrix becomes
\begin{align}
\label{ggmm1}
\Gamma_1 = I I Z. 
\end{align}
Here the tensor product of Pauli matrices are arranged according to   
$$\text{Majorana}\otimes \text{spin}\otimes \text{sub-lattice}.$$ In the presence of Hubbard $U$, there is the global charge-SU(2)
symmetry at half-filling. In the low energy theory, the charge-SU(2) transformation is generated by
\begin{align*}
	\text{Charge-$SU(2)$ generators}: T^a = (XYI, YII, ZYI)
\end{align*}
On the other hand, the spin-SU(2) transformations are generated by the following charge-SU(2) invariant matrices,
\begin{align*}
	\text{Spin-$SU(2)$ generators}: \Sigma^a = (YXI, IYI, YZI)
\end{align*}

\subsubsection{The charge-SU(2) confinement}
\hfill

Following the discussion in section \ref{NLSM2}, the massless free fermion theory is equivalent to the $O(4)$ level-$1$ WZW model. Gauging the charge-SU(2) symmetry of the sigma model, and integrating over the gauge field, amounts to imposing the Mott insulating constraint. 
Assume the system is in the charge-SU(2) confined phase, only charge-SU(2) singlet order parameters (mass terms) can exist at low energies. These mass terms satisfy
\begin{align*}
	&\{ \Gamma_1 , M \} = 0 \\
	&[T^a , M ] =0 \\
	&M^2 = I_{8}
\end{align*}
The most general mass $M$ satisfying the first two lines has the form
\begin{align}
	\label{mass1H}
	M =	n_0 \, IIY+ n_1 \, YXX+n_2 \, IYX  + n_3  \, YZX
\end{align}
Among the mass terms 
\begin{align*}
	IIY, YXX, IYX, YZX,
\end{align*}
the last three rotate into each other under the action of spin-SU(2) transformations, and the first one is invariant. They corresponds to the dimer and N\'eel order parameters respectively.
The condition that $M^2=I_8$ gives $$\sum_{i=0}^3 n_i^2=1.$$
\\

The non-linear sigma model describing the fluctuations of $\hat{n}$ has a WZW term because $\pi_3(S^3)=\mathbb{Z}$, namely, 
\begin{align}
	W[\hat{n}] = \frac{1}{4\pi} \int\limits_{\mathcal{M}} d^2 x \, \left( \partial_\mu \hat{n} \right)^2 - \frac{2\pi i }{12 \pi^2} \int\limits_{\mathcal{B}} \epsilon^{ijkl} \tilde{n}_i \, d\tilde{n}_j  \, d\tilde{n}_k \, d\tilde{n}_l .
	\label{o4}
\end{align}
This is the $SU(2)_1$ non-linear sigma model, known to be the effective theory of the Heisenberg spin chain \cite{Affleck1987}. \\

\subsection{$(2+1)$-D}
\label{Mott2D}
\hfill

\subsubsection{The analog of the  $\pi$-flux phase}
\hfill

\begin{figure}
	\begin{subfigure}{0.45\textwidth}
		\includegraphics[clip,trim=0cm 0cm 0cm 0cm,width=\linewidth]{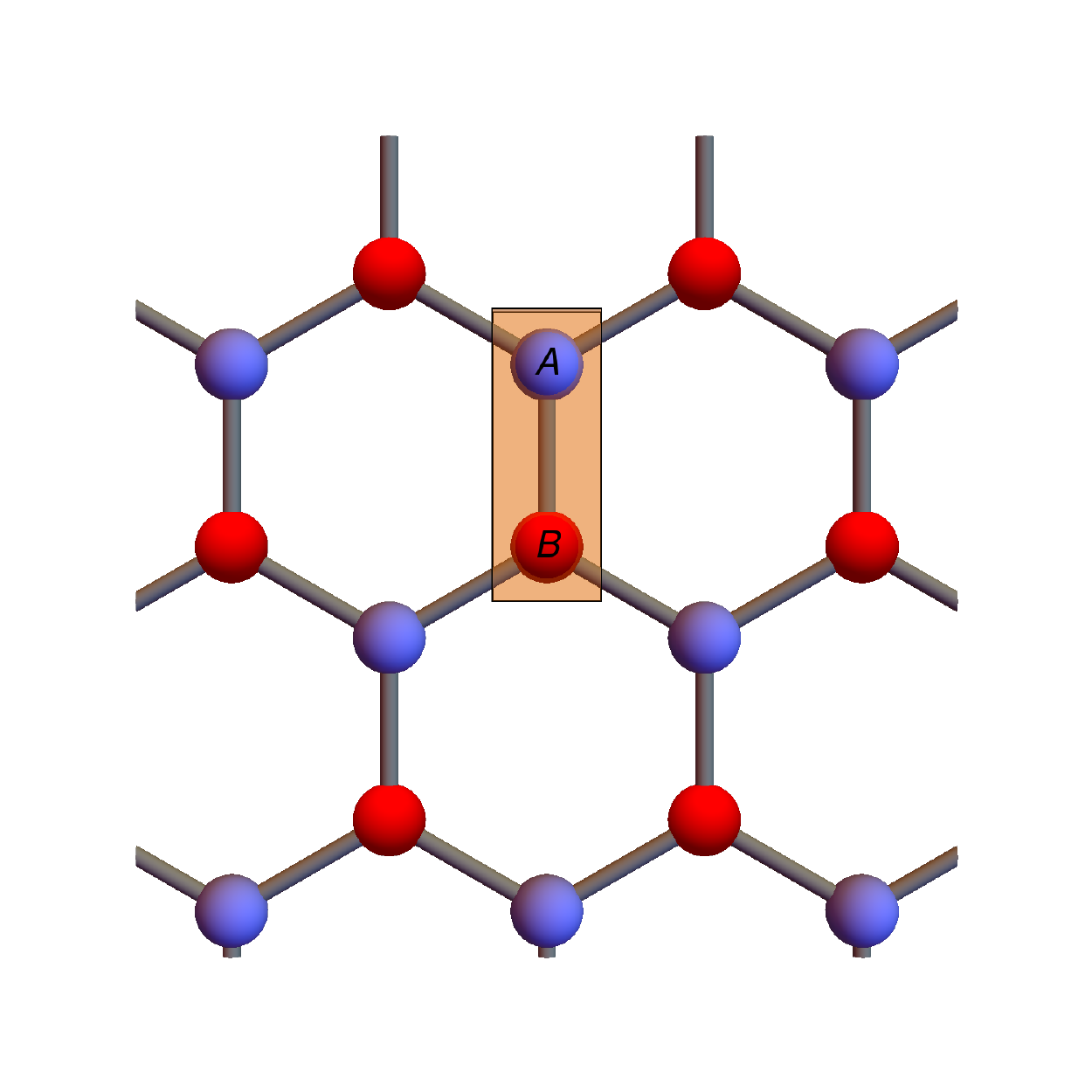}
		\caption{Real hopping} \label{2DTB_fig:1a}
	\end{subfigure}%
	\hspace*{0.3in}    % maximize separation between the subfigures
	\begin{subfigure}{0.45\textwidth}
		\includegraphics[clip,trim=0cm 0cm 0cm 0cm,width=\linewidth]{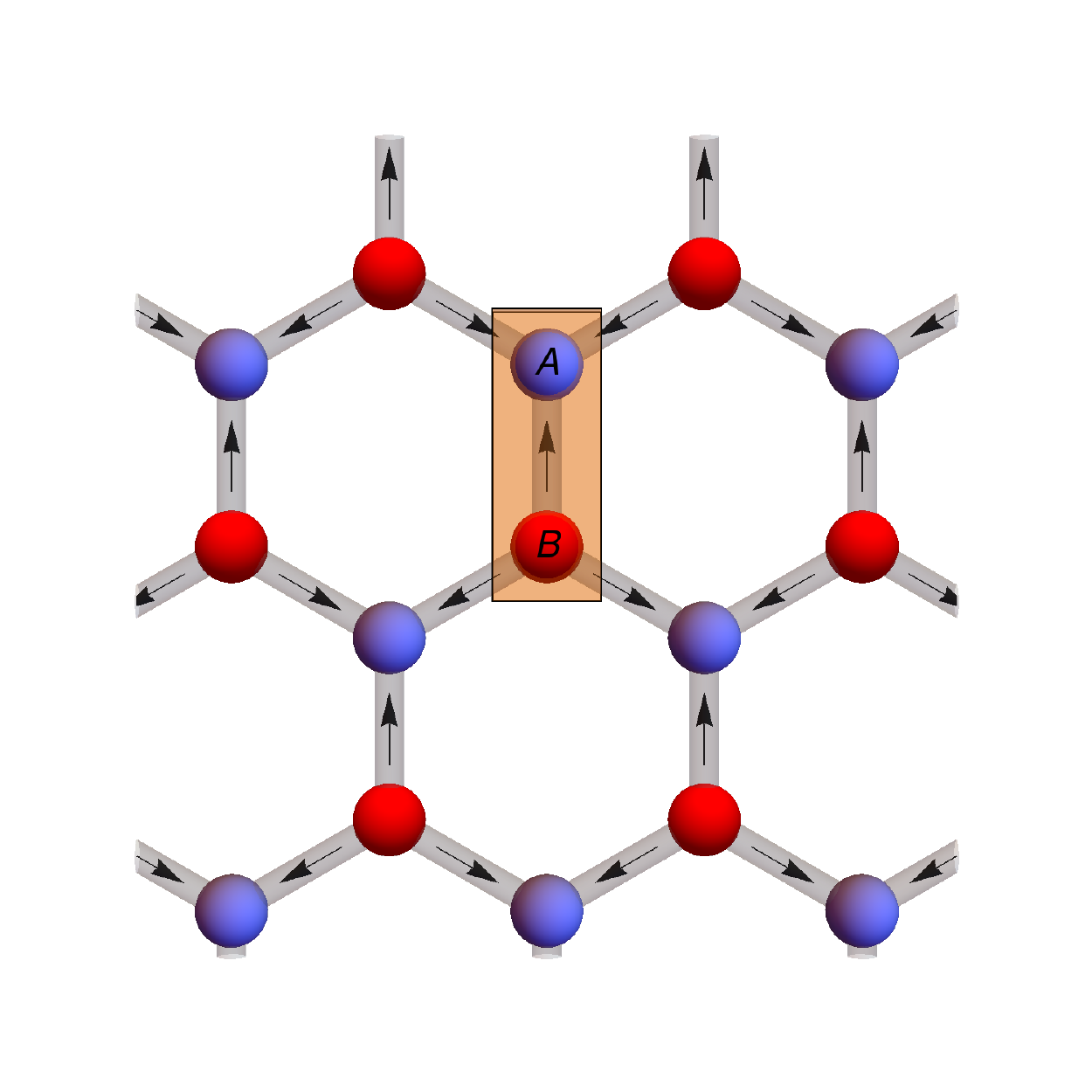}
		\caption{Imaginary hopping} \label{2DTB_fig:1b}
	\end{subfigure}%
	\caption{
		(a) The usual nearest neighbor tight-binding model on the honeycomb lattice with real hopping. (b) Upon the gauge transformation $(c^A_j ,c^B_{ j }) \rightarrow ( c^A_j, i \,c^B_{ j })$, hoppings become purely imaginary with alternating sign. The arrows point in the direction where the hopping is $+it$. The tight-binding Hamiltonian in panel (b) is charge-SU(2) invariant.
	}
	\label{2DTB}
\end{figure}

In $(2+1)$-D we use the honeycomb lattice to write down the tight-binding model. The lattice vectors in the real and momentum space are
\begin{align*}
	{\v a}_1 = \sqrt{3} a \left( \frac{1}{2}, \frac{\sqrt{3}}{2} \right), ~~~
	{\v a}_2 = \sqrt{3} a \left( -\frac{1}{2}, \frac{\sqrt{3}}{2} \right)
\end{align*}
and \begin{align}\label{reci}
	\v b_1 = \frac{4 \pi}{3a} \left(\frac{\sqrt{3}}{2}, \frac{1}{2} \right), ~~~
	\v b_2 = \frac{4 \pi}{3a} \left(-\frac{\sqrt{3}}{2}, \frac{1}{2} \right),
\end{align} 
respectively. 
%For the nearest neighbor tight-binding model with real hopping in 2D, one can break it into $A,B$ sub-lattice.
In the following we perform the gauge transformation $$(c^A_j ,c^B_{ j }) \rightarrow ( c^A_j, i \,c^B_{ j })$$ on the two sub-lattices, so that the nearest-neighbor hopping becomes purely imaginary (see figure \ref{2DTB}). %The reason for doing so is to make the hopping terms respect the global charge $SU(2)$ symmetry. 
The tight-binding Hamiltonian reads
\begin{align*}
	\hat{H} = &
	t \sum_{\v k} c_{\v k}^\dagger\left[  I \otimes \begin{pmatrix} 
		0	&	i + i e^{i {\v k} \cdot {\v a_1}} + i e^{i {\v k} \cdot {\v a_2}}	\\
		-i - i e^{-i {\v k} \cdot {\v a_1}}  - i e^{-i {\v k} \cdot {\v a_2}}	&	0		\\
	\end{pmatrix} \right] c_{\v k} \\
	=& 
	t \sum_{\v k} c_{\v k}^\dagger \Big[  
		-\left(\sin (\v k \cdot \v a_1) + \sin (\v k \cdot \v a_2) \right) IX 
		-\left( 1 +\cos (\v k \cdot \v a_1) + \cos (\v k \cdot \v a_2) \right) IY
		\Big] c_{\v k} 
\end{align*}
\noindent Here the Pauli matrices are arranged according to
$$ {\rm spin}\otimes \text{ sub-lattice}.$$

In the Majorana fermion basis,
\be
\hat{H}=t \sum_{\v k} \chi_{\v k}^\dagger \Big[  
	-\left(\sin (\v k \cdot \v a_1) + \sin (\v k \cdot \v a_2) \right) IIX 
	-\left( 1 +\cos (\v k \cdot \v a_1) + \cos (\v k \cdot \v a_2) \right) IIY
	\Big] \chi_{\v k}.
\label{latticeHoneycomb}\ee
\noindent Here the Pauli matrices are arranged according to   
$$\text{Majorana}\otimes\text{spin}\otimes\text{sub-lattice}.$$ \\

In the presence of repulsive Hubbard $U$, there is charge-SU(2) symmetry at half-filling. In the low energy theory the charge-SU(2) transformation is generated by
\begin{align*}
	\text{Charge-SU(2) generators}: T^a = (XYI, YII, ZYI)
\end{align*}
\noindent On the other hand, the spin-SU(2) transformations are generated by the following  matrices,
\begin{align*}
	\text{Spin-SU(2) generators}: \Sigma^a = (YXI, IYI, YZI)
\end{align*}
\noindent \Eq{latticeHoneycomb} is invariant under both the charge- and spin-SU(2). In momentum space
the Dirac points are located at $K$ and $\hat{K}$ points, i.e., $\pm k_0$ where $k_0 :=  \frac{1}{3} \left( \v b_1 - \v b_2 \right)$ (see \Eq{reci}). Note that in the Majorana fermion basis, the contribution of Hamtiltonian from $\v k$ and $-\v k$ are the same due to the constraint $\chi_{- \v k}^T = \chi^\dagger_{\v k}$. This means that one can take the fermion $\chi_{\v k_0 + \v q}$ around $k_0$ as the Fourier modes of complex fermion $\tilde{c}_{\v q}$ and discard the other node. We then break this complex fermion $\tilde{c}$ into real fermion by $\tilde{c} = \tilde{\chi}_1 + i \tilde{\chi}_2$ (in the following we shall refer to this 1 and 2 as the ``valley'' indices). In this final Majorana representation, the low energy Hamiltonian reads
\begin{align*}
 \hat{H}=\int d\v x ~\tilde{\chi}^T(\v x)~ \left[ -i \G_1 \partial_1 - i \G_2 \partial_2 \right]~\tilde{\chi}(\v x)
\end{align*}
\noindent where $\G_1 = IIIX$ and $\G_2 = IIYY$ \footnote{This is related to the gamma matrices in table \ref{tab:emergentSymm} by a basis transformation}. Here the Pauli matrices are arranged according to
$$\text{Majorana}\otimes\text{spin}\otimes\text{valley}\otimes\text{sub-lattice}.$$ In this basis, the symmetry generators are
\begin{align*}
	&\text{Charge $SU(2)$ generators}: T^a = (XYII, YIII, ZYII)\\
	&\text{Spin $SU(2)$ generators}: \Sigma^a = (YXII, IYII, YZII).
\end{align*}\\

\subsubsection{The charge-SU(2) confinement}
\hfill

Following the discussions in section \ref{bosonization2R}, the massless fermion theory is equivalent to the $\frac{O(8)}{O(4) \times O(4)}$ level-$1$ WZW model. Notice that the low energy fermion theory is identical to the $\pi$ flux phase spinon mean-field theory discussed in section \ref{Mott}. 
Imposing the Mott constraint  constraints the mass manifold. Specifically it requires the mass terms to commute with the charge-SU(2) generators. Under conditions the allowed mass terms satisfy
\begin{align*}
	&\{ \Gamma_i , M \} = 0 \\
	&[T^a , M ] =0 \\
	&M^2 = I_{16}
\end{align*}
The most general mass, $M \in \frac{O(8)}{O(4) \times O(4)}$, satisfying the first two equations has the form
\begin{align*}
	M =	n_1 \, YXIZ + n_2 \, IYIZ + n_3 \, YZIZ + n_4 \, IIXY + n_5 \, IIZY 
\end{align*}
\noindent Similar to the discussion in section \ref{Mott}, the first three of the masses in \Eq{NV} correspond to the N\'eel order parameters, while the last two to the valence bond solid (VBS) order parameters. The order parameter space forms an $S^4$. Plugging it into the $\frac{O(8)}{O(4) \times O(4)}$ level-$1$ WZW model, we arrive at the $O(5)$ WZW theory 
$$
W[n_i] = \frac{2}{\lambda_3} \int\limits_{\mathcal{M}} d^3 x \, \left( \partial_\mu n_i \right)^2 - \frac{2\pi i }{64 \pi^2} \int\limits_{\mathcal{B}} \epsilon^{ijklm} \tilde{n}_i \, d\tilde{n}_j  \, d\tilde{n}_k \, d\tilde{n}_l \, d\tilde{n}_m.
$$  Here we note that because $\pi_2(S^4)=0$ there is no soliton.
\\

\subsection{$(3+1)$-D}
\label{Mott3D}
\hfill

\subsubsection{The analog of the $\pi$-flux phase}
\hfill

%As a model for bipartite Mott insulator, we begin with a tight-binding model whose dispersion has Dirac nodes:% The following Hamiltonian is constructed using Wilson's regularization \cite{Wilson1977} %with purely imaginary hoppings
%\begin{align}
%	\hat{H} = \sum_{\v k} c_{\v k}^\dagger \left[ I \otimes \left(  
%			\thead{
%				\sin k_1 \, IXI + \sin k_2 \, IZI + \sin k_3 \, IYY \\
%				+\left( 3- \cos k_1 - \cos k_2 - \cos k_3 \right) IYZ
%			}
%	\right) \right] c_{\v k}.
%\end{align}
%This Hamiltonian is constructed using Wilson's regularization \cite{Wilson1977}, where the first Pauli matrix is associated with spin,  and the remaining three Pauli matrices in the parenthesis are associated with the sub-lattice degree of freedom. \\ 
% 
%  In the Majorana fermion representation the above Hamiltonian read,
%\begin{align}
%	\label{Mott3D_lattice}
%	\hat{H} = \sum_{\v k} \chi_{-\v k}^T \left[ II \otimes \left(  
%	\thead{
%		\sin k_1 \, IXI + \sin k_2 \, IZI + \sin k_3 \, IYY \\
%		+\left( 3- \cos k_1 - \cos k_2 - \cos k_3 \right) IYZ
%	}
%	\right) \right] \chi_{\v k}
%\end{align}
%\noindent where the Pauli matrices in the tensor product arranged according to
%\begin{align*}
%	\text{Majorana}\otimes \text{ spin}\otimes \text{ sub-lattice (which is $8 \times 8$)}
%\end{align*}

\begin{figure}
	\begin{subfigure}{0.5\textwidth}
		\includegraphics[clip,trim=4cm 3cm 2cm 1cm,width=\linewidth]{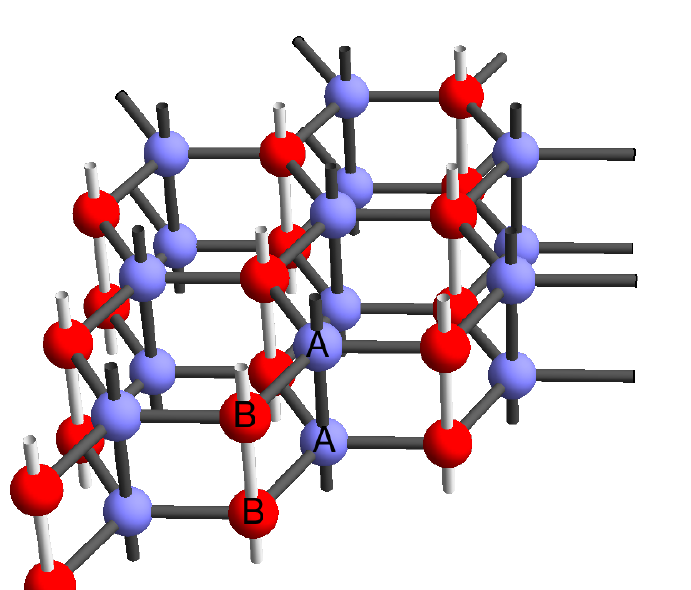}
		\caption{Real hopping} \label{3DTB_fig:1a}
	\end{subfigure}%
	%\hspace*{0.2in}    % maximize separation between the subfigures
	\begin{subfigure}{0.5\textwidth}
		\includegraphics[clip,trim=4cm 3cm 2cm 1cm,width=\linewidth]{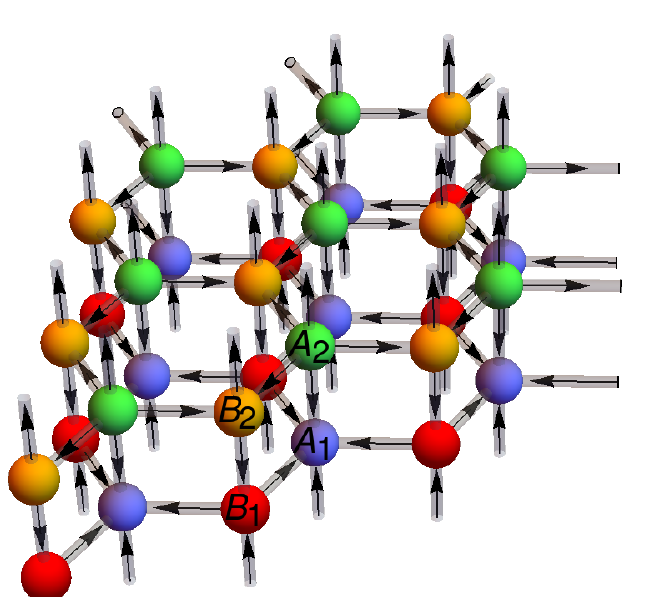}
		\caption{Imaginary hopping} \label{3DTB_fig:1b}
	\end{subfigure}%
	\caption{		
		(a) The tight-binding model on a stacked honeycomb lattice with real-valued nearest-neighbor hopping. Blue/red (A/B) mark the two sub-lattices of the honeycomb lattice respectively. The positive hoppings are drawn in black, while the negative hoppings in white.(b) After the gauge transformation $(c^{A_1}_j , c^{B_1}_j, c^{A_2}_j , c^{B_2}_j) \rightarrow (c^{A_1}_j , i c^{B_1}_j, i c^{A_2}_j , c^{B_2}_j)$,  a unit cell contains four sites. This is marked by blue/red/green/orange and labeled as $A_1/B_1/A_2/B_2$  respectively. The hoppings become purely imaginary. The arrows point in the direction where the hopping is $+it$. The tight-bonding Hamiltonian in panel (b) has charge-SU(2) symmetry.
	}
	\label{3DTB}
\end{figure}

As a model for bipartite Mott insulator, we begin with a 3-dimensional tight-binding model consists of stacked honeycomb lattice.  Here the lattice sites of each layer are stacked on top of those in the layer beneath. Within each layer, we have real hopping between the nearest-neighbor sites described in section \ref{Mott2D}. Between layers, the (real) hopping have the opposite sign for the $A$ and $B$ sub-lattice (see figure \ref{3DTB_fig:1a}). In order to make the hopping terms global charge-SU(2) invariant, we first enlarge the unit cell by grouping two adjacent layers to form $A_1$, $B_1$, $A_2$, $B_2$ sub-lattices as shown in figure \ref{3DTB_fig:1b}. We then perform the following gauge transformation,
\begin{align*}
(c^{A_1}_j , c^{B_1}_j, c^{A_2}_j , c^{B_2}_j) \rightarrow (c^{A_1}_j , i c^{B_1}_j, i c^{A_2}_j , c^{B_2}_j).
\end{align*}
\noindent Here the lattice vectors in the real and momentum spaces are
\begin{align*}
{\v a}_1 = \sqrt{3}  \left( \frac{1}{2}, \frac{\sqrt{3}}{2} ,0\right), ~~~
{\v a}_2 = \sqrt{3}  \left( -\frac{1}{2}, \frac{\sqrt{3}}{2} ,0\right), ~~~
{\v a}_3 = 3  \left( 0,0,1 \right)
\end{align*}
\noindent and
\begin{align}
\v b_1 = \frac{4 \pi}{3} \left(\frac{\sqrt{3}}{2}, \frac{1}{2} ,0\right), ~~~
\v b_2 = \frac{4 \pi}{3} \left(-\frac{\sqrt{3}}{2}, \frac{1}{2} ,0 \right),~~~
\v b_3 = \frac{2 \pi}{3} \left( 0 , 0 , 1 \right)
\label{reci3D}
\end{align} 
\noindent respectively. In the above we have assumed the magnitude of the hopping in the z-direction is the same as those within each layer. Moreover, we have tuned the lattice constant in the z-direction so that the Dirac cone is isotropic. The resulting tight-binding model reads
\be
&&\hat{H} = i \, t \sum\limits_{\v k} c^\dagger_{\v k}  \cdot I \otimes \begin{pmatrix} 
		0														&	S_{\rm xy}(\v k)	& S_{\rm z}(\v k)	&	0	\\
		-S_{\rm xy}^*(\v k)	&	0	&	0	&S_{\rm z}(\v k)	\\
		-S_{\rm z}^*(\v k)	&	0	&	0	&	-S_{\rm xy}(\v k)\\
		0	&-S_{\rm z}^*(\v k)	&	S_{\rm xy}^*(\v k)	&	0
	\end{pmatrix}c_{\v k}\nn
	&&=t \sum\limits_{\v k} c^\dagger_{\v k} I\otimes \left\{ 
	\thead{
	-\left[ \sin (\v k \cdot a_1 )+ \sin (\v k \cdot a_2)  \right] ZX  -\left[ 1+ \cos(\v k \cdot a_1 )+ \cos (\v k \cdot a_2)  \right] ZY \\
	+ \sin ( \v k \cdot a_3 ) XI - \left[1+\cos(\v k \cdot \v a_3) YI\right].
	}
	\right\} c_{\v k}\nn
\label{cpxhop}\ee
\noindent where the $S_{xy}$ and $S_z$ in \Eq{cpxhop} are defined as
\be &&S_{\rm xy}(\v k)= 1 + e^{i \v k \cdot \v a_1} +  e^{i \v k \cdot \v a_2}\nn
&&S_{\rm z}(\v k)=	1+ e^{-i \v k \cdot \v a_3},\nonumber\ee
and the Pauli matrices are arranged according to
$$ {\rm spin}\otimes \text{ sub-lattice } (4 \times 4).$$ It is simple to show that in the momentum space the Dirac points are located at $\pm \v k_0$, where $\v k_0 :=  \frac{1}{3} \left( \v b_1 - \v b_2 \right)$.\\

Converting \Eq{cpxhop} into the Majorana fermion basis, the Hamiltonian reads
\begin{align}
\hat{H}  =t \sum\limits_{\v k} \chi^T_{-\v k} II\otimes \left[ 
\thead{
	-\left( \sin (\v k \cdot a_1 )+ \sin (\v k \cdot a_2)  \right)ZX  -\left( 1+ \cos(\v k \cdot a_1 )+ \cos (\v k \cdot a_2)  \right)ZY \\
	+ \sin ( \v k \cdot a_3 ) XI - \left(1+\cos(\v k \cdot \v a_3) YI\right)
}
\right] \chi_{\v k}
\label{Mott3D_lattice}
\end{align}
\noindent where the first Pauli matrix $I$ acts in the Majorana space. The Hamiltonian in \Eq{Mott3D_lattice} is invariant under the global charge and spin $SU(2)$ transformations generated by
\begin{align}
&\text{Charge-SU(2): } T^a = (XYII, ~YIII, ~ZYII)\notag\\
&\text{Spin-SU(2): ~~~~} \Sigma^a = (YXII, ~IYII, ~YZII)
\label{tsu2g}
\end{align}
\\

When  performing the mode expansion near $\pm \v k_0$, because $\chi_{- \v k}^T = \chi^\dagger_{\v k}$, one can keep the complex fermion operator  $\tilde{c}_{\v q}=\chi_{\v k_0 + \v q}$ while disregard the mode expansion near $-\v k_0$. We subsequently break $\tilde{c}$ into real fermion operators $\tilde{c} = \tilde{\chi}_1 + i \tilde{\chi}_2$ (in the following we shall refer to this 1 and 2 as the ``valley'' indices). Omitting the tilde, in this final Majorana representation, the low energy theory of the Hamiltonian \Eq{Mott3D_lattice} is given by the $n=8$ real class massless fermion Hamiltonian
\be
	&&\hat{H}_{\rm eff} = \int d^3 x ~\chi^T \left[ - i  \sum_{i=1}^3  \Gamma_i \partial_i \right]\chi\nn
	&&\text{where } \Gamma_1 = IIZXI, ~ \Gamma_2 = IIZYY, ~ \Gamma_3 = IIXII,
	\label{3Dlow}
\ee
and \Eq{tsu2g} is given by
\begin{align}
&\text{Charge-SU(2): } T^a = (XYIII, ~YIIII, ~ZYIII)\notag\\
&\text{Spin-SU(2): ~~~~} \Sigma^a = (YXIII, ~IYIII, ~YZIII)
\label{tsu2g2}\end{align}
\noindent In this basis, the Pauli matrices correspond to
$$ \text{Majorana} \otimes \text{spin}\otimes \text{ sub-lattice } (4 \times 4) \otimes \text{valley}.$$\\

For the gamma matrices to be in the standard basis used in table \ref{tab:emergentSymm}, we can do the transformation $$\chi \rightarrow e^{i \frac{\pi}{4} IIZYI} \cdot e^{i \frac{\pi}{4} IIIXY}  \chi,$$ and then switch between the third and the fifth Pauli matrices. In the new basis, \Eq{3Dlow} becomes
\be
	&&\hat{H}_{\rm eff} = \int d^3 x ~\chi^T \left[ - i  \sum_{i=1}^3  \Gamma_i \partial_i \right]\chi\nn
	&&\text{where } \Gamma_1 = IIIZI, ~ \Gamma_2 = IIIXI, ~ \Gamma_3 = IIIYY,
	\label{3Dlow1}
\ee
while the symmetry generators in \Eq{tsu2g2} remain unchanged.  Upon bosonization, \Eq{3Dlow1} is equivalent to the $U(8)/O(8)$ nonlinear sigma model in \Eq{wzw3R}.\\

\subsubsection{The charge-SU(2) confinement}\label{3dmottspin}
\hfill

Following the discussion in section \ref{Mott_1D_2D}, the Mott  constraint  can be imposed by demanding the order parameter to be charge $SU(2)$ singlet. %fluctuating the charge $SU(2)$ gauge field. Assuming that the system is in the charge $SU(2)$ confined phase, the only the charge-$SU(2)$-singlet meson (mass) is allowed to fluctuate. 
%This greatly reduce the space of order parameter space. 
It is straightforward (but lengthy) to show that the following $Q^\mathbb{R}$ satisfies the charge-SU(2) singlet requirement 
\begin{align}
	\label{U(1)_S5}
	Q^\mathbb{R}(x) = e^{i \theta(x)} \left[ n_0(x) \, N_0 + i \sum_{i=1}^5 n_i(x) \, N_i   \right]:=e^{i \theta(x)}\cG_S(x),
\end{align} 
\noindent where
\begin{align*}
	&N_0 = III, ~N_1=IIZ, ~N_2=IIX, ~N_3=IYY, ~N_4=YZY, ~N_5=YXY\\
	 &\text{and}~\sum_{i=0}^5 n_i^2 = 1, \text{i.e.}, (n_0,n_1,n_2,n_3,n_4,n_5)\in S^5.
\end{align*}
In addition, in \Eq{U(1)_S5} $\cG_S$ is a symmetric special unitary $8\times 8$ matrix, namely,
$$\cG_S(x)\in {SU(8)\over O(8)}.$$\\

Substituting \Eq{U(1)_S5} into the bosonized nonlinear sigma model \Eq{wzw3R} and noting that 
$$\frac{1}{i}Q^{\mathbb{R}\dagger}\partial_\mu Q^{\mathbb{R}} = \frac{1}{i} \cG_S^\dagger\partial_\mu \cG_S + \partial_\mu \theta,$$  
the stiffness term becomes
\be
&&\frac{1}{4\lambda_4^2}   \int_\mathcal{M} d^4 x \, \tr\left[  \partial_{\mu} Q^{\mathbb{R}} \partial^{\mu}Q^{\mathbb{R}\dagger}\right] \nn
&&=\frac{8}{4\lambda_4^2}   \int_\mathcal{M} d^4 x \, \left[  \p_\mu\theta \p^{\mu}\theta\right] +\frac{1}{4\lambda_4^2}  \int_\mathcal{M} d^4 x \, \tr\left[  \partial_{\mu} \cG_S \partial^{\mu}\cG_S^{\dagger}\right] \nn
&&=\frac{2}{\lambda_4^2}   \int_\mathcal{M} d^4 x \, \left[  \p_\mu\theta \p^{\mu}\theta\right] +\frac{2}{\lambda_4^2}   \int_\mathcal{M} d^4 x \, \sum_{i=0}^5 (\partial_{\mu} n_i )^2
\label{stftinto}	\ee

The cross term vanishes because \be\frac{1}{i}\Tr[\cG_S^\dagger \partial_\mu \cG_S]=0.\label{notrace}\ee \Eq{notrace} is due to the fact that 
$\cG_S$ is a symmetric special unitary matrix hence $\in SU(n)$. As a result, the matrix part of $\frac{1}{i}\cG_S^\dagger \partial_\mu \cG_S$ can be decomposed into the generators $\{ t^a\}$ of $su(n)$, which are traceless. \\

As to the WZW term it's can be shown that 
\be &&-\frac{2\pi}{960\pi^3}    \int\limits_{\mathcal{B}}    \, \tr \Big[\left(\tilde{Q}^{\mathbb{R}\dagger} d\tilde{Q}^{\mathbb{R}} \right)^5 \Big]=-\frac{2\pi}{960\pi^3}    \int\limits_{\mathcal{B}}    \, \tr \Big[\left(\tilde{\cG_S}^{\dagger} d\tilde{\cG_S}\right)^5 \Big]\nn
&&=-\frac{2\pi i}{120\pi^3}    \int\limits_{\mathcal{B}}    \epsilon^{i_1i_2i_3i_4i_5i_6} \tilde{n}_{i_1}  d\tilde{n}_{i_2}  d\tilde{n}_{i_3}  d\tilde{n}_{i_4}  d\tilde{n}_{i_5} d\tilde{n}_{i_6}.\label{wzwtinto}\ee
(We shall prove this relation in appendix \ref{appendix:WZWdecompose}.) \\

Putting together \Eq{stftinto} and \Eq{wzwtinto}, the non-linear sigma model action is given by 
\be
W[\theta,\boldsymbol{\beta}] &&= \frac{2}{\lambda_4^2}   \int_\mathcal{M} d^4 x \, \left[  \p_\mu\theta \p^{\mu}\theta\right] +\frac{2}{\lambda_4^2}  \int_\mathcal{M} d^4 x \, \sum_{i=0}^5 (\partial_{\mu} n_i )^2  \nn
&&-\frac{2\pi i}{120\pi^3}    \int\limits_{\mathcal{B}}    \epsilon^{i_1i_2i_3i_4i_5i_6} \tilde{n}_{i_1}  d\tilde{n}_{i_2}  d\tilde{n}_{i_3}  d\tilde{n}_{i_4}  d\tilde{n}_{i_5} d\tilde{n}_{i_6}.\nn
\ee
Therefore unlike $(1+1)$- and $(2+1)$-D, the spin effective theory for $(3+1)$-D bipartite Mott insulator has an extra $U(1)$ mode!
\\

\subsubsection{Gapping out the $U(1)$ mode}
\hfill

In this subsection we show that there is a charge-SU(2) singlet fermion interaction term that gaps out the $\theta$ degree of freedom. For convenience, we use the basis in \Eq{3Dlow1}.
%we switch the order of Pauli matrices in \Eq{3Dlow} so that
%\be
%&&\Gamma_1 =  I_n \otimes XI  \nn
%&&\Gamma_2 =  I_n \otimes ZI \nn
%&&\Gamma_3 =  I_n \otimes YY  ,
%\label{3g}
%\ee
The emergent symmetry is $U(n)$ which includes a subgroup $U(1)$ (not to be confused with the extra $U(1)$ mode discussed earlier)  generated by $$Q_{U(1)} =  I_n \otimes IY.$$ We can use this $Q_{U(1)}$ to complexify the Majorana fermion
\footnote{
Note that although we have complexified the Majorana fermion using the emergent $U(1)$, this is different from the complex class because we allow the mass term to break this $U(1)$.
}
, namely,
\be
\psi_{i}^\alpha := \frac{1}{\sqrt{2}} \left( \chi_{\alpha i1} + i \,\chi_{\alpha i2} \right).\label{cpf}
\ee
\noindent Here the Majorana field $\chi_{\alpha i  a  }$ carries three indices: $\alpha=1,2,...,n$ is the flavor  index; % of the $n\times n$ identity matrix. 
 $i$ indexes the second Pauli matrix in \Eq{3Dlow1} and $a=1,2$ indexes the last Pauli matrix.  In terms of the complexified  fermion operators the mass term reads (see table\ref{tab:massManifold}) 

\be
&&\chi^T \left[    S_1 \otimes YX  +  S_2 \otimes YZ   \right] \chi \nn
&&= \left[    \psi_i^\alpha \, \left( i \, Y_{ij} \right)( S_1+ i \, S_2)_{\alpha\beta} \,\psi_j^\beta + h.c. \right]\nn
&&= \left[    \psi_i^\alpha \, E_{ij}  Q^\mathbb{R}_{\alpha\beta} \,\psi_j^\beta + h.c. \right]
\label{scm}
\ee
where $S_1$ and $S_2$ are symmetric real matrices. \\

Now we are ready to construct the desired interaction term to gap out the $U(1)$ mode in \Eq{U(1)_S5}
\begin{align}
\hat{H}_{\rm int}=- \frac{U_\theta}{2} \int d^4 x \Big[&E_{i_1 j_1} E_{i_2 j_2 }... E_{i_n j_n }\Big( \epsilon_{\alpha_1 \alpha_2 ... \alpha_n} \psi^{ \alpha_1}_{i_1 } \psi^{\alpha_2}_{i_2 } ... \psi^{ \alpha_n}_{i_n } \Big)\nn
&\times \Big( \epsilon_{\beta_1 \beta_2 ... \beta_n} \psi_{j_1 }^{\beta_1} \psi_{j_2 }^{\beta_2} ... \psi_{j_n }^{\beta_n} \Big) + {\rm h.c.} \Big].
\label{hitr}
\end{align}

First we note that \Eq{hitr} is a charge-SU(2) singlet, hence is unaffected by the Mott constraint. The proof goes as follows. When acted upon by the charge-SU(2) transformation, the fermion operator  in \Eq{cpf} transforms according to $$ \psi^{ \alpha}_{i}\ra u^{\alpha}_{\beta} \psi^{ \beta}_{i },$$ 	where $u^{\alpha}_{\beta}$ is the charge-SU(2) transformation matrix. Under such transformation, the term in each parenthesis of \Eq{hitr} transforms according to 		\begin{align*}
		 &\epsilon_{\alpha_1 \alpha_2 ... \alpha_n} \psi^{ \alpha_1}_{i_1 } \psi^{\alpha_2}_{i_2 } ... \psi^{ \alpha_n}_{i_n } \ra
			\epsilon_{\alpha_1 \alpha_2 ... \alpha_n} u^{\alpha_1}_{\beta_1} u^{\alpha_2}_{\beta_2}...u^{\alpha_n}_{\beta_n} \psi^{ \beta_1}_{i_1 } \psi^{\beta_2}_{i_2 } ... \psi^{ \beta_n}_{i_n }\\&=\left(\det{u}\right) \epsilon_{\beta_1 \beta_2 ... \beta_n}\psi^{ \beta_1}_{i_1 } \psi^{\beta_2}_{i_2 } ... \psi^{ \beta_n}_{i_n }= \epsilon_{\alpha_1 \alpha_2 ... \alpha_n} \psi^{ \alpha_1}_{i_1 } \psi^{\alpha_2}_{i_2 } ... \psi^{ \alpha_n}_{i_n } .
		\end{align*} 
Therefore \Eq{hitr} is charge-SU(2) invariant. \\

Next, we note, upon bosonization
\begin{align*}
E_{ij} \psi_i^\alpha \psi_j^\beta  \rightarrow Q^{\mathbb{R}}_{\alpha \beta} = \left( S_1 + i S_2\right)_{\alpha \beta},
\end{align*}
where $$Q^{\mathbb{R}}\in{U(8)\over O(8)}$$ is the order parameter of the nonlinear sigma model in \Eq{wzw3R}.
As the result, the action corresponds to $\hat{H}_{\rm int}$ is \be S_{\rm int}=- \frac{U_\theta}{2} \int d^4 x \left\{\det\left[Q^{\mathbb{R}}\right]+c.c\right\}.\label{stheta1}\ee
Substituting \Eq{U(1)_S5} into  \Eq{stheta1} we obtain 
 \be S_{\rm int}= - {U_\theta} \int d^4 x ~\cos(8\theta).\label{stheta}\ee\\

Naively, it might appear that the $\theta$-vacuum is $8$-fold degenerate, corresponding to $$\theta = \frac{2\pi l}{8} ~\text{with~} l=0,1,...7.$$ However, this is due to a redundancy in the splitting $U(8)/O(8) \ra  U(1) \times SU(8)/O(8)$. The transformation
\begin{align*}
e^{i \theta} \rightarrow&  e^{i (\theta + \frac{2\pi}{8}) } 
\end{align*}
can be absorbed by the following transformation of $\cG_S$ 
\begin{align*}
\cG_S \rightarrow& e^{i \frac{2 \pi}{8}} \cG_S.
\end{align*}
 Because $\left(e^{i \frac{2 \pi}{8}} \right)^8=1$, the transformed $\cG_S$ still belongs to $SU(8)/O(8)$. As a result,  the $8$ different $\theta$ vacua should be counted as one, as long as there is no spontaneous symmetry breaking in $\cG_S$ (i.e., when $\cG_S(x)$ fluctuates over all possible configurations in $SU(8)/O(8)$).\\

In the phase that the $\theta$ degrees of freedom are gapped out, we have
\begin{align}
	\label{S5}
	Q^\mathbb{R}(x) =  \left[ n_0(x) \, N_0 + i \sum_{i=1}^5 n_{i}(x) \, N_{i}   \right].
\end{align} 
\noindent Among the six order parameters, the first three are spin-SU(2) singlet and the last three are spin-SU(2) triplet. The latter can be interpreted as the anti-ferromagnetic order parameters. As to the first three, they break the lattice rotation symmetry, and can be identified as the VBS order parameters. The non-linear sigma model governing the $n_i$ degrees of freedom read
\be
W[n_i] = \frac{2}{\lambda_4^2}   \int_\mathcal{M} d^4 x \, \sum_{i=0}^5 (\partial_{\mu} n_i )^2 -\frac{2\pi i}{120\pi^3}    \int\limits_{\mathcal{B}}    \epsilon^{i_1i_2i_3i_4i_5i_6} \tilde{n}_{i_1}  d\tilde{n}_{i_2}  d\tilde{n}_{i_3}  d\tilde{n}_{i_4}  d\tilde{n}_{i_5} d\tilde{n}_{i_6},\nn
\ee

\noindent which is the $S^5$ (or $O(6)$) nonlinear sigma model with $k=1$ WZW term. Note that since $\pi_3(S^5)=0$, there is no soliton. This model is a natural generalization of the spin effective theory in $(1+1)$- and $(2+1)$-D. \\

Finally, the cautionary remarks in the summary of section \ref{Mott}, concerning the four-fermion interactions, induced by the charge-SU(2) gauge field fluctuations, also apply here.

\section{Twisted bi-layer graphene}\label{sect:TBLG}
\hfill

Another 2D system where relativistic electron dispersion comes into play is the twisted bilayer graphene (TBLG). When the twisting angle is close to the ``magic'' value, the relevant bands become very flat \cite{Bistritzer2011}, which suggests strong correlation. Under that condition, as a function of band filling $\nu$, a rich phase diagram emerges. This includes various insulating phases near integer filling and superconductivity when $\nu$ deviates from integer \cite{Cao2018,Yankowitz2019,Lu2019,Cao2018b,Sharpe2019,Serlin2020}. In the following, we shall hold the point of view that the essence of the TBLG physics is the fact that the interaction energy overwhelms the bandwidth, which does not require the bandwidth to be zero. Therefore we  restrict ourselves to twisting angles close but not exactly equal to the magic values. \\

In the non-interacting picture, the Fermi energy ($E_F$) only intersects the Dirac nodes at the charge neutral point $\nu=0$. However, by measuring the electronic compressibility, it is recently suggested that the coincidence of $E_F$ and Dirac nodes  reappears  at all integer filling factors \cite{Zondiner2020,Wong2020}.  Such ``Dirac revivals'' is interpreted as the evidence of the unequal filling of bands induced by the polarization of the flavor (including valley and spin) degrees of freedom. Therefore the relativistic massless fermions and bosonized non-linear sigma models discussed in Part I are good starting points to address the physics of TBLG.\\

\begin{figure}
	\begin{subfigure}{0.31\textwidth}
		\includegraphics[width=\linewidth]{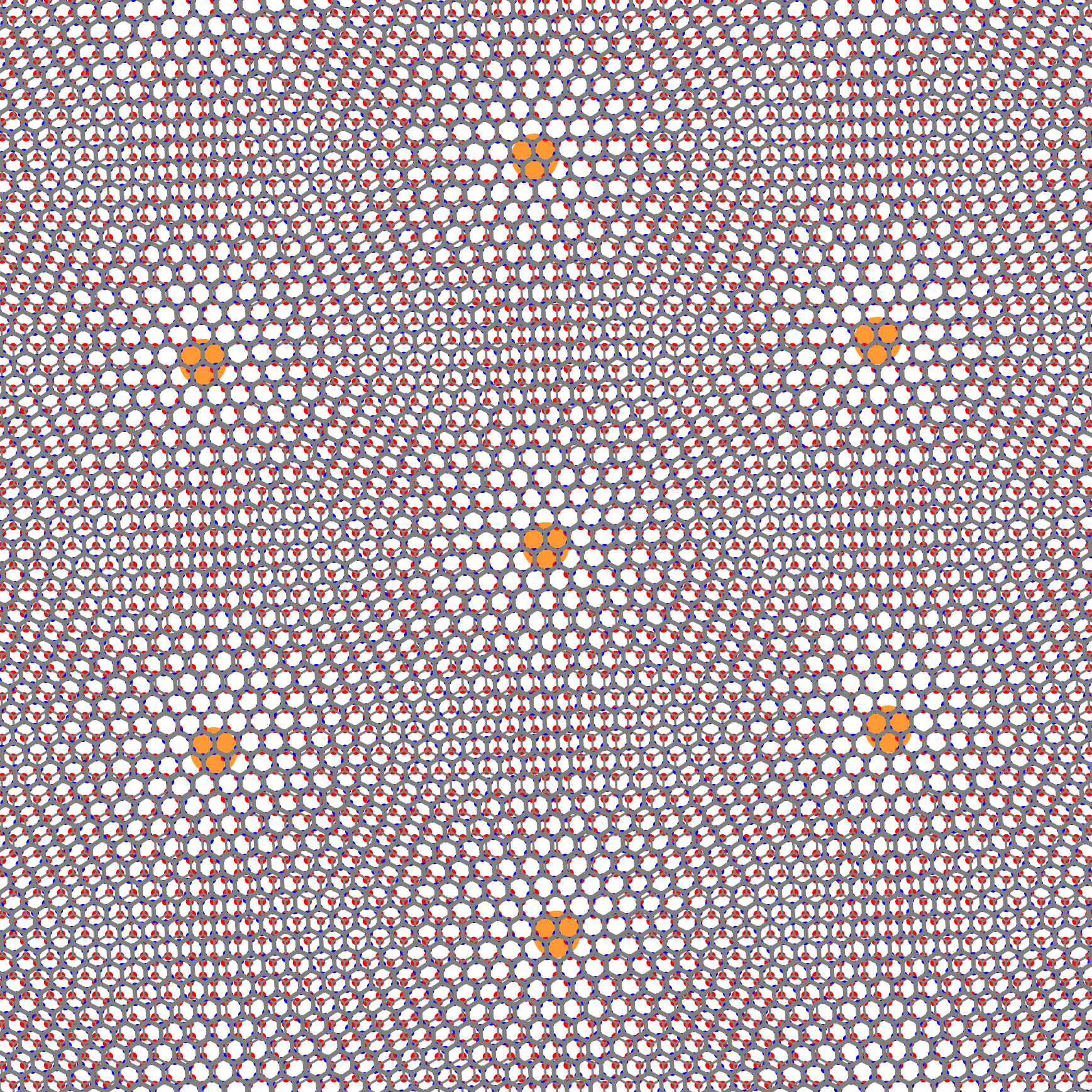}
		\caption{Real space} \label{fig:TBLGa}
	\end{subfigure}%
	\hspace*{\fill}   % maximize separation between the subfigures
	\begin{subfigure}{0.31\textwidth}
		\includegraphics[width=\linewidth]{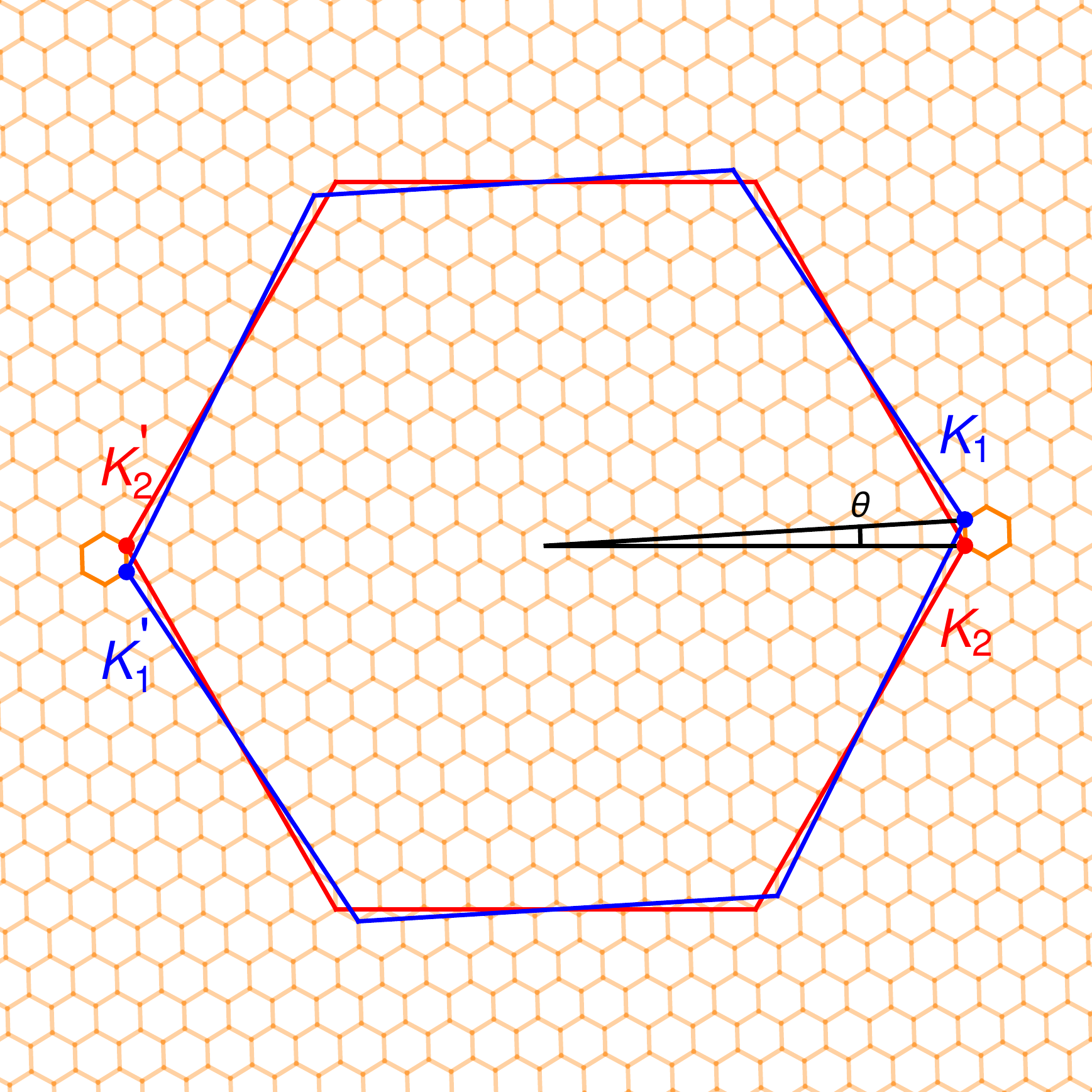}
		\caption{Momentum space} \label{fig:TBLGb}
	\end{subfigure}%
	\hspace*{\fill}   % maximizeseparation between the subfigures
	\begin{subfigure}{0.31\textwidth}
		\includegraphics[width=\linewidth]{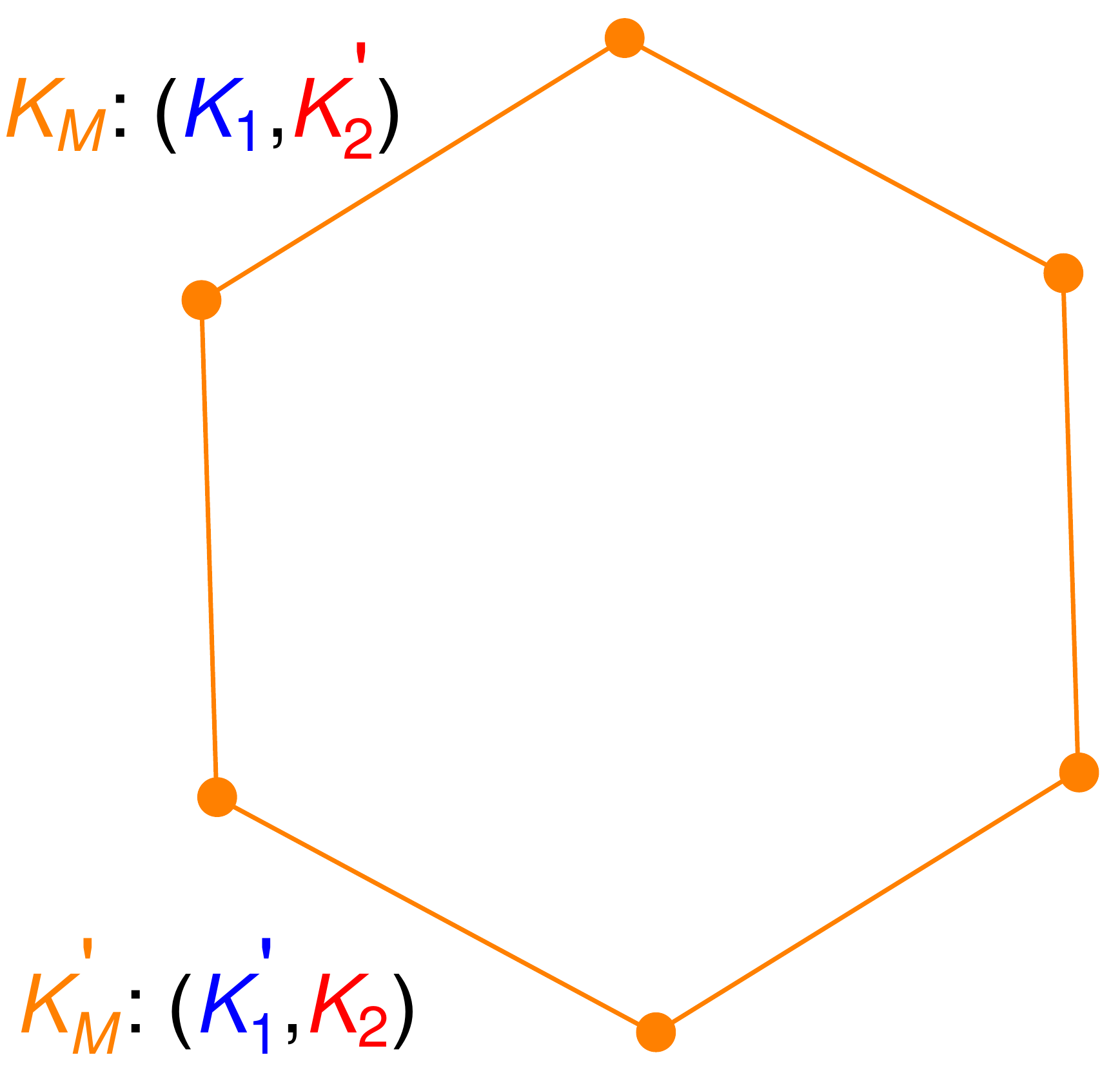}
		\caption{Moir\'e Brillouin zone} \label{fig:TBLGc}
	\end{subfigure}
	\caption{
		(a) A real space picture of twisted bilayer graphene. (b) Blue and red color the Brillouin zones of the first and second layer graphene. Orange colors the Brillouin zone of the Moir\'e superlattice.  (c) At $K_M$ there are the \B{$K_1$} of the first layer and \R{$K_2'$} of the first layer. At $K'_M$ there are the \B{$K_1'$} of the first layer and \R{$K_2$} of the second layer.
	}
	\label{TBLG}
\end{figure}

The real space structure of the TBLG is shown in \Fig{fig:TBLGa} for a certain small but commensurate twisting angle.  In \Fig{fig:TBLGb} we show the associated momentum space structure. The large blue and the red hexagons are the original graphene Brillouin zones for the two layers.
The small hexagons colored orange are the Brillouin zone of the Moir\'e superlattice. In \Fig{fig:TBLGc} we blow up one of the Moir\'e Brillouin zones. Here $K_M$ and $K_M^\prime$ labels the two valleys in the Moir\'e Brillouin zone, while the blue/red $K$ and $K^\prime$ labels the valleys of the graphene Brillouin zone. Note that each valley of the Moir\'e Brillouin zone consists of two opposite valleys of the graphene Brillouin zone,\\

\subsection{Charge neutral point $\nu=0$}
\hfill
 
In the presence of inter-layer hybridization, there are eight ``active''  graphene-like bands. We can label these eight ``flavors'' by the flavor index which represents $$\text{graphene valley, Moir\'e valley,  spin}$$ degrees of freedom. At the charge neutral point, the Fermi level crosses the Dirac points at $K_M$ and $K_M^\prime$.\\

In the momentum space we expand the band dispersion around $K_M$ and $K_M^\prime$, the resulting low energy Dirac-like band structure is described by the following continuum real-space Hamiltonian
\be
\hat{H}=\int d^2 x~\psi^\dagger(\v x)\left(-i\G_1\p_x-i\G_2\p_y\right)\psi(\v x),
\label{drc}\ee
where $\psi$ is an eight-component complex fermion field, and
\begin{align}
\G_1=XZII,~~\G_2=YIII.
\label{GammasTBLG}
\end{align}
Here the tensor product of Pauli matrices is arranged according to  $$ \text{ sub-lattice}\otimes {\rm graphene~valley}\otimes \text{Moir\'e valley}\otimes {\rm spin}.$$   
The reason we use the complex fermion (rather than Majorana) representation in \Eq{drc} is that at integer band fillings there is no evidence of superconductivity \cite{Lu2019}. Therefore \Eq{drc} belong to the complex class. \\

The massless free fermion Hamiltonian in \Eq{drc} has emergent $U(8)$ symmetry. After performing the  the basis transformation $$\psi \rightarrow  e^{i \frac{\pi}{4} XIII} \cdot \left[ \begin{pmatrix} I &  0 \\ 0 & Z \end{pmatrix} \otimes II \right] \psi$$ to cast the gamma matrices into the form used in table \ref{emsymm}, namely,
$$ \G_1=XIII, ~~\G_2=ZIII,$$ we can use our bosonization result (see appendix \ref{appendix:anomaliesb}).
%\footnote{
%Here we have done the basis transformation $\psi \rightarrow  \left[ \begin{pmatrix} I &  0 \\ 0 & Z \end{pmatrix} \otimes II \right]\cdot e^{i %\frac{\pi}{2} XIII}\psi$ to cast the gamma matrices into the form used in table \ref{emsymm} appendix \ref{appendix:anomaliesb}.
%}, 
In the presence of the electromagnetic  ($U(1)$) gauge field $A$, the massless fermion theory in \Eq{drc} is equivalent to the following gauged non-linear sigma model  
\be
&&W[Q^{\mathbb{C}}, A] ={1\over 2\lambda_3}  \int\limits_{\mathcal{M}} d^3 x \, \tr\Big[\left(\p_\mu Q^{\mathbb{C}}\right)^2\Big]-  \frac{ 2 \pi i }{256 \pi^2}   \Big\{   \int\limits_{\mathcal{B}} \tr \Big[\tilde{Q}^{\mathbb{C}}    \,\left( d \tilde{Q}^{\mathbb{C}} \right)^4 \Big] \nn
&&+ 8 \int\limits_{\mathcal{M}}  \tr \Big[ i A Q^{\mathbb{C}} (dQ^{\mathbb{C}})^2  - 2 AFQ^{\mathbb{C}} \Big]
\Big\},\label{gwzw}
\ee
where
 \be Q^{\mathbb{C}}\in ~
{U(8)\over U(4)\times U(4)}.\ee 
\\

As discussed in section \ref{IntFermion}, there exists a local interacting fermion model which respects all emergent symmetries %, and if the time-reversal symmetry is broken, 
and the phases (which might spontaneously break the continuous or discrete symmetries) are described by the effective theories given by \Eq{gwzw} but with 
\begin{align*}
Q^{\mathbb{C}}\in ~\bigcup_{l=0}^8~{U(8)\over U(l)\times U(8-l)}
\end{align*}

Among the last two terms of \Eq{gwzw}, the term  linear in $A_{\mu}$ measures the soliton current
$$
J^\mu=\frac{ i }{16 \pi} \e^{\mu\nu\rho} \tr \Big[Q^{\mathbb{C}}\p_\nu Q^{\mathbb{C}} \p_\rho Q^{\mathbb{C}}  \Big].
$$
The term proportional to $AF$ gives rise to a Chern-Simons term
$$- \frac{i }{8 \pi}     \int\limits_{\mathcal{M}} \tr[Q^{\mathbb{C}}] AF,$$ with the corresponding Hall conductance  
\be \s_{xy}={1\over 2}\tr[Q^{\mathbb{C}}]=l-4 .\label{sxy}\ee  
Therefore
only the $l=4$ mass manifold, ${U(8)\over U(4)\times U(4)}$, has $\s_{xy}=0$. Since so far there is no reported (non-zero) Hall conductivity at the charge neutral point \cite{Serlin2020}, we take it as implying the relevant mass manifold is  ${U(8)\over U(4)\times U(4)}$. 
\\

The resulting non-linear sigma model has two phases depending on the coupling constant $\lambda_3$ in the stiffness term. For $\lambda_3< \lambda_c$ there is a spontaneous breaking of the $U(8)$ symmetry, and the sigma model is gapped. We interpret this phase as the ``symmetry-breaking insulator''. For $\lambda_3>\lambda_c$, there is a gapless phase for the non-linear sigma model, and we interpret that as the semi-metal phase. As far as we know, it is still not totally clear whether the low-temperature phase at $\nu=0$ is a Dirac semimetal or a correlated charge insulator. \\

\subsection{$\nu=\pm 1, \pm 2, \pm 3$}
\label{TBLG:nu=2}
\label{oddi}
\hfill

Experimentally a sequence of asymmetric jumps in the electronic compressibility are observed near integer filling factors\cite{Zondiner2020,Wong2020}. In Ref.\cite{Zondiner2020} this is coined ``Dirac fermion revivals'', which is interpreted as due to ``flavor polarization''. In the following we shall assume this interpretation holds and regard the massless Dirac fermion as a good starting point for analyzing the low temperature phases. This point of view is also adopted in \cite{Christos2020}. \\

The mechanism of flavor polarization is likely due to a combination of Coulomb interaction and narrow bands, much like the occurrence of spin polarization (ferromagnetism) in narrow band metal. In the following, we shall assume the simplest flavor polarization mechanism. More complicated ones will not affect our discussions, as long as, after the polarization, the low energy spectrum forms Dirac cones and the number of active bands and the associated low-energy theory are captured correctly.\\

\begin{figure}
	\begin{subfigure}{0.35\textwidth}
		\includegraphics[clip,trim=1cm 0.cm 2cm 0cm,width=\linewidth]{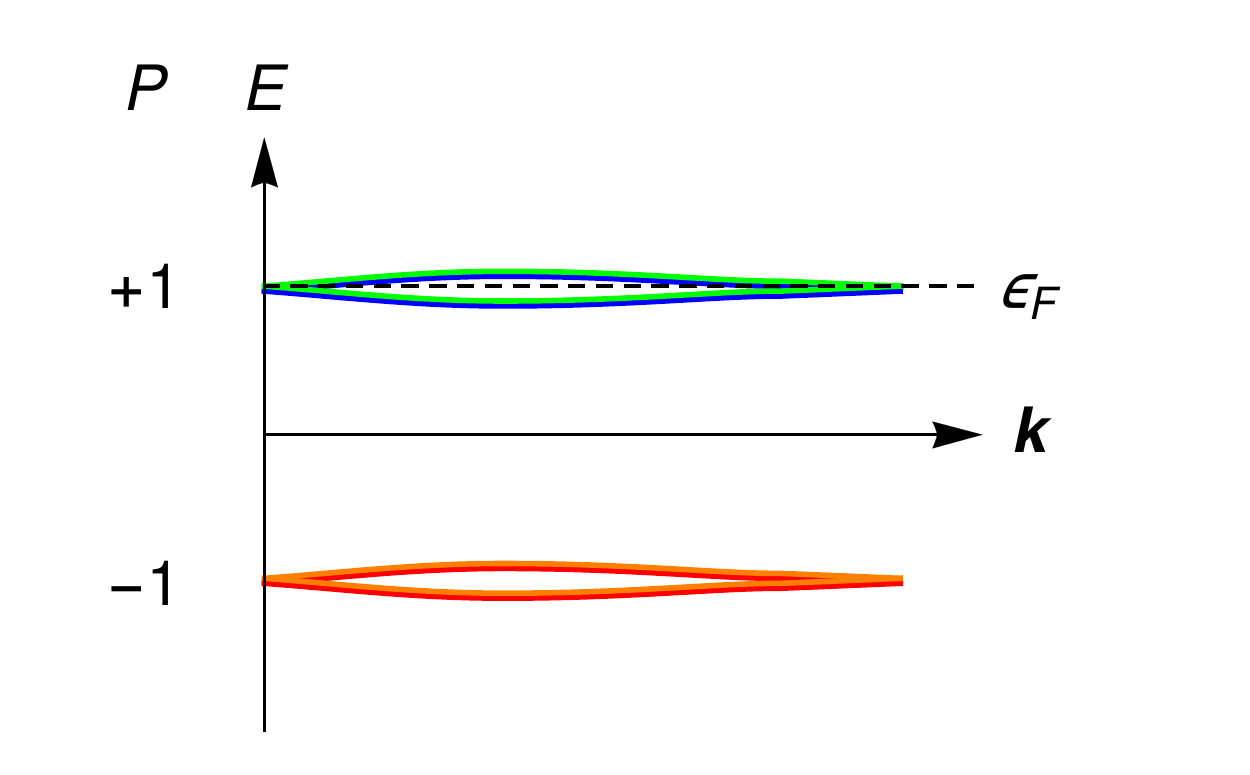}
		\captionsetup{font=normalsize,labelfont=normalsize}
		\caption{} \label{polarization_nu_2}
	\end{subfigure}%
	\hspace*{0.cm}   % maximize separation between the subfigures
	\begin{subfigure}{0.35\textwidth}
		\includegraphics[clip,trim=0cm 0cm 2cm 0cm,width=\linewidth]{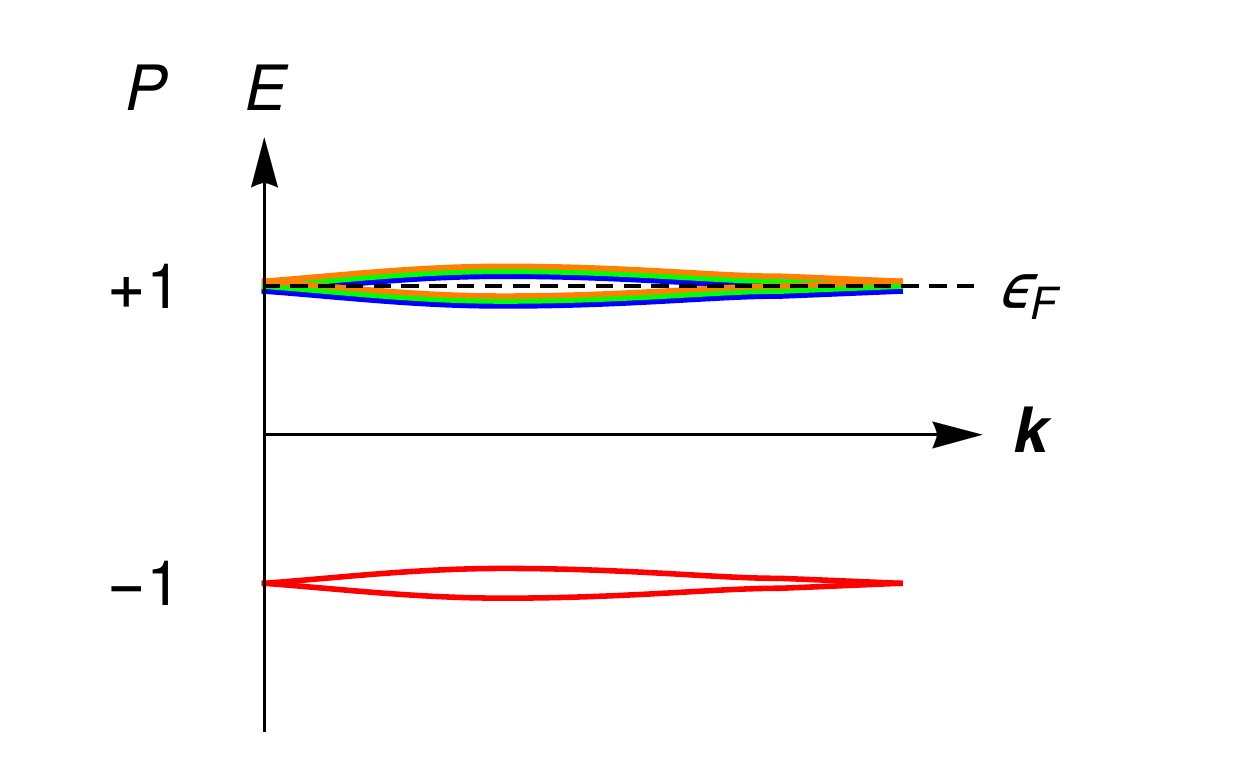}
		\captionsetup{font=normalsize,labelfont=normalsize}
		\caption{} \label{polarization_nu_3}
	\end{subfigure}%
	\hspace*{0.cm}   % maximize separation between the subfigures
	\begin{subfigure}{0.35\textwidth}
		\includegraphics[clip,trim=0cm 0cm 2cm 0cm,width=\linewidth]{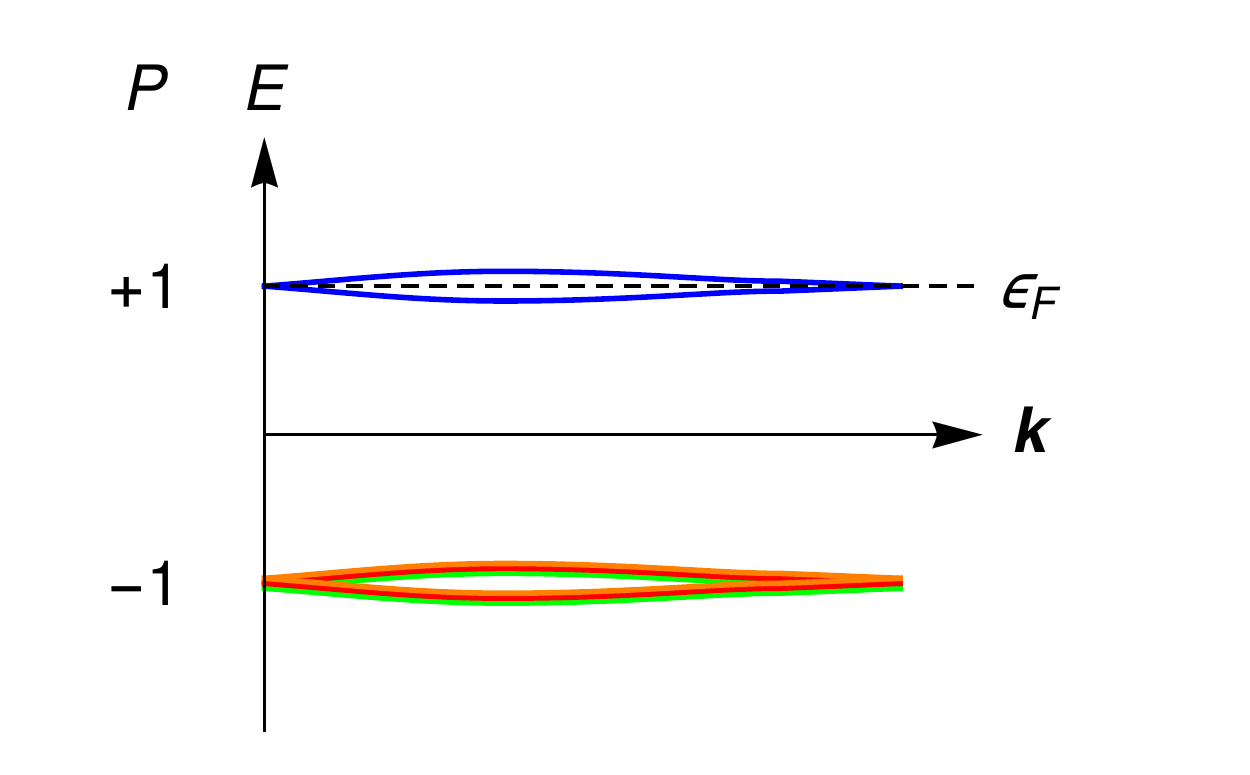}
		\captionsetup{font=normalsize,labelfont=normalsize}
		\caption{} \label{polarization_nu_1}
	\end{subfigure}%
	\caption{
		A caricature of the possible flavor polarization at (a) $\nu=2$ and (b) $\nu=1$ and (c) $\nu=3$. Note that as long as the Fermi level intersects bands with the right degeneracy, the bands below the Fermi energy can overlap without changing the filling factor. For $\nu=-2,-1,-3$ we simply reflect the figures with respect to the $x$-axis.
	}
	\label{polarization}
\end{figure}

For simplicity, we shall consider $\nu\geq 0$ in the following discussion. In the cases of $\nu= 1, 2, 3$ \footnote{$\nu=-1,-2,3$ can be mapped onto $\nu=1,2,3$ by flipping the signs of $\Delta_p$ in \Eq{polm} and $\e_F$ in \Fig{polarization}.}, the Fermi level crossing band number is reduced to $3,2,1$ respectively. This can be caused by a polarization operator 
\be\Delta_p\int d^2x~ \psi(\v x)^\dagger P~\psi(\v x)\label{polm}\ee where
$P$ is a hermitian matrix that commutes with $\Gamma_i$ and satisfies $P^2=I_{16}$. In addition, $P$ needs to be identity matrix for the Moir\'e valley degree of freedom. This leads to the space 
\be P \in \bigcup_{\nu=0}^4 \frac{U(4)}{U(4-\nu) \times U(\nu)},\label{pols}\ee 
\noindent Such a term will shift $4-\nu$ bands on the Moir\'e Brillouin zone upward and the remaining $\nu$ downward by the energy $\pm\Delta_p$.  For example, $P=IZII$ is one such polarization matrix for $\nu =2 $ causing the polarization of graphene valleys, with half of the bands are shifted upward/downward. The resulting spectrum for each $\nu$ is schematically shown in \Fig{polarization}.\\

After the polarization, the low energy free fermion Hamiltonian read 
\be
\hat{H}=\int d^2 x~\psi^\dagger(\v x)\left(- i\G_1^{(\nu)}\p_x - i\G_2^{(\nu)}\p_y\right)\psi(\v x),
\label{drc2}\ee
where, up to a flavor basis transformation,  
\be 
	\G_1^{(\nu)}= XII_{4-\nu}, ~~\G_2^{(\nu)}=YII_{4-\nu}.
\label{c1}
\ee  
Here $I_{4-\nu}$ is the identity matrix for size $4-\nu$. %In the presence of $U(4)$ symmetric electron-electron interaction, the non-linear sigma model that incorporates all possible spontaneous breaking of emergent symmetries has 
The order parameter associated with the Fermi-level crossing bands is
\be Q^{\mathbb{C}}\in \bigcup_{l=0}^{8-2 \nu} ~{U(8-2 \nu)\over U(l)\times U(8-2 \nu-l)},\label{tbmm2}\ee and the associated non-linear sigma model reads
\be
&&W[Q^{\mathbb{C}}, A] ={1\over 2\lambda_3}  \int\limits_{\mathcal{M}}d^3 x \, \tr\Big[\left(\p_\mu Q^{\mathbb{C}}\right)^2\Big]-  \frac{ 2 \pi i }{256 \pi^2}   \Big\{   \int\limits_{\mathcal{B}} \tr \Big[\tilde{Q}^{\mathbb{C}}    \,\left( d \tilde{Q}^{\mathbb{C}} \right)^4 \Big] \nn
&&+ 8 \int\limits_{\mathcal{M}} \tr \Big[ i A Q^{\mathbb{C}} (dQ^{\mathbb{C}})^2  - 2 AFQ^{\mathbb{C}} \Big]
\Big\}.\label{gwzw2}
\ee
where $A_\mu$ is the electromagnetic ($U(1)$) gauge field. Here, associated with each mass manifold the $\s_{xy}$ is given by 
$$\s_{xy}=l-(4-\nu).$$
\\

First consider we $\nu=1,2$. Since experimentally $\s_{xy}=0$ at $\nu=1,2$ \cite{Serlin2020} for $B=0$, we take it as implying the relevant mass manifold is  ${U(8-2\nu)\over U(4-\nu)\times U(4-\nu)}$. The resulting non-linear sigma model can have two phases. One of phases occurs for  $\lambda_3<\lambda_{\rm c}$,  where there is a spontaneous breaking of the $U(8-2\nu)$ symmetry and the sigma model is gapped. We interpret this phase as the ``symmetry-breaking correlated insulator''\footnote{Due to the flavor polarization, the original emergent symmetry is broken. Hence in principle, the low energy massless fermion theory can be regularized. If so there is the possibility that a Mott insulator phase exists.}. The other phase occurs for $\lambda_3>\lambda_{\rm c}$ where the sigma model remains gapless. We interpret that as the semi-metal phase.  \\

For $\nu=3$, the order parameter associated with the Fermi-level crossing bands is  
\be Q^{\mathbb{C}}\in \bigcup_{l=0}^2 ~{U(2)\over U(l)\times U(2-l)}.\label{tbmm1}\ee
The $l=2$ and $l=0$ mass manifolds break the time-reversal symmetry and yield $\s_{xy}=\pm 1$ (see appendix \ref{appendix:enlarge} for the details). Hence the phase corresponds to a quantum anomalous Hall state. This is consistent with the experimental observation of Ref.\cite{Serlin2020}. We stress that the non-zero $\s_{xy}$ associated with mass manifold $l=0$ or $2$ is independent of the choice of flavor polarization $P$ so long as it obeys \Eq{pols}.\\

The mass manifold associated with $l=1$ is $${U(2)\over U(1)\times U(1)}=S^2.$$ In that case $Q^{\mathbb{C}}$ can be replaced by a unit vector  
$\hat{n}\in S^2$. This leads to the bosonization of a small $n$ case (i.e., before the WZW term is stabilized). The resulting nonlinear sigma model was first derived in Ref. \cite{Abanov2000b} and reviewed in appendix \ref{appendix:enlarge}. The action is  given by  
$$W[\hat{n}]={1\over 2\lambda^\prime_3}\int_{\mathcal{M}} d^3x~\left(\p_\mu\hat{n}\right)^2+i\pi H[\hat{n}].$$ 
Here $H[\hat{n}]$ is the Hopf invariant of the $S^3\ra S^2$ mapping. In the presence of such Hopf term  the solitons are fermions \cite{Wilczek1983}. Depending on the value of $\lambda^\prime_3$ this non-linear sigma model can be gapless (preserving the $U(2)$ symmetry) for $\lambda_3'<\lambda_c$, or gapped (spontaneous symmetry breaking) for $\lambda_3'>\lambda_c$. In the latter case the fermionic solitons will be gapped.  In either case %$\tilde{T}$ is conserved and 
$\s_{xy}=0$. We viewed the gapped soliton phase a ``correlated insulator'' arising from symmetry breaking.

\part*{Conclusions} \label{sec:conclu}
\addcontentsline{toc}{part}{Conclusions}

In this paper we have (non-abelian) bosonized two classes of massless fermion theories, the real and complex class, in spatial dimensions 1, 2, and 3. 
The boson theories are non-linear sigma models with the level-1 Wess-Zumino-Witten terms. We have also included three examples showing how to apply the bosonization results.\\

Of course, the goal of bosonization is not simply writing down theories equivalent to that of massless free fermions. For example, the bosonized models manifest what are the ``nearby'' symmetry-breaking states. These symmetry-breaking states can be reached when anisotropy terms are added to the non-linear sigma models. The bosonized theories also allow one to include the effects of strong interaction such as the charge-SU(2) confinement discussed in the first two applications. Moreover, as we have discussed, the main idea of this bosonization is inspired by the physics of topological insulators and superconductors.  Indeed, the results discussed here can be applied to the boundary physics of such systems. \\

In this paper, when restoring the symmetries, we have restricted the bosonic order parameters to fluctuate smoothly. As the result, defect proliferation is not considered. In the literature, it is known that proliferation of symmetry-protected defects can lead to topological order (e.g., in Ref. \cite{Wang2013}). However, in that case, one is restricted to the boundary of topological insulators/superconductors (or more generally symmetry-protected topological states). 
%Because the (quantum) structure of defects, such as the presence of zero modes, which impacts the statistics of the defects, is sensitive to regularization. 
This is because defects are sensitive to short-distance physics, and the symmetries that protect the desired properties of defects can be broken by the regularization. Of course, unless the defects are on the boundary of an SPT, where regularization is provided by the bulk, and no symmetry breaking is necessary. An interesting question is how to reach a topological ordered state without invoking defects. 
%Another proposed application of the smooth order parameter fluctuation was for solving the chiral fermion problem in the standard model \cite{Wen2013}. 
These are directions that warrants more researches.\\

\part*{Acknowledgement} 
\addcontentsline{toc}{part}{Acknowledgement}

\noindent We thank Dr. Lokman Tsui for many helpful discussions. We also thank Professor Oskar Vafek and Professor Eres Berg for useful discussions on the ``Dirac revivals'' in twisted bilayer graphene. This work was primarily funded by the U. S. Department of Energy, Office of Science, Office of Basic Energy Sciences, Materials Sciences and Engineering Division under Contract No. DE-AC02-05-CH11231 (Theory of Materials program KC2301). This research is also funded in part by the Gordon and Betty Moore Foundation.\\

\begin{appendices}

\section{The emergent symmetries for $(2+1)$ and $(3+1)$-D}
\label{appendix:emergentSymm}
\hfill

In this appendix we derive the emergent symmetries of the massless fermion theory (see table \ref{tab:emergentSymm})   for spatial dimension $d=2,3$ (for $d=1$ the result has already been discussed in sections \ref{ES}).

\subsection{Complex class in $(2+1)$-D}
\hfill

In the complex fermion representation, the minimal size of the gamma matrices in two spatial dimensions is $2\times 2$. If the fermion has $n$ flavors, modulo a basis transformation, we have
\be
&&S_0 = \int d^3 x \, \psi^\dagger (\partial_0  - i  \sum_{i=1}^2 \Gamma_i \partial_i )\psi~~{\rm where}\nn
 &&\Gamma_1 =  Z  I_n, ~~~ \Gamma_2 = XI_n.
 \label{S02C}
\ee
It's easy to see that the full emergent symmetries include $U(n)$ transformations in the flavor degrees of freedom. %Here unitary transformation preserves the fermion anti-commutation relation. 
In addition, there are discrete symmetries, namely, charge conjugation and time-reversal symmetries. To summarize, \Eq{S02C} is invariant under
\begin{align}
&\text{$U(n)$ symmetry}: \notag\\
&U(n): \psi \rightarrow \left( I\otimes g \right) \psi  ~~~~\text{where }g \in U(n) \notag\\
&\text{Charge conjugation symmetry}:\notag\\
&C: \psi \rightarrow  \left( I \otimes I_n \right) (\psi^\dagger)^T \notag\\
&\text{Time reversal symmetry}:\notag\\
&T: \psi \rightarrow  \left( Y \otimes I_n \right) \psi . 
\label{symm2C}
\end{align}

\subsection{Real class in $(2+1)$-D}
\hfill

In the Majorana fermion representation, the minimal size of the gamma matrices in two spatial dimensions,  is $2\times 2$.
If the fermion has $n$ flavors, modulo a basis transformation, we have
\be
&&S_0 = \int d^3x \, \chi^T (\partial_0  - i  \sum_{i=1}^2 \Gamma_i \partial_i )\chi~~{\rm where}\nn
&& \Gamma_1 =  Z  I_n, ~~~ \Gamma_2 = XI_n.
\label{S02R}
\ee
It's easy to see that the full emergent symmetries include $O(n)$ transformations in the flavor degrees of freedom. %Here orthogonal transformation preserves the realness of the fermion operator and the fermion anti-commutation relation. 
In addition, there is time reversal symmetry. To summarize, \Eq{S02R} is invariant under
\begin{align}
&\text{$O(n)$ symmetry}:\notag\\
&O(n): \chi \rightarrow \left( I\otimes g \right) \chi  ~~~~\text{where }g \in O(n) \notag\\
&\text{Time reversal symmetry}:\notag\\
&T: \chi \rightarrow  \left( E \otimes I_n \right) \chi. 
\label{symm2R}
\end{align}

\subsection{Complex class in $(3+1)$-D}
\hfill

In the complex fermion representation, the minimal size of the gamma matrices in three spatial dimensions is $4\times 4$. If the fermion has $n$ flavors, modulo a basis transformation, we have
\be
&&S_0 = \int d^4x \, \psi^\dagger (\partial_0  - i  \sum_{i=1}^3 \Gamma_i \partial_i )\psi~~{\rm where}\nn
&& \Gamma_1 =  ZI  I_n, ~~~ \Gamma_2 = XII_n , ~~~ \Gamma_3 = YZI_n.
\label{S03C}\ee
Similar to the $(1+1)$-D case,  the chirality matrix $$\Gamma_5:=-i\Gamma_1 \Gamma_2 \Gamma_3=I Z I_n$$ commutes with the gamma matrices.
As a result, the full emergent include chiral $U(n)$ transformations, namely, $U_+(n) \times U_-(n)$ (see below). In addition, there are discrete symmetries, namely, charge conjugation, and time-reversal symmetries. To summarize, \Eq{S03C} is invariant under
\begin{align}
&\text{Chiral $U(n)$ symmetry}:\notag\\
&U_+(n)\times U_-(n): \psi \rightarrow \left( IP_+\otimes g_+ + IP_-\otimes g_- \right) \psi  ~~~~\text{where }g_\pm \in U_\pm(n) \notag\\
&\text{Charge conjugation symmetry}:\notag\\
&C: \psi \rightarrow  \left( IX \otimes I_n \right) (\psi^\dagger)^T \notag\\
&\text{Time reversal symmetry}:\notag\\
&T: \psi \rightarrow  \left( YZ \otimes I_n \right) \psi, 
\label{symm3C}
\end{align}
\noindent where $$P_{\pm}:={I\pm Z\over 2}.$$

\subsection{Real class in $(3+1)$-D}
\hfill

In the Majorana fermion representation, the minimal size of the gamma matrices in three spatial dimensions is $4\times 4$.
If the fermion has $n$ flavors, modulo a basis transformation, we have
\be
&&S_0 = \int d^4x \, \chi^T (\partial_0  - i  \sum_{i=1}^3 \Gamma_i \partial_i )\chi~~{\rm where}\nn
&& \Gamma_1 =  ZI  I_n, ~~~ \Gamma_2 = XII_n , ~~~ \Gamma_3 = YYI_n.
\label{S03R}\ee 
Although we can still define $\Gamma_1 \Gamma_2 \Gamma_3=I E I_n$, this is an anti-symmetric matrix with complex eigenvectors hence cannot be used to define chirality for Majorana (real) fermions.\\

\noindent To find the most general continuous unitary symmetry, notice that only $II$ and $IE$ commute with the first two Pauli matrices in $\G_{1,2,3}$.  Hence the symmetry transformation needs to be in the form  $$\chi \rightarrow \left( II\otimes g_1 - IE \otimes g_2 \right) \chi.$$ Here $g_1$ and $g_2$ are orthogonal matrices (which preserve the realness of the Majorana fermion operator and their anti-commutation relation). The condition of $g_1$ and $g_2$ being orthogonal matrices is equivalent to requiring $g_1+ i g_2 \in U(n)$\footnote{As an algebraic relation, $IE$ plays the role of $i$ here because $(IE)^2=-I_4$}. Thus, the unitary continuous symmetry is $U(n)$. In addition, there is time-reversal symmetry. To summarize, \Eq{S03R} is invariant under
\begin{align}
&\text{$U(n)$ symmetry}:\notag\\
&U(n): \chi \rightarrow \left( II\otimes g_1 - IE\otimes g_2 \right) \chi  ~~~~\text{where }g:= g_1 + i g_2 \in U(n) \notag\\
&\text{Time reversal symmetry}:\notag\\
&T: \chi \rightarrow  \left( EZ \otimes I_n \right) \chi 
\label{symm3R}
\end{align}

\section{The mass manifolds, homotopy groups and symmetry transformations }
\label{appendix:massManifold}
\hfill

In this appendix we derive the mass manifolds in table \ref{tab:massManifold}, and the transformation of $Q^{\mathbb{C}}$ and $Q^{\mathbb{R}}$ under the emergent symmetries in table \ref{tab:symmRestore} for $d=2,3$ (the $d=1$ case has been discussed in section \ref{massManifold1d} and \ref{restoreEmergent}). In addition, we discuss the relevant homotopy groups of the mass manifolds. For sufficiently large flavor number $n$, it turns out that the $\pi_{D+1}$, relevant to the existence of WZW term, are always equal to $\mathbb{Z}$. On the other hand, $\pi_{D-1}$, relevant to the existence of non-trivial soliton, are $\mathbb{Z}$ or $\mathbb{Z}_2$ depending on whether the class is complex or real. \footnote{Although we shall not further discuss it in this paper, the  $\mathbb{Z}$ or $\mathbb{Z}_2$ soliton classifications are originated from K-theory\cite{Kitaev2009} and the Bott periodicity \cite{Bott1957,Bott1959}. Therefore this statement holds true in even higher dimensions.}.\\

\subsection{Complex class in $(2+1)$-D}
\hfill

In $(2+1)$-D, complex fermion representation,  the gamma matrices in \Eq{S02C} are 
\begin{align}
\label{gamma2C}
\Gamma_1 = ZI_n,~~\Gamma_2 = XI_n. 
\end{align} 
The most general hermitian mass matrix $M$ satisfying $$\{M, \Gamma_i \}=0~{\rm and}~M^2 = I_{2n}$$ is of the form
\begin{align}
\label{mass2C}
M = Y \otimes H := Y\otimes Q^{\mathbb{C}} 
\end{align}
\noindent where $Q^{\mathbb{C}}= H$ is an $n \times n$ hermitian matrix satisfying $H^2=I_n$. This last condition requires the eigenvalues of $H$ to be $\pm 1$. Assuming $l$ of the eigenvalues are $+1$ and $n-l$ are $-1$, we have
\begin{align*}
Q^{\mathbb{C}} =W \cdot diag(\underbrace{+1,...,+1}_l,\underbrace{-1,...,-1}_{n-l}) \cdot W^{\dagger}.
\end{align*}
Different $Q^{\mathbb{C}}$ are characterized by the unitary matrix (whose columns are eigenvectors) $W \in U(n)$. However, not all $W$ will yield distinct $Q^{\mathbb{C}}$. Under the transformation 
\begin{align*}
W \rightarrow W \cdot \begin{pmatrix} \tilde{W}_1 & 0 \\ 0 & \tilde{W}_2 \end{pmatrix} \text{~~~where~~~} \tilde{W}_1 \in U(l) \text{~~~and~~~} \tilde{W}_2 \in U(n-l),
\end{align*}
$Q^{\mathbb{C}}$ is unchanged. Thus the mass manifold $\mathcal{M}$ is the union of the quotient spaces
\begin{align*}
\mathcal{M}= \bigcup_{l=0}^{n} \frac{U(n)}{ U(l) \times U(n-l)} .
\end{align*}
These quotient spaces are called ``complex Grassmannians''. Note that  $\mathcal{M}$ contains $n+1$ disconnected components.\\

Under the action of the emergent symmetries in \Eq{symm2C}, the order parameter $Q^{\mathbb{C}}$ transforms as
\begin{align*}
Q^{\mathbb{C}} &\xrightarrow{U(n)} g^\dagger \cdot Q^{\mathbb{C}} \cdot g \\
Q^{\mathbb{C}} &\xrightarrow{C}  \left(Q^{\mathbb{C}}\right)^T \\
Q^{\mathbb{C}} &\xrightarrow{T} - \left(Q^{\mathbb{C}} \right)^* 
\end{align*}
\noindent Among them, the time reversal transformation changes the signs of all eigenvalues and thus exchanges $l$ and $n-l$. Therefore only when $$Q^{\mathbb{C}}\in \frac{U(n)}{U(n/2) \times U(n/2)}~{\rm for~}n \in even$$ does the time reversal transformed $Q^{\mathbb{C}}$ stay in the same component of the mass manifold.
Only in this manifold, fluctuating $Q^{\mathbb{C}}$ can restore the {\it full} emergent symmetries.\\

Using the long exact sequence of the homotopy group corresponding to the fibration,
\begin{align*}
0 \rightarrow U(\frac{n}{2}) \times U(\frac{n}{2}) \rightarrow  U(n) \rightarrow \frac{U(n)}{ U(\frac{n}{2}) \times U(\frac{n}{2})} \rightarrow 0,
\end{align*}
 we can deduce the homotopy groups of the complex Grassmannian from the homotopy groups of $U(n)$ (see, e.g., \cite{Hatcher2001}). In table \ref{tab:homotopy2C} we list the results of the second, third, and fourth homotopy groups. They are relevant for determining the existence of solitons, $\theta$-term, and WZW term. These results are used in appendix \ref{appendix:fermionInt}.

\begin{table}
	\centering
	\begin{tabular}{ |c|c|c|c|c| }
		\hline
		$n$ (even)		&	Mass manifold	& 	\thead{$\pi_2$ \\	(soliton)}	&	$\thead{\pi_3\\ (\theta {\rm~ term})}$		&	\thead{$\pi_4$\\ (WZW)} \\
		\hline
		$\ge 4$ 	&   $\frac{U(n)}{U(n/2) \times U(n/2)}$			&			\framebox{$\mathbb{Z}$}	&		\framebox{$0$}	&		\framebox{$\mathbb{Z}$}	\\
		\hline
		$2$		&	$\frac{U(2)}{U(1) \times U(1)} = S^2$		&  		\framebox{$\mathbb{Z}$}		&		$\mathbb{Z}$	&		$\mathbb{Z}_2$	\\
		\hline
	\end{tabular}
	\caption{The homotopy groups of the complex Grassmannian $\frac{U(n)}{ U(n/2) \times U(n/2)}$. We box the homotopy group when it is stabilized, i.e., no longer changes with increasing $n$.}
	\label{tab:homotopy2C}
\end{table}

\subsection{Real class in $(2+1)$-D}
\hfill

The massless fermion Hamiltonian is given by \Eq{S02R}, where the gamma matrices are given by 
\begin{align}
\label{gamma2R}
\Gamma_1 = ZI_n~~\Gamma_2 = XI_n
\end{align}
The most general purely imaginary {\it antisymmetric} mass matrix (requirement due to hermiticity and Majorana condition) $M$ satisfying $$\{M, \Gamma_i \}=0~{\rm and~}M^2 = I_{2n}$$  is of the form
\begin{align}
\label{mass2R}
M = Y \otimes S := Y\otimes Q^{\mathbb{R}} 
\end{align}
\noindent where $Q^{\mathbb{R}}= S$ is an $n \times n$ real symmetric matrix satisfying $S^2=I_n$. This last condition requires the eigenvalues of $Q^{\mathbb{R}}$ to be $\pm 1$. Assuming $l$ of the eigenvalues are $+1$ and the rest are $-1$, we have
\begin{align*}
Q^{\mathbb{R}} =W \cdot diag(\underbrace{+1,...,+1}_l,\underbrace{-1,...,-1}_{n-l}) \cdot W^{\dagger}.
\end{align*}
Hence different $Q^{\mathbb{R}}$ are characterized  by the orthogonal matrix $W \in O(n)$. However, not all $W$  yield distinct $Q^{\mathbb{R}}$. Under the transformation 
\begin{align*}
W \rightarrow W \cdot \begin{pmatrix} \tilde{W}_1 & 0 \\ 0 & \tilde{W}_2 \end{pmatrix} \text{~~~where~~~} \tilde{W}_1 \in O(l) \text{~~~and~~~} \tilde{W}_2 \in O(n-l),
\end{align*}
$Q^{\mathbb{R}}$ is unchanged. Thus the mass manifold is the union of quotient spaces called ``real Grassmannians''
\begin{align*}
\mathcal{M}= \bigcup_{l=0}^n \frac{O(n)}{ O(l) \times O(n-l)}.
\end{align*}
Here,  $\mathcal{M}$ contains $n+1$ disconnected components.\\

Under the action of the emergent symmetries in \Eq{symm2R}, the order parameter $Q^{\mathbb{R}}$ transforms as
\begin{align*}
Q^{\mathbb{R}} &\xrightarrow{O(N)} g^T \cdot Q^{\mathbb{R}} \cdot g \\
Q^{\mathbb{R}} &\xrightarrow{T} -  Q^{\mathbb{R}}. 
\end{align*}
Among them, the time reversal transformation changes the signs of all eigenvalues and thus exchanges $l$ and $n-l$. Therefore only when $$Q^{\mathbb{R}}\in \frac{O(n)}{O(n/2) \times O(n/2)}~{\rm for~}n \in even$$ does the time reversal transformed $Q^{\mathbb{R}}$ stay in the same component of the mass manifold. Only in the mass manifold, fluctuating $Q^{\mathbb{R}}$  can restore the {\it full} emergent symmetries.\\

Using the long exact sequence of the homotopy group associated with the fibration,
\begin{align*}
0 \rightarrow O(\frac{n}{2}) \times O(\frac{n}{2}) \rightarrow  O(n) \rightarrow \frac{O(n)}{ O(\frac{n}{2}) \times O(\frac{n}{2})} \rightarrow 0,
\end{align*}
we can deduce the homotopy groups of the real Grassmannian from the homotopy groups of $O(n)$ (see e.g., \cite{Hatcher2001}). We list the results of the second, third, and fourth homotopy groups in table \ref{tab:homotopy2R}. They are relevant for determining the existence of solitons, $\theta$-term, and WZW term. These results are used in appendix \ref{appendix:fermionInt}.

\begin{table}
	\centering
	\begin{tabular}{ |c|c|c|c|c| }
		\hline
		%\hline\hline
		%\multicolumn{6}{|c|}{$(2+1)D$ with no constraint: $\frac{O(N)}{O(N/2)\times O(N/2)}$}  \\
		%\hline\hline
		$n$ (even) 		&	Mass manifold	& 	\thead{$\pi_2$ \\	(soliton)}	&		$\thead{\pi_3\\ (\theta {\rm~ term})}$			&	\thead{$\pi_4$\\ (WZW)} \\
		\hline
		$\ge 10$ 	&  	$\frac{O(n)}{O(n/2)\times O(n/2)}$		&			\framebox{$\mathbb{Z}_2$}		&		\framebox{$0$}	&		\framebox{$\mathbb{Z}$} 	\\
		\hline
		$2$		&	$S^1$			&	$0$	&		$0$	&		$0$	 	\\
		$4$		&	$\frac{ S^2 \times S^2 }{\mathbb{Z}_2}$			&	$\mathbb{Z}^2$	&		$\mathbb{Z}^2$	&		$\mathbb{Z}_2^2$	 	\\
		$6$		&				&	\framebox{$\mathbb{Z}_2$}	&	 $0$		&	$\mathbb{Z}$	 	\\
		$8$		&				&	\framebox{$\mathbb{Z}_2$}	&	$0$	&		$\mathbb{Z}^3$		 \\
		%$\geq 10$		&		&		\framebox{$\mathbb{Z}_2$}		&		\framebox{$0$}	&		\framebox{$\mathbb{Z}$}		\\
		\hline
	\end{tabular}
	\caption{The homotopy groups of the real Grassmannian $\frac{O(n)}{ O(n/2) \times O(n/2)}$. We box the homotopy group when it is stabilized , i.e., no longer changes with increasing $n$. }
	\label{tab:homotopy2R}
\end{table}

\subsection{Complex class in $(3+1)$-D}
\hfill

The massless fermion Hamiltonian is given by \Eq{S03C}, where the gamma matrices are given by 
\begin{align}
\label{gamma3C}
\Gamma_1 = ZII_n,~\Gamma_2 = XII_n,~\Gamma_3 = YZI_n
\end{align} 
The most general hermitian mass matrix $M$ satisfying $$\{M, \Gamma_i \}=0$$ is of the form
\begin{align}
\label{mass3C}
M= YX \otimes H_1 + YY \otimes H_2
\end{align}
\noindent Here $H_{1,2}$ are $n\times n$ hermitian matrices. It's easy to check that the extra condition on the mass matrix $$M^2 = I_{4n}$$ is equivalent to requiring
\begin{align*}
Q^{\mathbb{C}} := H_1 + i H_2 \in U(n)
\end{align*}
Thus, the mass manifold is $U(n)$. Here the mass manifold contains only a single component.\\

Under the action of the emergent symmetries in \Eq{symm3C}, the order parameter $Q^{\mathbb{C}}$ transforms as
\begin{align*}
Q^{\mathbb{C}} &\xrightarrow{U(n)\times U(n)} g_-^\dagger \cdot Q^{\mathbb{C}} \cdot g_+ \\
Q^{\mathbb{C}} &\xrightarrow{C}  \left(Q^{\mathbb{C}}\right)^T \\
Q^{\mathbb{C}} &\xrightarrow{T} \left(Q^{\mathbb{C}} \right)^* 
\end{align*}
\noindent Fluctuating $Q^{\mathbb{C}}$ within  $U(n)$ can restores the  {\it full} emergent symmetries.\\

We list the results of the second, third, and fourth homotopy groups in table \ref{tab:homotopy3C}. They are relevant for determining the existence of solitons, $\theta$-term, and WZW term. These results are used in appendix \ref{appendix:fermionInt}.

\begin{table}
	\centering
	\begin{tabular}{ |c|c|c|c|c|c|c| }
		\hline
		%\multicolumn{7}{|c|}{$(3+1)D$ with $Q$ constraint: $U(N)$} \\
		%\hline\hline
		$n$		&	Mass manifold					&	\thead{$\pi_3$ \\ (soliton) } & $\thead{\pi_4\\ (\theta {\rm~ term})}$	 & \thead{$\pi_5$ \\ (WZW) }\\
		\hline
		$\ge 3$ 	&   $U(n)$			&		\framebox{$\mathbb{Z}$}	&		\framebox{$0$}	&		\framebox{$\mathbb{Z}$} 	\\
		\hline
		$1$		&	$S^1$			&		$0$	&		$0$	&		$0$	\\
		$2$		&	$\frac{S^3\times S^1}{\mathbb{Z}_2}$			&		\framebox{$\mathbb{Z}$}	&		$\mathbb{Z}_2$	&		$\mathbb{Z}_2$	\\
		\hline
	\end{tabular}
	\caption{The homotopy groups of $U(n)$. We box the homotopy group when it is stabilized, i.e., no longer changes with increasing $n$. }
	\label{tab:homotopy3C}
\end{table}

\subsection{Real class in $(3+1)$-D}
\hfill

The massless fermion Hamiltonian is given by \Eq{S03R}, where the gamma matrices are given by 
\begin{align}
\label{gamma3R}
\Gamma_1 = ZII_n,~\Gamma_2 = XII_n,~\Gamma_3=YYI_n
\end{align}
The most general antisymmetric (to ensure hermiticity) mass matrix $M$ satisfying $$\{M, \Gamma_i \}=0$$ is of the form
\begin{align}
\label{mass3R}
M = YX \otimes S_1 + YZ \otimes S_2
\end{align}
Here $S_{1,2}$ are $n\times n$ real symmetric matrices. It's easy to check that the extra condition on the mass matrix $$M^2 = I_{4n}$$ is equivalent to requiring
\begin{align*}
Q^{\mathbb{R}} := S_1 + i S_2 \in \text{symmetric~} U(n).
\end{align*}
According to the ``Autonne decomposition'' (e.g., corollary 2.6.6 of \cite{Horn2012}), any symmetric unitary matrix can be decomposed into
\begin{align*}
Q^{\mathbb{R}}= U \cdot U^T
\end{align*}
\noindent where $U$ is a general $n\times n$ unitary matrix. However, not all $U$ will yield different $Q^{\mathbb{R}}$. The transformation 
$$U\ra U\cdot O,~\text{where}~O \in O(n)$$ leaves $Q^{\mathbb{R}}$ unchanged.
Thus the mass manifold is
\begin{align*}
\mathcal{M}_{m} = \frac{U(n)}{O(n)}.
\end{align*}
\noindent This mass manifold is called the ``real Lagrangian Grassmannian'', which contains a single component.\\

Under the action of the emergent symmetries in \Eq{symm3R}, the order parameter $Q^{\mathbb{R}}$ transforms as
\begin{align*}
&Q^{\mathbb{R}} \xrightarrow{U(n)}  g^T \cdot Q^{\mathbb{R}} \cdot g \\
&Q^{\mathbb{R}} \xrightarrow{T}  \left( Q^{\mathbb{R}} \right)^*  \\
\end{align*}
Fluctuating $Q^{\mathbb{R}}$  in $U(n)/O(n)$ can restore the {\it  full} emergent symmetries.\\

Using the long exact sequence of the homotopy group associated with  the fibration,
\begin{align*}
0 \rightarrow O(n)  \rightarrow  U(n) \rightarrow \frac{U(n)}{ O(n)} \rightarrow 0,
\end{align*}
we can deduce the homotopy groups of the  real Lagrangian Grassmannian from the homotopy group of $U(n)$ and $O(n)$ (see e.g., \cite{Hatcher2001}). We list the results of the second, third, and fourth homotopy groups in table \ref{tab:homotopy3R}. They are relevant for determining the existence of solitons, $\theta$-term, and WZW term. These results are used in appendix \ref{appendix:fermionInt}.

\begin{table}
	\centering
	\begin{tabular}{ |c|c|c|c|c| }
		\hline
		$n$ 		&	Mass manifold				&		\thead{$\pi_3$ \\ (soliton) } &$\thead{\pi_4\\ (\theta {\rm~ term})}$ & \thead{$\pi_5$ \\ (WZW) }\\
		\hline
		$\ge 6$ 	&  	$\frac{U(n)}{O(n)}$	&			\framebox{$\mathbb{Z}_2$}		&		\framebox{$0$}	&		\framebox{$\mathbb{Z}$} 	\\
		\hline
		$1$		&	$S^1$		&		$0$	&		$0$	 & $0$	\\
		$2$		&	$\frac{S^1 \times S^2}{\mathbb{Z}_2}$		&		$\mathbb{Z}$	&		$\mathbb{Z}_2$	 &	$\mathbb{Z}_2$ \\
		$3$		&	&	\framebox{$\mathbb{Z}_2$}	&	$0$ &  $\mathbb{Z}\times \mathbb{Z}_2$				\\
		$4$		&	&	\framebox{$\mathbb{Z}_2$}		&	$\mathbb{Z}$		&	$\mathbb{Z}\times \mathbb{Z}_2^2$	\\
		$5$		&	&		\framebox{$\mathbb{Z}_2$}		&		\framebox{$0$}	&		 $\mathbb{Z}\times \mathbb{Z}_2$	\\
		\hline
	\end{tabular}
	\caption{The homotopy groups of the real Lagrangian Grassmannian $\frac{U(n)}{ O(n)}$. We box the homotopy group when it is stabilized, i.e., no longer changes with increasing $n$. }
	\label{tab:homotopy3R}
\end{table}

\section{The anomalies of the fermion theories}
\label{appendix:anomaliesf}

In this section, we shall use the  heuristic method introduced in subsection \ref{Anomalyheuristic} to determine the anomalies associated with the emergent symmetries of the massless free fermion theory in $(1+1),(2+1)$ and $(3+1)$-D.  For each massless fermion theory, we shall determine 1) the largest subgroup of the continuous symmetry that is anomaly free, 2) whether the discrete symmetries are anomalous after imposing a regularization mass that is invariant under the  anomaly-free part of the continuous symmetry. Readers are referred to table \ref{emsymm} for the emergent symmetries of the massless free fermion theories;  table \ref{tab:massManifold} for the general form of mass terms ($Q^{\mathbb{C},\mathbb{R}}$) , and the topological space (mass manifold) they reside in;  %
table \ref{tab:symmRestore} for the transformations of $Q^{\mathbb{C},\mathbb{R}}$ under the emergent symmetries. \\ %Later, in subsection \ref{smano}, we shall determine the anomaly of the non-linear sigma model. This will allow us to make a comparison with the anomalies of the massless free fermion theories.\\

\subsection{The choice of discrete symmetry generators}\label{choice}
\hfill

Before discussing the free fermion anomalies, we would like to discuss the reason for making the particular choice for the time-reversal and charge conjugation generators in table \ref{tab:emergentSymm}. As mentioned in the main text, such  choice is not unique. Because we can generate new choices by compounding the $C$ and the $T$ in  table \ref{tab:emergentSymm} with other unitary symmetries.\\ %%For example, in the case of complex class in $(1+1)$-D, one can compound the $T$ and $C$ with a chiral $U_+(n) \times U_-(n)$ transformation. Had we done so, these discrete symmetries would also be anomalous.\\

In $(1+1)$-D and $(3+1)$-D we specifically choose the generators of $C$ and $T$ so that when a regularization mass is picked to commute with the maximal anomaly-free subgroup of the continuous symmetry, it is also charge conjugation and time reversal invariant. In $(2+1)$-D,  a regularization mass that commutes with the entire continuous symmetry group and $C$ exists. In contrast, it is impossible to pick a $T$ so that the chosen mass is also time reversal invariant.

\subsection{Complex class in $(1+1)$-D}
\hfill

As discussed in section\ref{Anomalyheuristic},  whether a  symmetry group is anomalous depends on whether there exists a regularization mass that is invariant under its action.   For continuous symmetries, the existence of an invariant regularization mass guarantees the possibility of gauging such  symmetries. \\
 
As shown in table \ref{emsymm} the continuous part of the emergent symmetries form  the $U_+(n) \times  U_-(n)$ group, and a general mass term has the following form \begin{align*}
M=X\otimes H_1 + Y \otimes H_2~~{\rm where}~~ H_1 + i H_2:=Q^\mathbb{C} \in U(n)
\end{align*}
Under the action of  $U_+(n) \times  U_-(n)$ these mass terms transform according to 
\begin{align*}
Q^\mathbb{C}\ra  g_-^\dagger \cdot Q^\mathbb{C} \cdot g_+,~~{\rm where}~~ (g_+, g_-) \in \, U_+(n) \times  U_-(n).
\end{align*} 
Since there is no (regularization) mass invariant under the action of  the entire $U_+(n) \times  U_-(n)$, it follows that $U_+(n) \times U_-(n)$ is anomalous. \\

The largest anomaly free subgroup is the diagonal $U(n)$, i.e., $g_+=g_-=g \in U(n)$. In this case we can choose  $Q^\mathbb{C} =I_n$ (i.e., $H_1=I_n$ and $H_2=0$), such that it is invariant under the diagonal $U(n)$. One can thus use 
\begin{align*}
M_{\rm reg}= X \otimes I_n
\end{align*}
\noindent as the regularization mass.\\

Note that this mass term is invariant under the time-reversal and charge-conjugation symmetries
\begin{align*}
Q^\mathbb{C}  \xrightarrow{T}& (Q^\mathbb{C})^T = I_n \\
Q^\mathbb{C}  \xrightarrow{C}& (Q^\mathbb{C})^* = I_n.
\end{align*}
\noindent Consequently there is no anomaly for these discrete symmetries after imposing the diagonal $U(n)$-invariant regularization mass.\\

\subsection{Real class in $(1+1)$-D}
\hfill

As shown in table \ref{emsymm} the continuous part of the emergent symmetries form  the $O_+(n) \times  O_-(n)$ group, and a general mass term has the following form \begin{align*}
M=Y\otimes S + X \otimes (iA) ~~{\rm where}~~S + A:=Q^\mathbb{R} \in O(n).
\end{align*}
Under the action of  $O_+(n) \times  O_-(n)$ these mass terms transform according to 
\begin{align*}
Q^\mathbb{R}\ra  g_-^T \cdot Q^\mathbb{R} \cdot g_+,~~{\rm where}~~ (g_+, g_-) \in \, O_+(n) \times  O_-(n).
\end{align*} 
Since there is no (regularization) mass invariant under the action of the entire $O_+(n) \times O_-(n)$, it follows that $O_+(n) \times O_-(n)$ is anomalous. \\

The largest anomaly free subgroup is the diagonal $O(n)$, i.e., $g_+=g_-=g \in O(n)$. In this case  $Q^\mathbb{R} =I_n$ (i.e., $S=I_n$ and $A=0$)  is invariant under the diagonal $O(n)$. One can thus use 
\begin{align*}
M_{\rm reg}= Y\otimes I_n
\end{align*}
\noindent as the regularization mass.\\

Since this mass term is invariant under the time-reversal, i.e.,
\begin{align*}
Q^\mathbb{R}  \xrightarrow{T}& (Q^\mathbb{R})^T = I_n, \\
\end{align*}
there is no anomaly for time-reversal symmetry after imposing the diagonal $O(n)$-invariant regularization mass.\\

\subsection{Complex class in $(2+1)$-D }
\hfill

As shown in table \ref{emsymm} the continuous part of the emergent symmetries form  the $U(n)$ group, and a general mass term has the following form \begin{align*}
M=Y \otimes Q^\mathbb{C}~~{\rm where}~~Q^\mathbb{C} \in  \bigcup_l\frac{U(n)}{U(l) \times U(n-l)}
\end{align*}
Under the action of  $U(n)$ these mass terms transform according to 
\begin{align*}
Q^\mathbb{C}\ra  g^\dagger \cdot Q^\mathbb{C} \cdot g,~~{\rm where}~~g\in \, U(n).
\end{align*} 

Because $Q^\mathbb{C}=\pm I_n$  is invariant under the action of $U(n)$ 
one can choose 
\begin{align*}
M_{\rm reg}=\pm Y \otimes I_n
\end{align*}
as the regularization mass. Hence the entire $U(n)$ is anomaly free.\\

It is easy to see that  $Q^\mathbb{C}= \pm I_n$ is the only  $U(n)$ preserving mass term. Although this mass term is invariant the charge-conjugation 
\begin{align*}
Q^\mathbb{C}  \xrightarrow{C}& (Q^\mathbb{C})^T = \pm I_n,
\end{align*}
it is odd under the time-reversal symmetry
\begin{align*}
Q^\mathbb{C}  \xrightarrow{T}& -(Q^\mathbb{C})^* = \mp I_n.
\end{align*}
Therefore the time-reversal symmetry is anomalous after imposing the $U(n)$-invariant regularization mass. Note that one cannot avoid this anomaly by compounding the time-reversal symmetry with other unitary symmetries, because all of the unitary symmetries leave the mass term invariant.\\

\subsection{Real class in $(2+1)$-D}
\hfill

As shown in table \ref{emsymm} the continuous part of the emergent symmetries form  the $O(n)$ group, and a general mass term has the following form \begin{align*}
M=Y \otimes Q^\mathbb{R}~~{\rm where}~~Q^\mathbb{R} \in  \bigcup_l\frac{O(n)}{O(l) \times O(n-l)}.
\end{align*}
Under the action of  $O(n)$ these mass terms transform according to 
\begin{align*}
Q^\mathbb{R}\ra  g^T \cdot Q^\mathbb{R} \cdot g,~~{\rm where}~~g\in \, O(n).
\end{align*} 

Because $Q^\mathbb{R}=\pm I_n$ is invariant under the action of $U(n)$, one can choose 
\begin{align*}
M_{\rm reg}=\pm Y \otimes I_n
\end{align*}
as the regularization mass. Hence the entire $O(n)$ is anomaly free.\\

It is easy to see that $Q^\mathbb{R}=\pm I_n$ is the only $O(n)$ preserving mass term. However, this mass term is odd under the time-reversal symmetry
\begin{align*}
Q^\mathbb{R}  \xrightarrow{T}& -Q^\mathbb{R} = \mp I_n.
\end{align*}
Therefore the time-reversal symmetry is anomalous after imposing the $O(n)$-invariant regularization mass. Note that one cannot avoid this anomaly by compounding the time-reversal symmetry with other unitary symmetries, because all of the unitary symmetries leave the mass term invariant.\\

\subsection{Complex class in $(3+1)$-D}
\hfill

As shown in table \ref{emsymm} the continuous part of the emergent symmetries form  the $U_+(n) \times  U_-(n)$ group, and a general mass term has the following form \begin{align*}
M=M_{\rm reg}=YX\otimes H_1 + YY \otimes H_2~~{\rm where}~~ Q^\mathbb{C}:= H_1 + i H_2 \in U(n)
\end{align*}
Under the action of  $U_+(n) \times  U_-(n)$ these mass terms transform according to 
\begin{align*}
Q^\mathbb{C}\ra  g_-^\dagger \cdot Q^\mathbb{C} \cdot g_+,~~{\rm where}~~ (g_+, g_-) \in \, U_+(n) \times  U_-(n).
\end{align*} 
Since there is no (regularization) mass invariant under the action of  the entire $U_+(n) \times  U_-(n)$, it follows that $U_+(n) \times U_-(n)$ is anomalous. \\

The largest anomaly free subgroup is the diagonal $U(n)$, i.e., $g_+=g_-=g \in U(n)$. In this case  $Q^\mathbb{C} =I_n$ (i.e., $H_1=I_n$ and $H_2=0$) is invariant under the diagonal $U(n)$. One can thus use 
\begin{align*}
M_{\rm reg}= YX \otimes I_n
\end{align*}
\noindent as the regularization mass.\\

Note that this mass term is invariant under the time-reversal and charge-conjugation symmetries,
\begin{align*}
Q^\mathbb{C}  \xrightarrow{T}& (Q^\mathbb{C})^* = I_n \\
Q^\mathbb{C}  \xrightarrow{C}& (Q^\mathbb{C})^T = I_n
\end{align*}
\noindent Consequently there is no anomaly for these discrete symmetries after imposing the diagonal $U(n)$-invariant regularization mass.\\

\subsection{Real class in $(3+1)$-D}
\hfill

As shown in table \ref{emsymm}, the continuous part of the emergent symmetries form  the $U(n)$ group, and a general mass term has the following form \begin{align*}
M=YX\otimes S_1 + YZ \otimes S_2~~{\rm where}~~ Q^\mathbb{R}:= S_1 + i S_2\in U(n)/O(n).
\end{align*}
Under the action of  $U(n)$, these mass terms transform according to 
\begin{align*}
Q^\mathbb{R}\ra g^T \cdot Q^\mathbb{R} \cdot g,~~{\rm where}~~ g \in \, U(n).
\end{align*} 
Since there is no (regularization) mass invariant under the action of the entire $U(n)$, it follows that $U(n)$ is anomalous. \\

The largest anomaly-free subgroup is $O(n)$, i.e., $g \in O(n)$. In this case  $Q^\mathbb{R} =I_n$ (i.e., $S=I_n$ and $S_2=0$) is invariant under the diagonal $O(n)$. One can thus use 
\begin{align*}
M_{\rm reg}=YX \otimes I_n
\end{align*}
\noindent as the regularization mass.\\

Note that this mass term is invariant under the time-reversal transformation,
\begin{align*}
Q^\mathbb{R}  \xrightarrow{T}& \, (Q^\mathbb{R})^* = I_n. \\
\end{align*}
\noindent Consequently there is no anomaly for time-reversal symmetry after imposing the  $O(n)$-invariant regularization mass.\\

\section{Fermion integration}
\label{appendix:fermionInt}
\hfill

In this section, we derive the nonlinear sigma models summarized in section \ref{NLSM} and \ref{NLSigma2D3D} by integrating out the gapped fermions.\\

\subsection{Integrating out real versus complex fermions}
\label{appendix:realComplexInt}
\hfill

For fermions in the real classes, we face integration of the following form
\be
&&Z[Q^{\mathbb{R}}(x)]=e^{-W[Q^{\mathbb{R}}(x)]} = \int D\chi(x) e^{-S[\chi(x),Q^{\mathbb{R}}(x)]}~~{\rm where}\nn
&&S[\chi,Q^{\mathbb{R}}(x)] = \int d^D x \, \chi^T \left\{ \partial_0 + \hat{H}[Q^{\mathbb{R}}(x)] \right\} \chi.
\label{mjef}
\ee
A convenient trick for doing such integration is to perform  the corresponding complex fermion integration and divide the resulting effective action by two. \\

Too see this, consider two copies of Majorana fermion $\chi_1$ and $\chi_2$ coupled to the {\it same} $Q^{\mathbb{R}}(x)$. After fermion integration, the result should be the square of that in \Eq{mjef}, namely,
\begin{align*}
\int D\chi_1 \, D\chi_2 \, e^{-\left\{ S[\chi_1(x),Q^{\mathbb{R}}(x)] + S[\chi_2(x),Q^{\mathbb{R}}(x)] \right\}}&=\left\{ Z[Q^{\mathbb{R}}(x)] \right\}^2=e^{-2W[Q^{\mathbb{R}}(x)]}\\&:=e^{-\tilde{W}[Q^{\mathbb{R}}(x)].}
\end{align*}
On the other hand, we can combine $\chi_{1,2}$ into a complex fermion field $$\psi = \chi_1 + i \chi_2,$$ so that the sum of the real fermion actions can be written as a complex fermion action,
\begin{align*}
&\chi_1^T \left[ \partial_0 + \hat{H}(Q^{\mathbb{R}}) \right] \chi_1 + \chi_2^T \left[ \partial_0 + \hat{H}(Q^{\mathbb{R}}) \right] \chi_2\\
&=\psi^\dagger \left[ \partial_0 + \hat{H}(Q^{\mathbb{R}}) \right] \psi.  
\end{align*}
Note that the cross terms cancel out, due to the anti-commutation relation between $\chi_1$ and $\chi_2$, and the fact that $$\Big[\partial_0 + \hat{H}(Q^{\mathbb{R}})\Big]^T=-\left[\partial_0 + \hat{H}(Q^{\mathbb{R}})\right].$$ 
Consequently if  $\tilde{W}[Q^{\mathbb{R}}(x)]$ is the effective action due to  the complex fermion integration, we have
\be
W[Q^{\mathbb{R}}(x)]  = \frac{1}{2} \tilde{W}[Q^{\mathbb{R}}(x)].
\label{crr}\ee
\\

Due to \Eq{crr}, we shall focus on the complex fermion integration in the following.

\subsection{Integrating out complex fermions}
\hfill

To make the action explicitly Lorentz invariant, we rewrite the fermion-boson action as
\begin{align}
S =& \int d\tau \, d{\v x}~  \psi^\dagger \Big[ \partial_0 - i   \sum_{i=1}^d \Gamma_i\partial_i +m \hat{M}(\tau, \v x) \Big] \psi \notag\\
=& \int d\tau \, d{\v x}~  \psi^\dagger (-i\g_0)\Big[(i\g_0) \partial_0 - i  (i\g_0) \sum_{i=1}^d \Gamma_i\partial_i +m (i \g_0)\hat{M}(\tau, \v x) \Big] \psi,
\label{actC}
\end{align}
\noindent where $\g^0$ is a $2n\times 2n$ hermitian matrix which anti-commutes with $\{\G_i \}$ and satisfying $(\g^0)^2=1$. In general we choose $\g^0$ to be identity matrix among the flavor degrees of freedom. We will write down $\gamma^0$ explicitly for each dimension later on. Here we also extract out the parameter $m$, which controls the size of the fermion gap. As discussed in section \ref{massManifold1d}, we will focus on the $\hat{M}$ belonging to the manifold manifold, i.e., satisfying $\hat{M}^2=1$. Now define 
\be
&&\g^\mu:=(\g^0,-i\g^0~\G_i)~{\rm where}~i=1,...,d\nn
&&\bar{\psi}:= \psi^\dagger (-i\gamma^0)\nn
&& \boldsymbol{\beta}(\t,\v x):=\g^0\hat{M}(\t,\v x) 
\ee
so that \Eq{actC} turns into
\begin{align*}
&S=\int d^D x~\bar{\psi} \left[ i \gamma^{\mu}   \partial_\mu + i m\, \boldsymbol{\beta}(x) \right] \psi
:=\int d^D x~\bar{\psi}~{\cal \hat{D}}~\psi\\
~~&{\rm where ~~} \hat{\mathcal{D}}:= i  \slashed{\partial} + i \, m \, \boldsymbol{\beta}(x).
\end{align*}
\noindent Using the anti-commutation relations between $\{\Gamma_i\}$ and $\gamma^0$, the $\g^\mu$ satisfies the Clifford algebra 
	$$\{\gamma^\mu, \gamma^\nu \}= 2 \delta^{\mu\nu}.$$ 
\noindent It's also easy to check that $\boldsymbol{\beta}(x)$, being a function of $Q^{\mathbb{C}}(x)$,  is a matrix-valued smooth function of space-time, satisfying 
$$\boldsymbol{\beta}(x)^\dagger \cdot \boldsymbol{\beta}(x) =1.$$ 
\noindent Note that $\boldsymbol{\beta}(x)$ is in general not hermitian. \\

%Similar to section \ref{NLSM1},  the complex fermion action we shall face has the following form,
%\noindent where $\gamma^\mu$s are the gamma matrices satisfying Clifford algebra $\{ \gamma^\mu , \gamma^\nu\} = 2 \delta^{\mu\nu}$ and $\bar{\psi} := \psi^\dagger (-i \gamma^0)$. 

Fermion integration generates the effective action
\begin{align*}
W=-\ln\det[{\cal \hat{D}}] = - \Tr \ln [{\cal \hat{D}} ].
\end{align*}

The variation of the effective action $W$ induced by a small variation in  $\delta \boldsymbol{\beta}$  (triggered by a small variation in $Q^{\mathbb{C}}$  subject to the constraint $\boldsymbol{\beta}(x)^\dagger \cdot \boldsymbol{\beta}(x) =I$) is given by
\begin{align*}
\delta W 
=& - \Tr\left[ \left(\delta \mathcal{\hat{D}} \right) \, \mathcal{\hat{D}}^{-1} \right] \\
=& - \Tr\left[ i  m \, \delta \boldsymbol{\beta} \, \mathcal{\hat{D}}^{-1} \right] \\
=& - \Tr\left[ i  m \, \delta \boldsymbol{\beta} \left(\mathcal{\hat{D}}^\dagger \mathcal{\hat{D}} \right)^{-1} \mathcal{\hat{D}}^\dagger \right] \\
=& - \Tr\left[ i  m \, \delta \boldsymbol{\beta} \left[G_0^{-1} - m (( \slashed{\partial} \boldsymbol{\beta})) \right]^{-1} \mathcal{\hat{D}}^\dagger \right] \\
=& - \Tr\left[ i  m \, \delta \boldsymbol{\beta} \left[G_0^{-1} \left( I - m G_0(( \slashed{\partial} \boldsymbol{\beta})) \right)\right]^{-1} \mathcal{\hat{D}}^\dagger \right] \\
=& -\Tr\left\{ i m \, \delta \boldsymbol{\beta} \left[ \sum\limits_{l=0}^{\infty} [ m G_0~((\slashed{\partial} \boldsymbol{\beta}))~  ]^l \right] G_0 \left( i \slashed{\partial} - i m \boldsymbol{\beta}^\dagger \right) \right\}
\end{align*}

\noindent Here the double parentheses in $((\slashed{\partial}\boldsymbol{\beta}))$ means that the derivative acts only on $\boldsymbol{\beta}$ and nothing afterward, and 
$$G_0:= (-\partial^2 + m^2)^{-1}.$$ One can thus express $\delta W$ in powers of $((\slashed{\partial} \boldsymbol{\beta}))$.  
In the following, we shall retain terms where the number of space-time derivatives is less or equal to $D$. Hence by dimension counting, each of these terms is either relevant or marginal.  The expansion is called the gradient expansion in the literature \cite{Abanov2000}. 
There are two types of terms having $\le D$ derivatives, namely, 
\begin{align}
\label{terms1}
 -\Tr\left\{ i m \, \delta \boldsymbol{\beta} \left[ \sum\limits_{l=0}^{D-1} [ m G_0~((\slashed{\partial} \boldsymbol{\beta}))~  ]^l \right] G_0 \left( i \slashed{\partial}\right) \right\}
\end{align}
and
\begin{align}
\label{terms2}
 -\Tr\left\{ i m \, \delta \boldsymbol{\beta} \left[ \sum\limits_{l=0}^{D} [ m G_0~((\slashed{\partial} \boldsymbol{\beta}))~  ]^l \right] G_0 \left( - i m \boldsymbol{\beta}^\dagger \right) \right\}
\end{align}
\\

It turns out that among all non-vanishing parts of Eq.\ref{terms1} and \ref{terms2} there is a unique pure imaginary term -- the WZW term. The rest  are real.  In $D=1+1$ and $2+1$ the only real term is the stiffness term. In $D=3+1$ there are several extra real terms in addition to the stiffness term. However, all of these extra terms contain four space-time derivatives. Hence they are irrelevant compared with the stiffness term. Therefore the non-linear sigma model with the WZW term contains the most relevant real and imaginary terms after the fermion integration. To avoid sidetracking, we shall leave these details in subsection \ref{appendix:otherTerms}.\\

Throughout this appendix, we shall adopt the following convention. $\Tr$ denotes the trace over both the space-time and the matrices in $\boldsymbol{\beta}$, $\delta\boldsymbol{\beta}$, and $\gamma^\mu$s. $\tr'$ denotes the trace over the matrices in $\boldsymbol{\beta}$, $\delta\boldsymbol{\beta}$, $\gamma^\mu$.  $\tr_\g$ denotes the trace over only the $\g$ matrices.  $\tr$ denotes the trace over the $n\times n$ matrices in $\boldsymbol{\beta},\delta\boldsymbol{\beta}$. According to the above convention
$$\tr'=\tr_\g\times \tr.$$ Moreover, we shall adopt the following short hand
$$\int {d^D k\over(2\pi)^D}:=\int_k$$

\subsubsection{The stiffness term}\label{stifft} 
\hfill

The first non-vanishing such term is the stiffness term,
\be
\delta W_{\rm stiff} =& -\Tr\left[ i m \, \delta \boldsymbol{\beta}  \left( m G_0 ((\slashed{\partial} \boldsymbol{\beta}))  \right)  G_0 \left( i \slashed{\partial} \right) \right]. 
\label{rsp}\ee
Fourier transforming \Eq{rsp}, we obtain
\be
&&\delta W_{\rm stiff} = -m^2 \int_p \int_q \, \tr' \left[ \delta \boldsymbol{\beta}_{-q} \frac{1}{(p+q)^2 + m^2 } \slashed{q} \boldsymbol{\beta}_q \frac{1}{p^2+m^2} \slashed{p} \right] \nn
&&\approx 2 m^2 \int_p \frac{1}{(p^2+m^2)^3}  \int_q (q\cdot p) \, \tr'\left[ \delta\boldsymbol{\beta}_{-q} \boldsymbol{\beta}^{\dagger}_q \slashed{q} \slashed{p} \right]\nn
&&= 2 m^2 \int_p \frac{p_\mu p_\nu}{(p^2+m^2)^3}  \int_q (q_\mu q_\lambda) \, \tr'\left[ \delta\boldsymbol{\beta}_{-q} \boldsymbol{\beta}^{\dagger}_q  \g^\nu \g^\lambda\right]
\label{msp}
\ee
As usual $$\slashed{p}:=\g^\mu p_\mu.$$ In passing to the second line in \Eq{msp} we have expanded the expression 
\begin{align}
\frac{1}{(p+q)^2 + m^2} = \frac{1}{p^2 + m^2} \sum\limits_{n=0}^\infty \left( -\frac{2 p \cdot q + q^2}{p^2+m^2} \right)^n
\label{qExpansion}
\end{align}
and keep the lowest order non-vanishing term. 
Because
\begin{align*}
\int_p \frac{1}{(p^2 + m^2)^3} p_\mu p_\nu =\frac{1}{D}\int_p\frac{p^2}{(p^2 + m^2)^3} \delta_{\mu\nu},
\end{align*}
\Eq{msp} turns into
\be 
\delta W_{\rm stiff}&&\approx {2 m^2\over D} \int_p \frac{p^2}{(p^2+m^2)^3}  \int_q  (q_\mu q_\lambda) \, \tr'\left[ \delta\boldsymbol{\beta}_{-q} \boldsymbol{\beta}^{\dagger}_q  \g^\mu \g^\lambda\right]\nn
&&={2 m^2\over D} \int_p \frac{p^2}{(p^2+m^2)^3}  \int_q q^2\, \tr'\left[ \delta\boldsymbol{\beta}_{-q} \boldsymbol{\beta}^{\dagger}_q \right]\label{msp2}\ee
In passing to the second line of \Eq{msp2} we have used the fact that $$q_\mu q_\lambda \g^\mu \g^\lambda=q^2 I.$$  
The $p$-integration in \Eq{msp2} converges for $(1+1)$-D and $(2+1)$-D, but diverges for $(3+1)$-D. We shall use dimensional regularization ($D=4-\epsilon$ with $\epsilon \rightarrow 0^+$),  which leads to 
\begin{align}
\label{deltaW}
\delta W_{\rm stiff}
\approx & 2 m^2 \frac{1}{D} \int_p \frac{p^2}{(p^2+m^2)^3}  \int_q  q^2 \, \tr'\left[ \delta\boldsymbol{\beta}_{-q} \boldsymbol{\beta}^{\dagger}_q \right] \notag\\
= &    \left[ \frac{\Gamma(2-\frac{D}{2})}{2(4\pi)^{D/2}}m^{D-2}\right]  \int_\mathcal{M} d^D x \, \tr'\left[  \partial_{\mu}(\delta \boldsymbol{\beta}) \partial^{\mu}\boldsymbol{\beta}^{\dagger} \right] 
\end{align}
Here $\Gamma(l)$ is the gamma function. For $(3+1)$-D, the dimension regularization is given by
$$\Gamma(2-\frac{D}{2}) = \Gamma(\frac{\epsilon}{2}) \approx \frac{2}{\epsilon} - \gamma + O(\epsilon)$$
\noindent where $\gamma$ is the Euler-Mascheroni constant. Thus, the term whose variation with respect to $\delta \boldsymbol{\beta}$ yields \Eq{deltaW} is
\begin{align}
W_{\rm stiff}[\boldsymbol{\beta}] =  & {1\over 2\lambda_D^{D-2}}  \int_\mathcal{M} d^D x \, \tr'\left[  \partial_{\mu} \boldsymbol{\beta} \partial^{\mu} \, \boldsymbol{\beta}^{\dagger} \right] \label{stiffnessg}
\end{align}
where $\lambda$ has the dimension of length. In the limit that the short-distance cutoff is zero,
\be{1\over\lambda_D^{D-2}}= \left[ \frac{\Gamma(2-\frac{D}{2})}{2(4\pi)^{D/2}} m^{D-2}\right].\label{lambdaD}\ee
\\

\subsubsection{The WZW (topological) term}\label{wzwt}
\hfill

The second type of non-vanishing term in the gradient expansion is topological in nature, namely, the Wess-Zumino-Witten term 
\be
&&\delta W_{\rm WZW} =  -\Tr\left[ i m \, \delta \boldsymbol{\beta}  \left( m G_0 ((\slashed{\partial} \boldsymbol{\beta}))  \right)^D G_0 \left( - i m \boldsymbol{\beta}^\dagger \right) \right]  \nn
&&\approx -m^{D+2} \left[ \int_p \frac{1}{(p^2+m^2)^{D+1}} \right] \int_\mathcal{M} d^D x \, \tr' \left[ \prod\limits_{a=1}^D(\gamma^{\mu_a} \partial_{\mu_a} \boldsymbol{\beta} )\,  \boldsymbol{\beta}^{\dagger} \, \delta\boldsymbol{\beta}\right] \nn
&&= - \left[ \frac{1}{(4\pi)^{D/2}} \frac{\Gamma(\frac{D}{2}+1)}{\Gamma(D+1)} \right]   \int_\mathcal{M} d^D x \, \tr' \left[ \prod\limits_{a=1}^D(\gamma^{\mu_a} \partial_{\mu_a} \boldsymbol{\beta} )\,  \boldsymbol{\beta}^{\dagger} \, \delta\boldsymbol{\beta}\right] 
\label{deltaWZW}
\ee
\Eq{deltaWZW} is the difference in the Berry phase between the order parameter configurations $\boldsymbol{\beta}(x)$ and $\boldsymbol{\beta}(x) +\delta\boldsymbol{\beta}(x)$. To determine the Berry phase for a specific $\boldsymbol{\beta}(x)$, we integrate \Eq{deltaWZW} from a reference configuration
$\boldsymbol{\beta}(x)={\rm constant~matrix}.$ The existence of a continuous retraction leading from $\boldsymbol{\beta}(x)$ to the reference configuration relies on $$\pi_{D}(\text{mass manifold})=0.$$ It turns out this is exactly the condition  when the WZW term exists (see later).
Under the condition that such continuous retraction exists, we can find a continuous family of configurations $\tilde{\boldsymbol{\beta}}(x,u)$ so that 
	\be 
	&&\tilde{\boldsymbol{\beta}}(x,u=1)=\boldsymbol{\beta}(x)\nn&&\tilde{\boldsymbol{\beta}}(x,u=0)={\rm constant~matrix}.
	\label{extu}
	\ee
We can integrate \Eq{deltaWZW} to yield
\be
&&W_{WZW}[\boldsymbol{\beta}]= - \left[ \frac{1}{(4\pi)^{D/2}} \frac{\Gamma(\frac{D}{2}+1)}{\Gamma(D+1)} \right]   \int\limits_{\mathcal{B}}   \, du  \, d^Dx\, \tr' \left[ \prod\limits_{a=1}^D(\gamma^{\mu_a} \partial_{\mu_a} \tilde{\boldsymbol{\beta}} )\,  \tilde{\boldsymbol{\beta}}^{\dagger} \, \partial_u \tilde{\boldsymbol{\beta}}\right]\notag\\ \label{WZWg}
\ee
\noindent As in the main text, $\mathcal{B}$ is the extension of space-time manifold $\mathcal{M}$, so that $$\partial \mathcal{B} = \mathcal{M}.$$

In summary, when the fermion flavor number, $n$, is sufficiently large so that the WZW term is stabilized, the non-linear sigma model action is
\begin{align}
W[\boldsymbol{\beta}]=&{1\over 2\lambda_D^{D-2}}  \int_\mathcal{M} d^D x \, \tr'\left[  \partial_{\mu} \boldsymbol{\beta} \partial^{\mu}\boldsymbol{\beta}^{\dagger} \right]\nn
 -& \left[ \frac{1}{(4\pi)^{D/2}} \frac{\Gamma(\frac{D}{2}+1)}{\Gamma(D+1)} \right]   \int\limits_{\mathcal{B}} \, du \, d^Dx \, \tr' \left[ \prod\limits_{a=1}^D(\gamma^{\mu_a} \partial_{\mu_a} \tilde{\boldsymbol{\beta}} )\,  \tilde{\boldsymbol{\beta}}^{\dagger} \, \partial_u \tilde{\boldsymbol{\beta}}\right]\nn
~~{\rm where ~}& {1\over\lambda_D^{D-2}}= \left[ \frac{\Gamma(2-\frac{D}{2})}{2(4\pi)^{D/2}} m^{D-2}\right].
\label{nls0}\end{align}
 In the following, we shall apply this result to $(1+1)$-D, $(2+1)$-D, and $(3+1)$-D \footnote{It turns out that it always contains a level-1 WZW term  in even higher dimensions, though we shall not discuss them in the present paper.}.

\subsection{The fermion integration results for sufficiently large $n$ so that the WZW term is stabilized }
\label{appendix:fermionIntLargen}
\hfill

In this subsection, we shall focus on the results of fermion integration when $n$ is sufficiently large so that $$\pi_{D+1}(\text{mass manifold})=\mathbb{Z}.$$ The case of small $n$, before the WZW term is stabilized, will be discussed in appendix \ref{appendix:enlarge}.

\subsubsection{Complex class in $(1+1)$-D}
\hfill

The fermion action for the complex class in $(1+1)$-D is given by \Eq{S1C}
\begin{align*}
S =&\int d^2x ~\psi^\dagger \left[ \partial_0 - i (Z  I_n)\partial_1 + m \left( X \otimes H_1 + Y \otimes H_2 \right) \right] \psi \\
=&\int d^2x~  \psi^\dagger (-i X I_n) \left[ i (X I_n )\partial_0 + i (-Y  I_n)\partial_1 + i m \left( I \otimes H_1 + i Z \otimes H_2 \right) \right] \psi \\
:=& \int d^2x~\bar{\psi} \left[ i \slashed{\partial} + i \, m  \, \boldsymbol{\beta} \right] \psi
\end{align*}

\noindent where \be &&\bar{\psi} =\psi^\dagger (-i X I_n)\nn
&&\gamma^0 = X  I_n, ~\gamma^1 = -Y I_n, ~\gamma^5 = Z  I_n\nn
&&\boldsymbol{\beta} = I \otimes H_1 + i \gamma^5 I\otimes H_2.\label{inp1}\ee \\

Plugging the above results into equation \Eq{stiffnessg} and \Eq{lambdaD}  the stiffness term is given by

\begin{align*}
W_{\rm stiff}[Q^{\mathbb{C}}] = &  \frac{1}{8\pi} \int d^2x\, \tr\left[ \partial_\mu(H_1+i H_2)  \partial^\mu(H_1-i H_2)  \right] \\
=&  \frac{1}{8\pi}   \int d^2x \, \tr\left[ \partial_\mu Q^{\mathbb{C}}  \partial^\mu Q^{\mathbb{C}\dagger} \right]
\end{align*}

where $Q^{\mathbb{C}}:= H_1 + i H_2\in U(n)$. Substitute \Eq{inp1} into \Eq{WZWg} we obtain WZW term as

\begin{align*}
W_{WZW}[Q^{\mathbb{C}}] =& - \frac{1}{8\pi}    \int\limits_{\mathcal{B}} \, du \, d^2x \, \tr'\left[ (\gamma^{\mu_1} \partial_{\mu_1} \tilde{\boldsymbol{\beta}} )(\gamma^{\mu_2} \partial_{\mu_2} \tilde{\boldsymbol{\beta}} )\,  \tilde{\boldsymbol{\beta}}^{\dagger} \, \partial_u \tilde{\boldsymbol{\beta}}\right] \\
=& - \frac{1}{8\pi}    \int\limits_{\mathcal{B}}  \, du\, d^2x  \, \tr' \Big[ \left( \gamma^{\mu_1}\gamma^{\mu_2} \right) \,  \left( I \otimes \tilde{H}_1 - i \gamma^5 I\otimes \tilde{H}_2 \right) \,  \partial_u \left( I \otimes \tilde{H}_1 + i \gamma^5 I\otimes \tilde{H}_2 \right)  \\  
&\times \partial_{\mu_1} \left( I \otimes \tilde{H}_1 - i \gamma^5 I\otimes \tilde{H}_2 \right) \,  \partial_{\mu_2} \left( I \otimes \tilde{H}_1 + i \gamma^5 I\otimes \tilde{H}_2 \right) \Big] \\
=& - \frac{1}{8\pi}    \int\limits_{\mathcal{B}}  \, du \, d^2x \, \tr' \Big[ \left( \gamma^{\mu_1}\gamma^{\mu_2} \gamma^5\right) \,  \left( I \otimes \tilde{H}_1 - i I\otimes \tilde{H}_2 \right) \,  \partial_u \left( I \otimes \tilde{H}_1 + i  I\otimes \tilde{H}_2 \right)  \\  
&\times \partial_{\mu_1} \left( I \otimes \tilde{H}_1 - i  I\otimes \tilde{H}_2 \right) \,  \partial_{\mu_2} \left( I \otimes \tilde{H}_1 + i  I\otimes \tilde{H}_2 \right) \Big] \\
=& - \frac{1}{8\pi}    \int\limits_{\mathcal{B}}  \, du \, d^2x \,  \left( -2 i \epsilon^{\mu_1 \mu_2}\right) \, \tr \Big[   \tilde{Q}^{\mathbb{C}\dagger} \,  \partial_u  \tilde{Q}^{\mathbb{C}} \,   \partial_{\mu_1} \tilde{Q}^{\mathbb{C}\dagger} \,  \partial_{\mu_2}  \tilde{Q}^{\mathbb{C}}  \Big] \\
=&  -\frac{i}{4\pi}    \int\limits_{\mathcal{B}}  \, du \, d^2x \,  \epsilon^{\mu_1 \mu_2} \, \tr \Big[   \left( \tilde{Q}^{\mathbb{C}\dagger}   \partial_u \tilde{Q}^{\mathbb{C}} \right) \left( \tilde{Q}^{\mathbb{C}\dagger} \partial_{\mu_1} \tilde{Q}^{\mathbb{C}} \right)\,  \left( \tilde{Q}^{\mathbb{C}\dagger}\partial_{\mu_2}\tilde{Q}^{\mathbb{C}} \right)  \Big] \\
=&  -\frac{i}{4\pi}\times  \frac{1}{3}    \int\limits_{\mathcal{B}}  \, du \, d^2x \,  \epsilon^{\tilde{\mu}_1 \tilde{\mu}_2 \tilde{\mu}_3} \, \tr \Big[   \left( \tilde{Q}^{\mathbb{C}\dagger}  \partial_{\tilde{\mu}_1}\tilde{Q}^{\mathbb{C}}  \right)  \left( \tilde{Q}^{\mathbb{C}\dagger} \partial_{\tilde{\mu}_2}\tilde{Q}^{\mathbb{C}}  \right)\,  \left(  \tilde{Q}^{\mathbb{C}\dagger}\partial_{\tilde{\mu}_3}\tilde{Q}^{\mathbb{C}}  \right)  \Big] \\
=&  -\frac{2\pi i}{24\pi^2}    \int\limits_{\mathcal{B}}   \, \tr \Big[   \left( \tilde{Q}^{\mathbb{C}\dagger}   d \tilde{Q}^{\mathbb{C}}  \right)^3    \Big]  
\end{align*}

\noindent where $$ \tilde{Q}^{\mathbb{C}}:=\tilde{H}_1 + i \tilde{H}_2\in U(n)$$ is the extension of $Q^{\mathbb{C}}= H_1 + i H_2$ into $\mathcal{B}$, and $\tilde{\mu}_a$ is extended space-time manifold index (i.e., they include $u$). This extension in \Eq{extu} is possible because $$\pi_2(U(N))=0.$$ The passing from the 2nd to the 3rd line is due to the fact that when $\tilde{H}_2 \rightarrow -\tilde{H}_2$ the whole expression changes sign, hence only the terms with an odd number of $\tilde{H}_2$ survive. %The notation $\tr$ denotes the trace over the last $n\times n$ matrix in \Eq{inp1} (similar meaning applies to $\tr$ in subsequent subsections). 
The final non-linear sigma model action, namely, 
\be
W[Q^{\mathbb{C}}]= \frac{1}{8\pi}   \int d^2x \, \tr\left[ \partial_\mu Q^{\mathbb{C}}  \partial^\mu Q^{\mathbb{C}\dagger} \right]-\frac{2\pi i}{24\pi^2}    \int\limits_{\mathcal{B}}    \, \tr \Big[   \left( \tilde{Q}^{\mathbb{C}\dagger}   d \tilde{Q}^{\mathbb{C}}  \right)^3    \Big]
\label{nls2c}\ee

 is the $U(n)_{k=1}$ WZW theory.\\

\subsubsection{Real class in $(1+1)$-D}
\hfill

The Majorana fermion action for the real class in $(1+1)$-D is given by \Eq{S1R}
\begin{align*}
S =& \int d^2x~\chi^T \left[ \partial_0 + i (Z  I_n)\partial_1 + m \left( X \otimes (iA) + Y \otimes S \right) \right] \chi ,\\
\end{align*}
where $S$ and $A$ are symmetric and anti-symmetric matrices, respectively.
Upon the complexification described in subsection \ref{appendix:realComplexInt}, the form of the action becomes exactly the same as the complex class action in the preceding section, except that $$H_1\ra iA~~H_2\ra S.$$ Following the discussion in subsection \ref{appendix:realComplexInt}, we can substitute $$Q^{\mathbb{C}} = H_1 + i H_2\ra i (A+S) := i Q^{\mathbb{R}}$$ into \Eq{nls2c}  and divide the result by $2$ to obtain the following non-linear sigma model action   
\be
&&W[Q^{\mathbb{R}}]= \frac{1}{16\pi}   \int d^2x \, \tr\left[ \partial_\mu Q^{\mathbb{R}} \partial^\mu (Q^{\mathbb{R}})^T \right]   -  \frac{2 \pi i}{48\pi^2}    \int\limits_{\mathcal{B}}  \, \tr \Big\{   \left[(\tilde{Q}^{\mathbb{R}})^T   d \tilde{Q}^{\mathbb{R}}\right]^3    \Big\} \nn 
\label{nls2r}\ee
This is the action of the $O(n)_{k=1}$ WZW theory.\\

\subsubsection{Complex class in $(2+1)$-D}
\hfill

The fermion action for complex class in $(2+1)$-D can be constructed from \Eq{gamma2C} and \Eq{mass2C},
\begin{align*}
S =&\int d^3x~ \psi^\dagger \left[ \partial_0 - i (Z  I_n)\partial_1 - i (X  I_n)\partial_2 + m \, Y \otimes Q^{\mathbb{C}} \right] \psi \\
=&\int d^3x~  \psi^\dagger (-i\, Y I_n) \left[ i (Y I_n )\partial_0 + i (X  I_n)\partial_1 + i (-Z  I_n)\partial_2 + i m \, I \otimes Q^{\mathbb{C}}  \right] \psi \\
:=& \int d^3x~\bar{\psi} \left[ i \slashed{\partial} + i m \boldsymbol{\beta} \right] \psi
\end{align*}
\noindent where \be &&\bar{\psi}= \psi^\dagger (-i Y I_n)\nn&&\gamma^0 = Y  I_n,~\gamma^1 = X I_n,~\gamma^2 = -Z I_n\nn&&\boldsymbol{\beta} = I \otimes Q^{\mathbb{C}}.\label{inp2}\ee 
\noindent Here $Q^\mathbb{C}(x)$ is an $n\times n$ hermitian-matrix-value function satisfying $\left( Q^\mathbb{C} \right)^2 = I_n$, forming the mass manifold $\bigcup_{l=0}^n \frac{U(n)}{U(l)\times U(n-l)}$ (see appendix \ref{appendix:massManifold}). As discussed earlier, the $l=n/2$ component is special because the full emergent symmetries of the fermion theory can be restored upon order parameter fluctuation. Hence as far as bosonization is concerned we will focus on $l=n/2$. However, the following derivation works for other values of $l$ too as long as both $l$ and $n-l$ are sufficiently large for the WZW term to be stabilized.\\
%The non-linear sigma model with $Q^\mathbb{C}(x)$ in other mass manifold component $l\ne n/2$ are used to describe  interacting fermion theory in section \ref{IntFermion} with spontaneous breaking of the time reversal symmetry.\\

Substitute \Eq{inp2} into \Eq{stiffnessg} and \Eq{lambdaD}, we obtain the following stiffness term
\begin{align*}
W_{\rm stiff}[Q^{\mathbb{C}}] = & {1\over 4\lambda_3}  \int d^3 x \, \tr'\left[  \partial_{\mu} \boldsymbol{\beta}  \partial^{\mu}\boldsymbol{\beta}^{\dagger} \right] \\
=& {1\over 2\lambda_3}   \int d^3 x \, \tr\left[  \partial_{\mu} Q^{\mathbb{C}} \partial^{\mu}Q^{\mathbb{C}}  \right], \\
\end{align*}
where $\lambda_3$ has the dimension of length and in the limit where the short-distance cutoff is zero, \be\lambda_3={8\pi\over m}.\label{lambda3}\ee

Substitution of \Eq{inp2} into equation (\ref{WZWg}) yields the WZW term

\begin{align*}
W_{\rm WZW}[Q^{\mathbb{C}}] =& - \left[ \frac{1}{(4\pi)^{3/2}} \frac{\Gamma(\frac{5}{2})}{\Gamma(4)} \right]   \int\limits_{\mathcal{B}}  \, du \, d^3x \, \tr' \left[ \prod\limits_{a=1}^D(\gamma^{\mu_a} \partial_{\mu_a} \tilde{\boldsymbol{\beta}} )\,  \tilde{\boldsymbol{\beta}}^{\dagger} \, \partial_u \tilde{\boldsymbol{\beta}}\right] \\
=& -  \frac{i}{32 \pi}      \int\limits_{\mathcal{B}}  \, du \, d^3x \,  \epsilon^{\mu_1 \mu_2 \mu_3} \, \tr \left[  \tilde{Q}^{\mathbb{C}}  \, \partial_u \tilde{Q}^{\mathbb{C}}  \, \partial_{\mu_1} \tilde{Q}^{\mathbb{C}}  \, \partial_{\mu_2} \tilde{Q}^{\mathbb{C}}  \, \partial_{\mu_3} \tilde{Q}^{\mathbb{C}} \right] \\
=& -  \frac{i}{128 \pi}      \int\limits_{\mathcal{B}}  \, du \, d^3x \,  \epsilon^{\tilde{\mu}_1 \tilde{\mu}_2 \tilde{\mu}_3 \tilde{\mu}_4} \, \tr \left[  \tilde{Q}^{\mathbb{C}}   \, \partial_{\tilde{\mu}_1} \tilde{Q}^{\mathbb{C}}  \, \partial_{\tilde{\mu}_2} \tilde{Q}^{\mathbb{C}}  \, \partial_{\tilde{\mu}_3} \tilde{Q}^{\mathbb{C}}  \, \partial_{\tilde{\mu}_4} \tilde{Q}^{\mathbb{C}} \right] \\
=& -  \frac{ 2 \pi i }{256 \pi^2}      \int\limits_{\mathcal{B}} \, \tr \left[\tilde{Q}^{\mathbb{C}}    \,\left( d \tilde{Q}^{\mathbb{C}} \right)^4 \right] .\\
\end{align*}
Here $\tilde{Q}^{\mathbb{C}}$ % \in \frac{U(n)}{U(n/2) \times U(n/2)}$$ 
is the extension of $Q^{\mathbb{C}} $ into $\mathcal{B}$, and $\tilde{\mu}_a$ extended space-time index. The extension in \Eq{extu} is possible because $$\pi_3\left(\frac{U(n)}{U(n/2) \times U(n/2)}\right)=0.$$ The existence of WZW is indicated by $$\pi_4\left(\frac{U(n)}{U(n/2) \times U(n/2)}\right)=\mathbb{Z},$$ with the topological invariant
\begin{align*}
\frac{1}{256 \pi^2}      \int\limits_{S^4} \, \tr \left[\tilde{Q}^{\mathbb{C}}    \,\left( d \tilde{Q}^{\mathbb{C}} \right)^4 \right] \in \mathbb{Z}.
\end{align*}
\noindent Comparing with the result of fermion integration, the WZW term is $2\pi i$ times the above topological invariant, implying the level, $k$, is 1. In summary, the non-linear sigma model action is 
\be
W[Q^{\mathbb{C}}]={1\over 2\lambda_3}   \int d^3 x \, \tr\left[  \partial_{\mu} Q^{\mathbb{C}} \partial^{\mu}Q^{\mathbb{C}}  \right] -  \frac{ 2 \pi i }{256 \pi^2}      \int\limits_{\mathcal{B}}  \, \tr \left[\tilde{Q}^{\mathbb{C}}    \,\left( d \tilde{Q}^{\mathbb{C}} \right)^4 \right] .
\label{nls2C}\ee

%This is the action of the $\frac{U(n)}{U(n/2) \times U(n/2)}$ non-linear sigma model with level one ($k=1$) WZW term.\\

\subsubsection{Real class in $(2+1)$-D}
\hfill

The fermion action for real class in $(2+1)$-D can be constructed from \Eq{gamma2R} and \Eq{mass2R},
\begin{align*}
S =&\int d^3x~ \chi^T \left[ \partial_0 + i (Z  I_n)\partial_1 + i (X I_n)\partial_2 + m \,  Y Q^{\mathbb{R}}   \right] \chi \\
\end{align*}
Note that the form of this action is the same as that in the preceding section, except that the fermions are Majorana and $Q^{\mathbb{R}} $ is real symmetric instead of hermitian. According to the discussion in subsection \ref{appendix:realComplexInt}, we can replace 
$$Q^{\mathbb{C}} \ra Q^{\mathbb{R}} \in \frac{O(n)}{O(n/2) \times O(n/2)}$$ in \Eq{nls2C} and divide the result by $2$. The resulting non-linear sigma model action is
\be
W[Q^{\mathbb{R}}]= {1\over 4\lambda_3}   \int d^3 x \, \tr\left[  \partial_{\mu} Q^{\mathbb{R}}  \partial^{\mu}Q^{\mathbb{R}}  \right]  -  \frac{2 \pi i}{512 \pi^2}      \int\limits_{\mathcal{B}} \tr \left[  \tilde{Q}^{\mathbb{R}}   \, (d \tilde{Q}^{\mathbb{R}})^4 \right].
\label{nls2R}
\ee
Here $\lambda_3$ has the dimension of length and in the  limit where the short-distance cutoff is zero $\lambda_3$ is given by \Eq{lambda3}. Moreover, $\tilde{Q}^{\mathbb{R}} $ is the extension of $Q^{\mathbb{R}} $ into $\mathcal{B}$. The extension in \Eq{nls2R} is possible because $$\pi_3\left(\frac{O(n)}{O(n/2) \times O(n/2)}\right)=0.$$ The existence of the WZW term is indicated by $$\pi_4\left(\frac{O(n)}{O(n/2) \times O(n/2)}\right)=\mathbb{Z},$$ with the topological invariant given by
\begin{align*}
\frac{1}{512 \pi^2}      \int\limits_{S^4} \, \tr \left[\tilde{Q}^{\mathbb{R}}    \,\left( d \tilde{Q}^{\mathbb{R}} \right)^4 \right] \in \mathbb{Z}.
\end{align*}
\noindent Comparing the WZW term with the topological invariant  we conclude \Eq{nls2R} is the action for the $\frac{O(n)}{O(n/2) \times O(n/2)}$ non-linear sigma model with $k=1$ WZW term.
\\

\subsubsection{Complex class in $(3+1)$-D}
\hfill

The fermion action for complex class in $(3+1)$-D can be constructed from \Eq{gamma3C} and \Eq{mass3C},
\begin{align*}
S =& \int d^4x~\psi^\dagger \left[ \partial_0 - i (ZI I_n)\partial_1 - i (XI I_n)\partial_2 - i (YZ  I_n)\partial_3 + m \left( YX \otimes H_1 + YY \otimes H_2 \right) \right] \psi \\
=&\int d^4x~  \psi^\dagger (-i YX I_n) \Big[ i (YX I_n )\partial_0 + i (XX I_n)\partial_1 + i (-ZX I_n)\partial_2 + i (-IY I_n)\partial_3 \\ 
&+ i m \left( II \otimes H_1 + i \, IZ \otimes H_2 \right) \Big] \psi \\
:=&\int d^4x~ \bar{\psi} \left[ i \slashed{\partial} + i m \boldsymbol{\beta} \right] \psi
\end{align*}
where \be &&\bar{\psi}=\psi^\dagger (-i YX I_n)\nn
&& \gamma^0 = YX I_n,~\gamma^1 = XX I_n~,\gamma^2 = -ZX I_n,~\gamma^3 = -IY I_n,~ \gamma^5 = IZ I_n\nn
&&\boldsymbol{\beta} = II \otimes H_1 + i \gamma^5 II\otimes H_2.\label{inp3}\ee \\

Substitute \Eq{inp3} into \Eq{stiffnessg} and \Eq{lambdaD}, the stiffness term read 
\begin{align*}
W_{\rm stiff}[Q^{\mathbb{C}}] = & {1\over 8\lambda_4^2}  \int d^4 x \, \tr'\left[  \partial_{\mu} \boldsymbol{\beta} \partial^{\mu}\boldsymbol{\beta}^{\dagger} \right] \\
= & {1\over 2\lambda_4^2}  \int d^4 x \, \tr\left[  \partial_{\mu} Q^{\mathbb{C}} \partial^{\mu}Q^{\mathbb{C}\dagger}\right] 
\end{align*}
where $$Q^{\mathbb{C}}= H_1 + i H_2 \in U(n).$$ The parameter $\lambda_4$ has the dimension of length and in the limit where the short-distance cutoff is zero,
\be{1\over \lambda_4^2}= \left[ \frac{\Gamma(0^+)m^{2}}{8 \pi^{2}} \right].\label{lambda4}\ee
In the case where the short distance cutoff is finite the coefficient $\Gamma (0^+)$ should be replaced by a cutoff dependent parameter. Substitution \Eq{inp3} into \Eq{WZWg} yields the WZW term
\begin{align*}
W_{\rm WZW}[Q^{\mathbb{C}}] =& - \left[ \frac{1}{(4\pi)^{2}} \frac{\Gamma(3)}{\Gamma(5)} \right]    \int\limits_{\mathcal{B}}  \, du \, d^4x  \, \tr' \left[ (\gamma^{\mu_1} \partial_{\mu_1} \tilde{\boldsymbol{\beta}} )(\gamma^{\mu_2} \partial_{\mu_2} \tilde{\boldsymbol{\beta}} )\,  (\gamma^{\mu_3} \partial_{\mu_3} \tilde{\boldsymbol{\beta}} )\,  (\gamma^{\mu_4} \partial_{\mu_4} \tilde{\boldsymbol{\beta}} )\,  \tilde{\boldsymbol{\beta}}^{\dagger} \, \partial_u \tilde{\boldsymbol{\beta}}\right] \\
=& -  \frac{1}{ 192\pi^{2}}     \int\limits_{\mathcal{B}}  \, du \, d^4x \, \tr' \Big[ \left( \gamma^{\mu_1}\gamma^{\mu_2} \gamma^{\mu_3}\gamma^{\mu_4} \right) \,  \left( I \otimes \tilde{H}_1 - i \gamma^5 I\otimes \tilde{H}_2 \right) \,  \partial_u \left( I \otimes \tilde{H}_1 + i \gamma^5 I\otimes \tilde{H}_2 \right)  \\  
&\times \partial_{\mu_1} \left( I \otimes \tilde{H}_1 - i \gamma^5 I\otimes \tilde{H}_2 \right) \,  \partial_{\mu_2} \left( I \otimes \tilde{H}_1 + i \gamma^5 I\otimes \tilde{H}_2 \right) \Big] \\
&\times \partial_{\mu_3} \left( I \otimes \tilde{H}_1 - i \gamma^5 I\otimes \tilde{H}_2 \right) \,  \partial_{\mu_4} \left( I \otimes \tilde{H}_1 + i \gamma^5 I\otimes \tilde{H}_2 \right) \Big] \\
=& -  \frac{1}{  192\pi^{2}}     \int\limits_{\mathcal{B}}  \, du \, d^4x \, \tr' \Big[ \left( \gamma^{\mu_1}\gamma^{\mu_2} \gamma^{\mu_3}\gamma^{\mu_4} \gamma^5 \right) \,  \left( I \otimes \tilde{H}_1 - i  I\otimes \tilde{H}_2 \right) \,  \partial_u \left( I \otimes \tilde{H}_1 + i  I\otimes \tilde{H}_2 \right)  \\  
&\times \partial_{\mu_1} \left( I \otimes \tilde{H}_1 - i  I\otimes \tilde{H}_2 \right) \,  \partial_{\mu_2} \left( I \otimes \tilde{H}_1 + i  I\otimes \tilde{H}_2 \right) \Big] \\
&\times \partial_{\mu_3} \left( I \otimes \tilde{H}_1 - i  I\otimes \tilde{H}_2 \right) \,  \partial_{\mu_4} \left( I \otimes \tilde{H}_1 + i  I\otimes \tilde{H}_2 \right) \Big] \\
=& -  \frac{1}{  192\pi^{2}}    \int\limits_{\mathcal{B}}  \, du \, d^4x \,  \left( 4  \epsilon^{\mu_1 \mu_2 \mu_3 \mu_4}\right) \, \tr \Big[   \tilde{Q}^{\mathbb{C}\dagger}\,  \partial_u \tilde{Q}^{\mathbb{C}} \,   \partial_{\mu_1}\tilde{Q}^{\mathbb{C}\dagger} \,  \partial_{\mu_2} \tilde{Q}^{\mathbb{C}} \, \partial_{\mu_3} \tilde{Q}^{\mathbb{C}\dagger} \,  \partial_{\mu_4} \tilde{Q}^{\mathbb{C}}  \Big] \\
=& -\frac{1}{  48\pi^{2}} \int\limits_{\mathcal{B}}  \, du \, d^4x \,  \epsilon^{\mu_1 \mu_2 \mu_3 \mu_4} \, \tr \Big[\left( \tilde{Q}^{\mathbb{C}\dagger}   \partial_u \tilde{Q}^{\mathbb{C}}  \right)  \left( \tilde{Q}^{\mathbb{C}\dagger}  \partial_{\mu_1}  \tilde{Q}^{\mathbb{C}}\right)\,  \left(\tilde{Q}^{\mathbb{C}\dagger} \partial_{\mu_2}\tilde{Q}^{\mathbb{C}}  \right)\\& \left( \tilde{Q}^{\mathbb{C}\dagger}  \partial_{\mu_3} \tilde{Q}^{\mathbb{C}}  \right)\left(  \tilde{Q}^{\mathbb{C}\dagger} \partial_{\mu_4}\tilde{Q}^{\mathbb{C}}  \right)  \Big] \\
=& -\frac{1}{  240 \pi^{2}}    \int\limits_{\mathcal{B}}  \, du \, d^4x \,  \epsilon^{\tilde{\mu}_1 \tilde{\mu}_2 \tilde{\mu}_3 \tilde{\mu}_4 \tilde{\mu}_5} \, \tr \Big[\left( \tilde{Q}^{\mathbb{C}\dagger} \partial_{\tilde{\mu}_1}\tilde{Q}^{\mathbb{C}} \right)\,\left( \tilde{Q}_3^{\mathbb{C}\dagger} \partial_{\tilde{\mu}_2}\tilde{Q}^{\mathbb{C}} \right)\left( \tilde{Q}_3^{\mathbb{C}\dagger} \partial_{\tilde{\mu}_3}\tilde{Q}^{\mathbb{C}} \right)\\&\left( \tilde{Q}_3^{\mathbb{C}\dagger} \partial_{\tilde{\mu}_4}\tilde{Q}^{\mathbb{C}} \right)\left( \tilde{Q}_3^{\mathbb{C}\dagger} \partial_{\tilde{\mu}_5}\tilde{Q}^{\mathbb{C}} \right)  \Big] \\
=& - \frac{2\pi}{480\pi^3}    \int\limits_{\mathcal{B}}    \, \tr \Big[\left(\tilde{Q}^{\mathbb{C}\dagger} d\tilde{Q}^{\mathbb{C}} \right)^5    \Big] 
\end{align*}
where $\tilde{Q}^{\mathbb{C}}$ is the extension of $Q^{\mathbb{C}} $ into $\mathcal{B}$, and $\tilde{\mu}_a$ is the coordinate index of the extended space-time manifold. The extension in \Eq{extu} is possible because $$\pi_4(U(n))=0.$$ In passing from the 2nd to the 3rd line is due to the fact that when $\tilde{H}_2 \rightarrow -\tilde{H}_2$, the entire expression changes sign, hence only terms with an odd number of $\tilde{H}_2$ survive. The existence of the WZW term is indicated by $$\pi_5(U(n))=\mathbb{Z},$$ with the topological invariant given by
\begin{align*}
 \frac{i}{480\pi^3}    \int\limits_{S^5}    \, \tr \Big[\left(\tilde{Q}^{\mathbb{C}\dagger} d\tilde{Q}^{\mathbb{C}} \right)^5    \Big] \in \mathbb{Z}.
\end{align*}
\noindent Comparing the WZW term with the topological invariant  we conclude the WZW term is at level $k=1$.
In summary, the non-linear sigma model action is given by

\be
W[Q^{\mathbb{C}}]={1\over 2\lambda_4^2}  \int d^4 x \, \tr\left[  \partial_{\mu} Q^{\mathbb{C}} \partial^{\mu}Q^{\mathbb{C}\dagger}\right]  - \frac{2\pi}{480\pi^3}    \int\limits_{\mathcal{B}}   \, \tr \Big[\left(\tilde{Q}^{\mathbb{C}\dagger} d\tilde{Q}^{\mathbb{C}} \right)^5    \Big] .
\label{nls3C}
\ee

\subsubsection{Real class in $(3+1)$-D}
\hfill

The fermion action for complex class in $(3+1)$-D can be constructed from \Eq{gamma3R} and \Eq{mass3R},
\begin{align*}
S =&\int d^4x~ \chi^T \left[ \partial_0 - i (XI I_n)\partial_1 - i (ZI  I_n)\partial_2  - i (YY  I_n)\partial_3  + m \left( YX \otimes S_1 +  YZ \otimes S_2 \right) \right] \chi 
\end{align*}
This action has the same form as that in the preceding section, except that the following differences. (i) The fermions are Majorana, (ii) an unitary change of the matrix basis, namely, rotation by $\pi/2$ generated by $IXI_n$ , and (iii) $H_1\ra S_1$ and $H_2\ra S_2$. According to the discussion in subsection \ref{appendix:realComplexInt}, we can use the result in the preceding section by substituting $Q^{\mathbb{C}} \ra Q^{\mathbb{R}}= S_1 + i S_2$  into \Eq{nls3C} and divide the final effective action by $2$. The resulting non-linear sigma model action is given by
\be
W[Q^{\mathbb{R}}]={1\over 4\lambda_4^2}  \int d^4 x \, \tr\left[  \partial_{\mu}Q^{\mathbb{R}} \partial^{\mu}Q^{\mathbb{R}\dagger} \right] -\frac{2\pi}{960\pi^3}    \int\limits_{\mathcal{B}}    \, \tr \Big[   \left(\tilde{Q}^{\mathbb{R}\dagger}    d \tilde{Q}^{\mathbb{R}}  \right)^5    \Big].\nn  
\label{nls3R}\ee
The existence of the WZW iterm is indicated by $$\pi_5(U(n)/O(n))=\mathbb{Z},$$ with the topological invariant given by
\begin{align*}
\frac{i}{960\pi^3}    \int\limits_{S^5}    \, \tr \Big[\left(\tilde{Q}^{\mathbb{R}\dagger} d\tilde{Q}^{\mathbb{R}} \right)^5    \Big] \in \mathbb{Z}.
\end{align*}
Again, we conclude that the WZW term is at level 1.\\

\subsection{The less relevant real terms originate from \Eq{terms1} and \Eq{terms2}}
\label{appendix:otherTerms}
\hfill

In this subsection, we provide the details which show that in $(1+1)$-D and $(2+1)$-D, the non-vanishing terms in \Eq{terms1} and \Eq{terms2} having  $\le D$ space-time derivatives are the stiffness and WZW terms. In $(3+1)$-D, there are extra real terms. In the following, we shall present a detailed analysis of these potential extra terms. 

\subsubsection{$(1+1)$-D}
\hfill

Given the fact that 
$$\boldsymbol{\beta} = I \otimes H_1 + i \gamma^5 I\otimes H_2,$$
the $l=0$ term in \Eq{terms1}  read
\begin{align*}
-\Tr\left[ i m \, \delta \boldsymbol{\beta} G_0 \gamma^\mu(i \partial_\mu) \right].
\end{align*}
Since $\boldsymbol{\beta}$ only contains $I$ or $\gamma^5$, it follows that the this term vanishes when we trace over the gamma matrices because both $$\tr_{\g}\left[\gamma^\mu\right]=0,~~ \tr_{\g}[\gamma^\mu \gamma^5]=0.$$ Similar argument applies to the $l=1$ term in \Eq{terms2}. \\

This leaves the $l=0$ term in \Eq{terms2} as the only term requiring further attention, namely, 
\begin{align*}
&- \Tr \left[ i  m \delta \boldsymbol{\beta} G_0 (-i \, m \, \boldsymbol{\beta}) \right] \\
&= - m^2 \int_{p,q}  \frac{1}{p^2+m^2}  \tr'\left[\boldsymbol{\beta}^\dagger_{-q} \delta \boldsymbol{\beta}_q \right] \\
&= - m^2 \left(\int_{p} \frac{1}{p^2+m^2} \right)\int d^2x \,\tr'\left[\boldsymbol{\beta}^\dagger \delta \boldsymbol{\beta}\right] \\
&= - 2m^2 \left(\int_{p} \frac{1}{p^2+m^2} \right)\int d^2x \, \delta \left\{ \tr\left[H_1^2 + H_2^2\right] \right\} \\
&= 0.
\end{align*}
\noindent In passing to the last line we used the constraint that $$Q^\mathbb{C}  = H_1 + i H_2 \in U(n)\Rightarrow H_1^2 + H_2^2 = I_n.$$ Hence the only non-vanishing terms are the stiffness and WZW terms in subsection \ref{stifft} and \ref{wzwt}.\\

\subsubsection{$(2+1)$-D}
\hfill

Given the fact that 
$$\boldsymbol{\beta}=I\otimes Q^{\mathbb{C}},$$ both
the $l=0$ term in \Eq{terms1} and the $l=1$ term in \Eq{terms2} vanishes under $\tr_\g$ because 
$$\tr_\g[\g^\mu]=0.$$
\\ 

The $l=2$ term in \Eq{terms1} gives
\begin{align*}
&-\Tr \left[ i m \, \delta \boldsymbol{\beta} \left( m G_0 ((\slashed{\partial} \boldsymbol{\beta} )) \right)^2 G_0 \left(i\slashed{\partial} \right)\right]\\
&=  m^3 \int_{p,q_1,q_2}\frac{1}{p^2+m^2} \frac{1}{(p+q_1)^2+m^2} \frac{1}{(p+q_1+q_2)^2+m^2} \tr'\Big[ \delta \boldsymbol{\beta}_{-q_1-q_2} (i \slashed{q_2} \boldsymbol{\beta}_{q_2} )(i \slashed{q_1} \boldsymbol{\beta}_{q_1} ) (i\slashed{p})\Big]\\
&\approx  m^3 \int_{p,q_1,q_2} \frac{\left(-2 p\cdot (2 q_1 + q_2) \right)}{(p^2+m^2)^4}  \left(2i \epsilon^{\mu\nu\rho} \right) q_2^\mu q_1^\nu p^\rho \, \tr\Big[ \delta Q^{\mathbb{C}}_{-q_1-q_2}  Q^{\mathbb{C}}_{q_2}  Q^{\mathbb{C}}_{q_1} \Big]\\
&=  \frac{-4 i m^3}{3} \int_{p,q_1,q_2} \frac{p^2}{(p^2+m^2)^4}   \epsilon^{\mu\nu\rho}  q_2^\mu q_1^\nu (2 q_1 + q_2)^\rho \, \tr\Big[ \delta Q^{\mathbb{C}}_{-q_1-q_2}  Q^{\mathbb{C}}_{q_2}  Q^{\mathbb{C}}_{q_1} \Big]\\
&= 0
\end{align*}
In passing to the third line we have traced over the $\g$ matrices, and in passing to the last line we have used the fact that  $q_2^\mu q_1^\nu (2 q_1 + q_2)^\rho$ is symmetric with respect to $(\nu ,\rho)$ or $(\mu ,\rho)$, while $\epsilon^{\mu\nu\rho}$ is totally anti-symmetric.\\

The $l=0$ term in \Eq{terms2} gives
\begin{align*}
&-\Tr \left[ i m \, \delta \boldsymbol{\beta}  G_0 \left(-i m \boldsymbol{\beta}^\dagger \right)\right]\\
&= -m^2 \int_p\frac{1}{p^2+m^2} \int d^3 x \, \tr'\left[\delta\boldsymbol{\beta} \boldsymbol{\beta}^\dagger \right]\\
&= -2m^2\int_p\frac{1}{p^2+m^2} \int d^3 x \, \tr\left[\delta Q^\mathbb{C} \, Q^\mathbb{C} \right]\\
&=0
\end{align*} 
In passing to the last line we noted that $$\left(Q^\mathbb{C}\right)^2 = I_n~\Rightarrow \delta Q^\mathbb{C} \, Q^\mathbb{C} = - Q^\mathbb{C} \delta Q^\mathbb{C}~\Rightarrow\tr\left[\delta Q^\mathbb{C} \, Q^\mathbb{C}\right]=-\tr\left[Q^\mathbb{C} \delta Q^\mathbb{C}\right]$$ Upon using the cyclic property of trace we conclude $$\tr\left[\delta Q^\mathbb{C} \, Q^\mathbb{C} \right] = 0.$$\\

The $l=2$ term in \Eq{terms2}  gives
\begin{align*}
&-\Tr \left[ i m \, \delta \boldsymbol{\beta} \left( m G_0 ((\slashed{\partial} \boldsymbol{\beta} )) \right)^2 G_0 \left(-i m \boldsymbol{\beta}^\dagger \right)\right]\\
&=-m^4\int_{p,q_1,q_2}\frac{1}{p^2+m^2}{1\over (p+q_1)^2+m^2}{1\over (p+q_1+q_2)^2+m^2}\tr'\left[\boldsymbol{\beta}_{-q_1-q_2-q_3}^\dagger \delta \boldsymbol{\beta}_{q_3}i \slashed{q_2}\boldsymbol{\beta}_{q_2}i \slashed{q_1}\boldsymbol{\beta}_{q_1}\right]\\
&\approx -m^4\int_{p}\frac{1}{(p^2+m^2)^3}\int_{q_1,q_2}\tr'\left[\boldsymbol{\beta}_{-q_1-q_2-q_3}^\dagger \delta \boldsymbol{\beta}_{q_3}i \slashed{q_2}\boldsymbol{\beta}_{q_2}i \slashed{q_1}\boldsymbol{\beta}_{q_1}\right]\\
&\approx -2m^4\int_p\frac{1}{(p^2+m^2)^3} \int d^3 x \, \tr\left[ Q^\mathbb{C}\, \delta Q^\mathbb{C} \, \partial_\mu Q^\mathbb{C} \, \partial_\mu Q^\mathbb{C} \right]\\
&= 0
\end{align*}
In passing from the second to the third line we have used the fact at most three $q_i$  are allowed (otherwise the term becomes irrelevant). Therefore at most we can expand the ${1\over (p+q_1)^2+m^2}{1\over (p+q_1+q_2)^2+m^2}$ to first order in $q_{1,2}$. However, such expansion inevitably comes with a $p$, and will vanish upon $p$ integration. Thus we can only keep the 0th order term ${1\over p^2+m^2}{1\over (p^2+m^2)^2}$.  In passing to the last line  we have used $\delta Q^\mathbb{C} \, Q^\mathbb{C} = - Q^\mathbb{C} \delta Q^\mathbb{C}$ three times to move $Q^\mathbb{C}$ to the end, and use the cyclic property to move it back to the front. In this way, we have proven that the quantity is the minus of itself, hence it is zero.\\

To summarize,  including all (the most and less relevant) terms, the non-linear sigma model is given by 
\be
W[Q^{\mathbb{C}}]={1\over 2\lambda_3}   \int d^3 x \, \tr\left[  \partial_{\mu} Q^{\mathbb{C}} \partial^{\mu}Q^{\mathbb{C}}  \right] -  \frac{ 2 \pi i }{256 \pi^2}      \int\limits_{\mathcal{B}}   \, \tr \left[\tilde{Q}^{\mathbb{C}}    \,\left( d \tilde{Q}^{\mathbb{C}} \right)^4 \right] .
\nonumber
\ee
\\

\subsubsection{$(3 +1)$-D}
\hfill

Given 
$$\boldsymbol{\beta} = I\otimes H_1 + i \gamma^5 I \otimes H_2,$$
\noindent  the $l=0,2$ terms of  \Eq{terms1} and the $l=1,3$ terms of \Eq{terms2} vanishes upon $\tr_\g$. This is because they contain either one or three $\g$ from $\slashed{p}$ or $\slashed{q_i}$. Because $\boldsymbol{\beta}$ contributes either $I$ or $\gamma^5$. These terms vanish due to the fact that $$\tr_\g[\gamma^\mu]=\tr_\g[\gamma^\mu \gamma^5]=\tr_\g[\gamma^\mu \gamma^\nu \gamma^\rho]=\tr_\g[\gamma^\mu \gamma^\nu \gamma^\rho \gamma^5]=0.$$\\

The $l=0$ term of \Eq{terms2} gives
\begin{align*}
&- \Tr \left[ i  m \delta \boldsymbol{\beta} G_0 (-i \, m \, \boldsymbol{\beta}) \right] \\
&= - m^2 \int_{p,q}\frac{1}{p^2+m^2}  \tr'\left[\boldsymbol{\beta}^\dagger_{-q} \delta \boldsymbol{\beta}_q \right] \\
&= - m^2 \left(\int_p\frac{1}{p^2+m^2} \right)\int d^4x \,\tr'\left[\boldsymbol{\beta}^\dagger \delta \boldsymbol{\beta}\right] \\
&= - 4m^2 \left(\int_p\frac{1}{p^2+m^2} \right)\int d^4x \, \delta \left\{ \tr\left[H_1^2 + H_2^2\right] \right\} \\
&= 0.
\end{align*}
In passing to the last line we use the same reasoning as the corresponding term in $(1+1)$-D.\\ 

While maintaining two space-time derivatives, the $l=1$ term of \Eq{terms1} gives rise to the variation  of the stiffness term $\delta W_{\rm stiffness}$  in subsection \ref{stifft}. Here we retain up to 4 space-time derivatives,
\begin{align}
&-\Tr\left[ i  m\delta\boldsymbol{\beta} \left( m G_0 ((\slashed{\partial} \boldsymbol{\beta} )) \right) G_0 (i \slashed{\partial}) \right]\notag\\
&=- m^2 \int_{p,q}\frac{1}{p^2+m^2}\frac{1}{(p+q)^2+m^2} \tr'\left[\delta \boldsymbol{\beta}_{-q} (\gamma^\mu q_\mu) \boldsymbol{\beta}_q (\gamma^\nu p_\nu) \right] \notag\\
&=- m^2 \int_{p,q}\frac{1}{p^2+m^2}\frac{1}{(p+q)^2+m^2} q_\mu p^\nu\tr'\left[\delta \boldsymbol{\beta}_{-q} \gamma^\mu \gamma^\nu \boldsymbol{\beta}_q^\dagger  \right]   \notag\\ 
&=\delta W_{\text{stiffness}} - m^2 \int_{p,q} \frac{1}{(p^2+m^2)^2} \left(\frac{4 q^2 (p\cdot q)}{(p^2+m^2)^2} - \frac{8 (p\cdot q)^3}{(p^2+m^2)^3} \right) (q\cdot p) \tr' \left[\delta \boldsymbol{\beta}_{-q}  \boldsymbol{\beta}_q^\dagger \right]\notag\\
&=\delta W_{\text{stiffness}}- m^2 \int_{p,q}\left(\frac{ p^2 q^4}{(p^2+m^2)^4} - \frac{ p^4 q^4}{(p^2+m^2)^5} \right)  \tr' \left[\delta \boldsymbol{\beta}_{-q}  \boldsymbol{\beta}_q^\dagger \right]\notag\\
&=\delta W_{\text{stiffness}} -\frac{1}{192 \pi^2} \int d^4 x \,  \tr' \left[\partial^2(\delta \boldsymbol{\beta})  \partial^2 \boldsymbol{\beta}^\dagger \right]\notag\\
&=\delta W_{\text{stiffness}} -\frac{1}{96 \pi^2} \int d^4 x \,  \tr \left[\partial^2(\delta Q^{\mathbb{C}})  \partial^2 Q^{\mathbb{C}\dagger} + \partial^2(\delta Q^{\mathbb{C} \dagger})  \partial^2 Q^{\mathbb{C}} \right]\notag\\
&=\delta W_{\text{stiffness}} -\frac{1}{96 \pi^2} \int d^4 x\, \delta \Big( \tr \left[\partial^2 Q^{\mathbb{C}}  \partial^2 Q^{\mathbb{C}\dagger}  \right]\Big)
\label{extra3D_1_1}
\end{align}
\noindent In passing from the 2nd to the 3rd line, we use the property $\boldsymbol{\beta} \gamma^\nu = \gamma^\nu \boldsymbol{\beta}^\dagger$ for $\mu=0,1,2,3$. From the 3rd to the 4th line we have used the fact that the trace is only non-zero if the $\g^\mu$ and $\g^\nu$ are the same. 
\noindent From the 4th to the 5th line, we used the fact that  rotational invariance allows the following replacement in the integrand of the $p$ integral  $$p^\mu p^\nu p^\rho p^\sigma\ra \frac{1}{D(D+2)} \left(\delta_{\mu\nu} \delta_{\rho\sigma} + \delta_{\mu \rho} \delta_{\nu \sigma} + \delta_{\mu \sigma} \delta_{\nu\rho}\right).$$ (The factor $\frac{1}{D(D+2)}$ can be fixed by taking trace on both sides.)  In the 5th line, only the terms in $\delta\boldsymbol{\beta}_{-q} \boldsymbol{\beta}_{q}^\dagger$ having an even number of $\gamma^5$ are non-zero. Moreover, since $\gamma^5$ is always accompanied by $H_2$, we can replace $\gamma^5$ with the identity matrix as long as we symmetrize the end result with respect to $H_2$. After the replacement, $\boldsymbol{\beta}$ becomes $II\otimes Q^\mathbb{C}$, the identity matrix can then be trace out, and the symmetrization amounts to sum over the terms with $Q^\mathbb{C}=H_1 + i H_2$ replaced by $Q^{\mathbb{C}\dagger}=H_1 - i H_2$. We will use this last trick several times in the following.\\

The $l=3$ term of \Eq{terms1} gives 
\begin{align}
&-\Tr \left[ i m \, \delta \boldsymbol{\beta} \left( m G_0 ((\slashed{\partial} \boldsymbol{\beta} )) \right)^3 G_0 \left(i\slashed{\partial} \right)\right]\notag\\
&=  m^4 \int_{p,q_1,q_2,q_3} \, \frac{1}{p^2+m^2} \frac{1}{(p+q_1)^2+m^2} \frac{1}{(p+q_1+q_2)^2+m^2} 
\frac{1}{(p+q_1+q_2 + q_3)^2+m^2}\notag\\&\times \tr'\Big[ \delta \boldsymbol{\beta}_{-q_1-q_2-q_3} (i \slashed{q_3} \boldsymbol{\beta}_{q_3} )(i \slashed{q_2} \boldsymbol{\beta}_{q_2} )(i \slashed{q_1} \boldsymbol{\beta}_{q_1} ) (i\slashed{p})\Big]\notag\\
&\approx m^4 \int_{p,q_1,q_2,q_3} \, \frac{-2 p\cdot (3q_1 + 2 q_2 + q_3)}{(p^2+m^2)^5} q_3^\mu q_2^\nu q_1^\rho p^\sigma \tr'\Big[ \gamma^\mu \gamma^\nu \gamma^\rho \gamma^\sigma \delta \boldsymbol{\beta}_{-q_1-q_2-q_3}  \boldsymbol{\beta}_{q_3}^\dagger  \boldsymbol{\beta}_{q_2} \boldsymbol{\beta}_{q_1}^\dagger \Big]\notag\\
&=-\frac{m^4}{2}\int_p \frac{p^2}{(p^2+m^2)^5} \int_{q_1,q_2, q_3} q_3^\mu q_2^\nu q_1^\rho (3q_1 + 2 q_2 + q_3)^\sigma
\left(\delta_{\mu\nu} \delta_{\rho \sigma} - \delta_{\mu \rho} \delta_{\nu\sigma} + \delta_{\mu\sigma} \delta_{\nu \rho} \right)\notag\\ &\times \tr'\left[\delta \boldsymbol{\beta}_{-q_1-q_2-q_3}  \boldsymbol{\beta}_{q_3}^\dagger  \boldsymbol{\beta}_{q_2} \boldsymbol{\beta}_{q_1}^\dagger \right]_{\text{even terms in} H_2}\notag\\
&=-\frac{1}{96\pi^2} \int_{q_1,q_2,q_3} \left( 3 q_1^2 (q_2 \cdot q_3) - 2 q_2^2 (q_1\cdot q_3) + q_3^2 (q_1 \cdot q_2 ) + 4 (q_1\cdot q_2)(q_2 \cdot q_3) \right) \notag\\
&~~~~\times \tr'\left[\delta Q^{\mathbb{C}}_{-q_1-q_2-q_3}  Q^{\mathbb{C}\dagger}  Q^{\mathbb{C}}_{q_2} Q^{\mathbb{C}\dagger}_{q_1} \right]_{\text{even } H_2} \notag\\
&=-\frac{1}{96\pi^2} \int d^4 x \, \frac{1}{2} \tr \left[ 
\thead{
	3 \, \delta Q^{\mathbb{C} } \partial_\mu Q^{\mathbb{C} \dagger} \partial Q^{\mathbb{C} } \partial^2 Q^{\mathbb{C} \dagger} + 3 \, \delta Q^{\mathbb{C}\dagger } \partial_\mu Q^{\mathbb{C} } \partial_\mu Q^{\mathbb{C} \dagger} \partial^2 Q^{\mathbb{C} }\\
	-2 \, \delta Q^{\mathbb{C} } \partial_\mu Q^{\mathbb{C} \dagger} \partial^2 Q^{\mathbb{C} } \partial_\mu Q^{\mathbb{C} \dagger} -2 \, \delta Q^{\mathbb{C}\dagger } \partial_\mu Q^{\mathbb{C} } \partial^2 Q^{\mathbb{C} \dagger} \partial_\mu Q^{\mathbb{C} }\\
	+ \, \delta Q^{\mathbb{C} } \partial^2 Q^{\mathbb{C} \dagger} \partial_\mu Q^{\mathbb{C} } \partial_\mu Q^{\mathbb{C} \dagger} +  \, \delta Q^{\mathbb{C}\dagger } \partial^2 Q^{\mathbb{C} } \partial_\mu Q^{\mathbb{C} \dagger} \partial_\mu Q^{\mathbb{C} }\\
	+4 \, \delta Q^{\mathbb{C} } \partial_\mu Q^{\mathbb{C} \dagger} \partial_\mu \partial_\nu Q^{\mathbb{C} } \partial_\nu Q^{\mathbb{C} \dagger} + 4 \, \delta Q^{\mathbb{C}\dagger } \partial_\mu Q^{\mathbb{C} } \partial_\mu \partial_\nu Q^{\mathbb{C} \dagger} \partial_\nu Q^{\mathbb{C} }\\
}
\right] \notag\\
&=\frac{1}{96\pi^2} \int d^4 x \, \tr \left[ Q^{\mathbb{C} \dagger} \delta Q^{\mathbb{C} }Q^{\mathbb{C} \dagger} \left( 
\thead{
-3 \,\partial_\mu Q^{\mathbb{C} } Q^{\mathbb{C} \dagger} \partial_\mu Q^{\mathbb{C} } Q^{\mathbb{C} \dagger} \partial^2 Q^{\mathbb{C} } \\
+2 \ \partial_\mu Q^{\mathbb{C} } Q^{\mathbb{C} \dagger} \partial^2 Q^{\mathbb{C} } Q^{\mathbb{C} \dagger} \partial_\mu Q^{\mathbb{C} } \\
- \partial^2 Q^{\mathbb{C} } Q^{\mathbb{C} \dagger} \partial_\mu Q^{\mathbb{C} } Q^{\mathbb{C} \dagger} \partial_\mu Q^{\mathbb{C} } \\
-4 \partial_\mu Q^{\mathbb{C} } Q^{\mathbb{C} \dagger} \partial_\mu \partial_\nu Q^{\mathbb{C} } Q^{\mathbb{C} \dagger} \partial_\nu Q^{\mathbb{C} } \\
+6 \partial_\mu Q^{\mathbb{C} } Q^{\mathbb{C} \dagger} \partial_\mu  Q^{\mathbb{C} } Q^{\mathbb{C} \dagger} \partial_\nu Q^{\mathbb{C} }  Q^{\mathbb{C} \dagger} \partial_\nu Q^{\mathbb{C} } \\
-2 \partial_\mu Q^{\mathbb{C} } Q^{\mathbb{C} \dagger} \partial_\nu  Q^{\mathbb{C} } Q^{\mathbb{C} \dagger} \partial_\nu Q^{\mathbb{C} }  Q^{\mathbb{C} \dagger} \partial_\mu Q^{\mathbb{C} } \\
+2 \partial_\mu Q^{\mathbb{C} } Q^{\mathbb{C} \dagger} \partial_\nu  Q^{\mathbb{C} } Q^{\mathbb{C} \dagger} \partial_\mu Q^{\mathbb{C} }  Q^{\mathbb{C} \dagger} \partial_\nu Q^{\mathbb{C} } 
}
\right)\right] 
\label{extra3D_1_3}
\end{align}
\noindent From the 3rd to the 4th line, we take the terms with even number of $\gamma^5$ (thus even number of $H_2$) from $\boldsymbol{\beta}$ and use the identity $\tr_{\gamma}\left[\gamma^\mu \gamma^\nu \gamma^\rho \gamma^\sigma \right] = 4 (\delta^{\mu \nu} \delta^{\rho \sigma} - \delta^{\mu\rho} \delta^{\nu \sigma} + \delta^{\mu\sigma} \delta^{\nu\rho})$. The terms with odd number of $\gamma^5$ vanish because $\tr_\gamma\left[ \g^\mu \g^\nu \g^\rho \g^\sigma \gamma^5\right] = 4 \epsilon^{\mu\nu\rho\sigma}$ is totally anti-symmetric, while $q_3^\nu q_2^\nu g_1^\rho (3q_1 + 2q_2 + q_3)^\sigma$ is symmetric with respect to either $(\mu,\sigma)$, $(\nu,\sigma)$, or $(\rho, \sigma)$. From the 4th line to the 5th line, we used the same trick as in \Eq{extra3D_1_1}. From the 6th to the 7th line, $\delta Q^{\mathbb{C} \dagger} = - Q^{\mathbb{C} \dagger} \delta Q^{\mathbb{C} } Q^{\mathbb{C} \dagger}$ is used repeatedly until  $\delta$ or $\partial$ act only on $Q^{\mathbb{C} }$.\\

The $l=2$ term of \Eq{terms2} gives
\begin{align*}
&-\Tr \left[ i m \, \delta \boldsymbol{\beta} \left( m G_0 ((\slashed{\partial} \boldsymbol{\beta} )) \right)^2 G_0 \left(-im \boldsymbol{\beta}^\dagger\right)\right]\\
&= - m^4 \int_{p,q_1,q_2,q_3} \frac{1}{(p+q_1 + q_2)^2+m^2} \frac{1}{(p+q_1)^2+m^2} \frac{1}{p^2+m^2}\\
&\times\tr'\left[ \boldsymbol{\beta}_{-q_1-q_2-q_3}^\dagger \delta \boldsymbol{\beta}_{q_3} (i \slashed{q_2} \boldsymbol{\beta}_{q_2} )(i \slashed{q_1} \boldsymbol{\beta}_{q_1} ) \right] \\
&\approx m^4 \int_{p,q_1,q_2,q_3} \frac{q_2^\nu q_1^\mu}{(p^2+m^2)^3} \Big[ 1 - \frac{q_1^2 + (q_1+q_2)^2}{p^2+m^2} + 4\frac{ (p \cdot q_1)^2 +(p \cdot (q_1+q_2))^2 + (p \cdot q_1) (p \cdot (q_1 + q_2)) }{(p^2+m^2)^2} \Big] \\
&\times \tr'\left[ \boldsymbol{\beta}_{-q_1-q_2-q_3}^\dagger \delta \boldsymbol{\beta}_{q_3} (  \gamma_\nu \gamma_\mu \boldsymbol{\beta}_{q_2}^\dagger   \boldsymbol{\beta}_{q_1} ) \right]\\
&= m^4 \int_{p,q_1,q_2,q_3}\frac{(q_1 \cdot q_2) }{(p^2+m^2)^3}\Big[ 1 - \frac{q_1^2 + (q_1+q_2)^2}{p^2+m^2} + \frac{ p^2 \left(q_1^2 + (q_1+q_2)^2 + q_1 \cdot (q_1+q_2) \right)}{(p^2+m^2)^2} \Big] \\
&\times \tr\left[ \boldsymbol{\beta}_{-q_1-q_2-q_3}^\dagger \delta \boldsymbol{\beta}_{q_3}  \boldsymbol{\beta}_{q_2}^\dagger   \boldsymbol{\beta}_{q_1}  \right]\\
&= m^4 \int_{q_1,q_2,q_3}(q_1 \cdot q_2)  \Big[ \frac{1}{32\pi^2 m^2} - \frac{1}{192\pi^2 m^4} \left( q_1^2 + q_2^2 + q_1 \cdot q_2\right) \Big] \tr'\left[ \boldsymbol{\beta}_{-q_1-q_2-q_3}^\dagger \delta \boldsymbol{\beta}_{q_3}  \boldsymbol{\beta}_{q_2}^\dagger   \boldsymbol{\beta}_{q_1}  \right]\\
&= \int d^4 x  \Big[ -\frac{m^2}{32 \pi^2}\tr'\left[ \boldsymbol{\beta}^\dagger \delta \boldsymbol{\beta} \partial_\mu \boldsymbol{\beta}^\dagger \partial_\mu \boldsymbol{\beta}\right]  - \frac{1}{192\pi^2}\tr'\left[ \boldsymbol{\beta}^\dagger \delta \boldsymbol{\beta}\left(  \partial_\mu \boldsymbol{\beta}^\dagger \partial_\mu \partial^2\boldsymbol{\beta} + \partial_\mu \partial^2 \boldsymbol{\beta}^\dagger \partial_\mu \boldsymbol{\beta}+\partial_\mu \partial_\nu \boldsymbol{\beta}^\dagger \partial_\mu \partial_\nu \boldsymbol{\beta} \right)\right]\Big]\\
&= \int d^4 x  \Big[ -\frac{m^2}{16 \pi^2}\tr\Big[ Q^{\mathbb{C}\dagger} \delta Q^{\mathbb{C}} \partial_\mu Q^{\mathbb{C}\dagger} \partial_\mu Q^{\mathbb{C}} + Q^{\mathbb{C}} \delta Q^{\mathbb{C}\dagger} \partial_\mu Q^{\mathbb{C}} \partial_\mu Q^{\mathbb{C}\dagger}\Big] \\
& - \frac{1}{96\pi^2}\tr\left[ 
\thead{
Q^{\mathbb{C}\dagger} \delta Q^{\mathbb{C}}\left(  \partial_\mu Q^{\mathbb{C}\dagger} \partial_\mu \partial^2 Q^{\mathbb{C}} + \partial_\mu \partial^2 Q^{\mathbb{C}\dagger} \partial_\mu Q^{\mathbb{C}}+\partial_\mu \partial_\nu Q^{\mathbb{C}\dagger} \partial_\mu \partial_\nu Q^{\mathbb{C}} \right)\\
+Q^{\mathbb{C}} \delta Q^{\mathbb{C}\dagger}\left(  \partial_\mu Q^{\mathbb{C}} \partial_\mu \partial^2 Q^{\mathbb{C}\dagger} + \partial_\mu \partial^2 Q^{\mathbb{C}} \partial_\mu Q^{\mathbb{C}\dagger}+\partial_\mu \partial_\nu Q^{\mathbb{C}} \partial_\mu \partial_\nu Q^{\mathbb{C}\dagger} \right)\\
}
\right]\Big]
\end{align*}
 In passing from the 6th to the last line we have used the symmetrization trick in arriving at \Eq{extra3D_1_1}.  Using $\delta Q^{\mathbb{C} \dagger} = - Q^{\mathbb{C} \dagger} \delta Q^{\mathbb{C} }Q^{\mathbb{C} \dagger} $, the first term in the last line gives zero. The second term can be evaluated using the same formula repeatedly. After some straight-forward expansion, most terms cancel out and we are left with
\begin{align}
& -\frac{1}{96\pi^2}\int d^4 x \,  \tr\Big[Q^{\mathbb{C} \dagger} \delta Q^{\mathbb{C} }Q^{\mathbb{C} \dagger} \left( \partial^2 Q^{\mathbb{C} } Q^{\mathbb{C} \dagger} \partial_\mu Q^{\mathbb{C} }Q^{\mathbb{C} \dagger} \partial_\mu Q^{\mathbb{C} } -   \partial_\mu Q^{\mathbb{C} }Q^{\mathbb{C} \dagger} \partial_\mu Q^{\mathbb{C} } Q^{\mathbb{C} \dagger} \partial^2 Q^{\mathbb{C} }\right) \Big]
\label{extra3D_2_2}
\end{align}\\

Summing over \Eq{extra3D_1_1}, \Eq{extra3D_1_3}, and \Eq{extra3D_2_2}, we obtain 
\begin{align}
\delta W_{\rm stiffness}+\delta \left( \frac{1}{92\pi^2} \int d^4 x \, \tr \left[ 
\thead{
 \partial_\mu Q^{\mathbb{C}\dagger} \partial_\mu Q^{\mathbb{C}} \partial_\nu Q^{\mathbb{C}\dagger} \partial_\nu Q^{\mathbb{C}}\\
-\frac{1}{2} \partial_\mu Q^{\mathbb{C}\dagger} \partial_\nu Q^{\mathbb{C}} \partial_\mu Q^{\mathbb{C}\dagger} \partial_\nu Q^{\mathbb{C}}\\
-\partial^2 Q^{\mathbb{C}} \partial^2 Q^{\mathbb{C}\dagger} \\
}
\right] \right).
\label{extraStiffness}
\end{align}
\noindent All these terms are real. At low energy and long wavelength they are dominated by the stiffness term. In appendix \ref{esnls} we shall refer to the stiffness term plus these extra terms as the ``generalized stiffness'' term.\\

To summarize, including all ``generalized stiffness'' terms, the non-linear sigma model is given by 
\begin{align}
W[Q^{\mathbb{C}}]&={1\over 2\lambda_4^2} \int_\mathcal{M} d^4 x \, \tr\left[  \partial_{\mu} Q^{\mathbb{C}} \partial^{\mu}Q^{\mathbb{C}\dagger}\right]  - \frac{2\pi}{480\pi^3}    \int\limits_{\mathcal{B}}     \, \tr \Big[\left(\tilde{Q}^{\mathbb{C}\dagger} d\tilde{Q}^{\mathbb{C}} \right)^5    \Big] \nn
&+ \frac{1}{92\pi^2} \int_\mathcal{M} d^4 x \, \tr \left[ 
\thead{
	\partial_\mu Q^{\mathbb{C}\dagger} \partial_\mu Q^{\mathbb{C}} \partial_\nu Q^{\mathbb{C}\dagger} \partial_\nu Q^{\mathbb{C}}\\
	-\frac{1}{2} \partial_\mu Q^{\mathbb{C}\dagger} \partial_\nu Q^{\mathbb{C}} \partial_\mu Q^{\mathbb{C}\dagger} \partial_\nu Q^{\mathbb{C}}\\
	-\partial^2 Q^{\mathbb{C}} \partial^2 Q^{\mathbb{C}\dagger} \\
}
\right]
\label{nls3Call}
\end{align}

\section{Emergent symmetries of the nonlinear sigma models}\label{esnls}
\hfill

In this appendix, we shall generalize the discussions in section \ref{symmNLSigma} to $(2+1)$-D and $(3+1)$-D, namely, showing the nonlinear sigma models respect the full emergent symmetries of the massless free fermion theories (see table \ref{tab:symmRestore}, or appendix \ref{appendix:emergentSymm} ).\\ 

As we explained in appendix \ref{appendix:fermionIntLargen}, the nonlinear sigma models in real classes can be derived from the complex classes by restricting $Q^{\mathbb{R}}$ to the appropriate sub-mass manifold of $Q^{\mathbb{C}}$. Similarly, for each space-time dimension the emergent symmetry group of the real class is a subgroup of the complex class (see table \ref{tab:symmRestore}). Hence, once we have matched the symmetries (between the nonlinear sigma models and fermion theories) for the complex class, it is straightforward to do the same for the real class. All we need to do is to restrict the order parameters to the appropriate sub-mass manifold and the symmetries to the appropriate subgroup.  Therefore we shall focus on the complex classes in the following.\\

\subsection{Complex class in $(2+1)$-D }
\hfill

The nonlinear sigma model is given by \Eq{nls2C}, namely,
\begin{align*}
W[Q^{\mathbb{C}}]= {1\over 2\lambda_3}   \int d^3 x \, \tr\left[  \partial_{\mu} Q^{\mathbb{C}} \partial^{\mu}Q^{\mathbb{C}}  \right] -  \frac{ 2 \pi i }{256 \pi^2}      \int\limits_{\mathcal{B}}  \, \tr \left[\tilde{Q}^{\mathbb{C}}    \,\left( d \tilde{Q}^{\mathbb{C}} \right)^4 \right] .
\end{align*}\\

{(i) \it Global $U(n)$\\}

 Using the cyclic invariance of trace, the action in \Eq{nls2C} clearly respects the $U(n)$ symmetry $$Q^\mathbb{C} \rightarrow g^\dagger \cdot Q^\mathbb{C} \cdot g.$$ \\

{(ii)\it Charge conjugation\\}

$Q^\mathbb{C}$ transforms under the charge conjugation as $$Q^\mathbb{C} \xrightarrow{C}  \left(Q^\mathbb{C}\right)^T.$$ Under such transformation the stiffness term becomes 
\be &&{1\over 2\lambda_3}    \int d^3 x \, \tr\left[  \partial_{\mu} \left(Q^{\mathbb{C}}\right)^T \partial^{\mu}\left(Q^{\mathbb{C}}\right)^T  \right]\nn
&&={1\over 2\lambda_3}      \int d^3 x \, \tr\left[ \partial^{\mu}Q^{\mathbb{C}} \partial_{\mu} Q^{\mathbb{C}}  \right]\nonumber\ee
Hence is invariant. In passing to the last line we have used the fact that the trace of a transposed matrix is the same as that of the original.\\

Under charge conjugation the WZW term transforms as
\begin{align*}
&-  \frac{ 2 \pi i }{256 \pi^2} \int_{\mathcal{B}}      \tr \left[\tilde{Q}^{\mathbb{C}}    \,\left( d \tilde{Q}^{\mathbb{C}} \right)^4 \right] 
\xrightarrow{C} -  \frac{ 2 \pi i }{256 \pi^2}   \int_{\mathcal{B}}    \tr \left[(\tilde{Q}^{\mathbb{C}})^T    \,\left( d (\tilde{Q}^{\mathbb{C}})^T \right)^4 \right] \\
&= -  \frac{ 2 \pi i }{256 \pi^2}   \int_{\mathcal{B}}   \tr \left[ \left( d \tilde{Q}^{\mathbb{C}} \right)^4 \, \tilde{Q}^{\mathbb{C}}     \right]\\ &=-  \frac{ 2 \pi i }{256 \pi^2}    \int_{\mathcal{B}}   \tr \left[\tilde{Q}^{\mathbb{C}}    \,\left( d \tilde{Q}^{\mathbb{C}} \right)^4 \right].
\end{align*}
In passing to the second line we have used the transposing invariance of the trace, and the fact the reordering caused by transposing results in an even number of exchanges between the differential $1$-forms, hence there is no sign change. The cyclic property of trace is used for the last equality. Therefore the WZW term is charge conjugation invariant.\\

{(iii)\it Time reversal\\}

Under time-reversal $Q^\mathbb{C}$ transforms as $$Q^\mathbb{C} \xrightarrow{T} - (Q^\mathbb{C})^*=-(Q^\mathbb{C})^T.$$ (Here we have used the fact that $Q^\mathbb{C}$ is hermitian). This results in the following transformation of the stiffness term
\begin{align*}
&{1\over 2\lambda_3}    \int d^3x~   \tr\left[  \partial_{\mu} Q^{\mathbb{C}} \partial^{\mu}Q^{\mathbb{C}}  \right] \\ 
\xrightarrow{T} &\left({1\over 2\lambda_3} \int d^3x~     \tr\left[ \partial_{\mu} (-Q^{\mathbb{C}*}) \partial^{\mu}(-Q^{\mathbb{C}*})  \right] \right)^* \\
= &{1\over 2\lambda_3}     \int d^3x~    \tr\left[  \partial_{\mu} Q^{\mathbb{C}} \partial^{\mu}Q^{\mathbb{C}}  \right]   \\
\end{align*}
In passing to the second line we have used the fact that in Euclidean space-time the Boltzmann weight needs to be complex conjugated under anti-unitary transformation. Therefore, the stiffness term is time reversal invariant.\\

The WZW term transforms as follows under time reversal
\begin{align*}
&-  \frac{ 2 \pi i }{256 \pi^2} \int_{\mathcal{B}}   \tr \left[\tilde{Q}^{\mathbb{C}}    \,\left( d \tilde{Q}^{\mathbb{C}} \right)^4 \right]  \xrightarrow{T} \left(  -\frac{ 2 \pi i }{256 \pi^2}  \int_{\mathcal{B}}     \tr \left[(-\tilde{Q}^{\mathbb{C}})^*    \,\left( d (-\tilde{Q}^{\mathbb{C}})^* \right)^4 \right] \right)^*\\
&=-  \frac{ 2 \pi i }{256 \pi^2}  \int_{\mathcal{B}}     \tr \left[\tilde{Q}^{\mathbb{C}}    \,\left( d \tilde{Q}^{\mathbb{C}} \right)^4 \right],
\end{align*}
\noindent where the five negative signs associated with transposing are canceled out by the negative sign arising from complex conjugation of $i$. Thus the WZW term is time reversal invariant. 
\\

In summary, the nonlinear sigma model respects the full emergent symmetries of the massless fermion theory (see table \ref{tab:symmRestore}).

\subsection{Complex class in $(3+1)$-D}
\hfill

The nonlinear sigma model  in \Eq{nls3Call} is given by  
\begin{align*}
W[Q^{\mathbb{C}}]=&{1\over 2\lambda_4^2}  \int_\mathcal{M} d^4 x \, \tr\left[  \partial_{\mu} Q^{\mathbb{C}} \partial^{\mu}Q^{\mathbb{C}\dagger}\right]  - \frac{2\pi}{480\pi^3}    \int\limits_{\mathcal{B}}   \, \tr \Big[\left(\tilde{Q}^{\mathbb{C}\dagger} d\tilde{Q}^{\mathbb{C}} \right)^5    \Big] \\
+& \frac{1}{92\pi^2} \int_\mathcal{M} d^4 x \, \tr \left[ 
\thead{
	\partial_\mu Q^{\mathbb{C}\dagger} \partial_\mu Q^{\mathbb{C}} \partial_\nu Q^{\mathbb{C}\dagger} \partial_\nu Q^{\mathbb{C}}\\
	-\frac{1}{2} \partial_\mu Q^{\mathbb{C}\dagger} \partial_\nu Q^{\mathbb{C}} \partial_\mu Q^{\mathbb{C}\dagger} \partial_\nu Q^{\mathbb{C}}\\
	-\partial^2 Q^{\mathbb{C}} \partial^2 Q^{\mathbb{C}\dagger} \\
}
\right]
\end{align*}\\

{(i) \it Global $U(n)\times U(n)$\\}

\Eq{nls3Call} is clearly invariant under the $U_+(n)\times U_-(n)$ transformations
$$Q^\mathbb{C} \rightarrow g_-^\dagger \cdot Q^\mathbb{C} \cdot g_+.$$ This is
because in \Eq{nls3Call} $Q^\mathbb{C}$ and $Q^{\mathbb{C}\dagger}$ appears sequentially.\\

{(ii)\it Charge conjugation\\}

Under charge-conjugation $Q^\mathbb{C}$ transforms as $$Q^\mathbb{C} \xrightarrow{C} (Q^\mathbb{C})^T.$$ Under such transformation the ``generalized stiffness'' terms transforms as
\be
&&{1\over 2\lambda_4^2}  \int d^4 x \, \tr\left[  \partial_{\mu} \left(Q^{\mathbb{C}}\right)^T \partial^{\mu}\left(Q^{\mathbb{C}\dagger}\right)^T\right] \nn&&+ \frac{1}{92\pi^2} \int d^4 x \, \tr \left[ 
\thead{
	\partial_\mu \left(Q^{\mathbb{C}\dagger}\right)^T \partial_\mu \left(Q^{\mathbb{C}}\right)^T \partial_\nu \left(Q^{\mathbb{C}\dagger}\right)^T \partial_\nu \left(Q^{\mathbb{C}}\right)^T\\
	-\frac{1}{2} \partial_\mu \left(Q^{\mathbb{C}\dagger}\right)^T \partial_\nu \left(Q^{\mathbb{C}}\right)^T\partial_\mu \left(Q^{\mathbb{C}\dagger}\right)^T \partial_\nu \left(Q^{\mathbb{C}}\right)^T\\
	-\partial^2 \left(Q^{\mathbb{C}}\right)^T \partial^2\left(Q^{\mathbb{C}\dagger}\right)^T\\
}
\right]\nn 
&&={1\over 2\lambda_4^2}  \int d^4 x \, \tr\left[  \partial_{\mu} Q^{\mathbb{C}} \partial^{\mu}Q^{\mathbb{C}\dagger}\right] + \frac{1}{92\pi^2} \int d^4 x \, \tr \left[ 
\thead{
	\partial_\mu Q^{\mathbb{C}\dagger} \partial_\mu Q^{\mathbb{C}} \partial_\nu Q^{\mathbb{C}\dagger} \partial_\nu Q^{\mathbb{C}}\\
	-\frac{1}{2} \partial_\mu Q^{\mathbb{C}\dagger} \partial_\nu Q^{\mathbb{C}} \partial_\mu Q^{\mathbb{C}\dagger} \partial_\nu Q^{\mathbb{C}}\\
	-\partial^2 Q^{\mathbb{C}} \partial^2 Q^{\mathbb{C}\dagger} \\
}
\right]\nn
\ee
In arriving at the final line we have used the transposing invariance of the trace. Therefore the ``generalized stiffness'' terms are charge conjugation invariant. \\

Under charge conjugation, the WZW term transforms as
\begin{align*}
&- \frac{2\pi}{480\pi^3} \int_{\mathcal{B}}    \tr \Big[\left(\tilde{Q}^{\mathbb{C}\dagger} d\tilde{Q}^{\mathbb{C}} \right)^5    \Big] \\
\xrightarrow{C} &- \frac{2\pi}{480\pi^3}  \int_{\mathcal{B}}    \tr \Big[\left( (\tilde{Q}^{\mathbb{C}})^* d (\tilde{Q}^{\mathbb{C}} )^T \right)^5    \Big]\\
=&- \frac{2\pi}{480\pi^3}  \int_{\mathcal{B}}   \tr \Big[\left( d\tilde{Q}^{\mathbb{C}}  \, \tilde{Q}^{\mathbb{C}\dagger} \right)^5    \Big] \\=&- \frac{2\pi}{480\pi^3}   \int_{\mathcal{B}}   \tr \Big[\left(\tilde{Q}^{\mathbb{C}\dagger} d\tilde{Q}^{\mathbb{C}} \right)^5    \Big] 
\end{align*}
In passing to the third line we have used the transposing invariance of the trace.  Note that there is no extra sign because the number of exchanges between 1-forms is even (10 times). In arriving at the last line, the last $\tilde{Q}^{\mathbb{C}\dagger}$ is moved to the front by the cyclic invariance of the trace. Thus the WZW term is charge conjugation invariant. \\

{(iii)\it Time reversal\\}

Under time-reversal $Q^\mathbb{C}$ transforms as $$Q^\mathbb{C} \xrightarrow{T} (Q^\mathbb{C})^*.$$
\begin{align*}
W\left[ Q^{\mathbb{C}}\right] \xrightarrow{T} \left(W\left[ \left( Q^{\mathbb{C}}\right)^*\right] \right)^* = W\left[ Q^{\mathbb{C}}\right].
\end{align*}
This is because all the coefficients (including those in front of the generalized stiffness terms and the WZW term) in the nonlinear sigma model are real, the complex conjugation of the Boltzmann weight cancels out with complex conjugation in $Q^\mathbb{C*}$.\\

To summarize, the nonlinear sigma model is invariant under the full emergent symmetries of the massless fermion theory (see table \ref{tab:symmRestore}).

\section{Anomalies of the nonlinear sigma models}
\label{appendix:anomaliesb}
\hfill

To reveal the 't Hooft anomalies of the non-linear sigma model we first need to gauge it.   
In this section, we shall extend the discussions in section \ref{tHooftWZW} to gauge the continuous symmetries of nonlinear sigma models in $(1+1)$-D, $(2+1)$-D, and $(3+1)$-D. We shall adopt Witten's trial-and-error method \cite{Witten1983b}.\\

We have discussed at the beginning of appendix \ref{esnls} that the mass manifold and emergent symmetries of the non-linear sigma model of real classes are the submanifold and sub-group of the corresponding sigma model of complex classes. Consequently, once one knows how to gauge the nonlinear sigma models in the complex classes,  one simply needs to restrict the order parameters ($Q^{\mathbb{R}}$) to the  submanifold, and the gauge group to the subgroup,  to derive the gauged non-linear sigma models of real classes.\\
 
\subsection{The ('t Hooft) anomalies associated with continuous symmetries}
\hfill

\subsubsection{Complex class in $(1+1)$-D}
\hfill

The discussion for gauging the nonlinear sigma model of complex class in $(1+1)$-D was already in section \ref{tHooftWZW}. We will not repeat the argument but just quote the result here:
\begin{align}
	W[Q^{\mathbb{C}},A_+,A_-]&=-\frac{1}{8\pi} \int\limits_{\mathcal{M}} d^2 x \, \tr\left[\left(Q^{\mathbb{C} \dagger} \left( \partial_\mu Q^{\mathbb{C}} - i Q^{\mathbb{C}} A_{+,\mu} + i A_{-,\mu} Q^{\mathbb{C}} \right) \right)^2 \right]\nn
	&-\frac{i}{12\pi} \int\limits_{\mathcal{B}} \tr \left[\left(Q^{\mathbb{C} \dagger} d Q^{\mathbb{C}} \right)^3 \right]
	- \frac{1}{4\pi} \int\limits_{\mathcal{M}} \tr\Big\{ A_+ \left( Q^{\mathbb{C} \dagger} d Q^{\mathbb{C} } \right)\nn& + A_- \left(d Q^{\mathbb{C} } Q^{\mathbb{C} \dagger}  \right)+ i A_+ Q^{\mathbb{C} \dagger}  A_- Q^{\mathbb{C} } \Big\} . 
	\label{gaugedWZW1C}
\end{align}
	
\noindent Under infinitesimal $U_+(n) \times U_-(n)$ gauge transformation,
\begin{align*}
&Q^{\mathbb{C}} \rightarrow  e^{-i \epsilon_-} Q^{\mathbb{C}} e^{i \epsilon_+}\nn
&A_\pm \rightarrow  A_\pm + d\epsilon_\pm + i [A_\pm, \epsilon_\pm],
\end{align*}
\Eq{gaugedWZW1C} acquires an addition piece  
\begin{align}
	\delta W = -\frac{i}{4\pi} \int\limits_{\mathcal{M}} \tr \left[ A_+  d\epsilon_+ - A_- d\epsilon_-   \right].
\label{gaugedWZWTrans1C}
\end{align}
Thus \Eq{gaugedWZW1C} is not gauge invariant, revealing the 't Hooft anomaly associated with $U_+(n) \times U_-(n)$. However, when one only gauges the diagonal $U(n)$, i.e., $A_+=A_- := A$ and $\epsilon_+ = \epsilon_- = \epsilon$, the non gauge invariant terms in  \Eq{gaugedWZWTrans1C} cancels out. Hence \Eq{gaugedWZW1C} is anomaly free with respect to the diagonal $U(n)$. This agrees with the free fermion anomaly.

\subsubsection{Real class in $(1+1)$-D}
\hfill

The gauged nonlinear sigma model for real class in $(1+1)$-D can be derived from the complex class by 1) restricting the order parameter $Q^\mathbb{C}\in U(n)$ to the subspace $Q^\mathbb{R}\in O(n)$, 2) restricting the gauge group  from $U_+(n) \times U_-(n)$ to $O_+(n) \times O_-(n)$, and 3) divide the nonlinear sigma model by a factor of two (see \ref{appendix:realComplexInt}). The result is

\begin{align}
W[Q^{\mathbb{R}},A_+,A_-] &= -\frac{1}{16\pi} \int\limits_{\mathcal{M}} d^2 x \, \tr \left[\left((Q^{\mathbb{R}})^T \left( \partial_\mu Q^{\mathbb{R}} - i Q^{\mathbb{R}} A_{+,\mu} + i A_{-,\mu} Q^{\mathbb{R}} \right) \right)^2 \right] \nn
	&+  \frac{2\pi i}{48\pi^2}    \int\limits_{\mathcal{B}}  \, \tr \Big[   \left( (\tilde{Q}^{\mathbb{R}})^T   d \tilde{Q}^{\mathbb{R}}  \right)^3    \Big]
+ \frac{1}{8\pi} \int\limits_{\mathcal{M}} \tr\Big\{ A_+ \left( d Q^{\mathbb{R}} (Q^{\mathbb{R}})^T\right)\nn
&+  A_- \left(  (Q^{\mathbb{R}})^T d Q^{\mathbb{R}} \right) - i A_+ (Q^{\mathbb{R}})^T A_- Q^{\mathbb{R}}\Big\}.
\label{gaugedWZW1R}
\end{align}

Here $A_\pm$ are the gauge fields associated with $O_+(n) \times O_-(n)$. 
\noindent Under the $O_+(n) \times O_-(n)$ gauge transformation,
\begin{align*}
&Q^{\mathbb{R}} \rightarrow  e^{-i \epsilon_-} Q^{\mathbb{R}} e^{i \epsilon_+}\nn
&A_\pm \rightarrow  A_\pm + d\epsilon_\pm + i [A_\pm, \epsilon_\pm],
\end{align*}
(here $\e_+$ and $\e_-$ are imaginary anti-symmetric matrices) \Eq{gaugedWZW1R} acquires an addition piece  
\begin{align*}
	\delta W = -\frac{i}{8\pi} \int\limits_{\mathcal{M}} \tr \left[ A_+  d\epsilon_+ - A_- d\epsilon_-   \right],
\end{align*}
manifesting the 't Hooft anomaly associated with $O_+(n) \times O_-(n)$. Again, when only the diagonal $O(n)$ is gauged, \Eq{gaugedWZW1R} is anomaly-free, consistent with the free fermion prediction.\\

\subsubsection{Complex class in $(2+1)$-D}
\hfill

In the following, we carry out Witten's method \cite{Witten1983b} for the non-linear sigma model.  The emergent continuous symmetry is $U(n)$ and under gauge transformation $Q^{\mathbb{C}}$ and $A$ change according to 
\be
	&& Q^{\mathbb{C}}\ra  Q^{\mathbb{C}}+\delta Q^{\mathbb{C}}~{\rm where}~\delta Q^{\mathbb{C}} = i [ Q^{\mathbb{C}}, \epsilon  ] \nn
	&&A\ra A+\delta A~{\rm where}~\delta A = d \epsilon + i [A, \epsilon].
	\label{gQA}
\ee
The gauge field enters stiffness term in \Eq{nls2C} via the minimal coupling,
\begin{align*}
	W_{\rm stiff}[Q^{\mathbb{C}}, A] ={1\over 2\lambda_3} \int\limits_{\mathcal{M}} d^3 x \, \tr \Big[\Big(\p_\mu Q^{\mathbb{C}}+i [A_\mu,Q^{\mathbb{C}}]\Big)^2\Big] 
\end{align*}
which is gauge invariant. \\

Following Witten's trial-and-error method, we now determine how gauge field enters through the WZW term. Under \Eq{gQA} the WZW term acquires an addition piece
\begin{align}
&\delta \left( \int\limits_\mathcal{B} \tr \left[ Q^\mathbb{C}( d Q^\mathbb{C})^4 \right] \right)\notag\\
=& \int\limits_\mathcal{B} \tr\left[ \delta Q^\mathbb{C}( d Q^\mathbb{C})^4 + 4Q d(\delta Q^\mathbb{C}) (dQ^\mathbb{C})^3 \right]\notag\\
=& \int\limits_\mathcal{B} \tr\left[ 5\delta Q^\mathbb{C}( d Q^\mathbb{C})^4 +d\left( 4 Q^\mathbb{C} \delta Q^\mathbb{C} (dQ^\mathbb{C})^3 \right) \right] \notag\\
=& 5\int\limits_\mathcal{B} \tr\left[  i \left(Q^\mathbb{C}\epsilon- \epsilon Q^\mathbb{C}\right)( d Q^\mathbb{C})^4  \right]  + \int\limits_\mathcal{M} \tr\left[  4 Q^\mathbb{C} i\left(Q^\mathbb{C}\epsilon- \epsilon Q^\mathbb{C}\right) (dQ^\mathbb{C})^3 \right] \notag\\
=&0 + 8 i \int\limits_\mathcal{M}\tr\left[  \epsilon (dQ^\mathbb{C})^3 \right] \notag\\
=&-8 i \int\limits_\mathcal{M} \tr\left[  d\epsilon \, Q^\mathbb{C}(dQ^\mathbb{C})^2 \right]
\label{gaugedWZW2C0}
\end{align}
In passing to the second line we used the constraint $Q^\mathbb{C}dQ^\mathbb{C} = - dQ^\mathbb{C} \, Q^\mathbb{C}$. An integration by part is done from the 2nd to the 3rd line. The 1st term in the 4th line vanishes because we can repeatedly use $Q^\mathbb{C}dQ^\mathbb{C} = - dQ^\mathbb{C} \, Q^\mathbb{C}$ and the cyclic invariance of the trace to show
$$
\tr\left[\epsilon Q^\mathbb{C}( d Q^\mathbb{C})^4  \right]=\tr\left[
Q^\mathbb{C}\epsilon ( d Q^\mathbb{C})^4  \right]. $$
\\

To cancel out the gauge dependent part of \Eq{gaugedWZW2C0}, we add an additional term 
\be\text{Added term}~~8 i \int\limits_\mathcal{M} \tr\left[ A \, Q^\mathbb{C}(dQ^\mathbb{C})^2 \right].\label{add1}\ee
Under the gauge transformation \Eq{gQA} this additional term transforms into
\begin{align}
&\delta \left(8i\int_\mathcal{M} \tr \left[A Q^\mathbb{C} (dQ^\mathbb{C})^2 \right] \right)\notag\\
=& 8i\int_\mathcal{M} \tr\left[\delta A Q^\mathbb{C} (d Q^\mathbb{C})^2 + \delta Q^\mathbb{C} (dQ^\mathbb{C})^2 A + d(\delta Q^\mathbb{C}) \left( dQ^\mathbb{C} \, A \, Q^\mathbb{C} + A \, Q^\mathbb{C} \, dQ^\mathbb{C} \right) \right]\notag\\
=& 8i\int_\mathcal{M} \tr\left[\delta A \, Q^\mathbb{C} (d Q^\mathbb{C})^2 + \delta Q^\mathbb{C} \left( 
\thead{
	(dQ^\mathbb{C})^2 A  +dQ^\mathbb{C} \, dA \, Q^\mathbb{C} - dQ^\mathbb{C} \, A \, dQ^\mathbb{C} \\
	- dA \, Q^\mathbb{C} \, dQ^\mathbb{C} + A \, dQ^\mathbb{C} \, d Q^\mathbb{C}
}
\right) \right]\notag\\
=& 8i\int_\mathcal{M} \tr\left[d\epsilon \, Q^\mathbb{C} (d Q^\mathbb{C})^2 + i[A,\epsilon] Q^\mathbb{C}(dQ^\mathbb{C})^2+ i[Q^\mathbb{C},\epsilon] \left( 
\thead{
	(dQ^\mathbb{C})^2 A  +dQ^\mathbb{C} \, dA \, Q^\mathbb{C} - dQ^\mathbb{C} \, A \, dQ^\mathbb{C} \\
	- dA \, Q^\mathbb{C} \, dQ^\mathbb{C} + A \, dQ^\mathbb{C} \, d Q^\mathbb{C}
}
\right) \right]\notag\\
=&8i \int_\mathcal{M} \tr\left[d\epsilon \, Q^\mathbb{C} (d Q^\mathbb{C})^2\right]- 8\int_\mathcal{M} \tr\left[\epsilon \, d\left( 
\thead{
	Q^\mathbb{C}\,dQ^\mathbb{C}\,A\,Q^\mathbb{C} + Q^\mathbb{C} \, A \, Q^\mathbb{C} \, dQ^\mathbb{C}\\
	Q^\mathbb{C} \, dA + dA \, Q^\mathbb{C}
}\right) \right]\notag\\
=&8i \int_\mathcal{M} \tr\left[d\epsilon \, Q^\mathbb{C} (d Q^\mathbb{C})^2\right]+ 8\int_\mathcal{M} \tr\left[d\epsilon \, \left( 
\thead{
	Q^\mathbb{C}\,dQ^\mathbb{C}\,A\,Q^\mathbb{C} + Q^\mathbb{C} \, A \, Q^\mathbb{C} \, dQ^\mathbb{C}\\
	Q^\mathbb{C} \, dA + dA \, Q^\mathbb{C}
}\right) \right]
\label{gaugedWZW2C1}
\end{align}
The first term of the final result cancels the gauge dependent term of \Eq{gaugedWZW2C0} by design. We continue to add the additional term 
\be\text{Added term}~~ -8 \int_\mathcal{M} \tr \left[ (AQ^\mathbb{C})^2 dQ^\mathbb{C}  \right]\label{add2}\ee in an attempt to cancel 
the term \begin{align}\label{attc} 8\int_\mathcal{M} \tr\left[d\epsilon \, \left( 
\thead{
	Q^\mathbb{C}\,dQ^\mathbb{C}\,A\,Q^\mathbb{C} + Q^\mathbb{C} \, A \, Q^\mathbb{C} \, dQ^\mathbb{C}
}\right) \right]\end{align}
in \Eq{gaugedWZW2C1}. Under the gauge transformation (\Eq{gQA}) the added term transforms as
\begin{align}
	&\delta \left( -8 \int_\mathcal{M} \tr \left[ (AQ^\mathbb{C})^2 dQ^\mathbb{C}  \right]\right)\notag\\
	=&-8 \int_\mathcal{M} \tr \left[ 
	\thead{ 
		\delta A \left( Q^\mathbb{C} \, A \, Q^\mathbb{C}\, dQ^\mathbb{C} + Q^\mathbb{C}\, dQ^\mathbb{C} \, A\, Q^\mathbb{C} \right)  \\
		+\delta Q^\mathbb{C} \left( 
		\thead{
			A \,Q^\mathbb{C}\, dQ^\mathbb{C} \,A + dQ^\mathbb{C}\, A \,Q^\mathbb{C} \,A \\
			-dA\, Q^\mathbb{C}\, A \, Q^\mathbb{C} + A \,dQ^\mathbb{C}\, A\, Q^\mathbb{C}\\
			+A\, Q^\mathbb{C}\, dA \, Q^\mathbb{C} - A \, Q^\mathbb{C}\, A\, dQ^\mathbb{C}
		}\right)
	}\right]\notag\\
	=&-8 \int_\mathcal{M} \tr \left[ 
	\thead{ 
		d\epsilon \left( Q^\mathbb{C} \, A \, Q^\mathbb{C}\, dQ^\mathbb{C} + Q^\mathbb{C}\, dQ^\mathbb{C} \, A\, Q^\mathbb{C} \right)  \\
		+ i \, \epsilon \, d\left( -A \, Q^\mathbb{C} \, A + Q^\mathbb{C} \, A \, Q^\mathbb{C} \, A \, Q^\mathbb{C} \right)
	}\right] \notag\\
	=& \int_\mathcal{M} \tr \left[ 
	\thead{ 
		-8 d\epsilon \left( Q^\mathbb{C} \, A \, Q^\mathbb{C}\, dQ^\mathbb{C} + Q^\mathbb{C}\, dQ^\mathbb{C} \, A\, Q^\mathbb{C} \right)  \\
		+ 8 i\, d\epsilon \, \left( -A \, Q^\mathbb{C} \, A + Q^\mathbb{C} \, A \, Q^\mathbb{C} \, A \, Q^\mathbb{C} \right)
	}\right] 
\label{gaugedWZW2C2}
\end{align}
\noindent The top line in the final result achieves canceling out \Eq{attc}. Now we focus on canceling out the terms in the bottom line %in the final result 
of \Eq{gaugedWZW2C2}. The term 
\begin{align}
\label{attc1} \int_\mathcal{M} \tr \left[ 
	\thead{ 
		+ 8 i\, d\epsilon \, \left( Q^\mathbb{C} \, A \, Q^\mathbb{C} \, A \, Q^\mathbb{C} \right)
	}\right] \end{align}
can be canceled out by adding the extra term
\be\text{Added term}~~-\frac{8i}{3} \int_\mathcal{M} \tr \left[ (A \,Q^\mathbb{C})^3 \right].\label{add3}\ee Under gauge transformation  the added term transforms as
\begin{align}
	&\delta \left(-\frac{8i}{3} \int_\mathcal{M} \tr \left[ (A \,Q^\mathbb{C})^3 \right] \right)\notag\\
	=&-8i \int_\mathcal{M} \tr \left[ \delta( A \, Q^\mathbb{C} )(A \,Q^\mathbb{C})^2 \right]\notag\\
	=&-8i \int_\mathcal{M} \tr \left[ \left( d\epsilon \, Q^\mathbb{C} + i [A,\epsilon]Q^\mathbb{C} + i A [Q^\mathbb{C},\epsilon] \right) (A \,Q^\mathbb{C})^2 \right] \notag\\
	=&-8i \int_\mathcal{M} \tr \left[ d\epsilon \, Q^\mathbb{C}(A \,Q^\mathbb{C})^2 \right]
\label{gaugedWZW2C3}
\end{align}
which indeed cancels \Eq{attc1}.
\noindent The remaining term
\begin{align}
\label{attc2}\int_\mathcal{M} \tr \left[ 
	\thead{ 
				+ 8 i\, d\epsilon \, \left( -A \, Q^\mathbb{C} \, A \right)
	}\right] 
\end{align}
in \Eq{gaugedWZW2C2} can be partially canceled by adding the extra term \be\text{Added term}~~ 8i \int_\mathcal{M}\tr \left[ A^3 Q^\mathbb{C} \right],\label{add4}\ee 
which transforms as 
\begin{align}
	&\delta \left(8i \int_\mathcal{M}\tr \left[ A^3 Q^\mathbb{C} \right] \right)\notag\\
	=&8i \int_\mathcal{M} \tr \left[ (d\epsilon + i[A,\epsilon]) \left( A^2 \, Q^\mathbb{C} + A\,Q^\mathbb{C}\,A + Q^\mathbb{C} \, A^2\right) + i[ Q^\mathbb{C}, \epsilon ]\left( A^3\right)\right]\notag\\
	=&8i \int_\mathcal{M} \tr \left[ d\epsilon \left( A^2 \, Q^\mathbb{C} + A\,Q^\mathbb{C}\,A + Q^\mathbb{C} \, A^2\right) \right]
	\label{gaugedWZW2C4}
\end{align}
under the gauge transformation. The second term in \Eq{gaugedWZW2C4} cancel \Eq{attc2}. \\

At this point, under the gauge transformation, the sum of the original WZW term and the added terms \Eq{add1},\Eq{add2},\Eq{add3},  \Eq{add4} acquires the extra piece
\begin{align}
	&\delta \left( \int\limits_\mathcal{B}  \tr \left[ Q^\mathbb{C}( d Q^\mathbb{C})^4 \right] +8 \int_\mathcal{M} \tr \left[
	\thead{
		i \, A \, Q^\mathbb{C} (dQ^\mathbb{C})^2- (AQ^\mathbb{C})^2 dQ^\mathbb{C}\\
		-\frac{i}{3}(A \,Q^\mathbb{C})^3 +i \, A^3 Q^\mathbb{C}
	}  
	\right] \right)\notag\\
	=& 8 \int\limits_\mathcal{M}  \tr \left[d\epsilon \left( Q^\mathbb{C} F + F Q^\mathbb{C}\right)\right]
	\label{gaugedWZW2C5}
\end{align}
\noindent where $F:= dA + iA^2$. This last non-gauge invariant term, \Eq{gaugedWZW2C5}, can also be canceled out by adding
\be
\text{Added term}~~-8 \int\limits_\mathcal{M}  \tr \left[ A \, Q^\mathbb{C} \, F + A \, F \, Q^\mathbb{C}  \right].
\label{add5}\ee
Indeed, under the gauge transformation the added term transforms as
\begin{align*}
	& \delta \left(-8 \int\limits_\mathcal{M}  \tr \left[ A \, Q^\mathbb{C} \, F + A \, F \, Q^\mathbb{C}  \right] \right) \\
	=&-8 \int\limits_\mathcal{M}  \tr \left[ 
	\thead{
		(d\epsilon + i [A,\epsilon]) ( Q^\mathbb{C}\, F + F \, Q^\mathbb{C}) \\
		+i \, [F,\epsilon] (A \, Q^\mathbb{C} + Q^\mathbb{C} \, A) \\
		+i \, [Q^\mathbb{C},\epsilon] (F \, A + A \, F)
	}
	\right]\\
	=&-8 \int\limits_\mathcal{M}  \tr \left[ 
		d\epsilon \,  \left( Q^\mathbb{C}\, F + F \, Q^\mathbb{C} \right)
	\right]
\end{align*}
which cancels  \Eq{gaugedWZW2C5}. Thus the entire  $U(n)$ symmetry can  be gauged without anomaly. This is consistent with the free fermion prediction.\\

In summary, the $U(n)$ gauged nonlinear sigma model in $(2+1)$-D is

\begin{align}
\label{gaugedWZW2C}
&W[Q^{\mathbb{C}},A] =  {1\over 2\lambda_3} \int\limits_{\mathcal{M}} d^3 x \, \tr \Big[\Big(\p_\mu Q^{\mathbb{C}}+i [A_\mu,Q^{\mathbb{C}}]\Big)^2\Big]-  \frac{ 2 \pi i }{256 \pi^2}\Big\{\int\limits_{\mathcal{B}} \tr \Big[\tilde{Q}^{\mathbb{C}}    \,\left( d \tilde{Q}^{\mathbb{C}} \right)^4 \Big]\notag\\
&+8 \int\limits_{\mathcal{M}} \tr \Big[ i A Q^{\mathbb{C}} (dQ^{\mathbb{C}})^2 - (A Q^{\mathbb{C}})^2 dQ^{\mathbb{C}} - \frac{i}{3} (A Q^{\mathbb{C}})^3 + i A^3 Q^{\mathbb{C}} -AQ^{\mathbb{C}}F - AFQ^{\mathbb{C}} \Big]
\Big\}.
\end{align}
\\

\subsubsection{Real class in $(2+1)$-D}
\hfill

The gauged nonlinear sigma model  can be derived from the results in preceding subsection by 1) restricting the order parameter $Q^\mathbb{C}\in \frac{U(n)}{U(n/2)\times U(n/2)}$ to the sub-manifold $Q^\mathbb{R}\in \frac{O(n)}{O(n/2)\times O(n/2)}$, 2) restricting the gauge group from $U(n) $ to $O(n)$, and 3) divide the effective action by a factor of two (see \ref{appendix:realComplexInt}). The resulting gauged nonlinear sigma model action is
\begin{align}
	\label{gaugedWZW2R}
	&W[Q^{\mathbb{R}},A] = {1\over 4\lambda_3} \int\limits_{\mathcal{M}} d^3 x \, \tr \Big[\Big(\p_\mu Q^{\mathbb{R}}+i [A_\mu,Q^{\mathbb{R}}]\Big)^2\Big]
	-  \frac{ 2 \pi i }{512 \pi^2} \Big\{ \int\limits_{\mathcal{B}} \tr \Big[\tilde{Q}^{\mathbb{R}}    \,\left( d \tilde{Q}^{\mathbb{R}} \right)^4 \Big] \notag \\
	&+ 8 \int\limits_{\mathcal{M}}  \tr \Big[ i A Q^{\mathbb{R}} (dQ^{\mathbb{R}})^2 - (A Q^{\mathbb{R}})^2 dQ^{\mathbb{R}} - \frac{i}{3} (A Q^{\mathbb{R}})^3 + i A^3 Q^{\mathbb{R}} -AQ^{\mathbb{R}}F - AFQ^{\mathbb{R}} \Big]
	\Big\}
\end{align}
Here $A$ is the gauge connection for the $O(n)$ gauge group. Again the entire  $O(n)$ symmetry is anomaly free, agreeing with the free fermion prediction.\\

\subsubsection{Complex class in $(3+1)$-D}
\hfill

The emergent symmetry is $U_+(n)\times U_-(n)$. The gauged WZW term was written down by Witten \cite{Witten1983b} with a minor correction in Ref.\cite{Kaymakcalan1984}. To simplify the notation, we will define $$\alpha_1 := dQ^{\mathbb{C}} \,  Q^{\mathbb{C}\dagger}, ~~\alpha_2 := Q^{\mathbb{C}\dagger} dQ^{\mathbb{C}}.$$ The derivation is rather long, so we shall not repeat it here. The result is \cite{Witten1983b}  
\begin{align}
	\label{gaugedWZW3C}
	&W[Q^{\mathbb{C}},A_+,A_-]\notag \\
	&=-{1\over 2\lambda_4^2}\int\limits_{\mathcal{M}} d^4 x \, \tr\left[\left(Q^{\mathbb{C}\dagger} \left( \partial_\mu Q^{\mathbb{C}} - i Q^{\mathbb{C}} A_{+,\mu} + i A_{-,\mu} Q^{\mathbb{C}} \right) \right)^2 \right]\notag\\
	& -\frac{2\pi}{480 \pi^3} \Big\{ \int_{\mathcal{B}}  \tr\left[(Q^{\mathbb{C}\dagger} dQ^{\mathbb{C}})^5\right] \notag\\
	&+ 5 \int_{\mathcal{M}}   \tr \left[ 
	\thead{
	-i \left(A_+ \alpha_2^3 + A_- \alpha_1^3 \right) 
	- \left( (dA_+ A_+ + A_+ d A_+)\alpha_2 +(dA_- A_- + A_- d A_-)\alpha_1    \right) \\
	+ dA_- dQ^{\mathbb{C}} A_+ Q^{\mathbb{C}\dagger} - dA_+ d(Q^{\mathbb{C}\dagger}) A_- Q^{\mathbb{C}} + A_+ Q^{\mathbb{C}\dagger} A_- Q^{\mathbb{C}} \alpha_2^2 - A_- Q^{\mathbb{C}} A_+ Q^{\mathbb{C}\dagger} \alpha_1^2 \\
	+ \frac{1}{2} \left( (A_- \alpha_1)^2 - (A_+ \alpha_2)^2 \right)- i \left( A_-^3 \alpha_1 + A_+^3 \alpha_2 \right) \\
	+i \left( \left( dA_+ A_+ + A_+ dA_+ \right) Q^{\mathbb{C}\dagger} A_- Q^{\mathbb{C}} - \left( dA_- A_- + A_- dA_-  \right)Q^{\mathbb{C}} A_+ Q^{\mathbb{C}\dagger} \right)\\
	- i \left( A_- Q^{\mathbb{C}} A_+ Q^{\mathbb{C}\dagger} A_- \alpha_1 +  A_+ Q^{\mathbb{C}\dagger} A_- Q^{\mathbb{C}} A_+ \alpha_2\right)\\
	+ \left( A_+^3 Q^{\mathbb{C}\dagger} A_- Q^{\mathbb{C}} - A_-^3 Q^{\mathbb{C}} A_+ Q^{\mathbb{C}\dagger} \right) + \frac{1}{2} (Q^{\mathbb{C}} A_+ Q^{\mathbb{C}\dagger} A_- )^2 
	}
	\right] \Big\}
\end{align}
\noindent Under the infinitesimal gauge transformation, the action transforms as
\begin{align*}
	\delta W =  \frac{2\pi i}{48 \pi^3}  \int\limits_{\mathcal{M}}  \tr \Big[ \epsilon_+  \Big(  (dA_+)^2 - \frac{i}{2}d( A_+^3)  \Big)  -\epsilon_-  \Big(  (dA_-)^2 - \frac{i}{2} d(A_-^3)  \Big) \Big]
\end{align*}
The situation is similar to the $(1+1)$-D case: there is an anomaly if we gauge $U_+(n)$ and $U_-(n)$ independently. However, there is no anomaly if we only gauge the diagonal part of $U(n)$. This is consistent with the free fermion prediction.\\

\subsubsection{Real class in $(3+1)$-D}
\hfill

The gauged nonlinear sigma model for the real class in $(3+1)$-D can be derived from the results of the preceding subsection by 1) restricting the order parameter $Q^\mathbb{C}\in U(n)$ to the submanifold $Q^\mathbb{R}\in \frac{U(n)}{O(n)}$ (the space of symmetric unitary matrix),  2) restricting the gauge group from $U_+(n)\times U_-(n) $ which transforms $Q^\mathbb{C}$ according to
\begin{align*}
	Q^\mathbb{C} \ra g_-^\dagger \cdot Q^\mathbb{C} \cdot g_+,
\end{align*}
\noindent to the sub-group $U(n)$  (the global symmetry group in the real class is $U(n)$ ), which transforms $Q^{\mathbb{R}}$ according to
\begin{align*}
	Q^\mathbb{R} \xrightarrow{u \in \, U(n)} u^T \cdot Q^\mathbb{R} \cdot u,
\end{align*}
 and 3) divide the action by a factor of two (see \ref{appendix:realComplexInt}). The resulting gauged nonlinear sigma model is
\begin{align}
	\label{gaugedWZW3R}
	&W[Q^{\mathbb{R}},A]\notag \\
	&=-{1\over 4\lambda_4^2} \int\limits_{\mathcal{M}} d^4 x \, \tr \left[\left(Q^{\mathbb{R}\dagger} \left( \partial_\mu Q^{\mathbb{R}} - i Q^{\mathbb{R}} A_{\mu} + i (-A_{\mu}^T) Q^{\mathbb{R}} \right) \right)^2 \right]\notag\\
	& -\frac{2\pi}{960 \pi^3} \Big\{ \int_{\mathcal{B}}  \tr\left[(Q^{\mathbb{R}\dagger} dQ^{\mathbb{R}})^5\right] \notag\\
	&+ 5 \int_{\mathcal{M}} \tr \left[ 
	\thead{
		-i \left(A \alpha_2^3 + (-A^T) \alpha_1^3 \right) 
		- \left( (dA A + A d A)\alpha_2 +(d(-A^T) (-A^T) + (-A^T) d (-A^T))\alpha_1    \right) \\
		+ d(-A^T) dQ^{\mathbb{R}} A Q^{\mathbb{R}\dagger} - dA d(Q^{\mathbb{R}\dagger}) (-A^T) Q^{\mathbb{R}} + A Q^{\mathbb{R}\dagger} (-A^T) Q^{\mathbb{R}} \alpha_2^2 - (-A^T) Q^{\mathbb{R}} A Q^{\mathbb{R}\dagger} \alpha_1^2 \\
		+ \frac{1}{2} \left( ((-A^T) \alpha_1)^2 - (A \alpha_2)^2 \right)- i \left( (-A^T)^3 \alpha_1 + A^3 \alpha_2 \right) \\
		+i \left( \left( dA A + A dA \right) Q^{\mathbb{R}\dagger} (-A^T) Q^{\mathbb{R}} - \left( d(-A^T) (-A^T) + (-A^T) d(-A^T)  \right)Q^{\mathbb{R}} A Q^{\mathbb{R}\dagger} \right)\\
		- i \left( (-A^T) Q^{\mathbb{R}} A Q^{\mathbb{R}\dagger} (-A^T) \alpha_1 +  A Q^{\mathbb{R}\dagger} (-A^T) Q^{\mathbb{R}} A \beta\right)\\
		+ \left( A^3 Q^{\mathbb{R}\dagger} (-A^T) Q^{\mathbb{R}} - (-A^T)^3 Q^{\mathbb{R}} A Q^{\mathbb{R}\dagger} \right) + \frac{1}{2} (Q^{\mathbb{R}} A Q^{\mathbb{R}\dagger} (-A^T) )^2 
	}
	\right] \Big\}.
\end{align}
Here we have used the definition
 $$\alpha_1 := dQ^{\mathbb{R}} \,  Q^{\mathbb{R}\dagger}, ~~\alpha_2 := Q^{\mathbb{R}\dagger} dQ^{\mathbb{R}}.$$
Under the infinitesimal gauge transformation,
	\begin{align*}
		&Q^{\mathbb{R}} \rightarrow  e^{i \epsilon^T} Q^{\mathbb{R}} e^{i \epsilon}\\
		&A \rightarrow  A + d\epsilon + i [A, \epsilon],
	\end{align*}
\noindent the gauged nonlinear sigma model acquires an addition piece
\begin{align}
\label{anr3}
	\delta W =  \frac{2\pi i}{96 \pi^3}  \int\limits_{\mathcal{M}}  \tr \Big[ d\epsilon \,  \Big( A dA - \frac{i}{2} A^3  \Big)  + d\epsilon^T \Big( (-A^T) d(-A^T) - \frac{i}{2} (-A^T)^3  \Big) \Big]
\end{align}
\noindent manifesting the 't Hooft anomaly of associated with $U(n)$.\\

However, if we only gauge the $O(n)$ subgroup of $U(n)$ 
$$\epsilon^T = -\epsilon, ~~A^T = -A.$$ Under such condition the two terms in \Eq{anr3}  cancel. 
Thus the $O(n)$ subgroup anomaly free. This agrees with the free fermion anomaly.

\subsection{Anomalies with respect to the discrete groups}
\hfill

After gauging the anomaly-free part of the continuous group, it is straightforward to determine how the resulting action transform under discrete symmetries. The necessary input is the transformation of the gauge field and the $Q^\mathbb{C,R}$. Here we simply state the results. In $(1+1)$-D and $(3+1)$-D there is no anomaly with respect to discrete symmetries after gauging the anomaly-free part of the continuous symmetries. In $(2+1)$-D, gauging the continuous symmetry breaks the time-reversal symmetry as discussed in subsection \ref{dsan}.\\

\section{Soliton's statistics}
\label{appendix:solitonStat}
\hfill

As discussed in subsection \ref{solitonClass} of the main text,  in $(2+1)$-D and $(3+1)$-D the mass manifolds for $Q^{\mathbb{C},\mathbb{R}}$ support solitons for sufficiently large $n$ (number of flavors).  In this appendix, we follow Ref.\cite{Witten1983a} to determine the statistics of soliton. This is achieved by computing the  Berry phase, arising from the WZW term, of an adiabatic self-rotating soliton. \\ 

Here is our strategy.  (1) We write down the  $Q^{\mathbb{C},\mathbb{R}}$ configuration corresponding to a static unit soliton.  (2) Based on the result of (1), we write down the $Q^{\mathbb{C},\mathbb{R}}$ configuration corresponding to an adiabatic self-$2\pi$-rotating soliton. (3) We plug the  $Q^{\mathbb{C},\mathbb{R}}$ configuration constructed in (2) into the WZW term to compute the Berry phase.\\

Because the space-time manifold $S^D$ is incompatible with the $Q^{\mathbb{C},\mathbb{R}}$ configuration  of a single soliton\footnote{On $S^D$, the infinite future corresponds to a single point. It follows that $Q^{\mathbb{C},\mathbb{R}}$ is a constant matrix at infinite future. This is incompatible with the single soliton configuration.}, in this section we shall follow Ref.\cite{Witten1983a} and use $$\mathcal{M}=S^{D-1}\times S^1$$ as the space-time manifold. Here $S^{D-1}$ is in the spatial manifold and $S^1$ is the loop in time. The extended manifold needed to define the WZW term is \cite{Witten1983a}  $$\mathcal{B}=S^{D-1} \times D^2,$$ where $D^2$ is a two-dimensional disk with the boundary $\partial D^2=S^1$ being the time loop.\\

\subsection{Complex class in $(2+1)$-D} 
\label{appendix:solitonStat2C}
\hfill

The mass manifold is $$\frac{U(n)}{U(n/2) \times U(n/2)}.$$ For $n\ge 4$ {\it both} homotopy groups $\pi_2$ (relevant to the existence of soliton)  and $\pi_4$ (relevant to the existence of the WZW term) are stabilized (see table \ref{tab:homotopy2C}). This is the situation we shall focus on in the following.\\

To write down a static soliton configuration, let us begin with $n=2$. This is because as far as $\pi_2$ (relevant to the existence of soliton) is concerned, it  
stabilizes at $n=2$, for which the mass manifold is $$\frac{U(2)}{U(1) \times U(1)}=S^2,$$ and $Q^{\mathbb{C}}$ is a $2\times 2$ hermitian matrix. Here a unit soliton is a degree 1 map from the spatial manifold $S^2$ to the mass manifold $S^2$. An example of such map is 
\be Q_{{\rm sol}}^{\mathbb{C}}(\theta,\phi)=\vec{n}\cdot \vec{\sigma} {\rm ~~where~~} \vec{n}=(\sin\theta\cos\phi,\sin\theta\sin\phi,\cos\theta),\label{unitsol}\ee
\noindent where $\theta$ and $\phi$ are the usual coordinates on $S^2$. This can be verified by computing the topological invariant associated with the soliton quantum number 
$$I_2= \frac{i}{16 \pi} \int \tr \left[ Q_{{\rm sol}}^{\mathbb{C}} (dQ_{{\rm sol}}^{\mathbb{C}})^2 \right]=1.$$ 
\\

For $n\ge 4$ we can write down a static unit soliton configuration as the {\it direct sum} of the $2\times 2 ~Q_{{\rm sol}}^{\mathbb{C}}(\theta,\phi)$ in \Eq{unitsol} with a number of  Pauli matrices  $Z$, i.e.,
\begin{align}
\label{gsolt}
Q_{\rm sol}^{\mathbb{C}}(\theta,\phi)=\vec{n}(\theta,\phi)\cdot\vec{\sigma}\oplus Z\oplus Z...
=\begin{pmatrix}
n_3			&	n_1- i n_2	&	0	&	0	\\
n_1+i n_2	&	-n_3	&	0	&	0	\\
0			&	0			&	1	&	0	\\
0			&	0			&	0	&	-1
\end{pmatrix}\oplus Z...
\end{align} 

To construct the configuration of a $2\pi$-self-rotating soliton (around, e.g., the $n_x$ axis) we introduce the following space-time dependent $Q^{\mathbb{C}}$, namely, 
\be
&&Q^{\mathbb{C}}(\theta,\phi,\tau) =R^T(\t)\cdot Q_{\rm sol}^{\mathbb{C}}(\theta,\phi) \cdot R(\t),~~{\rm where}\nn
&&R(\tau) =\left[
\begin{pmatrix}
e^{+i\frac{\tau}{2}}			&	0	&	0	&	0	\\
0	&	e^{-i\frac{\tau}{2}}		&	0	&	0	\\
0			&	0			&	1	&	0	\\
0			&	0			&	0	&	1
\end{pmatrix}\oplus I...\right]
\label{srans}\ee
Here $\tau$ ranges from $0$ to $2 \pi$ along the time loop, and $Q_{\rm sol}^{\mathbb{C}}(\theta,\phi)$ is given by \Eq{gsolt}. \\

As discussed earlier, in order to calculate the Berry phase arising from the WZW term, we needs to extend the space-time manifold from $\mathcal{M}=S^2\times S^1$ to $\mathcal{B} = S^2 \times D^2$. However, this extension is complicated by the fact that the orthogonal matrices $R(\t)$ in \Eq{srans}  is not single-valued as $\tau$ runs through the time loop (note, however, $Q^{\mathbb{C}}(\theta,\phi,\tau)$ is single valued). To overcome this difficulty, we use the algebraic fact observed by Witten \cite{Witten1983a}  that to reproduce the same $Q^{\mathbb{C}}(\theta,\phi,\tau)$, one can replace the $R(\t)$ in \Eq{srans} by the following single valued matrix 
\begin{align*}
R(\t)= 
\left[\begin{pmatrix}
1			&	0	&	0	&	0	\\
0	&	e^{-i \tau}		&	0	&	0	\\
0			&	0			&	e^{+i \tau}	&	0	\\
0			&	0			&	0	&	1
\end{pmatrix}\oplus I...\right].
\end{align*}
After such replacement, one can extend it to $S^2\times D^2$ by writing 
\be
&&\tilde{Q}^{\mathbb{C}}(\theta,\phi,\tau, u) =\tilde{R}^T(\t,u) \cdot Q_{\rm sol}^{\mathbb{C}}(\theta,\phi) \cdot 
\tilde{R}(\t,u),~~{\rm where}\nn
&&\tilde{R}(\t,u)=
\left[\begin{pmatrix}
0			&	0	&	0	&	0	\\
0	&	\sin u \, e^{-i \tau}		&	\cos u	&	0	\\
0			&	- \cos u			&	\sin u \, e^{+i \tau}	&	0	\\
0			&	0			&	0	&	1
\end{pmatrix}\oplus I...\right],\label{slfrsol}\ee
where  $u \in [0,\pi]$.\\

It's straightforward, though slightly tedious, to plug \Eq{slfrsol} in the WZW term \footnote{It can be checked that for the self-rotating soliton configurations we consider here and the rest of this appendix, where the space-time manifold is $S^{D-1}\times S^1$ and the extended space-time manifold is $S^{D-1}\times D^2$,  the WZW terms exist, and are  extension-independent up to $(2\pi i)~ \mathbb{Z}$. Thus, the soliton statistics is well-defined.} to obtain
\begin{align*}
W_{\rm WZW}[\tilde{Q}^{\mathbb{C}}] = & -  \frac{2\pi i}{256 \pi^2}      \int\limits_{\mathcal{B}} \tr \left[  \tilde{Q}^{\mathbb{C}}   \, (d \tilde{Q}^{\mathbb{C}})^4 \right]=i \pi.\\
\end{align*}
Therefore we conclude that the Berry's phase due to the self-rotation is $-1$, implying the unit soliton is a fermion. \\

\subsection{Real class in $(2+1)$-D}
\label{appendix:solitonStat2R}
\hfill

The relevant mass manifold is $\frac{O(n)}{O(\frac{n}{2}) \times O(\frac{n}{2})}$. From table \ref{tab:homotopy2R}, both $\pi_2$  and $\pi_4$ are stabilized for $n\ge 10$. In the following we shall restrict ourselves to such situation.\\

Unlike the case of complex class, the stabilized $\pi_2$ is $$\pi_2\left(\frac{O(n)}{O(\frac{n}{2}) \times O(\frac{n}{2})}\right)=Z_2,$$ rather than $Z$.  As a consequence, unlike the soliton in the preceding section, there is no integral form of topological invariant we can use to test whether a proposed $Q_{\rm sol}^{\mathbb{R}}$ configuration indeed corresponds to the non-trivial element of $Z_2$. The purpose of the following subsection is to establish such a testing method. \\

%The focus of the following discussion is to write down the $Q^{\mathbb{R}}$ configuration associated with the unit soliton which represents the  generator of $Z_2$. 
\subsubsection{How to test whether a proposed $Z_2$ soliton is trivial or not}
\hfill

Let's consider the ``fibration'' \be F\xrightarrow{i}  E\xrightarrow{p} B.\label{fibration}\ee Here $F$ stands for ``fiber space'', $E$ stands for ``total space'', and $B$ stands for ``base space''.  ``Fibration'' means that locally (i.e., in a small neighborhood of the base space $B$), the total space is the Cartesian product of the base space and the fiber space.  In \Eq{fibration} $i$ and $p$ stand for the inclusion and projection maps, respectively. They satisfies the property that {\it image} of $i$ is the {\it kernel} of $p$.
It is a non-trivial theorem that the fibration in \Eq{fibration} induces the following long exact sequence of mappings between homotopy groups (see, e.g., \cite{Hatcher2001})
\begin{align}
\label{longexact}
\begin{tikzcd}
...\pi_n(F)\xrightarrow{i_*}\pi_n(E) \xrightarrow{p_*} \pi_n(B)\ra\pi_{n-1}(F)\xrightarrow{i_*}\pi_{n-1}(E) \xrightarrow{p_*}\pi_{n-1}(B)...
\end{tikzcd} 
\end{align}
Here $i_*$, $p_*$ stand for the map between mapping classes induced by the inclusion and projection, respectively.
\Eq{longexact} has the property that for two consecutive mappings between homotopy groups, the {\it image} of the preceding map is equal to the {\it kernel} of the subsequent map. \\

In our case 
$$F=O\left(\frac{n}{2}\right)\times O\left(\frac{n}{2}\right),~~ E=O(n),~~B=\frac{O(n)}{O(\frac{n}{2}) \times O(\frac{n}{2})}.$$ The inclusion and projection maps in \Eq{fibration} are defined by 
\be
&&(O_1 , O_2 ) \xrightarrow{i} O:= \begin{pmatrix} O_1 & 0 \\ 0 & O_2 \end{pmatrix}~{\rm where}~~O_{1,2} \in O\left(\frac{n}{2}\right)~~{\rm and}~~O \in O(n)\nn
&&O \xrightarrow{p} S:= O \cdot diag(\underbrace{+1,...,+1}_{n/2},\underbrace{-1,...,-1}_{n/2}) \cdot O^T,~{\rm where}~S\in \frac{O(n)}{O(\frac{n}{2}) \times O(\frac{n}{2})}.\nn
\label{ipmaps}\ee

Our goal is to decide whether a given \be S^2 \xrightarrow{f_2} \frac{O(n)}{O(\frac{n}{2}) \times O(\frac{n}{2})}\label{notknow}\ee is topologically trivial or not.
To answer that we consider the following sub-sequence of \Eq{longexact}
\begin{align*}
\pi_2(O(n))\xrightarrow{p_*}\pi_2\left (\frac{O(n)}{O(\frac{n}{2}) \times O(\frac{n}{2})}\right)\xrightarrow{\beta_*}\pi_1\left(O\left(\frac{n}{2}\right) \times O\left(\frac{n}{2}\right)\right)\xrightarrow{i_*}\pi_1(O(n)).
\end{align*}
where it is known that
\begin{align*}
&\pi_2(O(n))=0\\ &\pi_2\left (\frac{O(n)}{O(\frac{n}{2}) \times O(\frac{n}{2})}\right)=\mathbb{Z}_2\\ &\pi_1\left(O\left(\frac{n}{2}\right) \times O\left(\frac{n}{2}\right)\right)=\mathbb{Z}_2\times\mathbb{Z}_2\\ &\pi_1(O(n))=\mathbb{Z}_2.
\end{align*}
\\

The map $$\pi_1\left(O\left(\frac{n}{2}\right) \times O\left(\frac{n}{2}\right)\right)\xrightarrow{i_*}\pi_1(O(n))$$
sends $$\mathbb{Z}_2\times\mathbb{Z}_2\ra\mathbb{Z}_2~~{\rm via}~~(s_1,s_2)\ra (s_1 + s_2 \mod 2).$$ % where $(s_1,s_2)\in \mathbb{Z}_2 \times \mathbb{Z}_2$. Let's denote this map by $i$. 
Hence the kernel of this map is $(0,0)$ and $(1,1)$. According to \Eq{longexact} these should be the image of the map
$$
\pi_2\left (\frac{O(n)}{O(\frac{n}{2}) \times O(\frac{n}{2})}\right)\xrightarrow{\beta_*}\pi_1\left(O\left(\frac{n}{2}\right) \times O\left(\frac{n}{2}\right)\right),
$$
or equivalently,
\be\mathbb{Z}_2\xrightarrow{\beta_*}\mathbb{Z}_2\times\mathbb{Z}_2.\label{z2z2toz2}\ee The requirement that the image of the map in \Eq{z2z2toz2} be 
${(0,0),(1,1)}$, implies that \be s\xrightarrow{\beta_*} (s,s).\label{demap}\ee Therefore the soliton configuration, which is an representative of the $s=1$ element of  $\pi_2\left (\frac{O(n)}{O(\frac{n}{2}) \times O(\frac{n}{2})}\right)$, is mapped to a configuration representative of the $(1,1)$ element of $\pi_1\left(O\left(\frac{n}{2}\right) \times O\left(\frac{n}{2}\right)\right)$ under $\beta$. Hence if we can tell whether a representative map of $\pi_1(O(n/2))$  is trivial or not, we can deduce whether the configuration in \Eq{notknow} is topologically non-trivial by applying $\beta$ to it.\\

But this requires us to know how to construct the $\beta$ map. To achieve that we consider the following commutative diagram (a diagram is commutative if different paths leading from the same initial space to the final space are the same map. The fact that the following diagram is commutative is by construction.)  
$$
\begin{tikzcd}
S^1 \arrow[dashed]{d}{f_1} \arrow{r}{\delta_1}		& 	D^2 \arrow[dashed]{d}{\lambda} \arrow{r}{\gamma_2} 	&	S^2 \arrow{d}{f_2}  \\
O(\frac{n}{2}) \times O(\frac{n}{2})  \arrow{r}{i}	& O(n) \arrow{r}{p}										&	\frac{O(n)}{O(\frac{n}{2}) \times O(\frac{n}{2})}	
\end{tikzcd}
$$
Here $\delta_1$ is the inclusion map which maps $S^1$ to the boundary of the 2-dimensional disk $D^2$; $\gamma_2$ is the map that compactifies the boundary of $D^2$ to single point; $f_2$ is the map in \Eq{notknow} and $f_1$ is the map obtained by applying $\beta$ to $f_2$, i.e., $\beta[f_2]=f_1$. By knowing whether $f_1$ is a non-trivial map representing $(1,1)$ in \Eq{demap} we can deduce whether $f_2$ is a non-trivial soliton configuration. In the commutative diagram $\lambda$ is the homotopy lift of $f_2 \circ \gamma_2$. The fact that such a lift exists is because $D^2$ is homeomorphic to a two dimensional cube, i.e., a square, hence by the homotopy lifting property $\lambda$ exists. \\

Because the image of $\gamma_2 \, \circ \, \delta_1$ is a point, so does the image of $ f_2 \,\circ \gamma_2 \, \circ  \, \delta_1 = p \, \circ\, \lambda \,\circ\, \delta_1$. This implies $(\lambda \,\circ\, \delta_1)[S^1]$ is  in the kernel of the map $O(n)\xrightarrow{p}\frac{O(n)}{O(\frac{n}{2}) \times O(\frac{n}{2})}$. Since $(\lambda \,\circ\, \delta_1)[S^1]$ is projected to a point in the base space, it must be contained entirely in a single fiber
$O\left({n\over 2}\right)\times O\left({n\over 2}\right).$ Therefore the sought-after $f_1$ is given by
$$f_1=\lambda \,\circ\, \delta_1.$$ The above arguments allows us take the map $f_2$ as input and produce the map $f_1$ as output, i.e., we have constructed $\beta$. 
 \\

In the following we apply the construction discussed above to the following proposed soliton configuration\footnote{It is illuminating to compare $ Q_{\rm sol}^{\mathbb{R}}$ with $Q_{\rm sol}^\mathbb{C}$ in \Eq{gsolt}, namely, 
$Q_{\rm sol}^{\mathbb{R}}=\Re [Q_{sol}^{\mathbb{C}}]\otimes I + \Im[Q_{sol}^{\mathbb{C}}]\otimes E$.}  
\begin{align}
\label{propsol}
f_2:(\theta,\phi)\ra Q_{\rm sol}^{\mathbb{R}}=& \left( n_1 XI + n_2 EE + n_3 ZI \right) \oplus Z \oplus Z ... 
\end{align}
where $$\vec{n}=(\sin\theta\cos\phi,\sin\theta\sin\phi,\cos\theta).$$

Because $Q_{\rm sol}^{\mathbb{R}}$ is a real symmetric matrix, it can be diagonalized by orthogonal transformation
\begin{align*}Q_{\rm sol}^{\mathbb{R}}=W \cdot \left[ \begin{pmatrix}
1	&	0	& 0	&	0	\\
0	& 	1	&	0	&	0	\\
0	&	0	&	-1	&	0	\\
0	&	0	&	0	&	-1								
\end{pmatrix} \oplus Z \oplus Z... \right]  \cdot W^T\end{align*} where
\begin{align*}
W(\theta,\phi)=& \begin{pmatrix}
\cos\frac{\theta}{2} \cos\phi	&	-\cos\frac{\theta}{2} \sin\phi	& -\sin\frac{\theta}{2}	&	0	\\
\cos\frac{\theta}{2} \sin\phi	&	\cos\frac{\theta}{2} \cos\phi	&	0	&	-\sin\frac{\theta}{2}	\\
\sin\frac{\theta}{2}			&	0								&	\cos\frac{\theta}{2} \cos\phi	&	\cos\frac{\theta}{2} \sin\phi	\\
0								&	\sin\frac{\theta}{2}			&	-\cos\frac{\theta}{2} \sin\phi	&	\cos\frac{\theta}{2} \cos\phi								
\end{pmatrix} \oplus I ...
\end{align*}
Naively, one might think $W(\theta,\phi)$ is a mapping between $S^2$ and $O(n)$. However, this is not true. To see it, let's inspect $W(0,\phi)$ and $W(\pi,\phi)$, 
\begin{align*}
W (0,\phi)=& \begin{pmatrix}
\cos\phi	&	-\sin\phi	& 0	&	0	\\
\sin\phi	&	\cos\phi	&	0	&0	\\
0			&	0								&	\cos\phi	&	\sin\phi	\\
0								&	0			&	-\sin\phi	&	 \cos\phi								
\end{pmatrix} \oplus I ...,
\end{align*}
\begin{align*}
W (\pi,\phi)=& \begin{pmatrix}
0	&	0& -1	&	0	\\
0	&0	&	0	&	-1\\
1		&	0								&0	&0	\\
0								&	1			&0&	0						
\end{pmatrix} \oplus I ...
\end{align*}
The fact that $W(0,\phi)$ depends on $\phi$ and $W(\pi,\phi)$ does not implies that we should view $W$ as a mapping between $D^2$ and $O(n)$, where $\theta=0$ corresponds to the boundary  while $\theta=\pi$ corresponds to the center of $D^2$ (i.e., the radius of $D^2$ is $\pi-\theta$). In fact, $W$ is the map $\lambda$ in the commutative diagram, namely, $$\lambda=W.$$ Hence the map $f_1$ is given by
\begin{align}
\label{dedf1} f_1(\phi)=
W (0,\phi)=& \begin{pmatrix}
\cos\phi	&	-\sin\phi	& 0	&	0	\\
\sin\phi	&	\cos\phi	&	0	&0	\\
0			&	0								&	\cos\phi	&	\sin\phi	\\
0								&	0			&	-\sin\phi	&	 \cos\phi								
\end{pmatrix} \oplus I ...\\=&\left\{\begin{pmatrix}
\cos\phi	&	-\sin\phi\\		
\sin\phi	&	\cos\phi\end{pmatrix}\oplus 1...\right\}\oplus \left\{\begin{pmatrix}
\cos\phi	&	\sin\phi\\		
-\sin\phi	&	\cos\phi\end{pmatrix}\oplus 1...\right\}. \notag
\end{align}
To summarize, given the map $f_2$ in \Eq{propsol}, we have obtained the  map $f_1$  in the commutative diagram via \Eq{dedf1}. \\

Now we are ready to determine whether \Eq{propsol} is a topological non-trivial soliton configuration. It is known that the following map from $S^1$ to $O(n/2)$ 
$$\tilde{f}_1(\phi)= \left\{\begin{pmatrix}
\cos\phi	&	\mp\sin\phi\\		
\pm\sin\phi	&	\cos\phi\end{pmatrix}\oplus 1...\right\}$$ is a representative of the generator of
$\pi_1(O(n/2))=\mathbb{Z}_2$.
Thus the mapping class of $f_1$ in \Eq{dedf1} is the $(1,1)$ element of $\pi_1(O(n/2)\times O(n/2))=\mathbb{Z}_2\times\mathbb{Z}_2$. It follows that  $f_2$ in \Eq{propsol} is a representative of the generator of 
$\pi_2(\frac{O(n)}{O(n/2)\times O(n/2)})=\mathbb{Z}_2$, i.e., it is a soliton configuration. \\ 

\subsubsection{The Berry phase of a self-rotating $\mathbb{Z}_2$ soliton}
\hfill

To calculate the Berry's phase due to a $2 \pi$ self-rotation of the soliton in \Eq{propsol}, we rotate the soliton configuration to produce $Q^\mathbb{R}(\theta,\phi,\tau)$ in the same way in the appendix \ref{appendix:solitonStat2C}. After all dust settles, we end up with

\begin{align*}
&Q^\mathbb{R}(\theta,\phi,\tau) = \Re [Q^{\mathbb{C}}(\theta,\phi,\tau) ]\otimes I + \Im[Q^{\mathbb{C}}(\theta,\phi,\tau) ]\otimes E= \\
&
\begin{pmatrix}
\cos \theta					&	0							&		\sin\theta \cos (\phi+\tau)	&		-\sin\theta \sin (\phi+\tau)		\\
0							&	\cos\theta					&		\sin\theta \sin (\phi+\tau)	&		\sin\theta \cos (\phi+\tau)			\\
\sin\theta \cos(\phi+\tau)	&	\sin\theta \sin(\phi+\tau)	&		-\cos\theta					&		0	\\
-\sin\theta \sin(\phi+\tau)	&	\sin\theta \cos(\phi+\tau)	&		0							&	-\cos\theta
\end{pmatrix} \oplus Z \oplus Z ...
\end{align*}
Similarly, we can extend the space-time to one extra dimension by defining $$\tilde{Q}^\mathbb{R}(\theta,\phi,\t,u):=\Re [\tilde{Q}^{\mathbb{C}}(\theta,\phi,\t,u)]\otimes I + \Im[\tilde{Q}^{\mathbb{C}}(\theta,\phi,\t,u)]\otimes E.$$ (It's easy to check it suffices all the properties we want for extension). Plugging the extended $\tilde{Q}_2^\mathbb{R}$ into the WZW term, we find

\begin{align*}
W_{WZW}[Q^\mathbb{R}] =& 2 \pi i \left( -  \frac{1}{512 \pi^2}      \int\limits_{\mathcal{B}} \tr \left[  \tilde{Q}^\mathbb{R}   \, (d\tilde{Q}^\mathbb{R})^4 \right] \right)
= i \pi.
\end{align*}
Hence the soliton is again a fermion.

\subsection{Complex class in $(3+1)$-D}
\label{appendix:solitonStat3C}
\hfill

The mass manifold in this situation is $U(n)$. $\pi_3$ (relevant to the existence of soliton) and the $\pi_5$ (relevant to the existence of the WZW term) are {\it both} stabilized for $n\ge 3$.  In the following, we shall restrict ourselves to such conditions.\\

The fact a unit soliton in this mass manifold is a fermion has already been discussed in \cite{Witten1983a}. We briefly repeat the argument here for completeness. To construct a static soliton we start from $n=2$ (as far as $\pi_3$ is concerned, it stabilizes for $n\ge 2$ with $\pi_3(U(n))=\mathbb{Z}$.)  Thus the $n=2$ unit soliton is just the degree one map of $S^3 \rightarrow 
SU(2)\sim S^3~~$\footnote{Note that $\pi_3(U(n))=\mathbb{Z}$ originates from the $SU(n)$ part of $U(n)$. Among other things, it means that we can limit ourselves to the $SU(n)$ WZW term for the Berry phase calculation.}. We can choose the unit soliton configuration to be
\begin{align*}
Q_{\rm sol}^\mathbb{C}(\vec{\Omega}) = \begin{pmatrix}
\Omega_0 + i \Omega_3		&		i(\Omega_1 - i \Omega_2)		\\
i(\Omega_1 + i \Omega_2)	&		\Omega_0 - i \Omega_3			\\
\end{pmatrix}
\end{align*}
where $$\Omega_0^2 +\Omega_1^2+\Omega_2^2+\Omega_3^2=1$$ are the coordinate on $S^3$. For $n\ge 3$  one can write the unit soliton as
\begin{align*}
Q_{\rm sol}^\mathbb{C}(\vec{\Omega})=\begin{pmatrix}
\Omega_0 + i \Omega_3		&		i(\Omega_1 - i \Omega_2)		\\
i(\Omega_1 + i \Omega_2)	&		\Omega_0 - i \Omega_3			\\
\end{pmatrix}\oplus 1\oplus 1...
\end{align*}\\

Next, we rotate the unit soliton in, say, the $\Omega_1$-$\Omega_2$ plane by $2 \pi$. The time-dependent soliton configuration can be written as 
\begin{align*}
Q^\mathbb{C}(\vec{\Omega},\t)= &\left[\begin{pmatrix}
e^{-i\frac{\tau}{2}}			&	0	&	0		\\
0	&	e^{+i\frac{\tau}{2}}		&	0		\\
0			&	0			&	1
\end{pmatrix}\oplus 1...\right] \cdot Q_{\rm sol}^\mathbb{C}(\vec{\Omega}) \cdot 
\left[\begin{pmatrix}
e^{+i\frac{\tau}{2}}			&	0	&	0		\\
0	&	e^{-i\frac{\tau}{2}}		&	0		\\
0			&	0			&	1		
\end{pmatrix}\oplus 1 ..\right] \\
=&\left[ \begin{pmatrix}
1			&	0	&	0		\\
0			&	e^{+i \tau}		&	0		\\
0			&	0			&	e^{-i \tau}	
\end{pmatrix}\oplus 1..\right] \cdot  Q_{\rm sol}^\mathbb{C}(\vec{\Omega})   \cdot 
\left[\begin{pmatrix}
1			&	0	&	0		\\
0	&	e^{-i \tau}		&	0		\\
0			&	0			&	e^{+i \tau}	
\end{pmatrix}\oplus 1..\right]
\end{align*}

\noindent where $\tau \in [0,2 \pi]= S^1$ is the time parameter. One can extend the configuration to $\mathcal{B} = S^3 \times D^2$, where $D^2$ is the two-dimensional disk with radius $u \in [0,\pi]$, by

\begin{align}
\label{3drtsol}
\tilde{Q}^\mathbb{C}(\vec{\Omega},\t,u)=& \left[\begin{pmatrix}
1			&	0	&	0		\\
0			&	\sin u \, e^{+i \tau}		&	-\cos u		\\
0			&	\cos u			&	\sin u \, e^{-i \tau}	
\end{pmatrix}\oplus 1...\right] \cdot  Q_{sol}^\mathbb{C}(\vec{\Omega}) \cdot 
\left[\begin{pmatrix}
1			&	0	&	0		\\
0	&	\sin u \, e^{-i \tau}		&	\cos u		\\
0			&	\cos u			&	\sin u \, e^{+i \tau}	
\end{pmatrix}\oplus 1..\right]
\end{align} 
Plugging \Eq{3drtsol} into the WZW term gives 
\begin{align*}
W_{WZW}[\tilde{Q}^\mathbb{C}] = &2 \pi i \left(  \frac{i}{480\pi^3}    \int\limits_{\mathcal{B}}  \,\tr \Big[   \left(\tilde{Q}^{\mathbb{C}\dagger}   d \tilde{Q}^\mathbb{C} \right)^5    \Big] \right)= i \pi.
\end{align*}
Hence the unit soliton is a fermion.\\

\subsection{Real class in $(3+1)$-D}
\label{appendix:solitonStat3R}
\hfill

The mass manifold is $U(n)/O(n)$. Here $\pi_3$ (relevant to the existence of soliton) and $\pi_5$ (relevant to the existence of the WZW term) are both stabilized for $n\ge 6$. To write down the degree one soliton in $U(n)/O(n)$, let's first look at the fibration \be O(n) \xrightarrow{i} U(n) \xrightarrow{p} U(n)/O(n).\label{ufib}\ee  Here the projection $p$ is defined by \be u \xrightarrow{p} u_S=u^T \cdot u,~{\rm where}~u\in U(n), ~u_S \in U(n)/O(n).\label{uprojj}\ee After the homotopy groups are stabilized, the long exact sequence associated with \Eq{ufib} is given by
\begin{align*}
... & \pi_4(U(n)/O(n)) \ra \pi_3(O(n))\xrightarrow{i_*} \pi_3(U(n)) \xrightarrow{p_*}\pi_3 (U(n)/O(n))\ra\pi_2(O(n))...\\
... &~~~~~~~~~~~0~~~~~~~~\ra~~~~~~\mathbb{Z}~~~~\xrightarrow{i_*}~~~~~\mathbb{Z}~~~~~\xrightarrow{p_*}~~~~~~~~~~~\mathbb{Z}_2~~~~~~~\ra ~~~~~~0~~~~~...
\end{align*}
This implies  that we can construct the unit soliton in $U(n)/O(n)$ by taking a unit soliton in $U(n)$, namely $Q_{\rm sol}^\mathbb{C}$ in appendix \ref{appendix:solitonStat3C}, and perform the projection map in \Eq{uprojj}, namely,  $$ Q_{\rm sol}^\mathbb{R}=\left(Q_{\rm sol}^\mathbb{C}\right)^T \cdot Q_{\rm sol}^\mathbb{C}.$$\\

The time dependent $Q^\mathbb{R}(\vec{\Omega},\tau)$ can be constructed by the similar projection of $Q^\mathbb{C}(\vec{\Omega},\tau)$, i.e., 
$$Q^\mathbb{R}(\vec{\Omega},\tau)=\left(Q^\mathbb{C}(\vec{\Omega},\tau)\right)^T\cdot Q^\mathbb{C}(\vec{\Omega},\tau).$$ 
The extended $\tilde{Q}^\mathbb{R}$  can also be constructed by the same projection 
$$\tilde{Q}^\mathbb{R}(\vec{\Omega},\tau,u)=\left(\tilde{Q}^\mathbb{C}(\vec{\Omega},\tau,u)\right)^T\cdot \tilde{Q}^\mathbb{C}(\vec{\Omega},\tau,u).$$ The result  $\tilde{Q}^\mathbb{R}(\vec{\Omega},\tau,u)$ can be substituted into the WZW term to obtain

\begin{align*}
W_{WZW}[Q^\mathbb{R}] =&  2 \pi i \left(  \frac{i}{960\pi^3}    \int\limits_{\mathcal{B}}   \, \tr \Big[   \left(\tilde{Q}^{\mathbb{R}\dagger}   d \tilde{Q}^\mathbb{R} \right)^5    \Big] \right)= i \pi
\end{align*}

Therefore  the unit soliton is again a fermion.\\

\section{Bosonization for small flavor number}
\label{appendix:enlarge}
\hfill

In this appendix we discuss the bosonization in cases when $n$, the number of flavors, is less than the value necessary to stabilize $\pi_{D+1}({\rm mass~manifold})$, or the WZW term.\\

In some cases, although the homotopy group $\pi_{D+1}({\rm mass~manifold})$ is not yet stabilized, it already contains $\mathbb{Z}$ as a subgroup. For instance, for real class in $(3+1)$-D, $\pi_5(U(3)/O(3)) = \mathbb{Z} \times \mathbb{Z}_2$. The nonlinear sigma model derived from fermion integration in appendix \ref{appendix:fermionInt} contains the level-$1$ WZW term, which is $2 \pi$ times the topological invariant of the $\mathbb{Z}$ part of $\pi_5$. This is also true for $n=6,8$ of the non-charge-conserved cases in $(2+1)$-D. In these cases the story is unchanged.\\

In other cases  $\pi_{D+1}({\rm mass~manifold})$  is a finite abelian group, e.g., $\mathbb{Z}_2$, or even $0$.  This requires a case-by-case study. Here, instead of attempting at studying all possible cases, we shall focus on the case that is relevant to the applications in section \ref{appi} of the main text, namely, the case of $n=2$ complex class in $(2+1)$-D (which is relevant to the discussions in subsection \ref{oddi} of the main text).

\subsection{Complex class in $2+1$ D with $n=2$}
\hfill

The mass manifold is $$\frac{U(2)}{U(1)\times U(1)} = S^2$$ and
\begin{align*}
	\pi_4( S^2) =  0 ,~~{\rm but}~~\pi_3(S^2)=\mathbb{Z}.
\end{align*} 
\noindent The generator of $\pi_3(S^2)$ is called Hopf map \cite{Hopf1931}. The question at hand is whether this signifies the presence of a topological term in the nonlinear sigma model. A similar situation occurs for, e.g., the non-linear sigma model describing the anti-ferromagnetic spin chains in $(1+1)$-D. There, the mass manifold is $S^2$ and $\pi_3(S^2)=0$ but $\pi_2(S^2)=\mathbb{Z}$. In the nonlinear sigma model, there is a topological term associated with the $\pi_2$ in the non-linear sigma model, the $\theta$ term, which is responsible for the difference between the integer and half-integer spin chains \cite{Haldane1983,Haldane1983b,Affleck1987}. \\

To answer the question posed above, the derivation in appendix \ref{appendix:fermionInt} is not applicable. This is because the Hopf term (or the $\theta$ term) is invariant under arbitrary infinitesimal deformation of $Q^{\mathbb{C}}$.\\

\subsection{Mass manifold enlargement}
\hfill

One way to proceed is to enlarge the mass manifold (or the target space of the order parameter). The idea \cite{Elitzur1984} is as follows. If two order parameter configurations cannot be deformed into each other, as in the case where configurations correspond to different elements of $\pi_D$ (in this case $\pi_3$), we can enlarge the mass manifold so that after the enlargement, one configuration can be continuously deformed to the other. One can then compute the Berry phase difference caused by infinitesimal order parameter variation using the method explained in \ref{appendix:fermionInt}, and integrate the result. However, it is important to note that enlarging the mass manifold requires adding extra fermion flavors. It is important to make sure that the initial and the final order parameters couple to the added fermion flavors in a trivial way (i.e., in the added flavor space, the order parameters are the same constant) so that the Berry phase difference is originated from the original fermions. Finally, one accounts for the Berry phase by picking the coefficient in front of the $\pi_D$ (here $\pi_3$) topological invariant.\\

Using this technique, Abanov \cite{Abanov2000b} enlarged $\frac{U(2)}{U(1)\times U(1)} = S^2$ to $$\frac{U(l+1)}{U(l)\times U(1)} = \mathbb{CP}^l$$ and showed that the nonlinear sigma model from the $n=2$ fermion integration contains a $\theta=\pi$ Hopf term. In the following we will choose an alternative enlargement, namely,  $$\frac{U(2)}{U(1)\times U(1)} = S^2\ra \frac{U(4)}{U(2)\times U(2)}.$$ We will show that the result is consistent with that of Abanov. Because of the Hopf term, the unit soliton has fermion statistics \cite{Wilczek1983}.\\

\subsection{The derivation of the Hopf term}
\label{appendix:Hopf}
\hfill

The Hopf map is a map from $S^3$ with coordinate $(\Omega_0,\Omega_1,\Omega_2,\Omega_3 )$ where $\sum_{i=0}^3\Omega_i^3=1$ to 
$S^2$ with coordinate $(n_1,n_2,n_3)$ where $\sum_{i=1}^3n_i^2=1.$ More explicitly,
\be
\label{zexpr}
	&&\vec{\Omega}\xrightarrow{\rm Hopf} \vec{n}= z^\dagger \sigma^a z 
	~\text{where } ~ z:=
		\begin{pmatrix} \Omega_0+ i \Omega_1 \\ \Omega_2 + i \Omega_3 \end{pmatrix} . 
		\ee
The $Q^{\mathbb{C}}$ of the non-linear sigma model is given by		
\begin{align}
\label{qzeig}
Q^\mathbb{C}	(\vec{\Omega}) = \sum_{a=1}^3n^a(\vec{\Omega})\cdot  \sigma^a 
				= 2 \, z(\vec{\Omega}) \, z(\vec{\Omega})^\dagger - I_{2},
\end{align}		where $z$ is given by \Eq{zexpr}. In writing down the 2nd equality we have used the identity $$\sum_{a=1}^3\sigma^a_{ij} \sigma^a_{kl} = 2\delta_{il} \delta_{jk} - \delta_{ij} \delta_{kl}. $$ In \Eq{qzeig} the $2\times 2$ matrix $Q^\mathbb{C}$ has eigenvalues $\pm 1$, and $z$ is the eigenvector associated with eigenvalue $+1$.\\ 

In the following, we enlarge the order parameter so that $Q^\mathbb{C}$ can be deformed to $\sigma_z$. To do so, we add two additional fermion flavors and enlarge the mass manifold to $$\frac{U(4)}{U(2)\times U(2)}.$$ \\

In the enlarged space the order parameter is given by 
\be
	\label{extendedHopf}
	&&{Q'}^{\mathbb{C}}(\vec{\Omega})={Q}^{\mathbb{C}}(\vec{\Omega})\oplus (-Z)
	= \begin{pmatrix} Q^{\mathbb{C}}(\vec{\Omega})& 0  \\  0  & -Z \end{pmatrix}\nn
	&&= \begin{pmatrix} 2 \, z(\vec{\Omega}) \, z(\vec{\Omega})^\dagger - I_2 & 0  \\  0  & -Z \end{pmatrix} 
\ee
where $z(\vec{\Omega})$ is given by \Eq{zexpr}. Here the fermions associated with extra flavors couple to the mass term $Y\otimes (-Z)$ (see table \ref{tab:massManifold}). Although the $Q^{\mathbb{C}}$ given by \Eq{qzeig} cannot be deformed to a constant configuration in the space $\frac{U(2)}{U(1)\times U(1)} = S^2$ (i.e., within the first $2 \times 2$ block) because 
$$\pi_3(S^2)=\mathbb{Z},$$
it is possible to deform the $4 \times 4$ ${Q'}^{\mathbb{C}}$ to a constant matrix.  This is because $$\pi_3\left( \frac{U(4)}{U(2)\times U(2)}\right)=0.$$ \\

Now we explicitly construct such a deformation. First we rewrite \Eq{extendedHopf} as 
\begin{align*}
	&{Q'}^{\mathbb{C}}(\vec{\Omega})  
	= \begin{pmatrix} 2 \, z'(\vec{\Omega}) \, {z'}(\vec{\Omega})^{\dagger} - I_3 & 0  \\  0  & +1 \end{pmatrix} \\
	&\text{where }  z'(\vec{\Omega}) := \begin{pmatrix} \Omega_0+ i \Omega_1 \\ \Omega_2 + i \Omega_3 \\0 \end{pmatrix}.
\end{align*}
We then write down a continuous deformation, as a function of $u$, as follows 
\be
	&&{\tilde{Q'}}^\mathbb{C}(\vec{\Omega},u)  
	= \begin{pmatrix} 2 \, \tilde{z}^\prime(\vec{\Omega},u)   \, \tilde{z}^{\prime\dagger}(\vec{\Omega},u)   - I_3 & 0  \\  0  & +1 \end{pmatrix} 
\nn
&& \tilde{z}^\prime(\vec{\Omega},u)  := 
	\begin{cases}	
		\begin{pmatrix} -\cos u \left( \Omega_0+ i \Omega_1 \right), -\cos u \left( \Omega_2 + i \Omega_3 \right) , \sin u \end{pmatrix}^T  ~\text{ for } u\in [\frac{\pi}{2}, \pi]\\
		\begin{pmatrix} \cos u, 0, \sin u \end{pmatrix}^T ~\text{ for } u\in [0,\frac{\pi}{2}).
	\end{cases}\nn
\label{qwext}\ee
\noindent This extends the configuration from the space-time $\mathcal{M}=S^3$ at $u=\pi$ to a four dimensional disk $\mathcal{B}=D^4$ with $u\in[0,\pi]$. Here $\partial{\mathcal{B}} = \mathcal{M}$ and with $u$ as the radial direction of $D^4$. For $u=\pi$ \Eq{qwext} reduces to  
\Eq{extendedHopf}, while for $u=0$ $${\tilde{Q'}}^\mathbb{C}(\vec{\Omega},u=0)= \begin{pmatrix} 1&0&0&0\\
0&-1&0&0\\0&0&-1&0\\ 0&0&0&+1 \end{pmatrix}=Z\oplus (-Z).$$ Therefore at $u=0$ and $u=\pi$ the fermions associated with the added flavors couples to exactly the same mass term $Y\otimes (-Z)$ according to table  \ref{tab:massManifold}.\\

 \Eq{qwext} constitutes an extension we need to define the WZW term (which is stabilized at $n=4$).
Now we can plug \Eq{qwext} into the WZW term of the $\frac{U(4)}{U(2)\times U(2)}$ non-linear sigma model. When all dust settles we obtain
\begin{align*}
	W_{\rm WZW}[\tilde{Q'}^{\mathbb{C}}] = & -  \frac{2\pi i}{256 \pi^2}      \int\limits_{\mathcal{B}} \tr\left[  \tilde{Q'}^{\mathbb{C}}   \, (d \tilde{Q'}^{\mathbb{C}})^4 \right]=i \pi.\\
\end{align*}
This result agrees with that of  Ref.\cite{Abanov2000b} and suggests the existence of a $\theta=\pi$ Hopf term.\\

\subsection{Gauging small $n$ non-linear sigma models}
\hfill

In appendix \ref{appendix:anomaliesb}, we have shown how to gauge the nonlinear sigma models. Recall that the non-trivial gauge coupling terms all originate from the WZW term. For small $n$, the WZW term does not exist. One might think we need to re-derive the gauging procedure. Fortunately, we can use the mass manifold enlargement idea discussed in the preceding subsection to derive the  coupling between $Q^{\mathbb{C},\mathbb{R}}$ and the gauge field. Without going into details we (1) add additional fermion flavors until the WZW term is stabilized. (2) Proceed as usual to gauge the continuous symmetries. (3) Restrict $Q^{\mathbb{C},\mathbb{R}}$ to the proper sub-mass manifold and the gauge group to the proper subgroup (so that the gauge field does not couple to the added fermion flavors). Following this procedure, we gauged the small $n$ nonlinear sigma model following the same try-and-error method.\\

As an example, we shall write down the charge-U(1) gauged nonlinear sigma model for $n=2$ in the $(2+1)$-D complex class. As shown in appendix \ref{appendix:Hopf}, the bosonized theory the $S^2$ nonlinear sigma model with the $\theta=\pi$ Hopf term. Plugging $Q^\mathbb{C} =n^a \sigma^a$ into the gauge coupling part in \Eq{gaugedWZW2C}, we arrive at
\begin{align}
W[\boldsymbol{\beta} ,A] = {1\over \lambda_3} \int\limits_{\mathcal{M}} d^3 x \, (\partial_\mu \hat{n})^2 + i\pi H(\hat{n})  +  \int\limits_{\mathcal{M}} d^3x \left[ i A_\mu \left( \frac{1}{8\pi} \epsilon^{abc} \epsilon^{\mu\nu\rho} n^a  \partial_\nu  n^b  \partial_\rho n^c \right) \right], \notag
\end{align}
\noindent where the last term makes the $S^2$ solitons carry $U(1)$ charge. For $Q^\mathbb{C}$ in the $l=0$ and $l=2$ component of te mass manifold, we have a constant configuration $Q^\mathbb{C} = \pm I \in \frac{U(2)}{U(2) \times U(0)}$. Plugging it into \Eq{gaugedWZW2C}, we get
\begin{align}
\pm \frac{i}{4\pi}  \int\limits_{\mathcal{M}} d^3 x \, \epsilon^{\mu\nu\rho} A_\mu \partial_\nu A_\rho,
\end{align}
 which gives $\sigma_{xy}=\pm 1$.

\section{Massless fermions as the boundary of bulk topological insulators/superconductors}
\label{appendix:SPT}
\hfill

The idea behind our bosonization is to fluctuate the bosonic order parameters ($Q^{\mathbb{C}}$ or  $Q^{\mathbb{R}}$) to restore the full emergent symmetries of the massless fermion theory.  These order parameters are chosen so that  when they are static, any $Q^{\mathbb{C},\mathbb{R}}(\v x)$ configuration will fully gap out the fermions. %All possible such order parameters form a topological space, the mass manifold, within which $Q^{\mathbb{C},\mathbb{R}}$  fluctuates. 
As shown in appendix \ref{appendix:massManifold}, a static $Q^{\mathbb{C},\mathbb{R}}(\v x)$ configuration breaks at least some of the emergent symmetries. Conversely, if the full emergent symmetries are unbroken the fermion spectrum should remain gapless. Putting it succinctly, the emergent symmetries protect the gapless fermions.\\

The above situation reminds us of the boundary gapless modes of SPTs. Therefore, it is natural to suspect that each of the gapless fermion theories can be realized at the boundary of certain emergent-symmetry-protected SPT. % (in field theory language, this is the anomaly-inflow picture \cite{Callan1985,Witten2016}). 
In this appendix, we show that this is indeed the case. Moreover, we shall construct the bulk SPT explicitly.\\

\subsection{The $\mathbb{Z}$ classification}
\hfill

As discussed in Ref.\cite{Kitaev2009}, the classification of free fermionic SPTs can be determined by checking how many copies of  the boundary theory can be ``stacked'' together before a symmetry allowed mass term emerges. For example, a $\mathbb{Z}_N$ classification implies, after stacking $N$ copies of the massless fermion theory, a mass term can be found without breaking any of the protecting symmetry (here the emergent symmetries). In the following, we show that the emergent-symmetry-protected SPT has  $\mathbb{Z}$ classification.\\

For the sake of generality, we shall use the the Majorana fermion representation, even for complex classes.  $N$ copies of the boundary theory is described by the gamma matrices and the matrices that execute symmetry transformations,
\begin{align*}
\Gamma_i^{(N)} 	=& \G_i \otimes I_N,~~i=1,...,d \\
T^{(N)} 		=& t \otimes I_N \\
U^{(N)}			=& u \otimes I_N     
\end{align*}
\noindent Here $t$ and $u$ are orthogonal matrices obeying $\{t,\G_i\}=[u,\G_i]=0$. The symbol $t$ and $u$ stand for anti-unitary and unitary, respectively. It is important to note that $t$ and $u$ represent the complete set of anti-unitary and unitary  transformation matrices, from the product of which all symmetry matrices can be constructed. In addition, $\G_i, t, u$  stand for the gamma and symmetry matrices for one copy of the massless fermion theory.\\

Existence of a mass term for the stacked massless fermion theory, implies that there exist a matrix $M^{(N)}$ that anti-commutes with all of the gamma matrices. The general form of $M^{(N)}$ is
$$M^{(N)}=m_s\otimes A_N+m_a\otimes S_N$$ where $m_s,S_N$ and $m_a, A_N$ are symmetric and anti-symmetric matrices, respectively.  Since  $\G_i^{(N)}=\G_i \otimes I_N,  ~T^{(N)} = t \otimes I_N, ~U^{(N)}= u \otimes I_N$  it follows that
\begin{align*}
\{m_{s,a} ,\G_i \} = \{m_{s,a},t\}=[m_{s,a},u]=0
\end{align*}
If such a $m_a\ne 0$ exists, we can use it as the mass term for the original massless fermion theory. This contradicts the statement that under the protection of emergent symmetry there is no mass term. Thus $m_a=0$ and $M^{(N)}$ reduces to \be M^{(N)}=m_s\otimes A_N.\label{MNA}\ee 

On the other hand, $m_{s}$ can then be used to construct an anti-unitary symmetry. By our assumption, such anti-unitary symmetry matrix $m_s$ must be the product $t$'s and $u$'s.  Thus the matrix
\begin{align*}
T^{(N)\prime} =m_{s} \otimes I_N
\end{align*} 
is an anti-unitary symmetry matrix of the stacked fermion theory. However such $T^{(N)\prime}$ commutes with  \Eq{MNA} which is a contradiction. 
(Recall that in Majorana fermion representation, a mass matrix must anti-commute with all anti-unitary symmetry matrices.) Therefore ${M}^{(N)}$ can not exist for any $N$. Consequently, the classification of the massless fermion theory must be $\mathbb{Z}$.

\subsection{Construction of the bulk SPT}
\hfill

To construct the bulk SPT, we follow the ``holographic construction'' in Ref.\cite{Tsui2019}. In the following, we just summarize the results.\\

For a massless fermion theory, with gamma matrices $\{\G_i | \, i=1,2,...,d \}$, anti-unitary symmetry $t$, and unitary symmetries $\{u\}$, we can construct the bulk matrices,
\begin{align*}
\G_i^{\rm(bk)}= &
\begin{cases}
	\G_i \otimes Z 				&\text{ for } i=1,...,d \\
	I_{\rm dim(\G_i)} \otimes X &\text{ for } i=d+1
\end{cases}\\
T^{\rm(bk)}= & t \otimes Z \\
U^{\rm(bk)} = & u \otimes I
\end{align*}
\noindent Here the label $\rm(bk)$ is for distinguishing the bulk from the boundary matrices. In \cite{Tsui2019}, it's shown that as long as the boundary massless fermion is irreducible, and  $t$,$u$ prohibit any mass term,  then there is single allowed bulk mass term which respects all the symmetries\footnote{Here the irreducibility means the gamma and the symmetry matrices cannot be  simultaneously block-diagonalized non-trivially. The fact that this is true for our case is because the inclusion of the full emergent symmetries. (Proof omitted.)}. Such mass term is given by
\begin{align*}
M^{\rm(bk)} = I_{\rm dim(\G_i)} \otimes Y.
\end{align*}
The above mass term can be used to regularize and gap out the fermion in the bulk. In Wilson's regularization, the SPT (single-particle) Hamiltonian in momentum space is given by 
\begin{align*}
h^{\rm(bk)}(\v k) = \sum\limits_{i=1}^{d+1} \sin k_i \, \Gamma_i^{\rm(bk)} + \left(d+1 + m_B - \sum\limits_{i=1}^{d+1} \cos k_i \right) M^{\rm(bk)}
\end{align*}
When $m_B<0$, and when the lattice is cut open in the $(d+1)$th direction (actually the gapless boundary modes exist when the cut is along any direction), the boundary low energy theory is that of the original massless fermions.

\section{The decoupling of the charge-SU(2) gauge field from the low energy non-linear sigma model after confinement}
\label{appendix:chargeSU(2)Decoupling}
\hfill

In this appendix, we  show that the charge-SU(2) gauge field is not coupled to \Eq{o5}. To recap, the charge-SU(2) singlet $Q^\mathbb{R}$ is given by
\be
&&Q^\mathbb{R}=n_i N_i~~{\rm where}\nn
&&N_i = \left(YXY, IYY, YZY, IIX, IIZ \right) .
\nonumber
\ee
Following appendix \ref{appendix:anomaliesb} after gauging the charge-SU(2) symmetry, the $\frac{O(8)}{O(4) \times O(4)}$ nonlinear sigma model with $k=1$ WZW term becomes

\begin{align}
W[Q^\mathbb{R}, a] =&\frac{1}{4 \lambda_3} \int\limits_{\mathcal{M}} d^3 x \, \tr\Big[\Big(\p_\mu Q^\mathbb{R}+i [a_\mu,Q^\mathbb{R}]\Big)^2\Big] 
-  \frac{ 2 \pi i }{512\pi^2}   \Big\{   \int\limits_{\mathcal{B}}  \tr \Big[\tilde{Q}^\mathbb{R}    \,\left( d \tilde{Q}^\mathbb{R} \right)^4 \Big] \notag\\
+& 8 \int\limits_{\mathcal{M}} \tr \Big[ i a Q^\mathbb{R} (dQ^\mathbb{R})^2 - (a Q^\mathbb{R})^2 dQ^\mathbb{R} - \frac{i}{3} (a Q^\mathbb{R})^3 + i a^3 Q^\mathbb{R} -a Q^\mathbb{R} f - a f Q^\mathbb{R} \Big]
\Big\}. 
\label{gzero}
\end{align}
Since all $N_i$ commute with the charge-SU(2) group, it follows that $Q^\mathbb{R}=n_i N_i$ commutes with charge-SU(2) gauge field $a$. Hence the gauge coupling term in the stiffness term vanishes. \\

To show this is also true for the gauged WZW term part, we shall take the $$\tr\left[a Q^\mathbb{R} (dQ^\mathbb{R})^2 \right]$$ term in \Eq{gzero} as an example. Plugging in $Q^\mathbb{R}=n_i N_i$, we obtain
$$  \tr\left[a Q^\mathbb{R} (dQ^\mathbb{R})^2 \right] = \sum_{i,j,k}  \tr[a \, N_i N_j N_k] n_i \, dn_j \, dn_k.$$ 
In the following we shall prove that each term in the sum vanishes, i.e.,  $$\tr[a \, N_i \, N_j \,  N_k]=0~~~~\forall (i,j,k).$$ 
To achieve that we insert a $N_l^2=1$ where $l\ne i,j,k$ into the trace and leave it invariant, i.e., 
$$\tr[a N_i N_j N_k]=\tr[N_l^2 a  N_i N_j N_k].$$ 
Due to the commutivity between $N_l$ and $a$ and the anti-commutivity between $N_l$ and each of the $N_{i,j,k}$, we can move one $N_l$ all the way to the right end and use the cyclic property of trace to put it back to the front
$$\tr[N_l^2  a N_i N_j N_k]=-\tr[N_l a  N_i N_j N_k N_l]=-\tr[N_l N_l a  N_i N_j N_k ]=-\tr[ a  N_i N_j N_k ].$$ Thus 
$$\tr[ a  N_i N_j N_k ]=-\tr[ a  N_i N_j N_k ]\Rightarrow \tr[ a  N_i N_j N_k ]=0.$$
This proof can be applied to all gauge coupling terms in \Eq{gzero} because there is an odd number of $Q^\mathbb{R}$s for every term that  couples to the gauge field. Therefore the charge-SU(2) gauge field is not coupled to $Q^\mathbb{R}=n_i N_i$.

\section{The WZW term in the  $(3+1)$-D real class non-linear sigma model}
\label{appendix:WZWdecompose}

In this section, we will show that upon the decomposition in  \Eq{U(1)_S5} of subsection \ref{3dmottspin}, namely, $$Q^\mathbb{R}(x)= e^{i\theta(x)} \cG_S(x),$$ the contribution of the WZW term is solely from the $\cG_S(x)$ part, i.e.,
namely
\begin{align*}
\tr\left[ \left( Q^{\mathbb{R}\dagger} dQ^{\mathbb{R}} \right)^5 \right] = \tr\left[ \left( \cG_S^\dagger d\cG_S \right)^5 \right] 
\end{align*}

\noindent First, note that one can at most choose $d\theta$ once in the expansion of $\tr\left[ \left( Q^{\mathbb{R}\dagger} dQ^{\mathbb{R}} \right)^5 \right]= \tr\left[(\cG_S^\dagger d\cG_S + i d\theta)^5\right]$, otherwise the differential form vanishes because $(d\theta)^2=0$. The only term that can possibly survive other than $\tr\left[(\cG_S^\dagger d \cG_S )^5\right]$ is then of the form
\begin{align*}
\tr\left[d \theta \left( \cG_S^\dagger d\cG_S \right)^4 \right] 
= d \theta  \left( \cG_S^\dagger d\cG_S \right)^a \left( \cG_S^\dagger d\cG_S \right)^b \left( \cG_S^\dagger d\cG_S \right)^c \left( \cG_S^\dagger d\cG_S \right)^d \tr\left[  t^a t^b t^c t^d\right]
\end{align*}
\noindent Here $\{ t^a\}$ are the complete basis for the generators of $SU(n)$ in the fundamental representation (note that $\cG_S$ are the symmetric special unitary matrices, which are special kind of unitary matrices). In the following, we will show that for every term from the trace $\tr\left[  t^a t^b t^c t^d\right]$, it is at least symmetric with respect to two of the  indices in $a,b,c,d$. If so, because the scalar valued one forms $\left( \cG_S^\dagger d\cG_S \right)^a$ anti-commute with each others $\tr\left[d \theta \left( \cG_S^\dagger d\cG_S \right)^4 \right]$ vanishes.\\

We shall choose the conventions
\begin{align}
\label{tatbOrtho}
\tr \left[ t^a t^a \right] = \frac{1}{2} \delta_{ab} \\
\label{tatbComm}
[t^a , t^b] = i f_{abc} t^c
\end{align}
\noindent where $f_{abc}$ is the structure constant for $SU(n)$. Here the Einstein summation convention is used.  $f_{abc}$ is real and totally anti-symmetric. We shall also define
\begin{align}
\label{dabc}
d_{abc} = 2 \, \tr\left[ \{t^a,t^b \}t^c \right]
\end{align}
It can be shown simply that due to the cyclic property of trace and the hermiticity of $t^a$, $d_{abc}$ is real and totally symmetric with respect to $a,b,c$.\\

As a pre-step, we would calculate $t^a  t^b$. Because the identity matrix $I_n$ together with $\{ t^a \}$ form a complete basis for all $n\times n$ complex matrices, we can decompose $t^a t^b$ in terms of them. The coefficients can be calculated making use of \Eq{tatbOrtho}, \Eq{tatbComm}, and \Eq{dabc},
\begin{align}
\label{tatb}
t^a t^b  =& \frac{1}{n}\tr \left[ t^a t^b\right]I_n +  \tr\left[\left(\{t^a,t^b \} + [t^a,t^b]\right)t^c \right] t^c\notag \\
=& \frac{1}{2} \left[ \frac{1}{n} \delta_{ab} I_n + (d_{abc} + i f_{abc}) t^c\right].
\end{align} 
\noindent The equation above implies
\begin{align}
\label{tatbAntiComm}
 \{t^a,t^b \} = \frac{1}{n} \delta_{ab} I_n + d_{abc} t^c
\end{align}
\noindent For later usage, we will derive another formula for the product of two $f_{abc}$s. By direct expansion, one can prove the following identity
\begin{align*}
[t^a , [ t^b, t^c]] = \{ \{t^a,t^b\},t^c\} - \{ \{t^a,t^c\},t^b\}. 
\end{align*}
Using of \Eq{tatbComm} and \Eq{tatbAntiComm} twice in the equation above, we  get
\begin{align}
\label{ff}
f_{abe} f_{cde} = \frac{2}{n} \left(\delta_{ac} \delta_{bd} - \delta_{ad} \delta_{bc} \right) + \left(d_{ace} d_{bde} - d_{ade} d_{bce} \right)
\end{align}\\

Now we can calculate $\tr \left[ t^a t^b t^c t^d\right]$  by applying \Eq{tatb} twice and carrying out the trace. After some algebra and the help of \Eq{ff}, we get
\begin{align*}
\tr \left[ t^a t^b t^c t^d\right]=& \frac{1}{4} \tr \left[\left( \frac{1}{n} \delta_{ab} I_n + (d_{abc} + i f_{abe}) t^e\right) \left( \frac{1}{n} \delta_{cd} I_n + (d_{cdf} + i f_{cdf}) t^f\right)\right]\\
=& \frac{1}{4} \Big[ 
\thead{
		\frac{1}{n} (\delta_{ab} \delta_{cd} - \delta_{ac} \delta_{bd} + \delta_{ad} \delta_{bc} ) \\
			+\frac{1}{2} (d_{abe} d_{cde} - d_{ace} d_{bde} + d_{ade}d_{bce} )\\
		+\frac{i}{2} ( f_{abe} d_{cde} + f_{cde} d_{abe} )
}
\Big]
\end{align*}
By the symmetry properties of $\delta_{ab}$ and $d_{abc}$, every term is least symmetric with respect to two indices. For example, $f_{abe} d_{cde}$ is symmetric with respect to $c,d$. This concludes our proof.

\end{appendices}

\newpage

\bibliographystyle{ieeetr}
\bibliography{bibs}
%\addcontentsline{toc}{part}{References}
\newpage

\end{document}